\documentclass[5p]{elsarticle}
\usepackage{xtab,booktabs}
\usepackage{listings}
\usepackage[dvips]{color}
\usepackage{amsmath}
\usepackage{amssymb}
\usepackage{comment}
\usepackage{ifthen}
\usepackage[breaklinks]{hyperref}
\usepackage{bold-extra}

\biboptions{authoryear}

\definecolor{lightgray}{rgb}{0.9,0.9,0.9}
\definecolor{darkgreen}{rgb}{0,0.4,0}

\newcommand{\va}{{\sc Vartools}}
\newcommand{\tabitem}{~~\llap{$-$}~~}

\newcommand{\forloop}[5][1]%
{%
\setcounter{#2}{#3}%
\ifthenelse{#4}%
	{%
	#5%
	\addtocounter{#2}{#1}%
	\forloop[#1]{#2}{\value{#2}}{#4}{#5}%
	}%
	{%
	}%
}% 

\newcommand{\ctbd}[1]{}

\newcommand\aj{AJ}% 
\newcommand\apj{ApJ}% 
\newcommand\apjl{ApJ}% 
\newcommand\apjs{ApJS}% 
\newcommand\ao{Appl.~Opt.}% 
\newcommand\apss{Ap\&SS}% 
\newcommand\aap{A\&A}% 
\newcommand\aapr{A\&A~Rev.}% 
\newcommand\mnras{MNRAS}% 
\newcommand\pasp{PASP}% 
\newcommand\nat{Nature}% 
\newcommand\iaucirc{IAU~Circ.}% 
\newcommand\planss{Planet.~Space~Sci.}% 
\newboolean{emulateapj}
\setboolean{emulateapj}{false}
\newboolean{astroph}
\setboolean{astroph}{true}
\lstdefinelanguage{vartools}
{alsoletter={-,\.},
 morekeywords={vartools},
 morekeywords={[2]-rms,-addnoise,-alarm,-aov,-aov\_harm,-autocorrelation,-binlc,-BLS,-BLSFixPer,-BLSFixDurTc,-changeerror,-changevariable,-chi2,-chi2bin,-clip,-converttime,-copylc,-decorr,-dftclean,-difffluxtomag,-ensemblerescalesig,-expr,-findblends,-fluxtomag,-GetLSAmpThresh,-if,-elif,-else,-fi,-Injectharm,-Injecttransit,-Jstet,-Killharm,-linfit,-LS,-MandelAgolTransit,-medianfilter,-microlens,-nonlinfit,-o,-Phase,-resample,-rescalesig,-restorelc,-restricttimes,-rmsbin,-savelc,-SoftenedTransit,-Starspot,-stats,-SYSREM,-TFA,-TFA\_SR,-wwz},
 morekeywords={[3]-i,-l,-basename,-example,-functionlist,-header,-headeronly,-help,-inlistvars,-inputlcformat,-jdtol,-quiet,-L,-listcommands,-log-command-line,-matchstringid,-nobuffer,-bufferlines,-noskipempty,-numbercolumns,-oneline,-parallel,-randseed,-readall,-readformat,-redirectstats,-showinputlcformat,-showinputlistformat,-skipmissing,-tab},
 emph={prompt},
 moredelim=**[is][\itshape]{@}{@},
 moredelim=**[is][\normalsize\bfseries]{~}{~},
    sensitive=true,
}

\begin{document}

\begin{frontmatter}

%% Titlepage
\title{\va{}: A Program for Analyzing Astronomical Time-Series Data}

%% Authors
\author{Joel~D.~Hartman}
\author{G\'asp\'ar~\'A.~Bakos}
\address{Princeton University, Department of Astrophysical Sciences,
	4 Ivy Lane, Princeton, NJ, 08544, USA; email: jhartman@astro.princeton.edu}

%% EOF authors

% #####################################################################
%% abstract
\begin{abstract}
%++++++++++++++++++++++++++++++++++++++++++++++++++++++++++++++++++++++
%++++++++++++++++++++++++++++++++++++++++++++++++++++++++++++++++++++++

\setcounter{footnote}{0}

This paper describes the \va{} program, which is an open-source
command-line utility, written in C, for analyzing astronomical time-series data,
especially light curves. The program provides a general-purpose set of
tools for processing light curves including signal identification,
filtering, light curve manipulation, time conversions, and modeling and
simulating light curves. Some of the routines implemented
include the Generalized Lomb-Scargle periodogram, the Box-Least
Squares transit search routine, the Analysis of Variance periodogram,
the Discrete Fourier Transform including the CLEAN algorithm, the
Weighted Wavelet Z-Transform, light curve arithmetic, linear and
non-linear optimization of analytic functions including support for
Markov Chain Monte Carlo analyses with non-trivial covariances,
characterizing and/or simulating time-correlated noise, and the TFA
and SYSREM filtering algorithms, among others. A mechanism is also
provided for incorporating a user's own compiled processing routines
into the program. \va{} is designed especially for batch processing of
light curves, including built-in support for parallel processing,
making it useful for large time-domain surveys such as searches for
transiting planets. Several examples are provided to illustrate the
use of the program.
\end{abstract}

% #####################################################################
%% keywords
\begin{keyword}
methods: data analysis --- methods: statistical --- time --- techniques: photometric
\end{keyword}

\end{frontmatter}

% ####################################################################
%% space velocity
\section{Introduction}

Measuring the variation over time in the apparent brightness of a
distant object is one of the primary techniques that an astronomer may
use to study the universe. A few classic examples include using
supernovae or pulsating variable stars as standard candles in setting
the cosmic distance ladder
\citep[e.g.][]{hubble:1929:m31,riess:1998,perlmutter:1999}, using
eclipsing binaries to measure the fundamental physical properties of
stars \citep[e.g.][]{kopal:1959,torres:2010:EBreview}, observing microlensing
events to constrain the population of MAssive Compact Halo Objects
(MACHOs) in the Galaxy or to identify extrasolar planets
\citep[e.g.][]{paczynski:1986,bond:2004}, and finding extrasolar
planets by searching for stars that show periodic transit events
\citep[e.g.][]{henry:1999,charbonneau:2000,konacki:2003}.

Developments in technology, together with scientific interest in
finding rare types of variability, have led to numerous massive time
domain surveys, especially in the optical and infrared bandpasses (see
for example \citealp{bakos:2012:hatsouth}, and references therein).
Adding the time dimension to a survey leads to a massive flow of data,
and a need for specialized software to analyze it. This need is
expected to grow even larger as new time-domain surveys (e.g., {\em LSST}, {\em TESS}, {\em PLATO}, {\em HATPI}, etc.) continue to
come online.

This paper describes \va{}, a program which provides a suite of tools
for processing light curves, and which is designed in particular
for handling large data sets. An early version of \va{}, which included
only a handful of period-finding tools, was briefly described in
\citet{hartman:m37:2} who used it in analyzing the data from a
variability survey of the open cluster M37. Since that time the
program has been significantly altered, motivating the more detailed
and up-to-date description provided here.

%; it has also been used in a
%number of additional time-series studies
%\citep[e.g.][]{hartman:m37:3,hartman:m37:4,hartman:2010:pleiades,hartman:2011:kmdwarf,nascimbeni:2012,barclay:2011,kim:2011,prsa:2011,shappee:2011,
%  henderson:2011,fleming:2010, nataf:2009}.
In addition to the numerous standalone software packages available
which perform particular analysis tasks on light curves
(e.g., transiting planet and eclipsing binary identification and
modeling software; \citealp{kovacs:2002:BLS}; \citealp{devor:2005}), there
are several other software suites available for general astronomical
light curve analysis. Notable examples include {\sc
  PyKE} \citep{still:2012}\footnote{\url{http://keplerscience.arc.nasa.gov/PyKE.shtml}}, the {\sc Peranso} period analysis program \citep{paunzen:2016}, the
{\sc Period04} program \citep{lenz:2005}, {\sc Wqed} which was
developed for the Whole Earth Telescope project \citep{thompson:2009},
and the open-source {\sc VStar} program provided by the American Association of Variable Star Observers\footnote{\url{http://www.aavso.org/vstar-overview}}.

The \va{} programs differs from these in a few key ways. First, it
provides more general purpose analysis routines than the other
programs. Second, unlike the other programs, it is designed
specifically for processing large numbers of light curves. This
includes the ability to seamlessly pass light curves between commands
without requiring intermediate temporary files, not requiring any
human interaction to process each light curve, and native support for
parallel processing of light curves.

General signal processing packages exist for data analysis platforms
such as {\sc matlab}, {\sc R}, {\sc idl}, or {\sc python}, but these are
typically not tailored for astronomical purposes, and in general are
collections of functions (like {\sc PyKE}), requiring the user to
solve a variety of bookkeeping problems, and in-so-doing to write a
fair amount of code, to apply them to large astronomical time domain
datasets. All of these platforms also allow executing system commands,
so \va{} may be incorporated into pipelines built for these platforms
as well.

The following section provides an overview of the program, including a
general description of its operation, and discussions of the
processing commands and control options that are provided; in
section~\ref{sec:examples} several examples are provided; performance
tests are presented in section~\ref{sec:performance}; possible areas
for future development are discussed in
section~\ref{sec:conclusion}. The appendices include comprehensive
listings of the input and output syntax and data formats.

\section{The \va{} Program}\label{sec:program}

The basic operation of \va{} is to read-in a light curve, apply one or
more ``commands'' (i.e.~processing routines) to the light curve, and
output resulting statistics to an ASCII table for later analysis. The
program is written in {\sc ansi c}, with regular updates provided on
the
web\footnote{\url{http://www.astro.princeton.edu/\~jhartman/vartools.html}}. This
article describes version 1.33 of the program, a static copy of which
is preserved on the website. To allow maximum portability, it can be
compiled with basic functionality without any non-standard external
libraries, however a number of features have library dependencies
(including the {\sc cfitsio} library,
\citealp{pence:1999}\footnote{\url{http://heasarc.gsfc.nasa.gov/fitsio/fitsio.html}};
the GNU scientific library, or GSL,
\citealp{galassi:2009}\footnote{\url{http://www.gnu.org/software/gsl/}};
and the JPL NAIF CSPICE library,
\citealp{acton:1996}\footnote{\url{http://naif.jpl.nasa.gov/naif/toolkit.html}}). Compilation
and installation are carried out using the GNU Build
System\footnote{\url{http://www.gnu.org/software/automake/manual/html_node/GNU-Build-System.html}}
and has been successfully carried out on a number of system
architectures (Linux, Mac and Windows). While \va{} may be used to
analyze an individual light curve of interest, it has been developed
primarily to be included in pipelines which conduct batch processing
of large numbers of light curves.

As a simple example, listing~\ref{lst:example1} shows the use of \va{} to calculate
the average, standard deviation, and expected standard deviation of
the magnitudes in the light curve stored in the file ``EXAMPLES/1''
(this ASCII text file is included in the distribution of the program).
\begin{lstlisting}[caption={A simple example of running \va{}. In this example, and others shown in this paper, the text following
``prompt$>$'' is typed by the user into the shell. In all examples in
this paper we assume a BASH environment. Note that the
``\textbackslash" character at the end of the line continues the
command over multiple lines; the ``-oneline'' argument, which causes
one output statistic to be printed per line, is thus also part of the
same command. The last five lines (beginning with ``Name'') are
the output of the program. Colors are used to highlight commands and options passed to \va{}.},label={lst:example1},frame=single,backgroundcolor=\color{lightgray},language=vartools]
prompt> vartools -i EXAMPLES/1 -rms \
         -oneline

@Name           = EXAMPLES/1@
@Mean_Mag_0     =  10.24745@
@RMS_0          =   0.15944@
@Expected_RMS_0 =   0.00101@
@Npoints_0      =  3122@
\end{lstlisting}
Alternatively, we can apply this to a list
of light curves stored in ``EXAMPLES/lc\_list'' (each line in this
list contains the name of a light curve file to process), and run this
using 32 parallel processes (listing~\ref{lst:example2}).
\begin{lstlisting}[caption={Processing a list of light curves in parallel. All quantities for a particular light curve are output on a
single line, the ``...'' indicates that additional output from the
program is not being shown here, and the parallel processing causes
the output light curves to be in a different order from the input
list.},label={lst:example2},frame=single,backgroundcolor=\color{lightgray},language=vartools,float=*]
prompt> vartools -l EXAMPLES/lc_list -rms \
        -header -numbercolumns -parallel 32

@#1_Name 2_Mean_Mag_0 3_RMS_0 4_Expected_RMS_0 5_Npoints_0@
@EXAMPLES/2  10.11802   0.03663   0.00102  3313@
@EXAMPLES/1  10.24745   0.15944   0.00101  3122@
@EXAMPLES/10  10.87781   0.00236   0.00143  3974@
@...@
\end{lstlisting}
A number of more complicated examples are provided in
Section~\ref{sec:examples}.

The next subsection describes the general operation of the
program, paying special attention to those features relevant for batch
processing. The processing algorithms which are included as
``commands'' in \va{} are discussed in Section~\ref{sec:commands},
built-in methods for extending \va{} with a user's own processing
routines are discussed in Section~\ref{sec:userlib}, and options which
may be used to control the operation of the program are discussed in
Section~\ref{sec:options}. Performance benchmark tests are provided in
Section~\ref{sec:performance}.

\subsection{General Operation}\label{sec:operation}

\subsubsection{Input}\label{sec:input}

A user may provide either a single light curve or an {\sc ascii} list
of light curve files as input to the program. Each light curve is
stored in a separate file, either in ascii text format (one
measurement per row, with separate white-space delimited columns for
the time, magnitude or flux, uncertainties, and other optional
information), or as a binary FITS table. The input list file may also
be used to provide light-curve dependent data (e.g., this could be the
right ascension and declination coordinates of each star) as needed by
the various processing commands invoked by the user. There is
flexibility for the user to specify the format of the input files. The user may also provide a string giving a set of commands that are passed to the shell and applied to the input light curve files before being read-in by the program (e.g., this may be used to process compressed light curve files without having to store the uncompressed files on the disk).

\subsubsection{Data Flow}\label{sec:dataflow}

The input light curves are then sent to the commands in the order that
they are invoked on the command-line. The user may choose from 51
built-in commands, as well as a potentially unlimited number of
user-developed commands. In principle an arbitrary number of commands
can be called in a single invocation of \va{}; though in practice a
machine-dependent limitation will be set by the amount of available
memory. These commands generally have parameters which control their
properties (e.g., the range of frequencies to search in a periodogram,
or the period, epoch, planet size, etc., for injecting a transit
signal into a light curve). Most commands allow a number of options for
specifying how the parameters are to be determined. Typically a
parameter can be specified on the command-line for all light curves,
it can be read-in from a column in the input list (so that a different
value can be used for each light curve), it can be set using the
results from a previously executed command (e.g., the period found by
one command can be used in subtracting a harmonic signal with another
command), it can be set to the value of a variable (see
Section~\ref{sec:variable}), or equal to the result from evaluating an analytic expression.

When batch processing hundreds of thousands (or more) of individual
light curve files, often the speed of the process is limited by
attempts to read-from or write-to the hard disk, rather than by
computation speed. In such cases there is a significant advantage to
eliminating redundant attempts to access the disk. This is one
advantage of writing a pipeline in \va{}, where a light curve is read
into memory once and output to the disk only when explicitly requested
(output is treated as a command, which may be invoked at any time in
the pipeline). When a similar pipeline is written using a shell script
to piece together many standalone processing programs, it is often the
case that the programs are not capable of transmitting the light
curves via memory, forcing redundant disk access\footnote{in some
  cases this problem can be overcome by writing temporary files to
  shared memory rather than a physical disk.}. Moreover, scripting a
pipeline around \va{} is generally simpler than using a variety of
different programs which often require different data formats,
necessitating messy, and potentially bug-prone data manipulation steps
within a script. While pipelines written in an interpretted language
such as {\sc matlab}, {\sc idl}, or {\sc python} may also avoid
redundant disk access, they generally have a slower execution than a
compiled program like \va{}. Moreover building a pipeline from scratch in an
interpreted language may well require more development effort for the
user than using \va{} where all of the necessary bookkeeping is
handled internally by the program.

\subsubsection{Parallelization}\label{sec:parallel}

One other significant feature of \va{} is its built-in support for
parallel-processing of light curves. In the present version this is
limited to shared-memory parallel processing, which is appropriate for
processing on a single multi-core machine. The parallelization is done
using the standard POSIX threads library, included with most $C$
compilers. \va{} creates a number of threads specified by the user and
assigns to each thread a light curve to fully process. Once that light
curve is processed, the parent thread is notified, and a new light
curve is assigned to the child thread. This is repeated until all
light curves have been processed. The parallelization of individual
commands (e.g. period searches) has not yet been implemented, but is
under development. For the common case where the number of light
curves exceeds the number of processors, the current parallelization
implementation should be adequate. Note that when run in parallel
mode, computed statistics are printed in the order that light curves
finish processing. We use locks to prevent simultaneously writing out
data for multiple light curves, with buffering of the output to
minimize occurrences of multiple threads waiting for the same lock. As
a result the ordering of light curves in the output table may not be
the same as in the input list, and in general will differ each time
the program is run.

\subsubsection{Variables}\label{sec:variable}  
\va{} supports the use of variables on the command-line to refer to
vector or scalar data read-in from the light curves or list files, or
computed by commands. These variables can also be used in evaluating
analytic expressions with the {\bf -expr} command, in controlling the
flow of the program with the {\bf -if}, {\bf -elif}, {\bf -else} and
{\bf -fi} commands, for processing multiple data columns within a
single light curve with the {\bf -changevariable} command, or in
fitting analytic functions to the light curves with the {\bf -linfit}
or {\bf -nonlinfit} commands. The special variable names $id$, $t$,
$mag$, and $err$ are reserved for a string giving the name of the
image that a measurement comes from, the time, magnitude (or flux) and
magnitude (flux) uncertainty, respectively. Additionally the special
variable $NR$ refers to the integer record number of a point in a
light curve (1 for the first observation, 2 for the second,
etc.). Finally a list of special names are reserved for analytic
functions, constants, and operators (these are listed in
\ref{sec:reservednames}). 

\subsection{Commands}\label{sec:commands}

Here the built-in processing commands supported by \va{} are described. A
detailed description of the expected input syntax is provided in
\ref{sec:syntax}, while a description of the quantities
output by each command is provided in \ref{sec:output}. The commands
can be separated into several categories: (1) finding periodic
signals; (2) calculating light curve statistics; (3) fitting models;
(4) filtering; (5) simulating light curves; (6) manipulating light
curves; and (7) controlling the flow of data through \va{}. In
addition to the built-in commands, \va{} provides support for the
dynamic inclusion of commands and functions developed by the user; this feature is described at the end of this section. Note that the choice
of commands built-in to the currrent version of \va{} has been
determined primarily by the authors' own research. As such there is
a bias towards specialized routines that are useful in dealing with
data from transiting planet surveys, while some general routines
useful for other types of time-series applications may not be present.

\subsubsection{Period-Finding Algorithms}\label{sec:periodograms}

Several of the commands included in \va{} are used to identify periodic
signals in light curves. Each of these routines generates a spectrum
giving the significance of a periodic signal as a function of
frequency or period. The user can optionally output the spectrum of
each light curve to a separate file, \va{} will also identify peaks in the
spectrum and report the period or frequency, a measure of the
significance, and other relevant information to the output ascii
table. Below is a synopsis of each of these commands:

\paragraph{\bf -LS}\label{cmd:ls}
Calculates the Generalized Lomb-Scargle (L-S) periodogram
of a light curve \citep{zechmeister:2009,lomb:1976,scargle:1982}
using the method of \citet{press:1989} to speed up the calculation of
various trigonometric sums \citep[we use the implementation of this
  ``extirpolation'' algorithm due to][]{press:1992}. The primary
motivation for using the L-S method as opposed to directly applying
the Fast Fourier Transform (FFT) to the light curve is that L-S
handles nonuniformily sampled data (in astronomy this is the rule
rather than the exception) without resorting to methods such as
zero-padding or interpolation to a uniform grid, which amounts to
making assumptions about the light curve at non-observed time
instances. For this reason, this tool is widely used for identifying
periodic signals in astronomy. The generalized periodogram extends the
traditional L-S periodogram by allowing for a floating mean and
non-uniform uncertainties (the user may optionally calculate the
traditional periodogram instead; this is calculated following
\citealp{press:1992}). The value of the periodogram at frequency $f$ (in
cycles per day) is given by equations 4 and 5 in
\citet{zechmeister:2009}, i.e.:
\begin{equation}
LS(f) = \frac{\chi^{2}_{0} - \chi^{2}(f)}{\chi^{2}_{0}}
\label{eqn:LSdef}
\end{equation}
where $\chi^{2}_{0}$ is the value of $\chi^2$ using the weighted mean
of the data as the model, and $\chi^{2}(f)$ is the value of $\chi^2$
using the best-fit sinusoidal signal with frequency $f$ as the
model. This statistic varies between 0 (for no signal present at all)
and 1 (for a perfectly sinusoidal signal).

In addition to the peak periodogram value, the \va{} implementation of
L-S outputs two measures of significance for peaks identified in the
periodogram. These are the logarithm of the formal false alarm
probability, and the signal-to-noise ratio (S/N).

There are several ways in which the false alarm probability may be
estimated. We follow the \citet{cumming:1999} method described in
section 3 of \citet{zechmeister:2009}. In this method the signal is
measured relative to the scatter in the residuals from the best-fit
model (as opposed to using the scatter in the original light curve, or
using the input measurement uncertainties, which are two other popular
methods). Namely we take
\begin{equation}
z(f) = \frac{N-3}{2}\frac{LS(f)}{1 - LS_{\rm best}}
\end{equation}
where $N$ is the number of observations in the light curve, and $LS_{\rm best}$ is the highest value of $LS$ found in the periodogram. The false alarm probability is then given by
\begin{equation}
{\rm FAP} = 1 - [1-{\rm Prob}(z > z0)]^{M}
\label{eqn:LSFAP}
\end{equation}
where the false alarm probability for an individual trial frequency is
\begin{equation}
{\rm Prob}(z > z0) = \left( 1 + \frac{2z_{0}}{N-3} \right)^{-(N-3)/2}
\end{equation}
and $M$ is the number of independent frequencies sampled (sometimes
called the ``bandwidth'' penalty). Here we differ slightly from
\citet{zechmeister:2009} and use the \citet{horne:1986} estimate of
\begin{equation}
M = 2f_{\rm max}T
\end{equation}
where $f_{\rm max}$ is the maximum frequency calculated in the
periodogram, and $T$ is the time spanned by the light curve. Note that
the false alarm probability will change if you scan a larger or
smaller frequency range---this is expected behavior. It is also
important to note that this false alarm estimate assumes Gaussian
white noise. For real data the noise is often correlated in time, or
does not follow a Gaussian distribution. In such cases the true false
alarm probability is often much higher than the calculated
probability, and should be determined through Monte Carlo
simulations. Also note that $M$ is a simple estimate that may be off
by as much as a few dex in cases where the sampling is highly
non-uniform. For applications requiring a false alarm probabilty
accurate to better than 1\,dex, it is necessary to conduct Monte Carlo
simulations applying L-S to light curves with simulated noise to
calibrate the FAP.

The S/N measure provided by \va{} is determined within the periodogram
as 
\begin{equation}
{\rm S/N} = (LS - \bar{LS})/\sigma_{LS}
\label{eqn:LSSNR}
\end{equation}
where $LS$ is the value of the periodogram for a given peak,
$\bar{LS}$ is the mean across the periodogram, and $\sigma_{LS}$ is
the standard deviation of the periodogram. The user may optionally
specify a clipping factor used to remove outlier points (peaks) from
the values used in determining $\bar{LS}$ and $\sigma_{LS}$. The S/N
value provides an alternative method for estimating the significance
of a detection, however it is important to keep in mind that the
``noise'' in the periodogram does not follow a normal distribution, so
the S/N itself does not follow a normal distribution (i.e., a
spectroscopic S/N of 10 does not correspond to a $10\sigma$
detection). The user specifies the number of peaks to identify in the
periodogram, these will be sorted based on the peak-height. If
searching for multiple peaks, the user has the option to iteratively
``whiten'' the light curve after finding a peak (i.e., subtract the
best-fit periodic sinusoidal signal from the light curve with the
period fixed to the value identified in the periodogram), recompute
the periodogram, and search for the next highest peak.

\paragraph{\bf -GetLSAmpThresh}\label{cmd:getlsampthresh}
Though not a method for identifying a periodic
signal per se, this command is closely related to {\bf -LS} and so it
is discussed in this section. For many applications it is important to
know not only that a signal is present in a light curve, but also what
the minimum amplitude the signal could have had and still have been
detected within the noisy observed data. This information helps in
characterizing the selection effects present in a survey, and is thus
vital for determining the completeness of a survey. This command fits
a multi-harmonic Fourier series to a light curve at the period found
by {\bf -LS}. The model is then subtracted from the light curve, its
amplitude is scaled by a factor $\alpha$, and then it is re-added to
the residual. The Lomb-Scargle {\rm FAP} is calculated at the fixed
period for this new light curve. The value of $\alpha$ is varied until
the resulting FAP is equal to a threshold value chosen by the user;
the resulting $\alpha$ times the observed amplitude is reported to the
user.

\paragraph{\bf -dftclean}\label{cmd:dftclean}
Calculates the Discrete Fourier Transform (DFT) of a
light curve, and optionally applies the CLEAN deconvolution algorithm
\citep{roberts:1987}. We follow the convention that the complex DFT of
the set of points ${(t_{j}, m_{j})}$, evaluated at frequency $f$ is
given by:
\begin{equation}
F(f) = \sum_{j=1}^{N}(m_{j}-\bar{m})\exp(i2\pi f (t_{j} - \bar{t}))
\label{eqn:dft}
\end{equation}
and use the method of \citet{kurtz:1985} to speed up the
calculation. The user may also optionally output the window function
(i.e.~the DFT of the set of points ${(t_{i}, 1)}$ where $t_{i}$ are
the times of observation, and the DFT is applied without subtracting
the average from the $y$ values). The DFT procedure takes two
parameters: $nbeam$ which determines the frequency sampling of the
spectrum via $\Delta f = 1/(T\times nbeam)$ where $T$ is the time
baseline of the observations, and the maximum frequency to calculate
the DFT up to (the default is the Nyquist frequency given by $1/(2
\Delta t_{\rm min})$ where $\Delta t_{\rm min}$ is the minimum time
separation between consecutive points in the light curve). The CLEAN
routine is controlled by the gain and the S/N threshold for selecting
peaks in the DFT. After the DFT is calculated the spectrum is searched
for the most significant peak above the S/N threshold. This peak,
times the gain, is added to the cleaned spectrum, while the window
function, scaled appropriately and shifted to the peak frequency, is
subtracted from the DFT. The procedure iterates until no peaks above
the S/N threshold remain.

\paragraph{\bf -aov}\label{cmd:aov}
Calculates the Analysis of Variance (AoV) periodogram of a
light curve, as defined by \citet{schwarzenbergczerny:1989}. The
periodogram statistic follows a Fisher distribution of the form $F(x;
r-1, n-r)$ where $x$ is the value of the AoV periodogram, there are
$n$ data points in the light curve, and $r$ bins are used in the fit.
As with all period searches, one must also take into account the
number of independent periods that are tried. We use the
\citet{horne:1986} estimate for the band-width penalty. Work requiring
a more exact determination of the false alarm probability should
conduct Monte Carlo simulations \citep[see for
  example][]{paltani:2004,schwarzenbergczerny:2012}. Using the Fisher
distribution to interpret the AoV statistic also naturally punishes
overfitting the data. While the previous two commands use a sinusoid
for the model signal, in this case the model which is fitted to the
light curve is a discrete set of step functions (i.e., it is the
phase-binned light curve). This method is thus comparable to the
classical Phase Dispersion Minimization technique
\citep{stellingwerf:1978}, but with a superior statistic. In principle
      {\bf -aov} is more sensitive than {\bf -LS} or {\bf -dftclean}
      to light curves with sharp features, such as eclipsing binaries,
      for which a sinusoid is a poor approximation. However, because
      the phase-binned model is quite general (a relatively large
      number of free parameters must be used to fit typical signals),
      for most classes of variables the significance of the detection
      is lower when using {\bf -aov} than another method such as {\bf
        -aov\_{}harm} or {\bf -LS} \citep[e.g.][]{kovacs:1980}.

\paragraph{\bf -aov\_{}harm}\label{cmd:aovharm}
Similar to {\bf -aov}, this command computes an
AoV periodogram for a light curve, however it uses a multi-harmonic
Fourier series model for the signal, rather than the phase-binning
model used by {\bf -aov}. The fit is done using a projection onto
orthogonal complex polynomials as described by
\citet{schwarzenbergczerny:1996,schwarzenbergczerny:2012}. Here one
may also use the AoV statistic to optimize the number of harmonics to
include in the fit.

\paragraph{\bf -BLS}\label{sec:BLS}\label{cmd:bls}
Computes a Box-Least Squares
(BLS; \citealp{kovacs:2002:BLS}) spectrum for a light curve. Unlike the
other period-finding commands, {\bf -BLS} does not search for signals
that are continuously variable over the full phase curve, but instead
searches for periodic box-shaped dips in a light curve. This algorithm
provides an efficient means of identifying detached eclipsing binary light
curves and transiting planet signals. The algorithm operates by
scanning through frequencies, phasing the light curve at each
frequency, and fitting to the phased light curve a model of the form
\begin{equation}
m(\phi) = \begin{cases} \bar{m}, & \phi < \phi_{0}~~{\rm or}~~\phi > \phi_{0}+q \\ \bar{m} + \delta_{m}, & \phi_{0} < \phi < \phi_{0}+q \end{cases}
\end{equation}
for $\phi_{0} + q \leq 1$, or
\begin{equation}
m(\phi) = \begin{cases} \bar{m}, & \phi_{0} + q - 1 < \phi < \phi_{0} \\ \bar{m} + \delta_{m}, & \phi < \phi_{0} + q - 1~~{\rm or}~~\phi > \phi_{0} \end{cases}
\end{equation}
for $\phi_{0} + q > 1$. Here $\phi$ is the phase, $\bar{m}$ is the
average magnitude of the light curve, and $\delta_{m}$, $\phi_{0}$ and
$q$ are free parameters representing the transit depth, starting phase
of the transit, and transit duration in phase, respectively. The fit
for the free parameters is done by binning the phased light curve into
$N_{\rm bin}$ bins, trying each of these bins for $\phi_{0}$ and
trying discrete values for $q$ of the form $i/N_{\rm bin}$ where $i$
is an integer that varies between $i_{\rm min} = {\rm floor}(q_{\rm
  min}N_{\rm bin})$ and $i_{\rm max} = {\rm ceil}(q_{\rm max}N_{\rm
  bin})$, and $q_{\rm min}$ and $q_{\rm max}$ are set by the user. The
user may optionally fit a simple ``trapezoid'' transit to the light
curve for each of the significant periods found in the BLS
spectrum. In this case the fractional duration of ingress and egress
(i.e., in units of the total transit duration) is an
additional free parameter in the fit. The trapezoid model is fit using the Downhill Simplex method
\citep[DHSX;][]{nelder:1965} with initial parameter values for the frequency, transit depth, out-of-transit magnitude, transit time and transit duration taken from the BLS peak.

The basic statistic computed by the BLS algorithm is the Signal
Residue (${\rm SR}$) as a function of trial transit frequency, defined
by \citet{kovacs:2002:BLS} as
\begin{equation}
{\rm SR}(f) = \max_{\phi_{0},q} \Big\{ \left[ \frac{s^{2}(\phi_{0},q,f)}{r(\phi_{0},q,f)(1-r(\phi_{0},q,f))} \right] \Big\}
\label{eqn:BLSSR}
\end{equation}
where 
\begin{equation}
s^{2}(\phi_{0},q,f) = \sum_{\rm transit}\bar{w_{i}}\bar{x_{i}}
\end{equation}
is the weighted sum of phased, weighted-mean-subtracted magnitudes $\bar{x_{i}}$ within a trial transit (i.e., points with phase $\phi$ for a given frequency $f$, such that $\phi_{0} < \phi < \phi_{0}+q$ if $\phi_0 + q \leq 1$, or $\phi < \phi_0 + q - 1$ or $\phi > \phi_0$ otherwise), $\bar{w_{i}}$ is the weight for point $i$ normalized such that the sum of all weights in the light curve is unity, and we have $r(\phi_{0},q,f) = \sum_{\rm transit}\bar{w_i}$.
The maximum in equation~\ref{eqn:BLSSR} is taken over all pairs of
$\phi_{0}$ and $q$ considered for a given transit frequency $f$. The
function ${\rm SR}(f)$ is also called the BLS
spectrum. \citet{kovacs:2002:BLS} also define the Signal Detection
Efficiency (${\rm SDE}$) for a given peak in the BLS spectrum, given
by
\begin{equation}
{\rm SDE} = \frac{{\rm SR}_{\rm peak} - \bar{\rm SR}}{\sigma_{\rm SR}}
\label{eqn:BLSSDE}
\end{equation}
where $\bar{\rm SR}$ is the average value of the BLS spectrum, and $\sigma_{\rm SR}$ is its standard deviation.

In many cases time-correlated noise leads to a BLS spectrum which
slowly rises with decreasing frequency. To account for this, by default the \va{} program ranks peaks by a modified version of the ${\rm SDE}$, called the spectroscopic signal to noise ratio (S/N), and calculated as follows. We first define
\begin{equation}
\tilde{\rm SR}(f) = {\rm mean}_{\phi_{0},q} \Big\{ \left[ \frac{s^{2}(\phi_{0},q,f)}{r(\phi_{0},q,f)(1-r(\phi_{0},q,f))} \right] \Big\}
\label{eqn:BLSSRtilde}
\end{equation}
where the mean is computed, for a given frequency, over all trial combinations of $\phi_{0}$ and $q$, and the quantity being averaged is the same quantity that is maximized in computing ${\rm SR}$. A $3\sigma$ clipping is first performed before taking the mean. One may interpret $\tilde{\rm SR}(f)$ as the average Signal Residue due to noise at a given frequency. The S/N is then taken to be
\begin{equation}
{\rm S/N}(f) = \frac{{\rm SR}(f) - \bar{\tilde{\rm SR}}(f)}{\sigma_{\tilde{\rm SR}}}
\label{eqn:BLSSN}
\end{equation}
where in this case $\bar{\tilde{\rm SR}}(f)$ is the average value of
$\tilde{\rm SR}$ computed over a frequency range from $f-100\Delta f$
to $f+100\Delta f$ with $\Delta f$ being the frequency sampling of the
spectrum, and $\sigma_{\tilde{\rm SR}}$ is the standard deviation of
$\tilde{\rm SR}$ taken over the full spectrum. We perform an
interative 3$\sigma$ clipping before computing the average as a
function of frequency and the standard deviation. This determination
of ${\rm S/N}$ is similar to the SDE, except that it is the
frequency-localized signal enhancement, relative to the global noise
in the spectrum. Note that because SR does not follow a Gaussian
distribution, this definition of ${\rm S/N}$ should not be interpreted
using such a distribution.  The user may optionally use $\bar{\rm SR}$
and $\sigma_{\rm SR}$ in equation~\ref{eqn:BLSSN} rather than
$\bar{\tilde{\rm SR}}(f)$ and $\sigma_{\tilde{\rm SR}}$, in which case the
${\rm S/N}$ is the same as the SDE, except for the clipping applied in
computing the average and standard deviation.

Another statistic which we calculate for each transit, and which may be used to select transit candidates, is the signal-to-pink noise ratio (S/N$_{\rm pink}$) defined, following \citet{pont:2006}, as:
\begin{equation}
{\rm S}/{\rm N}_{\rm pink} = \frac{\delta_{m}^{2}}{(\sigma_{w}^2/n_{t}) + (\sigma_{r}^{2}/N_{t})}
\label{eqn:BLSSNpink}
\end{equation}
where $\sigma_{w}$ is the ``white noise'' r.m.s., $\sigma_{r}$ is the ``red noise'' r.m.s., $n_{t}$ is the number of points in transit, and $N_{t}$ is the number of distinct transits sampled. We take $\sigma_{w}$ to be equal to the r.m.s.\ of the light curve after subtracting the transit model, while we estimate $\sigma_{r}$ using the expression
\begin{equation}
\sigma_{r}^{2} = \sigma_{\rm bin}^{2} - \sigma_{\rm bin,thy}^{2}
\label{eqn:BLSrednoise}
\end{equation}
where $\sigma_{\rm bin}$ is the r.m.s.\ of the residual light curve
after binning in time with a bin-size equal to the duration of a
transit, and $\sigma_{\rm bin,thy}$ is the expected r.m.s.\ of the
binned light curve if the noise were uncorrelated in time. In the
event that the estimate for $\sigma_{r}^2$ is less than zero, we take
$\sigma_{r} = 0$.

\paragraph{\bf -BLSFixDurTc}\label{cmd:blsfixdurtc}
Runs BLS on a light curve fixing the duration of
the transit and the reference time of transit. Optionally the depth
and the duration of ingress (for a trapezoidal-shaped transit) may be
fixed as well. This may be used, for example, in a case where a single
transit is observed with high precision (e.g., from space) and one
wishes to search lower-precision data from another instrument for
additional transits.

\paragraph{\bf -BLSFixPer}\label{cmd:blsfixper}
Runs BLS on a light curve fixing the period (i.e.,
it finds the most transit-like signal in the light curve at a
specified period). This may be useful for carrying out transit
injection/recovery simulations to determine the detection efficiency
where one does not wish to perform a full BLS search on the simulated
light curves.

\paragraph{\bf -wwz}\label{cmd:wwz}
Computes the weighted wavelet transform as defined by
\citet{foster:1996}. This tool may be used to characterize signals
with quasi-periodic behavior that evolve in time (in amplitude and/or
frequency). The weighted wavelet transform is a discrete approximation
to the continuous wavelet transform for the function $x(t)$ given by
\begin{equation}
W(\omega,\tau;x(t)) = \omega^{1/2}\int x(t)f^{*}(\omega(t-\tau))dt
\label{eqn:waveletdef}
\end{equation}
for frequency $\omega$, time-shift $\tau$ and wavelet kernel
$f(z)$. As shown by \citet{foster:1996} the weighted wavelet transform
is superior to the well known discrete wavelet transform for
non-uniformly sampled time-series. \citet{foster:1996} adopts the
abbreviated Morlet wavelet \citep{goupillaud:1984} for the kernel given
by
\begin{equation}
f(z) = e^{i\omega(t-\tau)-c\omega^{2}(t-\tau)^2}
\label{eqn:wavelet2}
\end{equation}
and approximates the integral using a weighted projection of the data
onto the sinusoidal basis functions ($e^{i\omega(t-\tau)}$) with
weights $e^{-c\omega^{2}(t-\tau)^2}$. That is, the data are modeled as
\begin{equation}
y(t) = \sum_{a} y_{a}\phi_{a}(t)
\end{equation}
where the basis functions are
\begin{eqnarray}
\phi_{1}(t) & = & 1 \\
\phi_{2}(t) & = & \cos(\omega (t-\tau)) \\
\phi_{3}(t) & = & \sin(\omega (t-\tau)) \\
\end{eqnarray}
and the coefficients $y_{a}$ are given by 
\begin{equation}
y_{a} = \sum_{b} S^{-1}_{ab}<\phi_{b}|x>
\end{equation}
where the weighted inner product of functions $f$ and $g$ is given by
\begin{equation}
<f|g> = \frac{\sum_{\alpha = 1}^N w_{\alpha}f(t_{\alpha})g(t_{\alpha})}{\sum_{\beta = 1}^N w_{\beta}}
\end{equation}
with weights
\begin{equation}
w_{\alpha} = e^{-c\omega^{2}(t_{\alpha}-\tau)^2}
\end{equation}
and the sums are over the observed times. The matrix $S_{ab}$ is given by
\begin{equation}
S_{ab} = <\phi_{a}|\phi_{b}>.
\end{equation}
\citet{foster:1996} defines the Weighted Wavelet Transform (WWT) as
\begin{equation}
WWT = \frac{(N_{\rm eff}-1)V_{y}}{2V_{x}}
\label{eqn:waveletT}
\end{equation}
which is a $\chi^2$ statistic with 2 degrees of freedom with expected value 1. 
\citet{foster:1996}, however, suggests using the $Z$ statistic (or
Weighted-Wavelet Z-Transform, or WWZ) given by
\begin{equation}
Z = \frac{(N_{\rm eff}-3)V_{y}}{2(V_{x}-V_{y})}
\label{eqn:waveletZ}
\end{equation}
to characterize the significance of a signal with frequency $\omega$ and time-shift $\tau$ in the observed data. In both of these relations
\begin{equation}
N_{\rm eff} = \frac{(\sum_{w_{\alpha}})^{2}}{(\sum w_{\alpha}^2)}
\label{eqn:waveletNeff}
\end{equation}
is the effective number of data-points contributing to the signal, and
\begin{equation}
V_{x} = <x|x> - <1|x>^2
\end{equation}
and
\begin{equation}
V_{y} = <y|y> - <1|y>^2
\end{equation}
are the weighted variations of the data and model. The amplitude of
the signal (dubbed the Weighted Wavelet Amplitude, or WWA) is given by
\begin{equation}
WWA = \sqrt{(y_{2})^2+(y_{3})^3}.
\label{eqn:waveletA}
\end{equation}

Given a value of $c$ (by default $c = 1/8\pi^{2}$ is assumed), the
{\bf -wwz} command scans through frequency and time-shift (in keeping
with the rest of the program we express frequency in cycles per day,
whereas the frequency $\omega$ appearing in the equations above is in
radians per day) computing the $y_{a}$ coefficients, WWT, WWZ, WWA and
$N_{\rm eff}$. The user may optionally output this full transform
(i.e., each of these quantities for every frequency and time-shift
combination that is tested) to a file (as an ascii table, or a
multi-extension fits image) for each light curve analyzed. For each
trial time-shift the frequency with the maximum value of $WWZ$ is also
identified (i.e., $f_{\rm max, \tau}$, and corresponding quantities
$WWZ_{\rm max, \tau}$, $WWA_{\rm max, \tau}$, $N_{\rm eff, max,
  \tau}$, and $y_{\rm a, max, \tau}$), and may be output to a separate
file (this gives effectively the maximum signal as a function of time
in the light curve). The maximum of $WWZ_{\rm max, \tau}$ across all
trial $\tau$ values is determined, and its associated frequency,
amplitude, $N_{\rm eff}$ and model coefficients are included in the
\va{} output ascii table. We also compute the median values of the max
quantities across $\tau$ and include these in the output ascii table
as well.

\subsubsection{Light Curve Statistics}\label{sec:lcstats}

There are eight built-in commands which calculate statistics useful
for charaterizing the variability of light curves.

\paragraph{\bf -alarm}\label{cmd:alarm}
Calculates the statistic proposed by \citet{tamuz:2006}
to detect variability. Let $\bar{m}$ be the weighted average magnitude
of the light curve defined by
\begin{equation}
\bar{m} = \frac{\sum_{i} m_{i}/\sigma_{i}^2}{\sum_{i} 1/\sigma_{i}^2}
\end{equation}
where the sum is over all points in the light curve. Then the alarm
$a$ is given by
\begin{equation}
a = \frac{\sum_{k} a_{k}^2}{\sum_{i} ((m_{i} - \bar{m})/\sigma_{i})^2} - 2.2732395
\end{equation}
where $a_{k}$ is given by
\begin{equation}
a_{k} = \sum_{j} (m_{j} - \bar{m})/\sigma_{j}
\end{equation}
the sum on $j$ is over the $k$th continuous set of points with the
same value of ${\rm sign}(m-\bar{m})$, the sum on $k$ is over all such
values of $a_{k}$, and the sum on $i$ is over all points in the light
curve. The constant term $2.2732395$ is chosen such that the expected
value of $a$ is zero for Gaussian white noise. A Large value of $a$
indicates a light curve exhibiting correlated variability.

\paragraph{\bf -autocorrelation}\label{cmd:autocorrelation}
Calculates the Discrete Autocorrelation
Function \citep[DACF;][]{edelson:1988} for a light curve. This variant of the
autocorrelation function handles non-uniformly sampled time series by
forming all pairs of points $m_{i}$ and $m_{j}$, calculating the correlation between the pairs via
\begin{equation}
C_{ij} = \frac{(m_{i} - \bar{m})(m_{j} - \bar{m})}{\sigma_i\sigma_j}
\label{eqn:autocorrcij}
\end{equation}
and then binning this in lag time $\tau$, such that the DACF at lag $\tau_{k}$ is given by
\begin{equation}
DACF(\tau_{k}) = \frac{1}{M_{k}}\sum C_{ij},
\end{equation}
where the sum is over all $ij$ pairs with lag $\tau_{ij} = |t_{i} - t_{j}|$ satisfying $|\tau_{ij} - \tau_{k}| < \Delta \tau/2$, for a fixed sampling $\Delta \tau$, and $M_{k}$ is the number of these pairs. Equation~\ref{eqn:autocorrcij} differs from the corresponding equation in \citet{edelson:1988} in that we use the product of the measurement uncertainties $\sigma_{i}\sigma_{j}$ in the denominator, rather than $(\sigma^2 - \epsilon^2)$ with $\sigma$ being the standard deviation of the time series, and $\epsilon$ being a ``measurement error'' associated with the data-set, and included to preserve the normalization. This modification allows for heteroscedastic data, but as a consequence, the autocorrelation at time-lag $\tau = 0$ in general will not equal 1. 

An uncertainty on this correlation is estimated by taking the standard error on the mean for the points in a bin:
\begin{equation}
\sigma_{DACF}(\tau_{k}) = \frac{1}{M_{k}-1} \Big\{ \sum \left[ C_{ij} - DACF(\tau_{k}) \right]^{2}\Big\}^{1/2}.
\end{equation}

The DACF will be written out to a separate file for
each light curve analyzed. This algorithm is sometimes used to
identify variability periods for non-stationary quasi-periodic signals
by searching for peaks in the DACF, however it is most useful in
characterizing the coherence time scale of features (either true
astrophysical signals, or correlated noise) in the light
curves. Currently it is left up to the user to separately analyze the
output autocorrelation function.

\paragraph{\bf -chi2}\label{cmd:chi2}
Calculates 
\begin{equation}
\chi^{2}/{\rm d.o.f.} = \frac{1}{N-1}\sum_{i=1}^{N}\left( \frac{m_{i}-\bar{m}}{\sigma_{i}} \right)^{2},
\end{equation}
the reduced $\chi^{2}$, for each light curve.

\paragraph{\bf -chi2bin}\label{cmd:chi2bin}
Calculates the reduced $\chi^{2}$ after applying a moving mean filter to each light curve. This filter operates by replacing each $m_{i}$ in the light curve by the average value of $m$ within a bin of size $\Delta t$. That is, $m_{i}$ is replaced with
\begin{equation}
\frac{\sum_{j}m_{j}}{\sum_{j} 1}
\end{equation}
where the sum is over points $j$ such that $|t_{j} - t_{i}| < \Delta
t$. The light curve uncertainties after applying the filter are reduced by the amount expected for white noise. That is
\begin{equation}
\sigma_{i}^{\prime} = \frac{\sqrt{\sum_{j}\sigma_{j}^{2}}}{(\sum_{j} 1) - 1}
\end{equation}
where $\sigma_{i}^{\prime}$ is the new value for the uncertainty, and
the sum is over the same points as above. One result of this is that
if the light curve has time correlated variations, the reduced
$\chi^{2}$ reported by {\bf -chi2bin} will increase as $\Delta t$ is
increased. For pure white noise, the reduced $\chi^{2}$ will be
constant (to within statistical uncertainties) as $\Delta t$ is
increased.

\paragraph{\bf -Jstet}\label{cmd:jstet}
Calculates the $J$ light curve variability statistic suggested
by \citet{stetson:1996}, together with the Kurtosis $K$ and the $L$ statistic also defined by \citet{stetson:1996}. The $J$ statistic is similar to {\bf -alarm} in that its value will be greater for a light curve that shows time-correlated variability than for a light curve with the same scatter, but for which the variations are not correlated in time. The $J$ statistic is given by the expression
\begin{equation}
J = \frac{\sum_{k=1}^{n}w_{k}{\rm sign}(P_{k})\sqrt{|P_{k}|}}{\sum_{k=1}^{n}w_{k}}
\label{eqn:jstet}
\end{equation}
where the sum in the numerator and denominator is of $n$ pairs of observations, $P_{k} = \delta_{i(k)}\delta_{j(k)}$ for $i(k) \neq j(k)$ or $P_{k} = \delta_{i(k)}^2 - 1$ for $i(k) = j(k)$, and where 
\begin{equation}
\delta_{i} = \sqrt{N/(N-1)}(m_{i} - \bar{m})/\sigma_{i}
\end{equation}
is the normalized residual for observation $i$, there are $N$ observations total, observations $i(k)$ and $j(k)$ form pair $k$, and $w_{k}$ is a weight to be assigned to a pair.

The Kurtosis of the light curve magnitudes (i.e., the fourth moment of the distribution) is estimated via
\begin{equation}
K = \frac{\frac{1}{N}\sum_{i=1}^{N}|\delta_{i}|}{\sqrt{\frac{1}{N}\sum_{i=1}^{N}\delta_{i}^2}}
\label{eqn:kurtosisstet}
\end{equation}
and has a value of $\sqrt{2/\pi} \approx 0.798$ when $N$ is large and the magnitude values are drawn from a Gaussian distribution.

The $L$ statistic is then defined as
\begin{equation}
L=\frac{JK}{0.798}\frac{\sum w}{w_{\rm all}}
\label{eqn:lstet}
\end{equation}
where $w_{\rm all}$ is the maximum value of $\sum w$ that any light
curve from the survey could have (i.e., a light curve with
measurements from all images in a survey). This statistic provides
enhanced significance over the $J$ statistic for light curves with
non-Gaussian magnitudes, and reduces the significance for light curves
with missing observations.

\paragraph{\bf -rms}\label{cmd:rms}
Calculates the standard root-mean-square (RMS) scatter of the light curve given by:
\begin{equation}
{\rm RMS} = \sqrt{\sum_{i=1}^{N}(m_{i} - \bar{m})^{2}/(N-1)}
\end{equation}
Additionally this command will compute the expected RMS of the light curve given the input photometric uncertainties. This is given by
\begin{equation}
{\rm \bar{RMS}} = \sqrt{\sum_{i=1}^{N}(\sigma_{i}^2)/N}
\label{eqn:expectedrms}
\end{equation}

\paragraph{\bf -rmsbin}\label{cmd:rmsbin}
Calculates the RMS after applying a moving mean filter
to each light curve. The filter is the same as used for the {\bf
  -chi2bin} command.

\paragraph{\bf -stats}\label{cmd:stats}
Calculates one, or more, basic statistics for one, or more,
light-curve vectors (i.e., variables read-in from the light curve). The
available statistics include: the mean; the mean weighted by the input
photometric uncertainties; the median; the weighted median; the
standard deviation calculated with respect to the mean; the standard
deviation calculated with respect to the median; the median of the
absolute deviations from the median (medmeddev); the MAD ($1.483
\times$medmeddev, which for a Gaussian distribution equals the
standard deviation in the limit of large numbers, but is robust
against outliers); the kurtosis; the skewness; percentile values (any
floating point number between 0 and 100 may be used); weighted
percentile values; the maximum value; the minimum value; and the sum
of all elements.

\subsubsection{Model-Fitting}\label{sec:models}

There are seven built-in commands which can be used to fit model
signals to light curves.

\paragraph{\bf -decorr}\label{cmd:decorr}
Fits a model that is linear in its parameters to a
light curve. This command is deprecated in the current version of {\sc
  vartools}, being replaced by the more flexible {\bf -linfit}
command, so it will not be discussed in detail here.

\paragraph{\bf -Killharm}\label{cmd:killharm}
Fits a harmonic series to a light curve. The model has the form
\begin{align}
\label{eqn:harm}
& m_{0} + \sum_{i=1}^{N_{P}}\Big\{ a_{i,1}\sin(2\pi f_{i}t) + b_{i,1}\cos(2\pi f_{i}t) \\
& + \sum_{k=2}^{N_{\rm harm,i}+1}\left[ a_{i,k}\sin(2\pi k f_{i}t) + b_{i,k}\cos(2\pi k f_{i}t)\right] \nonumber \\
& + \sum_{k=2}^{N_{\rm subharm,i}+1}\left[ c_{i,k}\sin(2\pi f_{i}t/k) + d_{i,k}\cos(2\pi f_{i}t/k)\right] \Big\} \nonumber
\end{align}
where there are $N_{p}$ fixed frequencies $f_{i}$ in the fit, each having
$N_{\rm harm,i}$ higher-order harmonics and $N_{\rm subharm,i}$ subharmonics, and
the parameters $m_{0}$, $a_{i,k}$, $b_{i,k}$, $c_{i,k}$ and $d_{i,k}$ are
fitted by the procedure. Note that this same model can also be fit
through the more general-purpose {\bf -linfit} command; the separate
{\bf -Killharm} command is provided for convenience.

By default the best-fit values for the parameters $m_{0}$, $a_{i,k}$, $b_{i,k}$, $c_{i,k}$ and $d_{i,k}$ are reported in the output table. The user may optionally output amplitudes, phases, or relative amplitudes and phases instead.  Here the amplitude of harmonic $k$ is given by
\begin{equation}
A_{i,k} = \sqrt{a_{i,k}^{2}+b_{i,k}^{2}}
\label{eqn:harmamp}
\end{equation}
while the amplitude for sub-harmonic $k$ is given by
\begin{equation}
B_{i,k} = \sqrt{c_{i,k}^{2}+d_{i,k}^{2}}.
\label{eqn:subharmamp}
\end{equation}
The amplitudes may also be specified relative to the amplitude of the fundamental frequency via
\begin{equation}
R_{i,k,1} = A_{i,k}/A_{i,1}
\label{eqn:harmrelamp}
\end{equation}
and
\begin{equation}
Q_{i,k,1} = B_{i,k}/A_{i,1}.
\label{eqn:subharmrelamp}
\end{equation}
The phases are calculated as
\begin{equation}
\phi_{i,k} = {\rm atan2}(-b_{i,k}, a_{i,k})
\label{eqn:harmphase}
\end{equation}
and
\begin{equation}
\psi_{i,k} = {\rm atan2}(-d_{i,k}, c_{i,k})
\label{eqn:subharmphase}
\end{equation}
where ${\rm atan2(y,x)}$ is the standard 4-quadrant inverse tangent function with output in radians. The relative phases of the harmonics and sub-harmonics are given by
\begin{equation}
\phi_{i,k,1} = \phi_{i,k} - k\phi_{i,1}
\label{eqn:harmrelphase}
\end{equation} and
\begin{equation}
\psi_{i,k,1} = \psi_{i,k} - \frac{1}{k}\phi_{i,1}.
\label{eqn:subharmrelphase}
\end{equation}
Using relative amplitudes and phases for the harmonics and sub-harmonics is a common way to characterize the shape of a signal independently of the overall phase and amplitude. Such parameters are useful, for example, in classifying variable star light curves based on their morphology.

The {\bf -Killharm} command also provides a measurement of the
peak-to-peak amplitude of each harmonic series that is fit to the data
(a separate amplitude is supplied for each period).  To determine this
amplitude, the model harmonic series is evaluated at $100 \times
(N_{\rm harm,i} + N_{\rm subharm,i} + 1)$ phase steps. The minimum and
maximum values of the series at the evaluated points are
determined. Refined measurements are then performed near these points
using the {\sc dbrent} algorithm \citep{press:1992}.

\paragraph{\bf -linfit}\label{cmd:linfit}
Fits an analytic model that is linear in its parameters to a light
curve. The user provides on the command-line an analytic expression
together with a list of the names of variables in that expression that
are to be varied. The fit is done by singular value decomposition
(SVD) as implemented in the {\sc svdfit} function of Numerical Recipes
\citep{press:1992}. This function includes a correction which
suppresses poorly constrained parameter combinations, with the goal of
providing greater numerical stability. Because this can lead to
differences from other tools which solve the linear least squares
problem by direct inversion of the so-called design matrix, we provide
a summary of the {\sc svdfit} method below.

Let the light curve magnitudes be represented as a column vector $\vec{m}$ of length $N$, and suppose these are to be fit with a model $\vec{mod}$ which can be expressed as
\begin{equation}
\vec{mod} = \sum_{i}^{N_{\rm param}}a_{i}\vec{x}_{i},
\end{equation}
i.e., the linear combination of $N_{\rm param}$ column vectors $\vec{x}_{i}$. These vectors are supplied by the user, and may be the times of observations (perhaps raised to some power), other columns read-in from the light curve (e.g., the $X$ and/or $Y$ coordinates of the star on each image), or any analytic expression involving such terms. The $a_{i}$ parameters are the $N_{\rm param}$ variables to be fit by this command by minimizing $\chi^2$, which is given by
\begin{equation}
\chi^{2} = |{\bf A}\vec{a}-\vec{b}|^2.
\end{equation}
Here the $N \times N_{\rm param}$ design matrix ${\bf A}$ has components
\begin{equation}
A_{ij} = a_{j}x_{ji}/\sigma_{i}
\end{equation}
with $x_{ji}$ being the $i$th component of vector $\vec{x}_{j}$ and $\sigma_{i}$ is the $i$th measurement uncertainty in the light curve,
and the vector $\vec{b}$ has components
\begin{equation}
b_{i} = m_{i}/\sigma_{i}.
\end{equation}
Applying SVD (via the {\sc svdcmp} function in Numerical Recipes) to
{\bf A} yields an $N \times N_{\rm param}$ column-orthogonal matrix
{\bf U}, a $N_{\rm param} \times N_{\rm param}$ orthogonal matrix {\bf
  V}, and a $N_{\rm param} \times N_{\rm param}$ diagonal matrix ${\bf
  W}$ with positive or zero elements, which together satisfy
\begin{equation}
{\bf A} = {\bf U} {\bf W} {\bf V}^{T}
\end{equation}
and which can be used to determine the least-squares solution for $\vec{a}$ via
\begin{equation}
\vec{a} = \sum_{i=1}^{N_{\rm param}} \frac{{\bf U}_{i} \cdot \vec{b}}{w_{i}} {\bf V}_{i}
\label{eqn:svdlinfitsol}
\end{equation}
where ${\bf U}_{i}$ and ${\bf V}_{i}$ are the $i$th columns of each
matrix, and $w_{i}$ is the $i$th diagonal element of ${\bf
  W}$. The evaluation of equation~\ref{eqn:svdlinfitsol} can be done
efficiently by back-substitution (the {\sc svbksb} function in
Numerical Recipes). In cases where degeneracies exist between the different $\vec{x}_{i}$ vectors, very small values of $w_{i}$ may be found leading to numerical instabilities in evaluating equation~\ref{eqn:svdlinfitsol}. This problem is overcome by setting $\frac{1}{w_{i}} = 0$ in cases where $w_{i} < {\rm TOL} \times w_{\rm max}$, for some tolerance factor ${\rm TOL}$ and with $w_{\rm max}$ being the maximum element in ${\bf W}$. We adopt ${\rm TOL} = 10^{-9}$ in \va{}.

Formal
uncertainties on the parameters are also determined based on the
covariance matrix--these uncertainties are accurate only if the
measurement uncertainties supplied in the light curve are correct, and
if the noise is uncorrelated in time and drawn from a normal
distribution.

\paragraph{\bf -MandelAgolTransit}\label{cmd:mandelagoltransit}
Fits the analytic model for the transit of a
dark spherical planet across a limb-darkened spherical star described
by \citet{mandel:2002}. This model is parameterized by the orbital
period, an initial epoch of transit center, the ratio of the planet
radius to the stellar radius, the ratio of the semi-major axis to the
stellar radius, the normalized impact parameter (the minimum projected
distance between the stellar and planet centers divided by the sum of
the planet and stellar radii), the orbital eccentricity, the argument
of periastron, and either two or four limb darkening coefficients (two
for a quadratic limb-darkening law, and four for a non-linear
limb-darkening law). Note that this choice of the impact parameter
differs from some implementations where the normalization is done only
to the stellar radius. Normalizing to the sum of the radii has the
benefit of limiting the range of possible values to between $0$ and
$1$ to ensure transit. The user may also simultaneously fit a
Keplerian orbit to a set of radial velocity data, input as a separate
file. In this case the orbital semi-amplitude and center-of-mass
velocities are additional parameters. Because the transit model is
nonlinear in its parameters, it is necessary to choose initial values
from which to begin the fit. The DHSX is then used to search for a local $\chi^{2}$
minimum near the initial position. The user may either fix the initial
values on the command line, or use the results from the {\rm -BLS}
command to automatically initialize the parameters. Note that this
routine is not intended to produce publication-quality fits, as it
provides no mechanism for error determination, and does not support
simultaneous fits of multiple light curves. It is instead useful for
rapidly processesing many candidate transiting planet systems to
identify initial parameter estimates (as might be done in a transit
search pipeline).

\paragraph{\bf -microlens}\label{cmd:microlens}
Fits a simple microlensing model due to
\citet{wozniak:2001} to a light curve. The model is given by
\begin{equation}
M(t) = f_{0} + f_{1}(A(t)-1)
\label{eqn:microlens1}
\end{equation}
where
\begin{equation}
A(t) = \frac{u^{2}+2}{u\sqrt{u^{2}+4}}
\label{eqn:microlens2}
\end{equation}
and
\begin{equation}
u^{2} = u_{0}^2 + \left( \frac{t-t_{\rm max}}{t_{0}} \right) ^{2}
\label{eqn:microlens3}
\end{equation}
The free parameters in the model include: the zero-point flux $f_{0}$;
the lensed flux $f_{1}$; the impact parameter $u_{0}$ in units of the
Einstein radius; the Einstein ring crossing time $t_{0}$; and the time
of maximum lensing $t_{\rm max}$. The model is fit using DHSX.

\paragraph{\bf -nonlinfit}\label{cmd:nonlinfit}
Fits an analytic model to a light curve (non-analytic functions may
also be fit through the use of user-developed functions, see
Section~\ref{sec:userfunc}). The model may be non-linear in its
parameters. The user provides on the command-line an analytic
expression together with a list of the names of variables in that
expression that are to be varied, initial estimates for their best-fit
values, and step-sizes to use in initializing the search
algorithms. The user may optionally include a separate list of
parameters that enter linearly into the model, and which will be
optimized using linear least squares at each step in the non-linear
model fitting (via eq.~\ref{eqn:svdlinfitsol}). In addition to
specifying the model and the free parameters, the user may also
provide prior probability estimates for the parameters and/or
constraints on the parameters (a list of analytic inequalities that
must be satisfied by the parameters). The routine will attempt to
minimize the expression:
\begin{equation}
X^{2} = \ln(\det({\bf \Sigma})) + (\vec{m} - \vec{mod})^{T}{\bf \Sigma}^{-1}(\vec{m} - \vec{mod}) + \sum_{j=1}^{N_P} P_{j}
\label{eqn:nonlinfitx2}
\end{equation}
where $\vec{m}$ and $\vec{mod}$ are column vectors containing the
measured and model magnitudes, respectively, ${\bf \Sigma}$ is the
covariance matrix for the observational data, and $P_{j}$ are the
$N_{P}$ priors which are analytic functions of the free parameters. By
default ${\bf \Sigma}$ is taken to be a diagonal matrix with
components $\Sigma_{ii} = \sigma_{i}^2$, where $\sigma_{i}$ is the
magnitude uncertainty for observation $i$. In this case the term
$\ln(\det({\bf \Sigma}))$ is ignored by the routine as it is
independent of the model parameters. The user may optionally specify
an analytic expression for the uncertainties (this expression may
include parameters which are varied in the fit), in which case the
term will be included. Options for specifying a non-diagonal
covariance matrix are discussed below. The value of $X^{2}$ is
minimized over the region satisfying the parameter constraints. The
quantity $X^{2}$ is related to the likelihood $L$ via $X^{2} = -2 \ln
L + C$ for some constant $C$ and is equal to $\chi^{2}$ when using a
diagonal covariance matrix and no priors. Note that the $P_{j}$
functions specified by the user are actually equal to $-2$ times the
natural logarithm of functions proportional to the prior probability
densities, the $P_{j}$ functions are not the prior probability density
functions themselves. For example, if the user wishes to specify a
Gaussian prior of the form:
\begin{equation}
P \propto \exp \left[ -(a - \bar{a})^{2}/(2\sigma_{a}^2) \right]
\end{equation}
for some parameter $a$, they would express it as $(a -
\bar{a})^{2}/(\sigma_{a}^2)$.  

We allow three different forms for describing non-diagonal elements in
the covariance matrix (i.e., allowing for time-correlated
noise). These are the squared exponential, exponential and Mat\'ern noise
models. In the squared exponential model the covariance between times
$i$ and $j$ is given by
\begin{equation}
\Sigma_{ij} = \sigma_{i}^2\delta_{ij} + a\exp(-\frac{(t_{i}-t_{j})^2}{2\rho^2})
\label{eqn:squareexp}
\end{equation}
where $\sigma_{i}$ is the formal uncertainty for measurement $i$ (if
the user specifies an analytic expression to use for the uncertainty,
then $\sigma_{i}$ is the expression evaluated at time $i$),
$\delta_{ij}$ is the Kronecker delta function, and $a>0$, and $\rho >
0$ are parameters of the noise model. The exponential model is similar, with 
\begin{equation}
\Sigma_{ij} = \sigma_{i}^2\delta_{ij} + a\exp(-\frac{|t_{i}-t_{j}|}{\rho}).
\label{eqn:expnoise}
\end{equation}
The Mat\'ern model
\citep{handcock:1993} is given by
\begin{equation}
\Sigma_{ij} = \sigma_{i}^2\delta_{ij} + a\frac{2^{1-\nu}}{\Gamma(\nu)}\left( \frac{\sqrt{2\nu}|t_{i}-t_{j}|}{\rho}\right) ^{\nu} K_{\nu}\left( \frac{\sqrt{2\nu}|t_{i}-t_{j}|}{\rho}\right)
\label{eqn:matern}
\end{equation}
where $\Gamma$ is the standard gamma function, $K_{\nu}$ is the modified Bessel
function of the second kind, and $a > 0$, $\rho >0$ and $\nu > 0$ are
free parameters. This model allows for covariances that fall off more
slowly with separation than the squared exponential model. For $\nu =
1/2$ the Mat\'ern model is proportional to
$\exp(-|t_{i}-t_{j}|/\rho)$, while for $\nu \rightarrow \infty$ it
converges to the squared exponential. When non-diagonal covariances
are used it is necessary to compute the inverse of the covariance
matrix, which as an $\mathcal{O}(N^{3})$ operation becomes
computationally expensive when analyzing time series with tens to
hundreds of thousands of points. To speed up the calculation we make
use of the fact that for the noise models assumed we have $\Sigma_{ij}
> \Sigma_{ik}$ for $j \geq i$ and $k > j$ if the light curve is sorted
in time. Moreover, because the elements of ${\bf \Sigma}$ approach zero very rapidly when
moving away from the diagonal, the covariance matrix is typically a
sparse, near-diagonal matrix. We therefore compute the expression ${\bf
  \Sigma}^{-1}({\bf m}-{\bf mod})$ using Cholesky decomposition and
back-substitution, truncating the loops within the algorithm when
$C_{ij} < \epsilon$ for an assumed tiny value of $\epsilon$. The logarithm of the determinant of ${\bf \Sigma}$ is computed from this same decomposition.

To optimize the non-linear parameters the user may choose between
using the DHSX method as implemented in the {\sc amoeba} routine by
\citet{press:1992}, or running a Differential Evolution Markov Chain Monte Carlo (DEMCMC) procedure.

The {\sc amoeba} routine performs a greedy search for the $X^2$ minimum,
and will tend to find the local minimum nearest to the initial
starting position. The search is carried out using a set of $N_{\rm
  par}+1$ vectors of parameters (the so-called simplex, here $N_{\rm
  par}$ is the number of free parameters). In the \va{} implementation
the components of each vector are initially set to the initial
parameter values specified by the user, except that for vector $j$ the
component $j-1$ is set to the initial value plus the specified
step-size (for the first vector no component is adjusted). The routine
evaluates $X^{2}$ for each vector of parameters, and then repeatedly
adjusts the components of each vector following a prescribed method in
an attempt to find the minimum value of $X^{2}$. The routine stops
once it achieves:
\begin{equation}
\frac{2|X_{\rm max}^2 - X_{\rm min}^2|}{|X_{\rm max}^2 + X_{\rm min}^2| + \epsilon} < {\rm TOL}
\label{eqn:amoebaconvergence}
\end{equation}
where $X_{\rm max}^2$ is the maximum value of $X^{2}$ amongst the
vectors in the simplex, $X_{\rm min}^2$ is the minimum value,
$\epsilon = 10^{-10}$ is a small number used to prevent division by
zero, and ${\rm TOL}$ is the convergence tolerance (by default equal
to $10^{-10}$, but the user may specify a different value on the
command line).

The second option is to run a DEMCMC procedure
\citep{terbraak:2006}. This procedure runs $2N_{\rm par}$ simultaneous
Markov chains (a minimum of three chains are used). By default \va{}
will first run the {\sc amoeba} downhill simplex to find a local minimum
about which the DEMCMC will be executed, though the user may skip this
step. One of the chains is then initialized to the best-fit parameter
values, for the other chains we add to each parameter a Gaussian
random number multiplied by the step-size. The initial user-supplied
step sizes are iteratively adjusted to determine the step-size such
that $\Delta \chi^2 = 1$ for each adjusted parameter, holding all
other parameters fixed. We cycle through each of the chains proposing
a new link $N$ in the chain $j$ using
\begin{equation}
\vec{a}_{j,N} = \vec{a}_{j,N-1} + \gamma (\vec{a}_{k,N-1} - \vec{a}_{l,N-1}) + \epsilon {\bf U}\times \vec{r}
\label{eqn:mcmcdef}
\end{equation}
where $k$ and $l$ are the indices of two randomly selected chains, $\gamma =
1.7/\sqrt{2N_{\rm par}}$, $\epsilon$ is a small number ($0.1$ by
default, but can be modified by the user), $\vec{r}$ is a vector of Gaussian
random numbers and ${\bf U}$ is an upper triangular matrix determined from
the Cholesky decomposition of the parameter covariance matrix ${\bf \Sigma}$
\citep[e.g.][]{press:1992}. We estimate ${\bf \Sigma}$ from the existing DEMCMC chains,
recomputing ${\bf U}$ every 10000 links. We calculate $\Delta X^{2}$, the
difference in $X^{2}$ between the proposed link and the prior link in
the chain, and then accept the proposal with probability $\exp(-\Delta
X^{2}/2)$ (if $\Delta X^{2} < 0$, the proposed link is always
accepted). If the proposed link is rejected, the previous link is
copied to link $N$. The process terminates once either the total
number of links reaches a user specified limit, or the total number of
accepted links reaches a limit. 

The $N_{c}$ chains are stored in a
single array such that element $i$ in this array is associated to
chain $i_{\rm chain} = i\,{\rm modulo}\,N_{c}$. The user has the
option of writing this array out to a file. By default all links are
written, but the user may write out only a fraction of the links using
the ``printevery'' option. To allow the user to interrupt the process
without losing all of the results the links are written to the file
buffer immediately after each MCMC step. However, we do not
explicitely flush the file buffer, so the data will only be written to
disk once the buffer is filled (this may be every few hundred links,
but depends on the number of free parameters, and on the specific
computer system used). 

The full chain is also stored in memory and is
used to compute statistics included in the output ascii table. By
default these are the median and standard deviation of each free
parameter. These may be interpretted as the most likely values and
uncertainties for each parameter. The values corresponding to the
lowest $X^{2}$ model are also included in the output ascii table. The
user may change the statistics computed from the chain and may also
specify different combinations of the parameters on which to compute
the statistics. By default the first 10\% of the chain is excluded
from calculating the output statistics, however the user may
optionally specify a different burn-in fraction. 

A limit is also
provided on the total amount of memory allowed to be allocated to a
particular MCMC command. Once this limit is reached, each time a new
link is added to the chain the earliest link in the chain is
forgotten. The default memory limit is 4\,GB. Note that if \va{} is
run in parallel mode, then the limit for each thread is the total
limit divided by the number of threads.

\paragraph{\bf -SoftenedTransit}\label{cmd:softenedtransit}
Fits an approximate transit model due to
\citet{protopapas:2005} to a light curve. The model is given by:
\begin{equation}
M(t) = M_{0} + \frac{1}{2}\delta \Big\{ 2 - \tanh[c(t^{\prime}+0.5)] + \tanh[c(t^{\prime}-0.5)] \Big\}
\label{eqn:softenedtransit1}
\end{equation}
with
\begin{equation}
t^{\prime}(t) = \frac{P}{\pi\eta}\sin(\pi(t-T_{0})/P)
\label{eqn:softenedtransit2}
\end{equation}
where $M_{0}$ is the out of transit magnitude, $\delta$ is the depth
of the transit, $c$ is a constant controlling the ``sharpness'' of the
transit, $P$ is the period, $T_{0}$ is the reference time of mid
transit, and $\eta$ is the transit duration.

\paragraph{\bf -Starspot}\label{cmd:starspot}
Fits a single, circular, uniform temperature spot
model given by \citet{dorren:1987} to a light curve. This model is
given by: 
\begin{equation}
M(t) = M_{0} - 2.5\log_{10}\left[ 1 - (aA(\alpha,\beta(t))+bB(\alpha,\beta(t)))\right]
\label{eqn:starspot1}
\end{equation}
where $M_{0}$ is the out of transit magnitude, $A$ and $B$ are functions given by \citet{dorren:1987} that depend on $\alpha$, the spot angular radius, and $\beta$, the angle between the line of sight and the surface normal at the spot center, and $a$ and $b$ are quantities given by
\begin{align}
a = & \frac{(1-\mu_{\star})-(1-\mu_{s})\frac{F_{s}}{F_{\star}}}{\pi(1-\mu_{\star}/3)} \\
b = & \frac{\mu_{\star}-\mu_{s}\frac{F_{s}}{F_{\star}}}{\pi(1-\mu_{\star}/3)} \\
\end{align}
with $\mu_{\star}$ and $\mu_{s}$ being the linear limb darkening
coefficients for the unspotted photosphere of the star and for the
spot, respectively, and $F_{s}/F_{\star}$ being the spot to
photosphere flux ratio. \citet{dorren:1987} suggest values of
$a = 0.0298$ and $b = 0.08745$ for the Sun at 5000\,\AA. The angle
$\beta$ is itself a function of the time of the observation, the
rotation period of the star, the inclination of the stellar rotation
axis, the latitude of the spot, and the longitude of the spot center
at a reference time. This procedure is deprecated in the current
version of \va{}, users should instead use the {\bf -macula} model
which is distributed with \va{} as a user-developed command.

\subsubsection{Filtering}\label{sec:filters}

The built-in commands listed below can be used to filter light
curves. Additionally, filtering may be done through the procedures
described in the previous subsection (e.g., {\bf -Killharm} to filter
harmonic series from the data, or {\bf -linfit} to apply a linear
filter).

\paragraph{\bf -clip}\label{cmd:clip}
Removes points from the light curve. Each point is
removed from all vectors associated with the light curve (e.g. the
time, magnitude, magnitude uncertainty, and other data read-in from
the light curve). The default method is to $\sigma$-clip outlier
magnitude values. In other words, given an average magnitude $\bar{m}$
measured RMS magnitude $\sigma_{m}$, and specified clipping factor
$\alpha$, any point $i$ with magnitude $m_{i}$ satisfying $|m_{i} -
\bar{m}| > \alpha \sigma_{m}$ will be removed from the light
curve. The user may optionally run this in iterative mode, in which
case $\bar{m}$ and $\sigma_{m}$ are both recalculated after clipping,
and the clipping is repeated with the new values; this can either
be repeated for a specified number of iterations, or until no further
points are removed. Finally the user may also simply remove points
with formal errors $\leq 0$ or undefined magnitude values. This later
capability can be used in combination with an {\bf -expr} command to
perform a customized clipping procedure (e.g., listing~\ref{lst:clipexample}).

\begin{lstlisting}[caption={example of using the {\bf -clip} command to remove points with magnitude values greater than 10.},label={lst:clipexample},frame=single,backgroundcolor=\color{lightgray},language=vartools]
vartools -i inputfile.txt \
 ... \
 -expr 'err=(mag>10)*(-1)+(mag<=10)*err' \
 -clip -1 0 \
 ...
\end{lstlisting}

\paragraph{\bf -medianfilter}\label{cmd:medianfilter}
For each point $i$, determines the median
magnitude value of all points $j$ that satisfy $|t_{i} - t_{j}| <
\Delta t$ for fixed $\Delta t$, and then either subtracts
this value from the magnitude of point $i$ (high-pass filtering the
light curve), or replaces the magnitude of point $i$ with this value
(low-pass filtering the light curve). The user may optionally use the
average, or the uncertainty-weighted average instead of the median.

\paragraph{\bf -restricttimes}\label{cmd:restricttimes}
Filter observations from the light curve(s)
based on the time values. The user either indicates observations to
include or exclude. This can be done by giving a range of time values,
a list of time values to filter, or a list of image $id$ values to filter.

\paragraph{\bf -SYSREM}\label{cmd:sysrem}
Performs the trend-filtering procedure suggested by
\citet{tamuz:2005}, which effectively amounts to determining the principle components of the light curve dataset, and filtering these from the light curves. The filter operates by
attempting to model the light curves indexed by $i$ consisting of
points indexed by $j$ as:
\begin{equation}\label{eqn:sysrem}
m_{i,j} = \epsilon_{i,j} + \sum_{k=1}^{N_{trend}}c_{k,i}a_{k,j}
\end{equation}
where $\epsilon_{i,j}$ is a gaussian random number drawn from a normal
distribution with mean $\bar{m_{i}}$ and standard deviation
$\sigma_{i,j}$; and there are $N_{trend}$ trend vectors $a_{k,j}$,
called ``airmass-like'' terms by \citet{tamuz:2005}, to which each
light curve has a linear response parameterized by $c_{k,i}$, called
``color-like'' terms by \citet{tamuz:2005}. Initial airmass and/or
color-like vectors are assumed (these might be the airmasses of the
images, or the x/y pixel positions of the stars, for example). For
each assumed airmass-like (color-like) vector the associated
color-like (airmass-like) vector is solved for by linear least squares
minimization. The so-determined color-like (airmass-like) vector is
then fixed, and the airmass-like (color-like) vector is then
redetermined by linear least squares minimization. The procedure is
repeated for the next vector until all $N_{trend}$ trends are
fitted. To prevent outliers from dominating the determination of
trends, the user specifies two $\sigma$-clipping parameters. The first
for excluding points from individual light curves in determining the
average magnitudes of the light curves, the second for excluding
points from the fits for the airmass-like or color-like terms.

\paragraph{\bf -TFA}\label{cmd:tfa}
Applies the Trend Filtering Algorithm suggested by
\citet{kovacs:2005:TFA}. This is similar to SYSREM in that the light
curves are modeled as a linear combination of trend vectors as
described in equation~\ref{eqn:sysrem}, however in this case the
$a_{k,j}$ ``airmass-like'' terms are taken to be $N_{trend}$ template
light curves chosen randomly from among the full ensemble of observed
light curves, and only the $c_{k,i}$ ``color-like'' terms are
determined by fitting. One other practical difference between SYSREM
and TFA is that the latter typically uses more trend vectors in
modelling the light curves (e.g.~for a field containing $\gtrsim 10000$
stars, each containing $\sim 10000$ observations, it is not uncommon
to use $\sim 1000$ template light curves in TFA, whereas typically
applications of SYSREM may only use a few dozen trend vectors). When
the same set of templates is used to filter numerous light curves the
filter may be applied efficiently by a combination of Singular Value
Decomposition (or alternatively Lower-Upper Diagonal Decomposition)
applied once to the matrix of templates, followed by the much faster
Singular Value Back Substitution procedure applied to each light curve
to be filtered.
 
\paragraph{\bf -TFA\_{}SR}\label{cmd:tfasr}
Applies the Trend Filtering Algorithm in
``signal-reconstruction'' mode. In this case the assumed light curve
model is modified from that given in equation~\ref{eqn:sysrem} to:
\begin{equation}
m_{i,j} = \epsilon_{i,j} + S_{i}(t_{j}) + \sum_{k=1}^{N_{trend}}c_{k,i}a_{k,j}
\end{equation}
where $S_{i}(t_{j})$ is a light-curve-dependent signal. In \va{} the allowed
signals include: (1) a periodic step-function of the form
\begin{eqnarray}
S(t) & = & \sum_{u=1}^{N_{\rm bin}}\alpha_{u}\Theta(\phi - (u-1)/N_{\rm bin})\Theta(u/N_{\rm bin} - \phi) \nonumber \\
\phi & = & t/P - {\rm floor}(t/P)
\end{eqnarray}
where $\Theta(X)$ is the Heavside-step function ($=0$ for $x < 0$;
$=1$ for $x > 0$), $P$ is the fixed period of the signal, there are
$N_{\rm bin}$ steps used, and the free parameters are the $\alpha_{u}$ terms;
(2) a fixed signal form evaluated at the times of observation (and
read-in from a file); or (3) a harmonic series of the form given in
equation~\ref{eqn:harm}. Additionally the user may include a number of
light-curve-specific vectors among the series of trend vectors used in
the fit. In practice this slows down the filter as it requires a
re-inversion of the design matrix used in the linear fit.

\subsubsection{Light Curve Simulation}\label{sec:lcsim}

The following built-in commands may be used in simulating light curves.

\paragraph{\bf -addnoise}\label{cmd:addnoise}
Adds noise with optional time-correlation to the light curve. Five
different methods may be selected from. The first is simple
uncorrelated Gaussian white noise, in which case the user specifies
the standard deviation of the noise.

The second, third and fourth methods simulate a Gaussian process with a fixed covariance matrix ${\bf \Sigma}$. These methods differ in the form of the covariance matrix. The second method assumes a squared exponential model where the elements of the covariance matrix are given by Eq.~\ref{eqn:squareexp}. The third method assumes an exponential covariance with elements given by Eq.~\ref{eqn:expnoise}. The fourth method uses the Mat\'ern class of covariance matrices with elements given by Eq.~\ref{eqn:matern}. In all three cases the Cholesky decomposition of ${\bf \Sigma}$ is computed and then multiplied against a vector of random numbers drawn from a normal distribution to produce a light curve drawn from the specified covariance matrix. 

In the fifth method the noise is modeled as
\begin{equation}
\epsilon_{w} + \epsilon_{r}
\end{equation}
where $\epsilon_{w}$ is a white-noise component that is a random
number drawn from a Gaussian distribution with zero mean and standard
deviation $\sigma_{w}$ (specified by the user), while $\epsilon_{r}$
is a red-noise component drawn from a distribution with standard
deviation $\sigma_{r}$ and time-correlations such that the
power-spectral density of the noise scales as $1/f^{\gamma}$ (both
$\gamma$, which must satisfy $-1 < \gamma < 1$, and $\sigma_{r}$ are
specified by the user; taking $\gamma = 0$ yields uncorrelated
noise). We use the method of \citet{mccoy:1996} to generate the
red-noise. The \citet{mccoy:1996} procedure for simulating a series of
$N = 2^{p}$ points uniformly sampled in time, described in Section~4.2
of that paper, works as follows: (1) generate a set of points
$d_{m,k}$ with $m = 1,\ldots,p$ and $k=1,\ldots,2^{p-m}$ each drawn
from a Gaussian distribution with zero mean and variance $S_{m}$, and
generate an additional point $x_{p},1$ with zero mean and variance $S_{p+1}$
(\citealp{mccoy:1996} gives a prescription for determining the
variances $S_{m}$ that depends on $\sigma_{r}$ and $\gamma$); (2)
apply the inverse discrete wavelet transform (we use the
implementation in the GNU scientific library) to the vector
$(x_{p,1},d_{p,1},d_{p-1,1},d_{p-1,2},\ldots,d_{1,1},\ldots,d_{1,2^{p-1}})$. To
generate non-uniformly sampled observations, we first simulate a
series of $N = 2^{p}$ uniformly sampled points as above, taking
\begin{equation}
p = \min(20,{\rm ceil}(\lg(T/\delta_{t})))
\end{equation}
where $T$ is the time spanned by the non-uniformly sampled
observations, and $\delta_{t}$ is the minimum time between successive
observations. We then linearly interpolate this sequence onto the
non-uniformly sampled observation times.

\paragraph{\bf -copylc}\label{cmd:copylc}
Makes copies of the light curve under analysis. Data
read in from the input list file, as well as data computed by commands
executed prior to the copy command are also replicated. Each copy is
then independently processed through the subsequent commands issued to
\va{}. This command is useful, for example, for carrying out
simulations in which different signals are injected into the same
light curve to estimate the recovery efficiency, or different
pure-noise light curves with the same time sampling are simulated to
determine the bandwidth correction for period finding algorithms such
as {\bf -LS}.

\paragraph{\bf -Injectharm}\label{cmd:injectharm}
Adds to the light curve a harmonic series of the form 
\begin{align}
& A_{1}\cos(2\pi(t/P + \phi_{1})) \nonumber \\
& + \sum_{k=2}^{N_{\rm harm}+1}(A_{k} \cos(2\pi(kt/P + \phi_{k}))) \nonumber \\
& + \sum_{k=2}^{N_{\rm subharm}+1}(B_{k} \cos(2\pi(t/(kP) + \psi_{k})))
\label{eqn:injectharm}
\end{align}
where $P$ is the period of the signal, $A_{1}$, $A_{k}$, and $B_{k}$
are the amplitudes of the fundamental, harmonic, and subharmonic
terms, respectively, and $\phi_{1}$, $\phi_{k}$, and $\psi_{k}$ are
the phases of the fundamental, harmonic, and subharmonic terms,
respectively. The user indicates the number of desired harmonic and
subharmonic terms ($N_{\rm harm}$ and $N_{\rm subharm}$) and how each
of these are to be determined (their values may be fixed on the
command-line, read-in from the input list of light curves, or randomly
drawn from uniform or uniform-log distributions). The amplitudes and
phases of the harmonic and subharmonic terms may also be specified
relative to the amplitude and phase of the fundamental (i.e.~the input
is $R_{i1} = A_{i}/A_{1}$ rather than $A_{i}$, and/or $\phi_{k1} =
\phi_{k} - k\phi_{1}$ rather than $\phi_{k}$). This latter option is
useful if one wants to inject into the light curve a signal with a
fixed shape (e.g. a saw-tooth like pulsation), but random overall
amplitude and phase, which someone might want to do to determine the
completeness of a variability survey to a particular class of variable
stars. In this case one would fix the relative phases and amplitudes
of the harmonic and/or subharmonic terms, but allow the amplitude and
phase of the fundamental to vary.

\paragraph{\bf -Injecttransit}\label{cmd:injecttransit}
Adds a transit signal to the light curve. \va{}
uses the \citet{mandel:2002} model for the transit of a nonluminous
spherical object in front of a limb-darkened spherical star. The user
may specify the model parameters, including: the orbital period, the
radius of the planet, the mass of the planet, the phase of the transit
at time $T=0$, $\sin i$ ($i$ is the orbital inclination), the
eccentricity and argument of periastron or $e \cos \omega$ and $e \sin
\omega$, the mass of the star, the radius of the star, quadratic or
non-linear limb darkening coefficients for the star, and an optional
light dilution factor. Alternatively some of these parameters may be
drawn from random distributions. For the period this can be a uniform
distribution in $P$, $\log P$, $f = 1/P$ or $\log f$. For the planet
mass and/or radius this can be a uniform distribution or a log-uniform
distribution. For $\sin i$ it is a uniform distribution of orbit
orientations in space, subject to the constraint that there must be a
transit, which corresponds to $\cos i$ being drawn from a uniform
distribution between $0$ and $C$ with
\begin{equation}
C = \frac{R_{\rm star} + R_{\rm pl}}{a}\frac{1 + e \cos (\pi - \omega)}{1 - e^{2}}.
\end{equation}

\subsubsection{Light Curve Manipulation}\label{sec:lcmanip}

The following commands can be used to perform common light curve manipulation tasks: 

\paragraph{\bf -binlc}\label{cmd:binlc}
Bins the light curve in time. The user either specifies the number of
bins to use, or the size of the bins in units of the time coordinate,
and may also control the starting time of the first bin (by default it
is the time of the first observation in the light curve). The user may
either use the average, median or error-weighted average for the
resulting binned value. All vectors associated with this light curve
(i.e.~columns read-in from the light curve file including the
magnitudes), except for the time and uncertainty, vectors storing
string or character data, or any vectors explicitly indicated by the
user, are binned as follows:
\begin{equation}
\bar{x}_{i} = \begin{cases} \sum_{j} x_{j} / N_{i}, & \mbox{average} \\ {\rm med}({x_{j}}), & \mbox{median} \\ \sum_{j} (x_{j}\sigma_{j}^{-2})/\sum_{j} \sigma_{j}^{-2}, & \mbox{weighted average} \end{cases}
\end{equation}
while the uncertainties are binned following:
\begin{equation}
\bar{\sigma}_{i} = \begin{cases} \sqrt{\sum_{j} (\sigma_{j}^2) / N_{i}^2}, & \mbox{average} \\ 1.253\sqrt{\sum_{j}(\sigma_{j}^2) / N_{i}^2}, & \mbox{median} \\ \sqrt{(\sum_{j}(\sigma_{j}^{-2}))^{-1}}, & \mbox{weighted average} \end{cases}
\end{equation}
where $\bar{x}_{i}$ is the $i$th binned value of vector $x$, and
$\bar{\sigma}_{i}$ is the $i$th binned uncertainty, the sum on $j$ in
each case is over the unbinned points with times that fall within bin
$i$, $N_{i}$ is the number of points that contribute to bin $i$, and
the uncertainties are the standard errors on the binned quantities
assuming Gaussian uncorrelated noise. Bins for which $N_{i} = 0$ are
excluded from the binned light curve. The time reported for each bin
is either the time at the center of the bin, the average of the times
that fall within the bin (i.e.~$\bar{t}_{i} = \sum_{j}t_{j}/N_{i}$),
the median of the times that fall within the bin, or the user may
choose not to shrink the size of the light curve, in which case all
points within a bin are replaced by their binned value, but the light
curve times are not changed. The user may also optionally provide a
list of variables which will not be binned in the default manner. For
each variable in the list the user also indicates which statistic to
use in calculating the binned value. Options are the same as for the
{\bf -stats} command.

\paragraph{\bf -changeerror}\label{cmd:changeerror}
Sets the magnitude uncertainties $\sigma_{j}$
 in a light curve equal to the measured r.m.s.\ scatter of the light curve magnitudes.

\paragraph{\bf -converttime}\label{cmd:converttime}
Converts the time-system of the light curves between julian date (JD),
modified julian date (MJD$=$JD$-2400000.5$), helio-centric julian date
(HJD; i.e., JD approximately corrected to the center of the solar
system assuming observations are made at the Earth-Moon barycenter,
and that this position follows an elliptical orbit about the center of
the solar system with linear perturbations to the orbital elements),
and bary-centric julian date (BJD; i.e., time corrected to the center
of the solar system using the NASA JPL ephemeris to determine the
location of the observer relative to the barycenter at a given time);
constants may be added to or subtracted from the input and/or output
times. 

Time conversions between a UTC-based system (i.e., JD values
having been calculated directly from UTC without accounting for
leap-seconds; this is not formally correct, but a very common
practice) and a TDB-based system (barycentric dynamical time, i.e.~the
UTC time is corrected to a leap-second-free time system before
converting to JD) are also supported. In the year 2015 there is
approximately a one minute difference between the two systems
\citep[see the discussion in][]{eastman:2010}. 

Conversions to/from BJD
and between UTC and TDB use the NASA JPL NAIF cspice library
\citep{acton:1996} following the expressions given by
\citet{eastman:2010}; we do not allow for Shapiro or Einstein
time-delays in converting to BJD (i.e., gravitational redshifts are
ignored). 

Differences between BJD and HJD are due mostly to the orbits
of Jupiter and Saturn (which are ignored for HJD), and can be as large
as $4.2$ seconds for observations made between 1980 and 2020. 

The user
may specify an observatory where the data were collected, or the
latitude, longitude, and altitude. If this is not done then the center
of the Earth is assumed, which can introduce up to a 21 millisecond
error in the BJD correction. The procedure optionally accounts for
proper motion and/or coordinate precession. 

Based on comparing time
conversions made with \va{} to those computed through the JPL HORIZONS
interface\footnote{\url{http://ssd.jpl.nasa.gov/horizons.cgi}} we
conclude that this command has an internal precision of $\sim 0.1$
milliseconds for conversions near J2000.0. At present this command
does not support time conversions for observations made off of the
surface of the Earth.

\paragraph{\bf -difffluxtomag}\label{cmd:difffluxtomag}
Converts the light curves from differential flux
units (which are output, for example, by the popular ISIS image
subtraction procedure; \citealp{alard:1998}) into magnitudes. The expression for converting from differential flux $df$ to magnitude $m$ is given by:
\begin{equation}
m = m_{0}+\Delta_{m} - 2.5*\log_{10}(f_{\rm ref} - df)
\end{equation}
where $m_{0}$ is the magnitude of a source yielding a flux of 1, $\Delta_{m}$ is an offset that the user may give to switch to a different magnitude system, and the reference flux $f_{\rm ref}$ is determined from the input reference magnitude $m_{\rm ref}$ via:
\begin{equation}
f_{\rm ref} = 10^{-0.4(m_{\rm ref} - m_{0})}.
\end{equation}

\paragraph{\bf -ensemblerescalesig}\label{cmd:ensemblerescalesig}
Transforms the magnitude uncertainties of
the light curves by the expression
\begin{equation}
\label{eqn:ensemblerescaletrans}
\sigma_{i,j}^{\prime} = \sqrt{a\sigma_{i,j}^{2}+b}.
\end{equation}
Where $\sigma_{i,j}^{\prime}$ is the new uncertainty for light curve
$i$, observation $j$, after the transformation. Here $a$ and $b$ are
determined by fitting a linear relation of the form:
\begin{equation}
y_{i} = ax_{i} + b
\label{eqn:ensemblerescalefitrelation}
\end{equation}
where 
\begin{equation}
y_{i} = (\chi_{i}^{2}/{\rm dof})\bar{RMS_{i}}^2
\label{eqn:ensemblerescalefitrelationyi}
\end{equation}
 for light curve
$i$, and
\begin{equation}
x_{i} = \bar{RMS_{i}}^2
\label{eqn:ensemblerescalefitrelationxi}
\end{equation}
for light curve $i$, and $\bar{RMS}$ is the
expected RMS of the light curve (equation~\ref{eqn:expectedrms}). The
result of this transformation is that, for many datasets, the
$\chi_{i}^{2}/{\rm dof}$ values are distributed about unity. The
justification for this is given in
\ref{sec:ensemblerescaletransjust}. The parameter $a$ is related to the ratio of the gain and
the effective gain, while $b$ is related to a constant magnitude
error-term \citep[e.g.][]{hartman:2005:ngc6791}. This routine requires
all of the light curves to be read-in simultaneously, and thus is
incompatible with the {\bf -parallel} option in the current version of
\va{}.

\paragraph{\bf -expr}\label{cmd:expr}
Sets a variable equal to an analytic expression. For example, one
might give the command ``-expr mag=mag/2'' which evaluates ${\rm
  mag}/2$ at each point in the light curve and replaces the magnitude
with the result. If the variable on the left-hand-side of the equality
has not previously been defined, it will be created. Variables which
appear on the right-hand-side can be the name of a vector which is
read-in from the light curve, a scalar or vector created by another
command (e.g.~the fitting parameters, or the output model vector,
created by the {\bf -linfit} command), or any output parameter from a
previously executed command (such as the light curve RMS computed with
the {\bf -rms} command). \va{} uses an implementation of the
precedence climbing algorithm \citep{richards:1979} for parsing and
evaluating analytic expressions. The list of recognized functions is
provided in \ref{sec:reservednames}, while a method for users to
define their own functions which may then be used in such expressions
is discussed in Section~\ref{sec:userfunc}.

\paragraph{\bf -fluxtomag}\label{cmd:fluxtomag}
Converts a light curve from fluxes into
magnitudes. The uncertainties are also converted from flux
uncertainties into magnitude uncertainties.

\paragraph{\bf -Phase}\label{cmd:phase}
Replaces the time coordinate of a light curve with its
phase (from $0$ to $1$) and sorts the light curve by the phase. The
user must specify the period to use for phasing the light curve, this
may either be fixed on the command line for all light curves, read-in
from the input light curve list, or taken from a previous command
(e.g.~the period found by {\bf -LS}). The user may also optionally
control the reference time for phase $0$.

\paragraph{\bf -resample}\label{cmd:resample}
Resamples the light curve onto a new time
base. Several methods are allowed for interpolating or extrapolating
the data, including setting resampled points to the value of the
observation that is closest in time, performing linear interpolation
between points, cubic spline interpolation \citep[e.g.,][]{press:1992}, cubic spline interpolation with a
constraint that the interpolating function be monotonic between points
\citep{steffen:1990}, and Basis-spline interpolation (implemented
using GSL routines). The user may optionally adopt different interpolation methods for resampled points that are close to observations and those that are far from observations.

\paragraph{\bf -rescalesig}\label{cmd:rescalesig}
Rescales the magnitude uncertainties for a light
curve (i.e.~$\sigma = \alpha\sigma$) such that $\chi^2 / {\rm dof} =
1$ for that light curve.

\subsubsection{Control of Data Flow}\label{sec:controlflow}

The following commands control the flow of data through the pipeline:

\paragraph{\bf -changevariable}\label{cmd:changevariable}
Changes the variable used for the time, magnitude, magnitude
uncertainty, or image identifier in subsequent commands. For example,
the command ``-changevariable mag mag2'' would cause subsequent
commands to \va{} to operate on the data stored in the variable $mag2$
for the light curve magnitudes. The data stored in the original
variable $mag$ remains unchanged, and may be used in analytic
expressions.

\paragraph{\bf -if, -elif, -else, -fi}\label{cmd:if}
Can be used to make
 execution of commands conditional upon the evaluation of expressions,
 following the typical logic of ``if'', ``else if'', ``else'' and
 ``end if''-type conditional statements. The use of the terms {\bf
   -elif} and {\bf -fi} are as in the Bourne-Again SHell. \va{}
 supports nested conditionals, but all conditional constructs are
 ignored by commands which process all light curves simultaneously
 (e.g.~{\bf -SYSREM}) as well as by the {\bf -savelc} and {\bf
   -restorelc} commands.

\paragraph{\bf -o}\label{cmd:o}
Outputs the light curve in its present state to a file (or to {\sc
  stdout} if specified by the user). The user can specify a rule for
naming the output file, and can also control the format of the output
data. By default the light curve will be output as an ascii text file,
but may also be output as a binary FITS table.

\paragraph{{\bf -restorelc} and {\bf -savelc}}\label{cmd:restorelc}
Are used to save a light curve
(including all associated vectors) and later restore it to this
previously saved state.

\subsubsection{Miscellaneous Commands}\label{sec:misc}

The commands listed here do not fall under the other categories.

\paragraph{\bf -findblends}\label{cmd:findblends}
This command may be used to determine if a signal detected in a given
light curve is likely due to blending with a nearby variable
source. The command follows the procedure described by
\citet{hartman:2011:kmdwarf}. Each light curve in the input list is
spatially matched to another list of light curves (by default the
input light curve list is matched to itself, but the user may
optionally provide a different list for the matching). A harmonic
series with a specified period (either read-in from the input list
file, fixed on the command line, or set equal to the output from a
previously executed command) is then fit to both the input light
curve, and to all matching light curves. The peak-to-peak amplitude of
the fitted harmonic series is then determined for each source and
converted to flux units (by default the routine assumes the input is
in magnitudes).  The light curve with the highest flux amplitude is
reported. If this is not the same as the input light curve, then the
neighboring object is the most likely source of the
variability.

\subsection{User-Developed Extensions}\label{sec:userlib}

While \va{} includes a wide-range of built-in processing routines, it
falls far short of encompassing the full range of routines that one
might wish to apply to a light curve. It is therefore essential to
enable a user to incorporate his/her own light curve processing
algorithms within the program. To that end two mechanisms have been
included: one for dynamically loading at run-time a compiled library
defining a user-developed command; and another for dynamically loading
a library defining a user-developed function which can be used in
analytic expressions on the command-line. Each of these are described in turn.

\subsubsection{User-Developed Commands}\label{sec:usercommand}

Each library implementing a user-developed command must contain five
functions, with standardized names and calling syntaxes, which are
used to interface the command with \va{}. These functions perform such
tasks as initializing variables defining standard properties of the
command, parsing the command-line when the user calls the command,
displaying the syntax and help information for the command, and
executing the command on a light curve. These required functions
should be written in $C$, while the algorithm itself may be written in
any language callable from $C$. Templates are provided of the
functions required in each library, together with a Makefile for
compiling and linking the code into a dynamically loadable shared
object library. A template is also provided illustrating how to
include a processing algorithm written in $FORTRAN$ in \va{}. 

\va{} uses the GNU Libtool
package\footnote{\url{http://www.gnu.org/software/libtool}} for
compiling and dynamically loading libraries in a system-independent
fashion. When \va{} encounters an unrecognized term on the
command-line, it will search for an appropriate library in a data
directory defined during the installation of \va{}, load it if
available, and execute the command. So long as the libraries are
installed in the correct directory, they may be called by the user in
the same way that all other commands are executed. To avoid performing
a directory search, the user may also use the {\bf -L} option to
explicitly load a library file.

A number of example libraries are distributed with \va{} which provide additional commands. These include:

\paragraph{\bf -fastchi2}\label{cmd:fastchi2}
Calculate the Fast $\chi^{2}$ periodogram following
\citet{palmer:2009}. This routine scans through a range of frequencies
fitting a harmonic series to the data at each trial frequency and
reporting the minimum $\chi^2$. It uses the fast Fourier transform to
do the fit quickly in $\mathcal{O}(N \log N)$ time. The user specifies
the number of harmonics to include in the series, the minimum and
maximum frequencies to search, the order and reference time of a
polynomial in time which is removed from the light curve before
performing the search, the time-span to use in computing the Nyquist
frequency (if not calculating it automatically), an over-sampling
factor for determining the resolution of the periodogram, and a
tolerance parameter used in performing a higher resolution search
around the peaks in the periodogram.

\paragraph{\bf -jktebop}\label{cmd:jktebop}
Fit or inject a detached eclipsing binary light curve
model [in]to a light curve. This command uses the code from the
JKTEBOP program due to \citet{southworth:2004}, converted into a
callable library. The JKTEBOP code is in turn based on the Eclipsing
Binary Orbit Program \citep{popper:1981,etzel:1981} and implements the
model of \citet{nelson:1972}. The JKTEBOP-based library is written in
$FORTRAN$, so this command can be used as a template for incorporating
a $FORTRAN$ algorithm into \va{}.  The basic parameters for the model
include the orbital period of the binary, a reference primary eclipse
epoch, the normalized sum of the component radii ($(R_{1}+R_{2})/a$),
the ratio of the radii ($R_{2}/R_{1}$), the mass ratio
($M_{2}/M_{1}$), the surface brightness ratio ($J_{2}/J_{1}$), the
orbital inclination or normalized impact parameter, the eccentricity
parameters $e\sin \omega$ and $e\cos\omega$, limb darkening
coeffections, gravity darkening coeffections, reflection effect
coeffections, third light, and a tidal lead/lag angle. The fit is
performed using DHSX.

\paragraph{\bf -macula}\label{cmd:macula}
Fit or inject a Macula spot model [in]to a light
curve. The model and $FORTRAN$ code implementing it are both due to
\citet{kipping:2012:macula}. This model allows for an arbitrary number
of starspots which are allowed to evolve over time with a linear
growth/decay law. The parameters in the model include: the equatorial
rotation period; the inclination of the star; quadratic and quartic
differential rotation coefficients; four limb-darkening coefficients
for the stellar photosphere and four for the spots; and a blend
parameter. Additionally each spot is parameterized by: its longitude
at the time of maximum size; its latitude at the time of maximum size;
its maximum angular size; the spot-to-star flux ratio; the time of
maximum size; the lifetime; the
duration of ingress (time to appear); and the duration of egress (time
to disappear). The model can be fit either with DHSX, or with the
Levenberg-Marquardt algorithm \citep{press:1992}. The latter makes use
of partial derivatives that are provided by
\citet{kipping:2012:macula}.

\paragraph{\bf -magadd}\label{cmd:magadd}
Adds a constant to a light curve. This is provided as a
trivial example of a user-developed command, and may be used as a
template.

\subsubsection{User-Developed Functions}\label{sec:userfunc}

The mechanism for incorporating a user's own functions into \va{} is
similar to the mechanism for incorporating user-developed
commands. Again the user creates a dynamically loadable library
written in $C$. The library must contain an initialization function
with a standardized name which is called when the library is
opened. This function is used to indicate the names of all new
analytic functions supplied by the library, the number of
double-precision arguments expected by each function (at present all
functions input and output double-precision variables), and pointers
to the $C$ functions in the library (i.e., variables storing their
location in memory) which execute the computations associated with
each analytic function. These $C$ functions are expected to take a
vector of double-precision numbers, with each component in the vector
representing a separate argument to the function, and to output a
single double-precision quantity as the result. The functions may in
turn be wrappers to routines written in other languages callable from
$C$.

In order to use the functions provided by a library, the user must
provide the {\bf -F} option on the command-line, followed by the name
of the library (excluding filename extensions). An example library is
included with the \va{} distribution, which may be used as a template.

\subsection{Options}\label{sec:options}

In addition to the commands mentioned above, a number of command-line
options are available to change the behavior of \va{}. These options
may be given at any point on the command-line (except for within the
list of parameters supplied to a command). The available options
include:

{\bf -basename}: only the base filename of each light curve is
included in the output table, rather than the full path.

{\bf -bufferlines}: use this to adjust the number of lines buffered
within \va{} before being written to standard out during parallel
processing. This buffering is separate from the system buffering done
on stdout itself, which may be turned off with the {\bf-nobuffer}
option. The {\bf-bufferlines} option only has an effect if the
{\bf-parallel} option is used, and will help speed up the processing
by preventing threads from waiting on each other to output
results. The tradeoff with using a larger buffer is that more memory
is used, and results will be output less frequently.

{\bf -example}: used in conjunction with the name of a command to print out an example of how that command is used.

{\bf -functionlist}: shows the list of function names that are
recognized in analytic expressions.

{\bf -header}: include column headers in the output ascii table.

{\bf -headeronly}: prints the output header that a call to \va{} would
produce and exits without processing any light curves. This is useful for finding the column names or numbers to use when values computed by one command are used to set parameter values for subsequent commands.

{\bf -help}: shows detailed help for a command.

{\bf -inlistvars}: used to specify a format for the input light curve
list, and store columns from this list as vector variables.

{\bf -inputlcformat}: used to specify the format of the input light
curves. One may use this option to read-in in more than just the
default JD, magnitude, and magnitude uncertainty values from the light
curve.

{\bf -jdtol}: sets the tolerance for considering two observations to
have come from the same epoch. The default tolerance is
$10^{-5}$\,days. This is used by commands such as {\bf -TFA} or {\bf
  -SYSREM} which match points from different light curves.

{\bf -quiet}: process the light curves, but do not output the ascii
table of statistics.

{\bf -L}: load a user-compiled library defining a new \va{} command.

{\bf -listcommands}: shows a terse list of the available commands.

{\bf -log-command-line}: includes the command-line in the header of
the output ascii table.

{\bf -matchstringid}: use image string identifiers, rather than the
time, to match points from different light curves.

{\bf -nobuffer}: outputs each line in the ascii table as soon as it is
generated, rather than buffering the output.

{\bf -noskipempty}: by default empty light curves are skipped and not
included in the output table. To not skip these give this option. This
option has no effect if the {\bf -readall} option is used.

{\bf -numbercolumns}: pre-pend column numbers to the header names in
the output ascii table.

{\bf -oneline}: display each computed statistic on a separate line for
each light curve, rather than in a table format.

{\bf -parallel}: process multiple light curves in parallel.

{\bf -randseed}: specifies a seed for the random number generator. If
this is not used, then every call to \va{} will produce the same set
of random numbers. The system time may be used as a seed to produce
quasi-random numbers.

{\bf -readall}: read-in all of the light curves at once, rather than
as they are processed.

{\bf -readformat}: This option is deprecated in the current version of
\va{}, users are encouraged to use the {\bf -inputlcformat} option
instead. It is provided to allow compatibility for older processing scripts.

{\bf -redirectstats}: output the ascii table of statistics to a file
rather than to stdout.

{\bf -showinputlcformat}: displays the expected format of the input
light curves, given a call to \va{}. No processing will occur if this
option is given, it is provided for assistance in developing scripts
using \va{}.

{\bf -showinputlistformat}: displays the expected format of the input
list of light curves, given a call to \va{}. No processing will occur if this
option is given, it is provided for assistance in developing scripts
using \va{}.

{\bf -skipmissing}: do not abort if a missing or unreadable light
curve file is encountered. Instead skip the light curve and proceed
with others in the list.

{\bf -tab}: outputs the ascii table as a starbase table (tab-delimited
columns).

\section{Examples}\label{sec:examples}

\subsection{Searching a set of light curves for periodic variable stars with Lomb-Scargle}\label{sec:example1}

For the first example, suppose we have a collection of light curves
that we wish to search for periodicity using the L-S method. Further, suppose
each light curve is stored in a separate file with the format shown in listing~\ref{lst:example1lcformat}, and that we created a list of light curves as shown in listing~\ref{lst:example1listformat}. These light curves may be searched for periodic signals using the Generalized Lomb-Scargle periodogram as shown in listing~\ref{lst:example1GLS}.

\begin{lstlisting}[caption={light curve format for example~\ref{sec:example1}},label={lst:example1lcformat},frame=single,backgroundcolor=\color{lightgray},language=vartools]
> cat 1.txt

# Time[BJD]  Mag    Err
53725.173920 10.44080 0.00136
53725.176540 10.43881 0.00166
53725.177720 10.44024 0.00145
...
\end{lstlisting}

\begin{lstlisting}[caption={light curve list format for example~\ref{sec:example1}},label={lst:example1listformat},frame=single,backgroundcolor=\color{lightgray},language=vartools]
> cat lclist.txt

# FileName
1.txt
2.txt
3.txt
...
\end{lstlisting}

\begin{lstlisting}[caption={running the Generalized Lomb-Scargle algorithm on a list of light curves as discussed in example~\ref{sec:example1}},label={lst:example1GLS},frame=single,backgroundcolor=\color{lightgray},language=vartools,float=*]
prompt> vartools -l lclist.txt \
    -LS 0.01 100.0 0.1 1 0 \
    -header -numbercolumns

@#1_Name 2_LS_Period_1_0 3_Log10_LS_Prob_1_0 4_LS_Periodogram_Value_1_0 5_LS_SNR_1_0@
@1.txt    77.76775250 -5709.91013    0.99392  386.76802@
@2.txt     1.23440877 -3999.59411    0.99619  558.03142@
@3.txt    18.29829471  -25.09202    0.03822   38.31823@
...
\end{lstlisting}

Here we search for periods between 0.01 days and 100.0 days sampling
the periodogram for each light curve using frequency steps of $0.1/T$,
where $T$ is the total time spanned by the light curve. We only output
one peak, which is the highest one found, and we do not store the
periodograms.  The options ``-header'' and ``-numbercolumns'' are used
to provide the header given in the output, and to also prepend the
column number to the name of each column in the header.

\begin{figure*}[]
\includegraphics[scale=0.7]{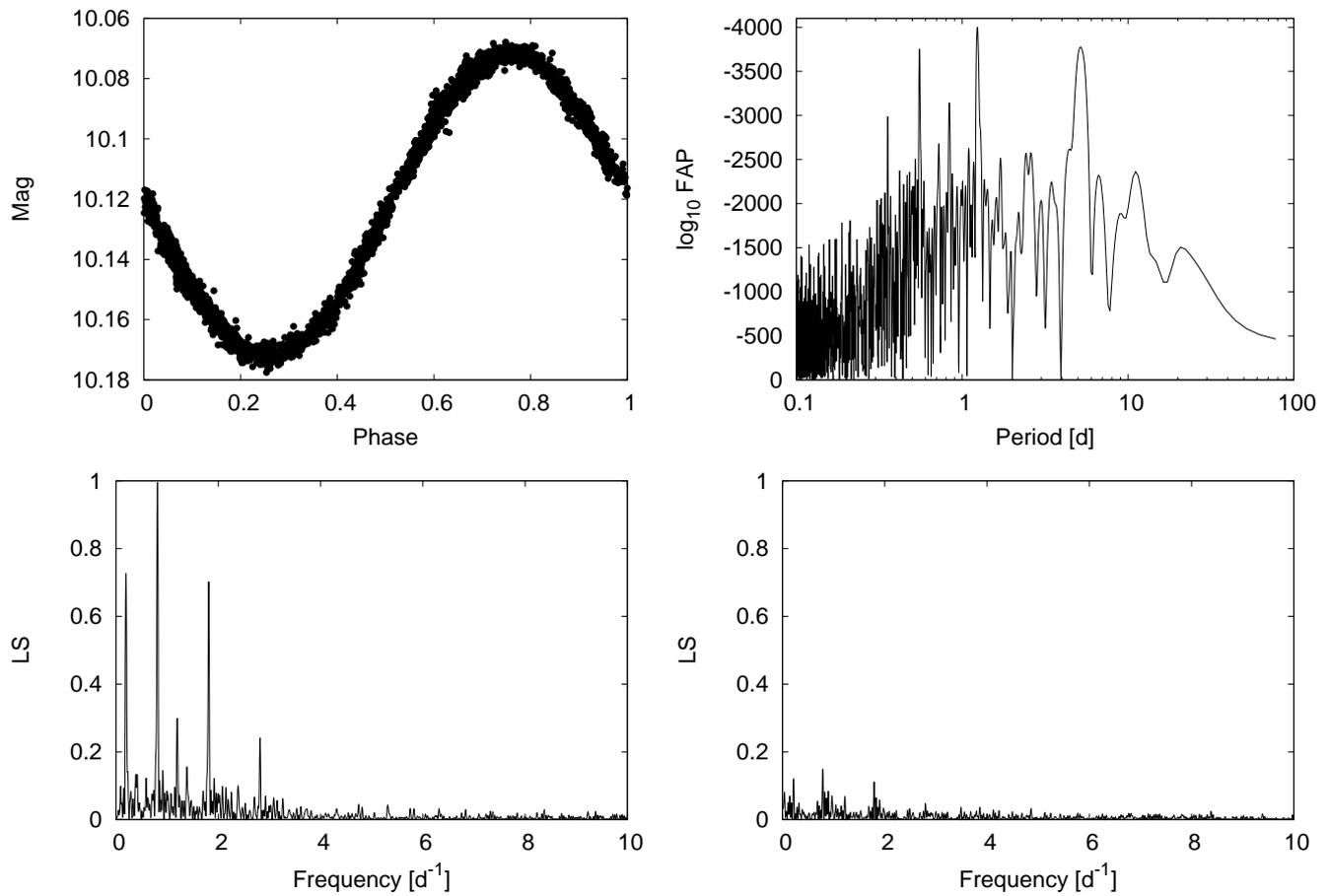}
\caption[]{ Periodogram computed in example~\ref{sec:example1} (listing~\ref{lst:example1GLSwithPER}) for light curve
  ``2.txt''. The upper left panel shows the light curve phase-folded
  at the period 1.2344\,d found by the {\bf -LS} command. The
  upper right panel shows the periodogram using period on the
  horizontal axis and the logarithm of the false alarm probability on
  the vertical axis. This same periodogram is plotted again in the
  lower left panel, this time using frequency as the horizontal axis,
  and the LS statistic (equation~\ref{eqn:LSdef}) on the vertical
  axis. The lower right panel shows the periodogram after one
  whitening cycle.
\label{fig:Periodogram1}
}
\end{figure*}

As seen in the output the first light curve has a periodic signal of
$77.7776$\,days, detected with false alarm probability
$10^{-5710}$ and spectroscopic S/N $387$.  In fact, this particular
example light curve has a strong linear trend which is well-fit by a
periodic signal with a period longer than the time-span of the
data. The second light curve also has a periodic signal, this time
with a period of $1.2344$\,days, detected with false alarm probability
$10^{-3999.6}$. This example light curve has a strong sinusoidal
signal (Figure~\ref{fig:Periodogram1}).

Listing~\ref{lst:example1GLSwithPER} shows a command which will also output the periodograms for each light
curve to the directory ``PER'' and identify 5 peaks in the
periodogram, in each case whitening the light curve (i.e., fitting and
removing a harmonic function from the light curve) at the previously
identified peak before finding the next one. For clarity we truncate
the output shown below, the actual output contains 21 columns, with
the period, false alarm probability, periodogram value, and S/N
reported for each of the five peaks. Figure~\ref{fig:Periodogram1} shows an example of the periodogram
output for the second light curve, while the format of the periodogram file produced by
\va{} is shown in listing~\ref{lst:example1PERformat}.

\begin{lstlisting}[caption={running the Generalized Lomb-Scargle algorithm on a list of light curves, while outputing the periodogram files, and identifying 5 peaks with whitening, as discussed in example~\ref{sec:example1}},label={lst:example1GLSwithPER},frame=single,backgroundcolor=\color{lightgray},language=vartools,float=*]
prompt> vartools -l lclist.txt \
    -LS 0.01 100.0 0.1 5 1 PER whiten \
    -header -numbercolumns

@#1_Name 2_LS_Period_1_0 3_Log10_LS_Prob_1_0 4_LS_Periodogram_Value_1_0@ \
@ 5_LS_SNR_1_0 6_LS_Period_2_0@ ...
@1.txt    77.76775250 -5709.91013    0.99392  683.48184    22.21935786@ ...
@2.txt     1.23440877 -3999.59411    0.99619  825.42315     0.55747493@ ...
@3.txt    18.29829471  -25.09202    0.03822   38.97469     1.15639781@ ...
...
\end{lstlisting}

\begin{lstlisting}[caption={Format of a periodogram file produced by the command in listing~\ref{lst:example1GLSwithPER}. We truncate the initial header (columns 6 through 11 repeat
P(omega) and the logarithm of the FAP for whitening cycles 2 through
4).},label={lst:example1PERformat},frame=single,backgroundcolor=\color{lightgray},language=vartools,float=*]
prompt> cat PER/2.txt.ls

@# Column 1 = Frequency in cycles per input light curve time unit.@
@# Column 2 = Unnormalized P(omega) (equation 5 of Zechmeister &@
@#            K\"urster 2009, A&A, 496, 577). Whitening Cycle 0.@
@# Column 3 = Logarithm of the false alarm probability.@
@#             Whitening Cycle 0.@
@# Column 4 = Unnormalized P(omega) (equation 5 of Zechmeister &@
@#            K\"urster 2009, A&A, 496, 577). Whitening Cycle 1.@
@# Column 5 = Logarithm of the false alarm probability.@
@#             Whitening Cycle 1.@
...
@0.012858800310578017 0.0034929037863222533 -463.7738154443843@ \
@0.03192079715170023 -22.681157234188063 0.025954476868188435@ \
@-16.539118701220865 0.032410425800583273 -20.616443696388661@ \
@0.0022737099428519927 0@
...
\end{lstlisting}

\subsection{Processing a FITS-format light curve from {\em Kepler}}\label{sec:exampleKepler}

\begin{lstlisting}[caption={Processing a light curve from the {\em Kepler} mission, as discussed in example~\ref{sec:exampleKepler}},label={lst:exampleKepler},frame=single,backgroundcolor=\color{lightgray},language=vartools,float=*]
prompt> vartools -i kplr001429092-2009166043257_llc.fits \
             -inputlcformat t:1,pdcsap_flux:8,pdcsap_flux_err:9 \
             -changevariable mag pdcsap_flux \
             -changevariable err pdcsap_flux_err \
             -fluxtomag 25.0 0 \
             -rms \
             -LS 0.1 30. 0.1 4 0 \
             -o tmp.lc columnformat t,pdcsap_flux:%.10f,pdcsap_flux_err:%.10f \
             -oneline

@Mean_Mag_3               =  14.48415@
@RMS_3                    =   0.00346@
@Expected_RMS_3           =   0.00037@
@Npoints_3                =  1624@
@LS_Period_1_4            =     8.16381396@
@Log10_LS_Prob_1_4        = -264.61584@
@LS_Periodogram_Value_1_4 =    0.53223@
@LS_SNR_1_4               = 7938.63608@
@LS_Period_2_4            =     4.46288496@
@Log10_LS_Prob_2_4        = -120.94203@
@LS_Periodogram_Value_2_4 =    0.19710@
@LS_SNR_2_4               = 2939.26680@
@LS_Period_3_4            =     3.84731462@
@Log10_LS_Prob_3_4        =  -11.73784@
@LS_Periodogram_Value_3_4 =    0.01976@
@LS_SNR_3_4               =  293.78100@
@LS_Period_4_4            =     2.88548597@
@Log10_LS_Prob_4_4        =   -2.35786@
@LS_Periodogram_Value_4_4 =    0.00694@
@LS_SNR_4_4               =  102.53805@

prompt> head -3 tmp.lc
131.51272409615922 14.4798143174 0.0003752172
131.53315879162255 14.4799784985 0.0003757022
131.55359338685957 14.4793099117 0.0003756885
\end{lstlisting}

Listing~\ref{lst:exampleKepler} shows an example of processing the
quarter~1 binary FITS-format light curve from the NASA {\em Kepler}
mission \citep{borucki:2010} for the star KIC~1429092\footnote{This
  light curve may be downloaded from \url{https://archive.stsci.edu/kepler/publiclightcurves.html}}. The {\bf
  -inputlcformat} option indicates the table columns to use. The first
column is $BJD-2454833$, the eigth column is the PDC-corrected simple
aperture photometry flux, and the ninth column is the uncertainty on
this flux. We use the {\bf -changevariable} commands to tell \va{} to
use the $pdcsap\_flux$ variable for the light curve brightness
measurements, and the $pdcsap\_flux\_err$ variable for the
uncertainties. Of course in this example we could have read columns 8
and 9 directly into the $mag$ and $err$ variables, but it can be
helpful to keep track of what quantities are actually being input when
there are different possibilities. The {\bf -fluxtomag} command
converts the flux into magnitudes. We then calculate the
r.m.s.\ scatter of the magnitudes and the L-S periodogram. The
magnitude-converted light curve is output to the file ``tmp.lc'' using
the specified format. Lines~32 through 35 of
listing~\ref{lst:exampleKepler} show the format of the output light
curve.

\subsection{Running the BLS algorithm on a light curve to find transits and eclipses}\label{sec:example2}

\begin{lstlisting}[caption={Running the BLS algorithm on a light curve, as discussed in example~\ref{sec:example2}},label={lst:example2BLS},frame=single,backgroundcolor=\color{lightgray},language=vartools,float=*]
prompt> vartools -i EXAMPLES/3.transit \
                 -inputlcformat t:1,mag:2,err:3 \
                 -BLS q 0.002 0.05 0.5 100. 100000 500 0 5 1 PER \
                      1 MODEL 0 fittrap \
                      nobinnedrms ophcurve MODEL -0.5 0.5 0.001 \
                 -oneline

@Name                         = EXAMPLES/3.transit@
@BLS_Period_1_0               =     2.12316764@
@BLS_Tc_1_0                   = 53727.297790176432@
@BLS_SN_1_0                   =   4.42764@
@BLS_SR_1_0                   =   0.00238@
@BLS_SDE_1_0                  =   4.36608@
@BLS_Depth_1_0                =   0.01229@
@BLS_Qtran_1_0                =   0.03609@
@BLS_Qingress_1_0             =   0.20215@
@BLS_OOTmag_1_0               =  10.16687@
@BLS_i1_1_0                   =   0.98229@
@BLS_i2_1_0                   =   1.01838@
@BLS_deltaChi2_1_0            = -24267.20922@
@BLS_fraconenight_1_0         =   0.43182@
@BLS_Npointsintransit_1_0     =   166@
@BLS_Ntransits_1_0            =     4@
@BLS_Npointsbeforetransit_1_0 =   128@
@BLS_Npointsaftertransit_1_0  =   145@
@BLS_Rednoise_1_0             =   0.00151@
@BLS_Whitenoise_1_0           =   0.00489@
@BLS_SignaltoPinknoise_1_0    =  14.54094@
@BLS_Period_2_0               =     1.85908563@
...
\end{lstlisting}

Listing~\ref{lst:example2BLS} shows an example of running the BLS
algorithm on a light curve containing a transit signal. This example
light curve is included with the \va{} distribution, and is produced
in a manner similar to that described below in
example~\ref{sec:example3}.  

The parameters are as follows: we
consider fractional transit durations between 0.002 and 0.05 in phase,
search for periods between 0.5 days and 100 days, use 100,000
frequency steps and 500 phase bins. We use 0 for the timezone (this is
used to assign points to an individual night in calculating the
fraction of $\Delta \chi^2$ that comes from one night), and report the
top five peaks in the periodogram. We output the periodograms to the
directory PER, output the best-fit transit model to the directory
MODEL, and we do not subtract the model from the light curve. A
trapezoid-shaped transit is fit to each peak to refine the transit
parameters compared to those found in the full frequency scan. We use
the ``nobinnedrms'' keyword to use $\bar{\rm SR}$ and $\sigma_{\rm
  SR}$ in equation~\ref{eqn:BLSSN} for the output BLS\_SN columns,
rather than $\bar{\tilde{\rm SR}}(f)$ and $\sigma_{\tilde{\rm SR}}$
(see the discussion in section~\ref{sec:BLS}), and we output a
uniformly sampled phase curve for the best-fit transit model to the
directory MODEL, to use for plotting. 

\begin{figure}[]
\includegraphics[scale=0.7]{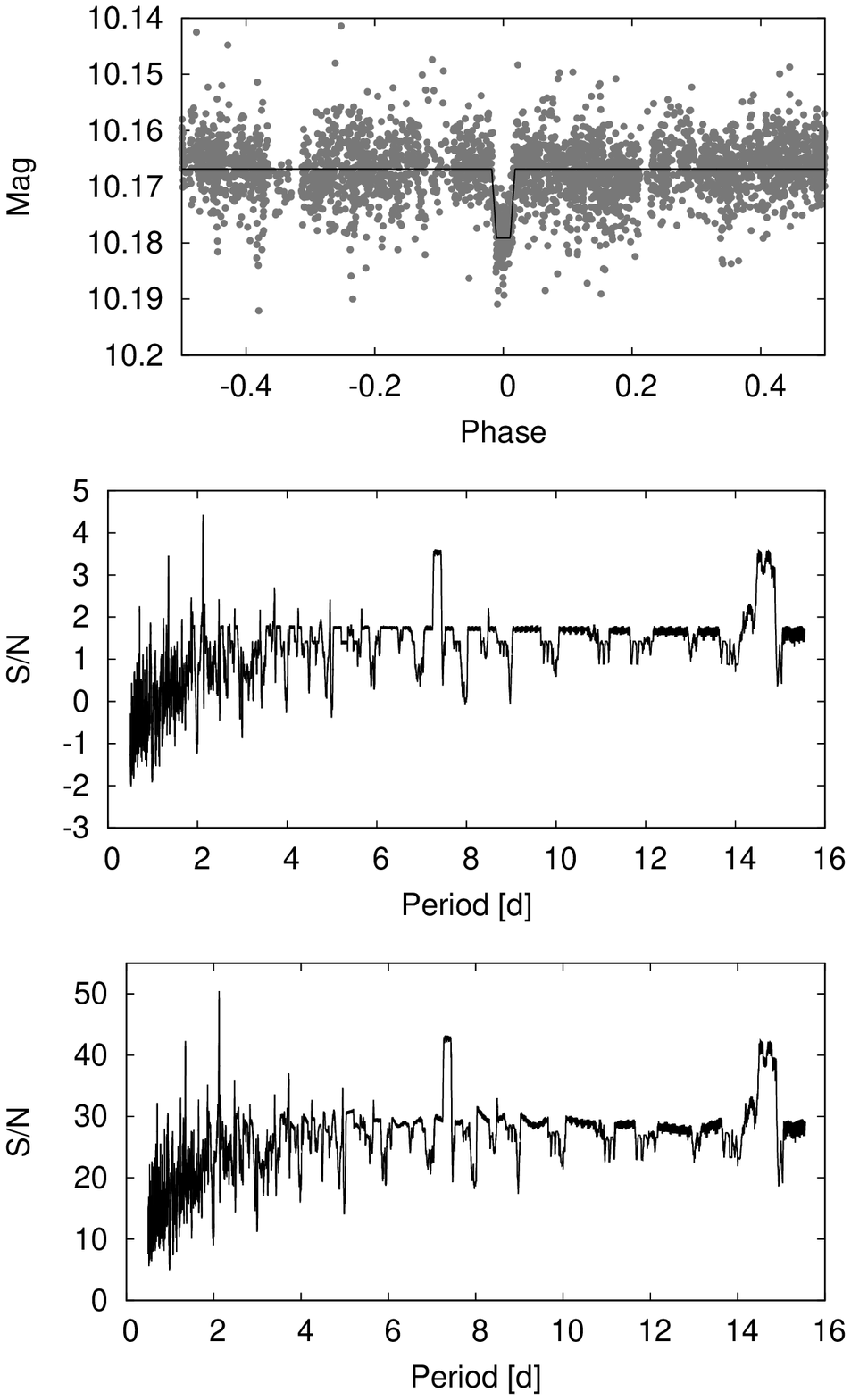}
\caption[]{ Phase-folded light curve and BLS spectra for
  example~\ref{sec:example2} (listing~\ref{lst:example2BLS}). Top:
  Phase-folded light curve as output by the {\bf -BLS} command,
  together with the model phase curve (output using the ``ophcurve''
  keyword). Middle: The BLS spectrum using the spectroscopic S/N when
  the ``nobinnedrms'' keyword is given (i.e., this is equivalent to
  SR, with the average value subtracted, and divided by the standard
  deviation). Bottom: The BLS spectrum using the spectroscopic S/N
  when the ``nobinnedrms'' keyword is not given
  (Eq.~\ref{eqn:BLSSN}). The primary difference is a change in scale
  due the difference between $\bar{\rm SR}$ (Eq.~\ref{eqn:BLSSR}) and
  $\bar{\tilde{\rm SR}}$ (Eq.~\ref{eqn:BLSSRtilde}).
\label{fig:BLS2}
}
\end{figure}

The results are shown in
figure~\ref{fig:BLS2}. The transit is recovered with a period of 2.1232
days, with a spectroscopic S/N of 4.4, and with ${\rm S}/{\rm N}_{\rm
  pink} = 14.5$ (Eq.~\ref{eqn:BLSSNpink}). We note that the latter is
probably the most useful of the metrics calculated here for selecting
transit candidates.

\subsection{Simulating and Recovering a Transiting Planet Light Curve}\label{sec:example3}

In this example we simulate a light curve with red-noise, add a
transit signal to it, recover the transit with {\bf -BLS}, and then
fit a \citet{mandel:2002} transit model to the light curve. A similar
procedure might be repeated millions of times to determine the
detection efficiency of a transit survey. The commands are shown in
listing~\ref{lst:example3simtransit}. 

In this case only the times and
magnitude uncertainties are read-in from the light curve file, which
we use to define the time-base for our simulated observations, and we
then initialize all magnitudes to a value of $10$ using the {\bf
  -expr} command. 

We use the {\bf -addnoise} command to add
time-correlated noise to the light curve, adopting the
exponential noise model (Eq.~\ref{eqn:expnoise}) with $\rho = 0.05$
days, $\sigma_{r} = 0.0015$ mag, and $\sigma_{w} = 0.0015$ mag. The
{\bf -changerror} command is called to set the magnitude uncertainties
equal to the r.m.s.\ scatter of the light curve, and we use the {\bf
  -o} command to output the simulated light curve to the file
``MODEL/1.simnoise''. 

Next comes the {\bf -Injecttransit} command
which is used to add the transit signal to the light curve. Here we
use the ``fix'' version of each parameter to specify the value to use
on the command line, if one were to carry out a large transit
injection/recovery simulation with \va{}, then the ``list'' or
``rand'' versions may be used. The injected signal has a period of
$2.123456$ days, a planet radius of 1.0\,$R_{J}$, a planet mass of
1.0\,$M_{J}$, a random phase, a random inclination, $e$ and $\omega$
fixed to zero, a stellar mass of 1.0\,$M_{\odot}$, a stellar radius of
1.0\,$R_{\odot}$, and quadratic limb darkening coefficients of 0.3 and
0.2. We do not output the model, but use the following {\bf -o}
command to output the light curve with the transit injected to the
file ``MODEL/1.injecttransit''. 

The {\bf -BLS} command is then used in a similar fashion to the
previous example to recover the transit, though in this case we only
identify 1 peak in the spectrum. We then use the {\bf
  -MandelAgolTransit} command to fit a \citet{mandel:2002} transit
model to the light curve, initializing the parameters based on the
previous {\bf -BLS} command. We use the same limb darkening
coefficients as for the injection, and vary all parameters except for
$e$, $\omega$, and the limb darkening coefficients. We do not attempt
to fit an RV curve, and we do not subtract the best-fit model. We
output the model together with the model phase curve to the directory
``MODEL''. The phase curve (for plotting) runs from phase $-0.5$ to
$0.5$ with a phase step of $0.001$.

\begin{figure}[]
\includegraphics[scale=0.7]{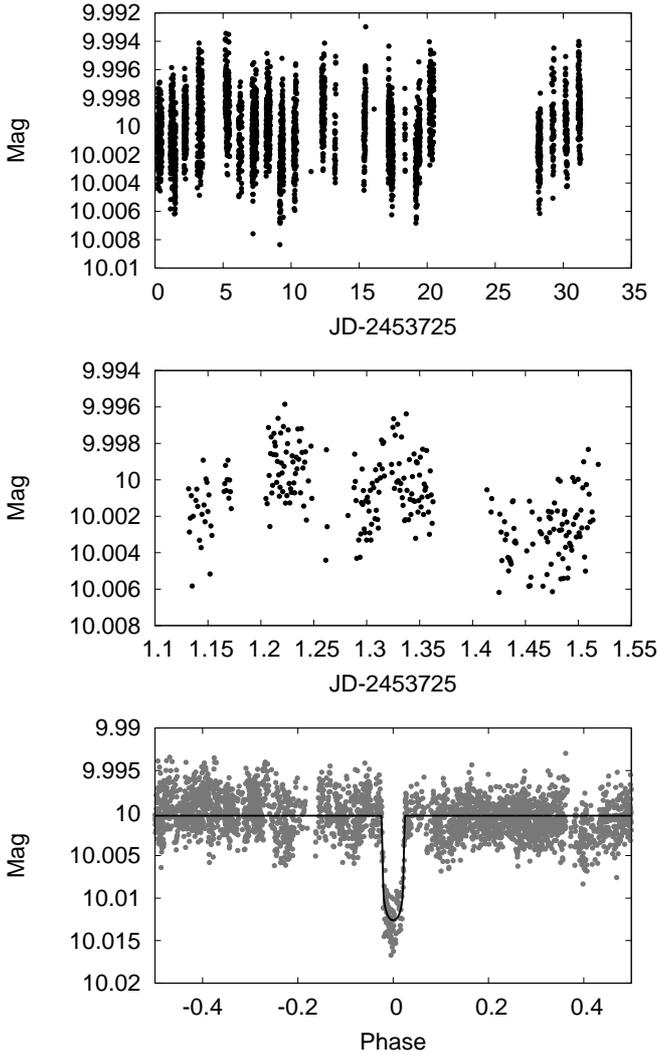}
\caption[]{ Top: a simulated light curve with time correlated noise
  generated in example~\ref{sec:example3}. This is the light curve
  produced by the {\bf -o} command on line~7 of
  listing~\ref{lst:example3simtransit}. Middle: the same light curve
  in the top panel, zooming in on a single night. Bottom: the
  phase-folded light curve after injecting the transit signal,
  recovering it with {\bf -BLS} and modelling it with the {\bf
    -MandelAgolTransit} command. The best-fit transit model is
  over-plotted. This is the light curve and model output through the
  {\bf -MandelAgolTransit} command on line~16 of
  listing~\ref{lst:example3simtransit}.
\label{fig:example3}
}
\end{figure}

A selection of the output
from this routine is shown in the listing. Figure~\ref{fig:example3}
shows the simulated noise-only light curve, together with the
phase-folded light curve with the transit signal and
\citet{mandel:2002} model overplotted.

\begin{lstlisting}[caption={Simulating a light curve with red-noise, adding a transit signal, recovering it with BLS and fitting a \citet{mandel:2002} transit model to it, as discussed in example~\ref{sec:example3}},label={lst:example3simtransit},frame=single,backgroundcolor=\color{lightgray},language=vartools,float=*]
prompt> vartools -i EXAMPLES/1 \
                 -inputlcformat t:1,err:3 \
                 -expr 'mag=10' \
                 -addnoise exp rho fix 0.05 sig_red fix 0.0015 \
                     sig_white fix 0.0015 \
                 -changeerror \
                 -o MODEL/1.simnoise \
                 -Injecttransit Pfix 2.123456 Rpfix 1.0 Mpfix 1.0 phaserand \
                     sinirand eomega efix 0. ofix 0. Mstarfix 1.0 \
                     Rstarfix 1.0 quad ldfix 0.3 0.2 0 \
                 -o MODEL/1.injecttransit \
                 -BLS q 0.002 0.05 0.5 100. 100000 500 0 1 1 PER \
                      1 MODEL 0 fittrap \
                      nobinnedrms ophcurve MODEL -0.5 0.5 0.001 \
                 -MandelAgolTransit bls quad 0.3 0.2 \
                      1 1 1 1 0 0 1 0 0 0 0 1 MODEL \
                      ophcurve MODEL -0.5 0.5 0.001 \
                 -oneline

@Name                         = EXAMPLES/1@
@Mean_Mag_2                   =  10.00029@
@RMS_2                        =   0.00222@
@Npoints_2                    =  3122@
@Injecttransit_Period_4       =     2.12345600@
...
@Injecttransit_phase_4        =   0.34066@
@Injecttransit_sin_i_4        =   0.99940@
...
@BLS_Period_1_6               =     2.12739225@
@BLS_Tc_1_6                   = 53726.949086475011@
@BLS_SN_1_6                   =   3.95738@
...
@BLS_Depth_1_6                =   0.01207@
...
@BLS_Npointsintransit_1_6     =   120@
@BLS_Ntransits_1_6            =     3@
@BLS_Npointsbeforetransit_1_6 =   169@
@BLS_Npointsaftertransit_1_6  =    69@
@BLS_Rednoise_1_6             =   0.00113@
@BLS_Whitenoise_1_6           =   0.00221@
@BLS_SignaltoPinknoise_1_6    =  17.60859@
@BLS_Period_invtransit_6      =     6.45330708@
@BLS_deltaChi2_invtransit_6   = -327.05444@
@BLS_MeanMag_6                =  10.00068@
@MandelAgolTransit_Period_7   =     2.12519667@
@MandelAgolTransit_T0_7       = 53726.95486357@
@MandelAgolTransit_r_7        =   0.09941@
@MandelAgolTransit_a_7        =   6.80453@
@MandelAgolTransit_bimpact_7  =   0.26979@
@MandelAgolTransit_inc_7      =  87.72772@
...
@MandelAgolTransit_chi2_7     =   0.99461@
\end{lstlisting}

\subsection{Searching a set of light curves for variable stars, while filtering noise}\label{sec:example4}

\begin{lstlisting}[caption={Light curve format for example~\ref{sec:example4}.},label={lst:example4lcformat},frame=single,backgroundcolor=\color{lightgray},language=vartools,float=*]
prompt> cat 1.txt

@# Image  Time[BJD]  Mag    Err    X     Y    FWHM   Ellipticity  PA  Airmass@
....
....
\end{lstlisting}

In this example we perform a general search for variability on a set
of light curves after applying two different noise-filtering
techniques. Listing~\ref{lst:example4lcformat} shows the format of the
light curves for this example. In addition to the usual time,
magnitude and uncertainty values provided in the light curve,
auxiliary information is provided for each observation including the
filename of the image from which the observation is generated, the X
and Y coordinates of the source on the image, the full-width at half
maximum (FWHM), ellipticity and position angle of a Gaussian fit to
the point spread function (PSF), and the airmass of the observation.

Listing~\ref{lst:example4processing} shows the procedure. We will
calculate five variability statistics: (1) the r.m.s.\ scatter of the
light curve; (2) the Generalized L-S periodogram; (3) the BLS
spectrum; (4) Stetson's $J$ statistic; and (5) the harmonic AoV
periodogram. This will be done three times for each light curve: on
the raw (un-filtered) input light curve; on the light curve after
decorrelating against the external parameters X, Y, FWHM, Ellipticity,
PA and Airmass; and on the light curve after applying TFA filtering in
addition to the decorrelation. To speed up the processing, this is run
in parallel on the light curves.

\begin{lstlisting}[caption={\va{} procedure for carrying out a general search for variability on a set of light curves, including different levels of filtering, as discussed in example~\ref{sec:example4}},label={lst:example4processing},frame=single,backgroundcolor=\color{lightgray},language=vartools,float=*]
prompt> vartools -l inputlist.txt \
    -inputlcformat id:1,t:2,mag:3,err:4,x:5,y:6,fw:7,el:8,pa:9,air:10 \
    -rms \
    -LS 0.1 100. 0.1 1 0 \
    -BLS q 0.002 0.05 0.5 100. 100000 500 0 5 0 0 0 fittrap \
    -Jstet 0.5 dates.txt \
    -aov_harm  1 0.1 100. 0.1 0.01 1 0 \
    -linfit \
      'a0+a1*x+a2*y+a3*fw+a4*el+a5*pa+a6*air+a7*sin(2.0*pi*x)+a8*sin(2.0*pi*y)' \
      a0,a1,a2,a3,a4,a5,a6,a7,a8 correctlc \
    -rms \
    -expr 'mag=mag+(Mean_Mag_0-Mean_Mag_6)' \
    -o LCEPD \
    -LS 0.1 100. 0.1 1 0 \
    -BLS q 0.002 0.05 0.5 100. 100000 500 0 5 0 0 0 fittrap \
    -Jstet 0.5 dates.txt \
    -aov_harm  1 0.1 100. 0.1 0.01 1 0 \
    -TFA trendlist.txt readformat 0 1 3 imagelist.txt 10 xycol 2 3 1 0 0 \
    -o LCTFA \
    -rms \
    -LS 0.1 100. 0.1 1 0 \
    -BLS q 0.002 0.05 0.5 100. 100000 500 0 5 0 0 0 fittrap \
    -Jstet 0.5 dates.txt \
    -aov_harm  1 0.1 100. 0.1 0.01 1 0 \
    -parallel 16 \
    -header -numbercolumns -matchstringid
\end{lstlisting}

Parameters are chosen for the commands as follows. For L-S we scan
periods from 0.1 to 100\,days with a subsample factor of 0.1. We only
report the top peak, and we do not output the periodogram. For BLS we
consider fractional transit durations between 0.002 and 0.05 in phase,
search for periods between 0.5\,days and 100\,days, use 100,000
frequency steps and 500 phase bins. We use 0 for the timezone (this is
used to assign points to an individual night in calculating the
fraction of $\Delta \chi^2$ that comes from one night), and report the
top five peaks in the periodogram. We do not output the periodogram or
the best-fit transit model, and we do not subtract the model from the
light curve. A trapezoid-shaped transit is fit to each peak to refine
the transit parameters compared to those found in the full
frequency scan. For Stetson's $J$ we use a timescale of 0.5\,days to
distinguish between ``near'' and ``far'' observations, and
``dates.txt'' is a file listing all of the dates in the light
curves. For the harmonic AoV we use a single harmonic model (i.e., a
simple sinusoid), search for periods between 0.1\,days and 100\,days
with a sub-sample factor of 0.1. A higher resolution scan, with a
subsample of 0.01 is performed around each peak. We only report the
top peak in the periodogram, and we do not output the
periodogram. Note that the parameters chosen for this example are just
to illustrate the use of the program, and are not meant to be best or
default choices. In practice all parameters need to be carefully
considered based on the properties of the light curves being analyzed.

After calculating the statistics for the raw light curve, we fit a
model that is a linear function of all of the external parameters (in
practice this is unlikely to be a very good model, one should inspect
the light curves to determine the best function to use for
filtering). This is then subtracted from the light curve using the
'correctlc' keyword. The linear coefficients which we optimize are
labeled $a0$ through $a8$. The terms 'sin(2.0*pi*x)' and
'sin(2.0*pi*y)' give a term that is periodic in one pixel. Light
curves generated through aperture photometry commonly have systematic
variations that are correlated with terms such as these. Following the '-linfit'
call, we use '-expr' to set the light curve mean back to its
pre-filtered value (subtracting the linear model causes the mean to be
close to zero). This is achieved by adding the difference between the
mean magnitude computed by '-rms' before '-linfit' and the mean
magnitude computed after it. These are referenced by their output
column names (``Mean\_Mag\_0'' and ``Mean\_Mag\_6''). The two calls to
``-o'' are used to output the light curves after each filtering
step. The decorrelated light curves are output to the directory LCEPD,
while the EPD+TFA-filtered light curves are output to the directory LCTFA.

To run TFA on the light curves we provide a list of files to be used
as TFA templates in 'template.txt'. These light curves are assumed to
have the same format as the input light curves, so we use the
readformat option to indicate that column 1 will be used for matching
points (this is the image id) and column 3 for the magnitudes. To use
the image ids for matching points in the light curves to points in the
templates, rather then the JD values, the '-matchstringid' option is
given at the end of the call to \va{}. A list of all image ids is
given in the file 'imagelist.txt'. Any template star that is within 10
pixels of the light curve being filtered will be removed from the
template for that light curve. The xycol option is used to indicate
that the average X and Y positions of the sources are read from
columns 2 and 3 of the input light curve list file. The light curves
passed on from this command will have the filtering applied, but we do
not output the TFA coefficients or the model trend signals.

Running this command on a set of light curves produces an ascii table
with 381 columns, which may be daunting to read. For convenience,
Listing~\ref{lst:example4columns} shows a method for writing out the
column names only, in a human-readable list. In this case one would
fill in the '...' with the full command given in
listing~\ref{lst:example4processing}, and the '-headeronly' option
causes \va{} to output the header and quit without processing any
light curves. The calls to the standard command-line utilities 'sed'
and 'awk' are used to reformat the names into a readable list,
including blank lines between the columns produced by each command. We
show the first few column names below the command. A simple selection
of variables might be to select any light curve with $\log_{10}{\rm
  FAP}_{\rm LS} < -100$, a BLS signal-to-pink-noise greater than 10,
$J_{\rm stet} > 1.5$, or $\log_{10}{\rm FAP}_{\rm AoV} < -100$.  This
can be done, for example, by applying a simple 'awk' script to the
output. In practice these thresholds would need to be adjusted based
on the noise properties of the survey (for example, the threshold of
$\log_{10}{\rm FAP}_{\rm LS} < -100$ is an extremely low false alarm
probability which is a reasonable threshold for highly time-correlated
noise, but far too conservative a threshold if the light curves
actually have Gaussian white noise).

\begin{lstlisting}[caption={A technique for printing out the column names for a \va{} procedure in a human-readable list. In this case the call to \va{} is the same as in listing~\ref{lst:example4processing}, but with with the ``-headeronly'' option used to output only the command headers. The procedure is discussed in example~\ref{sec:example4}.},label={lst:example4columns},frame=single,backgroundcolor=\color{lightgray},language=vartools,float=*]
prompt> vartools -l inputlist.txt \
    -inputlcformat id:1,t:2,mag:3,err:4,x:5,y:6,fw:7,el:8,pa:9,air:10 \
    -rms \
    ...
    -header -numbercolumns -matchstringid \
    -headeronly | \
sed 's|#||g' | \
awk '{c = ""; 
      for(i=1; i <= NF; i += 1) {
          n = split($i,s,"_");
          if(s[n] != c) print "";
          c = s[n];
          print $i;
      }}'

@1_Name@

@2_Mean_Mag_0@
@3_RMS_0@
@4_Expected_RMS_0@
@5_Npoints_0@

@6_LS_Period_1_1@
...
\end{lstlisting}

\subsection{Trend-filtering while preserving stellar variability with signal-reconstruction TFA}\label{sec:exampleTFASR}

In this example we show the use of signal-reconstruction-mode TFA ({\bf -TFA\_SR}) to filter systematic noise from a light curve while preserving real variability. Here the variability signal is modeled using a harmonic series. The command is shown in listing~\ref{lst:exampleTFASRlisting}.  A periodic signal is first identified with the {\bf -LS} command. We then save the light curve using the {\bf -savelc} command, to allow us to try different filtering techniques in the same call to \va{}. The {\bf -Killharm} command is used to subtract the signal from the light curve, and the call to {\bf -rms} calculates the scatter of the residuals. Calling {\bf -restorelc 1} restores the light curve to its state at the {\bf -savelc} command. We next run {\bf -TFA} to apply the TFA method in non-reconstructive mode. The trends are read from the file ``EXAMPLES/trendlist\_tfa'', the list of dates from the file ``EXAMPLES/dates\_tfa'', we use a separation of 25.0 pixels to remove close neighbors from the trend list, we specify the columns in the input list from which to read the X and Y pixel positions of the source, and we subtract the model from the light curve but do not output the trend coefficients or model signal.  After {\bf -TFA} we output the filtered light curve, and then subtract the signal with {\bf -Killharm} and calculate the scatter of the residuals with {\bf -rms}. Finally we restore the light curve again to its original state, and then filter using {\bf -TFA\_SR}. We use mostly the same options as for {\bf -TFA}, but now we output the coefficients and model trends to the directory ``EXAMPLES/OUTDIR1''. We do not apply TFA first, we use a stopping threshold of 0.001 and a maximum number of iterations of 100 (in this case, where we are using a harmonic series to model the signal, the trend and model are fit simultaneously to the data so these parameters have no affect), we use a ``harm''onic series to model the signal with zero higher-order harmonics and no sub-harmonics, and we take the period from the prior {\bf -LS} command. Again we output the light curve, subtract the signal, and compute the scatter of the residuals.

\begin{figure}[]
\includegraphics[scale=0.7]{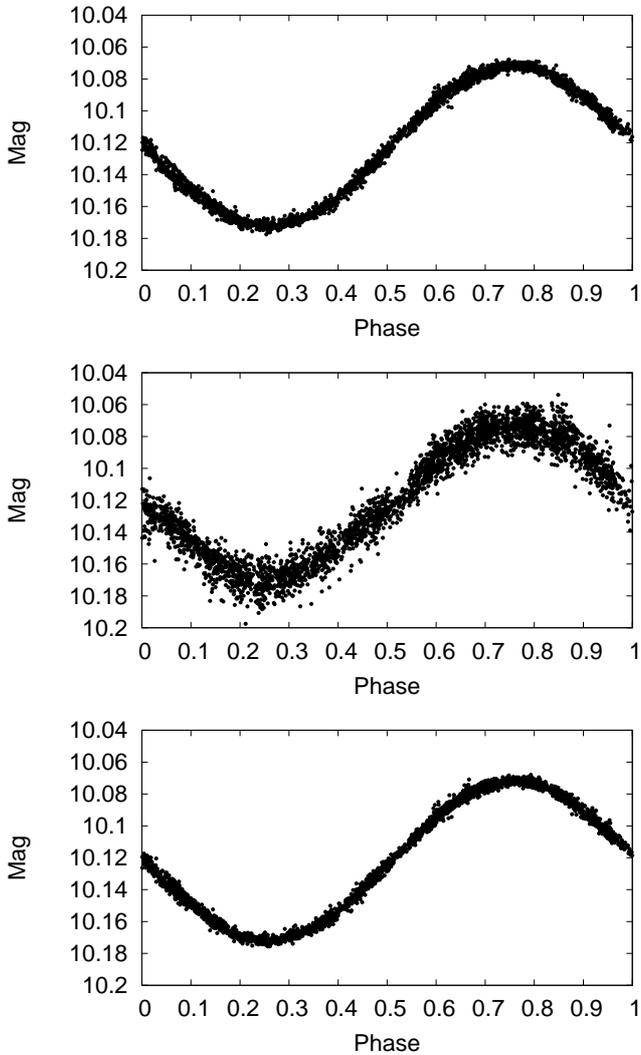}
\caption[]{ Top: phase-folded light curve of a simulated variable
  star before applying noise filtering as discussed in
  example~\ref{sec:exampleTFASR}. This is the input light curve
  processed in listing~\ref{lst:exampleTFASRlisting}. Middle: the
  phase-folded light curve after processing with the
  non-reconstructive-mode {\bf -TFA} command (i.e., that output on
  line 8 of listing~\ref{lst:exampleTFASRlisting}). Bottom: the
  phase-folded light curve after processing with the
  reconstructive-mode {\bf -TFA\_SR} command (i.e., that output on
  line 15 of listing~\ref{lst:exampleTFASRlisting}). Using the {\bf
    -TFA\_SR} command does a better job of filtering the noise while
  preserving the signal shape than does the {\bf -TFA} command.
\label{fig:exampleTFASR}
}
\end{figure}

By comparing the values for ``RMS\_4'', ``RMS\_8'' and ``RMS\_13'' one
can see the affect on the scatter in the residuals from the different
filtering techniques. Applying {\bf -TFA} in non-reconstructive model
increases the scatter from 2.3\,mmag to 7.2\,mmag, but when the
filtering is run in reconstructive mode the residual scatter decreases
to 2.1\,mmag. Figure~\ref{fig:exampleTFASR} compares the phase-folded
light curves after {\bf -TFA} and after {\bf -TFA\_SR} to the original
light curve, where it can be seen that the signal-reconstruction
correctly preserves the signal shape. Please note that this example is
concocted primarily to illustrate the syntax and operation of the {\bf
  -TFA\_SR} command. The noise-filtering in this example is not
terribly effective, but in practice one would likely use a much larger
set of template light curves, chosen in an intelligent fashion, which,
if the light curves share systematic trends, would reduce the scatter
of the residuals even further.

\begin{lstlisting}[caption={Example of filtering a light curve with TFA in non-reconstructive and reconstructive modes, as discussed in example~\ref{sec:exampleTFASR}},label={lst:exampleTFASRlisting},frame=single,backgroundcolor=\color{lightgray},language=vartools,float=*]
prompt> vartools -l EXAMPLES/lc_list_tfa_sr_harm -oneline -rms \
        -LS 0.1 10. 0.1 1 0 \
        -savelc \
        -Killharm ls 0 0 0 \
        -rms -restorelc 1 \
        -TFA EXAMPLES/trendlist_tfa EXAMPLES/dates_tfa \
            25.0 xycol 2 3 1 0 0 \
        -o EXAMPLES/OUTDIR1 nameformat 2.test_tfa_nosr \
        -Killharm ls 0 0 0 \
        -rms -restorelc 1 \
        -TFA_SR EXAMPLES/trendlist_tfa EXAMPLES/dates_tfa \
            25.0 xycol 2 3 1 \
            1 EXAMPLES/OUTDIR1 1 EXAMPLES/OUTDIR1 \
            0 0.001 100 harm 0 0 period ls \
        -o EXAMPLES/OUTDIR1 nameformat 2.test_tfa_sr_harm \
        -Killharm ls 0 0 0 \
        -rms \
        -oneline

@Name                                  = EXAMPLES/2@
@Mean_Mag_0                            =  10.11802@
@RMS_0                                 =   0.03663@
@Expected_RMS_0                        =   0.00102@
@Npoints_0                             =  3313@
@LS_Period_1_1                         =     1.23440877@
@Log10_LS_Prob_1_1                     = -4000.59209@
@LS_Periodogram_Value_1_1              =    0.99619@
@LS_SNR_1_1                            =   45.98308@
@Killharm_Mean_Mag_3                   =  10.12217@
@Killharm_Period_1_3                   =     1.23440877@
@Killharm_Per1_Fundamental_Sincoeff_3  =   0.05008@
@Killharm_Per1_Fundamental_Coscoeff_3  =  -0.00222@
@Killharm_Per1_Amplitude_3             =   0.10026@
@Mean_Mag_4                            =  10.11176@
@RMS_4                                 =   0.00231@
@Expected_RMS_4                        =   0.00102@
@Npoints_4                             =  3313@
@TFA_MeanMag_6                         =  10.11766@
@TFA_RMS_6                             =   0.03555@
@Killharm_Mean_Mag_7                   =  10.12211@
@Killharm_Period_1_7                   =     1.23440877@
@Killharm_Per1_Fundamental_Sincoeff_7  =   0.04802@
@Killharm_Per1_Fundamental_Coscoeff_7  =  -0.00268@
@Killharm_Per1_Amplitude_7             =   0.09620@
@Mean_Mag_8                            =  10.11169@
@RMS_8                                 =   0.00725@
@Expected_RMS_8                        =   0.00102@
@Npoints_8                             =  3313@
@TFA_SR_MeanMag_10                     =  10.11788@
@TFA_SR_RMS_10                         =   0.03642@
@Killharm_Mean_Mag_12                  =  10.12210@
@Killharm_Period_1_12                  =     1.23440877@
@Killharm_Per1_Fundamental_Sincoeff_12 =   0.04986@
@Killharm_Per1_Fundamental_Coscoeff_12 =  -0.00237@
@Killharm_Per1_Amplitude_12            =   0.09984@
@Mean_Mag_13                           =  10.11166@
@RMS_13                                =   0.00210@
@Expected_RMS_13                       =   0.00102@
@Npoints_13                            =  3313@
\end{lstlisting}

\subsection{Using the Weighted Wavelet Z-Transform to Characterize a Quasi-Periodic Signal}\label{sec:examplewwz}

\begin{figure}[]
\includegraphics[scale=0.7]{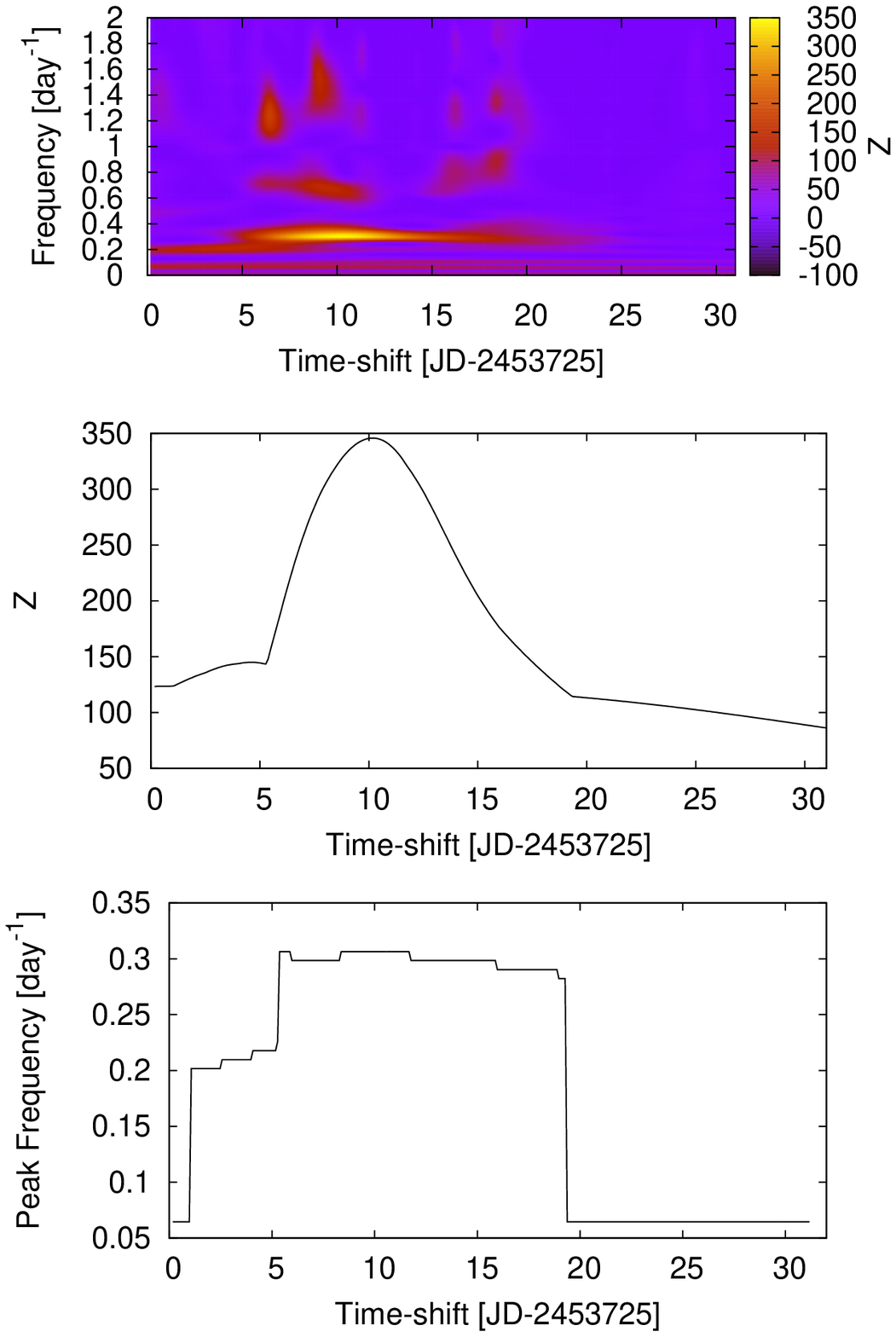}
\caption[]{ Top: Weighted Wavelet Z-Transform calculated in example~\ref{sec:examplewwz} (listing~\ref{lst:examplewwz}) for the light curve shown in figure~\ref{fig:examplewwzLC}. Middle: The maximum $Z$-transform as a function of time-shift. A strong signal is seen near a time of $JD-2453725 = 10.174$). Bottom: the peak frequency in the transform as a function of time-shift. The signal near $JD-2453725 = 10.174$ has a frequency of $0.3065$\,days$^{-1}$.
\label{fig:examplewwz}
}
\end{figure}

\begin{figure}[]
\includegraphics[scale=0.7]{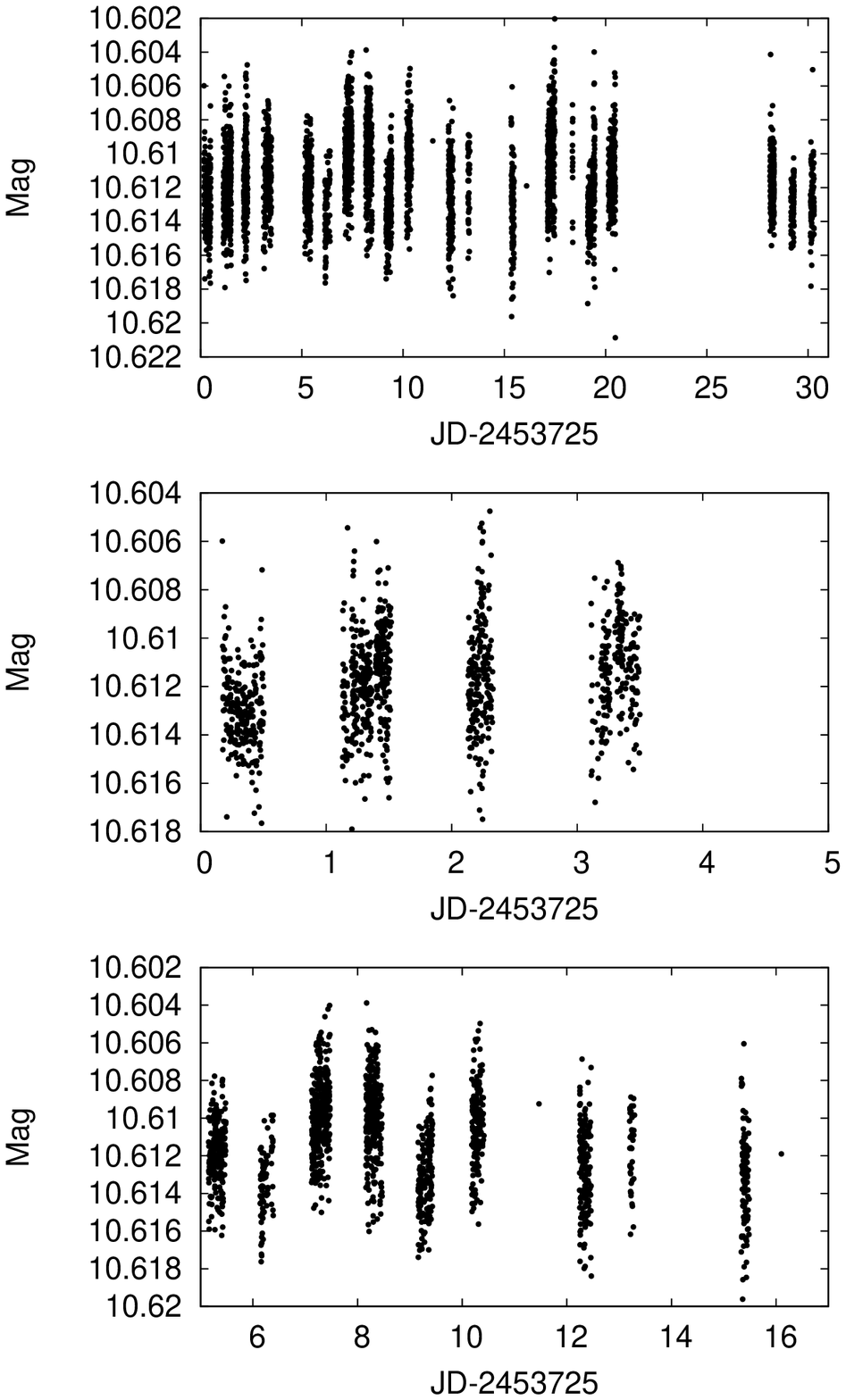}
\caption[]{ Top: the light curve analyzed in example~\ref{sec:examplewwz} (listing~\ref{lst:examplewwz}). Middle: the same light curve zoomed-in on early times when no strong signal is detected (though a signal with a frequency of $\sim 0.2$\,day$^{-1}$ may be present).  Bottom: the same light curve zoomed-in on times where a significant signal with a frequency of $0.3065$\,day$^{-1}$ is detected.
\label{fig:examplewwzLC}
}
\end{figure}

Listing~\ref{lst:examplewwz} shows an example of using the {\bf -wwz}
command to calculate the Weighted Wavelet Z-Transform of a light
curve. The transform is calculated up to a maximum frequency of
2.0\,d$^{-1}$ with a frequency step of $0.25/T$ where $T$ is the time
base-line of the light curve.  We use the ``auto'' keyword for both
``tau0'' and ``tau1'' to consider time-shifts running from the first
observation to the last, and use a time-shift step of 0.1\,days.  The
full transform is output to the file ``EXAMPLES/OUTDIR1/8.wwz'' while
the projection of the wavelet onto the time-shift axis is output to
the file ``EXAMPLES/OUTDIR1/8.mwwz''. These are plotted in
Figure~\ref{fig:examplewwz}. We also show the light curve in
Figure~\ref{fig:examplewwzLC}. For the full 2-d transform we use the
``pm3d'' keyword to output the file in a format that is convenient for
plotting with the {\sc gnuplot} program. The commands for plotting
this are shown at the bottom of Listing~\ref{lst:examplewwz}. The data
has a signal with a frequency of $0.3065$\,day$^{-1}$ ($P =
3.2632$\,days) centered at the time $JD - 53725 = 10.174$. The signal
is not present at later times, while a lower significance signal with
a frequency of $\sim 0.2$\,day$^{-1}$ may be present at earlier
times. This particular light curve comes from the MMT photometric
survey of M37 presented by \citet{hartman:m37:2}, and may be
a spotted star with an evolving surface brightness distribution.

\begin{lstlisting}[caption={Example of calculating the Weighted Wavelet Z-Transform for a light curve, as discussed in example~\ref{sec:examplewwz}},label={lst:examplewwz},frame=single,backgroundcolor=\color{lightgray},language=vartools,float=*]
prompt> vartools -i EXAMPLES/8 -oneline \
        -wwz maxfreq 2.0 freqsamp 0.25 tau0 auto tau1 auto dtau 0.1 \
         outfulltransform EXAMPLES/OUTDIR1/ pm3d \
         outmaxtransform EXAMPLES/OUTDIR1

@Name                = EXAMPLES/8@
@MaxWWZ_0            = 345.87167006071132@
@MaxWWZ_Freq_0       = 0.30645161290322581@
@MaxWWZ_TShift_0     = 53735.173920000001@
@MaxWWZ_Power_0      = 243.91899737448088@
@MaxWWZ_Amplitude_0  = 0.0019655476933700114@
@MaxWWZ_Neffective_0 = 1651.192187564548@
@MaxWWZ_AverageMag_0 = 10.611015587079519@
@Med_WWZ_0           = 135.53839042847693@
@Med_Freq_0          = 0.20967741935483872@
@Med_Power_0         = 116.72893440099938@
@Med_Amplitude_0     = 0.00096335163669205139@
@Med_Neffective_0    = 2188.0750923512305@
@Med_AverageMag_0    = 10.611486630808924@

prompt> gnuplot
gnuplot> set pm3d map
gnuplot> unset key
gnuplot> splot ``EXAMPLES/OUTDIR1/8.wwz'' u 1:2:3
\end{lstlisting}

\subsection{Performing an MCMC fit of a non-linear function to a light curve}\label{sec:examplemcmc}

Listing~\ref{lst:nonlinfitmcmc} shows an example of fitting an
analytic function that is non-linear in its parameters to a light
curve, and using an MCMC procedure to determine the uncertainties. The
initial {\bf -stats} command, together with the following {\bf -expr}
commands, are used to add a Gaussian function to a light curve. The
{\bf -stats} command is used to find the minimum and maximum times in
the light curve, which are stored in the variables $STATS\_t\_MIN\_0$
and $STATS\_t\_MAX\_0$. The first two calls to {\bf -expr} define the
variables $t1$ and $Dt$ in terms of the minimum and maximum values,
and then the third call to {\bf -expr} adds the Gaussian with
amplitude $0.1$\,mag, standard deviation $0.05Dt$ and peak time
$t1+Dt*0.2$. For our illustrative call to {\bf -nonlinfit} we provide
the analytic function to fit to the light curve in terms of the free
parameters $a$, $b$, $c$ and $d$. We then set the initial values and
uncertainty steps for each of these parameters on
line~7 of listing~\ref{lst:nonlinfitmcmc}. We use the ``mcmc'' keyword
to indicate that the DEMCMC procedure should be run, stopping it after
a total of 10000 links (in practice one would typically run more
links, a relatively small number is used here to minimize the
computing time for this simple example). The full MCMC chain will be
output to the file ``EXAMPLES/OUTDIR1/3.mcmc''.  The output table from
\va{} gives the best-fit (maximum likelihood) values for the free
parameters, together with the value of $\chi^2$ for those
parameters. Following this, the median and standard deviation of the
MCMC chain for each parameter is provided. These can be taken as
measures of the most likely values for each parameter, together with
their $1\sigma$ uncertainties. Following the output from \va{}, lines
28 through 33 of listing~\ref{lst:nonlinfitmcmc} show the format of
the file containing the full MCMC chain.

\begin{lstlisting}[caption={Running an MCMC procedure to fit a function that is non-linear in its parameters to a light curve, as discussed in example~\ref{sec:examplemcmc}.},label={lst:nonlinfitmcmc},frame=single,backgroundcolor=\color{lightgray},language=vartools,float=*]
prompt> vartools -i EXAMPLES/3 \
        -stats t min,max \
        -expr t1=STATS_t_MIN_0 \
        -expr 'Dt=(STATS_t_MAX_0-STATS_t_MIN_0)' \
        -expr 'mag=mag+0.1*exp(-0.5*((t-(t1+Dt*0.2))/(Dt*0.05))^2)' \
        -nonlinfit 'a+b*exp(-(t-c)^2/(2*d^2))' \
            'a=10.167:0.0002,b=0.1:0.0008,c=(t1+Dt*0.2):(0.005),d=(Dt*0.05):(0.016)' \
            mcmc Nlinkstotal 10000 outchains EXAMPLES/OUTDIR1/ \
        -oneline

@Name                     = EXAMPLES/3@
@STATS_t_MIN_0            = 53725.173920000001@
@STATS_t_MAX_0            = 53756.281021000003@
@Nonlinfit_a_BestFit_4    = 10.16733595824326@
@Nonlinfit_b_BestFit_4    = 0.10068199797234947@
@Nonlinfit_c_BestFit_4    = 53731.405627923843@
@Nonlinfit_d_BestFit_4    = 1.4997732382607727@
@Nonlinfit_BestFit_Chi2_4 = 90077.83071503813@
@Nonlinfit_a_MEDIAN_4     = 10.167339206527549@
@Nonlinfit_a_STDDEV_4     = 2.1634712351800478e-05@
@Nonlinfit_b_MEDIAN_4     = 0.10071313795015273@
@Nonlinfit_b_STDDEV_4     = 7.5501039462404719e-05@
@Nonlinfit_c_MEDIAN_4     = 53731.40592424116@
@Nonlinfit_c_STDDEV_4     = 0.0010974050260291169@
@Nonlinfit_d_MEDIAN_4     = 1.4987003116627102@
@Nonlinfit_d_STDDEV_4     = 0.001632159684888244@

prompt> head -3 EXAMPLES/OUTDIR1/3.mcmc
@# a b c d -2ln(L)@
@10.167347849788445 0.10070890704157852 53731.406660238252 1.4983339944179459@ \
@90080.167413138683@
@10.167330685303858 0.10072766325485784 53731.406153148055 1.5002348151196629@ \
@90081.07747374264@
\end{lstlisting}

\subsection{Converting the times in a light curve from UTC to BJD}\label{sec:exampleconverttime}

Listing~\ref{lst:converttime} shows an example of converting the times
in a light curve from UTC to TDB-corrected BJD. Lines 1--4 show the
format of the input light curve file ``EXAMPLES/1.UTC'', where the
first column gives the UTC times of the observations. By default \va{}
will attempt to read the time column directly into a double-precision
variable, which it interprets as the JD. To convert the input time
from UTC to JD, we use the call to {\bf -inputlcformat} on line~7 of
the listing. Here we indicate that the time will be read from the
first column, that it is in ``UTC'' string format, and we then
indicate how the string should be parsed, where ``\%Y'' is
interpretted as the year, ``\%M'' is the month, ``\%D'' is the day,
``\%h'' is the UT hour, ``\%m'' is the minute, and ``\%s'' are the
seconds. Finally we read the magnitude and uncertainties from columns
2 and 3, respectively. By default these are read in as
double-precision variables, so we do not need to specify a variable
type or format for these quantities. Lines 8 through 14 of the listing
execute the time conversion. We first need to specify the input time
system, which we do on line 8 (it is JD, and we use the
``inputsys-utc'' keyword to indicate that the input times have not
been corrected for leap-seconds). We then indicate the output time
system on line 9. Here we will convert to BJD, and will subtract
2400000 from the result, and we will convert to the TDB system to
correct the times for leap-seconds. To convert to BJD we need to tell
\va{} the coordinates of the star, which we do on line~10. We also
need to provide a set of files from JPL which provide the data for
calculating positions in the solar system (``ephemfile''), the list of
leap-seconds that have occurred since 1970 (``leapsecfile'') and the
detailed position of locations on the surface of the Earth with
respect to the center of the Earth-Moon system as a function of time
(``planetdatafile'')\footnote{The relevant files may be obtained from
  \url{ftp://naif.jpl.nasa.gov/pub/naif/generic_kernels/}}. Line~14
indicates the observatory where the observations were made (in this
case, Fred Lawrence Whipple Observatory in Arizona). Finally on
line~15 we output the light curve to the file
``EXAMPLES/OUTDIR1/1.bjdtdb'', whose format may be seen on lines
17--20 of the listing.

\begin{lstlisting}[caption={Converting the times in a light curve from UTC to BJD, as discussed in example~\ref{sec:exampleconverttime}.},label={lst:converttime},frame=single,backgroundcolor=\color{lightgray},language=vartools,float=*]
prompt> head -3 EXAMPLES/1.UTC
@2005-12-20T16:10:26.69 10.085 0.00119@
@2005-12-20T16:14:13.06 10.0847 0.00144@
@2005-12-20T16:15:55.01 10.0825 0.00123@

prompt> vartools -i EXAMPLES/1.UTC -quiet \
        -inputlcformat 't:1:utc:%Y-%M-%DT%h:%m:%s,mag:2,err:3' \
        -converttime input jd inputsys-utc \
            output bjd outputsubtract 2400000. outputsys-tdb \
            radec fix 88.079166 32.5533 \
            ephemfile CSPICEKERNELS/de432s.bsp \
            leapsecfile CSPICEKERNELS/naif0010.tls \
            planetdatafile CSPICEKERNELS/pck00010.tpc \
            observatory flwo \
        -o EXAMPLES/OUTDIR1/1.bjdtdb

prompt> head -3 EXAMPLES/OUTDIR1/1.bjdtdb
@  53725.180285235  10.08500   0.00119@
@  53725.182905252  10.08470   0.00144@
@  53725.184085226  10.08250   0.00123@
\end{lstlisting}

\section{Performance Tests}\label{sec:performance}

Here we present performance tests to compare the time required for I/O
between ASCII and Binary format data, and to also determine the
execution time for various period-finding commands. We focus here on
the periodic signal detection routines as these are generally the
time-limiting components of processing pipelines for large
datasets.

The following tests were carried out on a machine with 32 AMD Opteron
6140 64-bit CPUs, each with a maximum clock speed of 2.6\,GHz. The
machine runs Debian GNU/Linux version 7.8 with Linux kernel version
3.2.0-4-amd64. \va{} was compiled on this machine with {\sc gcc}
version 4.7.2 using $O2$ optimization, and was linked against all
supported external libraries. Here we are primarily interested in the
processing time for different routines, so to minimize I/O overhead
light curves were read-in from shared memory (/run/shm/ in the
listings below; this is a tmpfs file-system on this machine, and we
confirmed that there was sufficient free memory during our tests that
the files placed here were stored in RAM rather than swap space) and
all output was ignored (redirected to /dev/null; we do not use the
{\bf-quiet} option, so time is still spent generating the ASCII
table). Except where stated otherwise, all processing was performed on
the light curve stored in ``EXAMPLES/2.fits'', which is a binary FITS
table containing 3313 observations, and which is included in the \va{}
distribution.  

\subsection{Test 1: ASCII vs.\ Binary I/O}

There are four processes which contribute to the time required to
input and output a light curve: (1) reading data from disk into
memory; (2) parsing the raw data chunks into the appropriate variables
within \va{}; (3) transforming data from the vector variables back
into the appropriate format for storage on disk; (4) writing the data
to disk. If the numerical data are stored in ASCII format then steps
(2) and (3), which involve parsing ASCII strings into binary data and
vice versa, can be time consuming. If the data is stored in binary
format then steps (2) and (3) will be significantly faster (or not
applicable, depending on the data organization). Storing the data in
binary format also requires less total space than storing in ASCII to
the same precision, reducing the time required for steps (1) and (4)
as well. The primary benefit to using ASCII format is that the data
are easily human-readable with widely available software tools, may be
edited with simple text editors, and may be processed using a wide
range of tools which expect ASCII format data (e.g., standard shell
tools such as {\sc awk} or {\sc sed}).

To help in evaluating whether the time and storage savings from using
binary data are worth the possible tradeoffs with data transparency,
we carry out tests to evaluate the I/O performance of \va{} for
handling both types of data.

The time required to read and write data from/to disk is highly
system-dependent, and will vary in time as well, depending, for
example, on whether the data have been recently accessed and are
stored in cache memory. Regardless of the system, the time difference
between ASCII and binary format data for steps (1) and (4) should be
proportional to the difference in file-size. Here we focus on testing
steps (2) and (3) and attempt to minimize the time required for steps
(1) and (4) by using shared memory for the file storage. We first
generate an ASCII light curve using the command:
\begin{lstlisting}[frame=single,backgroundcolor=\color{lightgray},language=vartools]
prompt> \
  echo 0 | \
  awk ' \
    function gauss(u,v) { \
      u = sqrt(-2*log(rand())); \
      v = 4*atan2(1,0)*rand(); \
      return(u*cos(v)); \
    } \
    {
      for(i=1; i <= 1000000; i += 1) { \
        printf ``%.17g %.17g %.17g\n'', \
        2450000 + i/100000, \
        gauss(), gauss(); \
      } \
    }' \
  > /run/shm/test_ascii.txt

promp> head -3 /run/shm/test_ascii.txt
2450000.0000100001 -0.43249305127101612 \
0.33363186465509398
2450000.0000200002 0.35977598533128757 \
0.29950390190544596
2450000.0000300002 0.69568490442241193 \
-1.4140794262294938
\end{lstlisting}
Where the $gauss()$ function used in the {\sc awk} script is a simple routine
for generating Gaussian random numbers, and where we use ``\%.17g'' to
write the ASCII data to full double-precision. 

To determine the time for ASCII input only we carry out the following:
\begin{lstlisting}[frame=single,backgroundcolor=\color{lightgray},language=vartools]
prompt> time \
        vartools -i /run/shm/test_ascii.txt \
            -if 0 -rms -fi > /dev/null

real    0m2.340s
user    0m2.240s
sys     0m0.080s
\end{lstlisting}
Here the expression ``-if 0 -rms -fi'' is included to allow \va{} to
execute without performing any processing of the light curve (if no
commands are given to \va{} then it will terminate without reading the
light curve, with this expression the light curve will be read-in and
parsed into the appropriate variables, but the conditional placed in
front of ``-rms'' will prevent it from executing). This test indicates
that it takes \va{} $\sim 2.3$\,s of user+system time on our test
machine to parse a 3-column ASCII table with 1,000,000 rows and with
each column containing 17 digits. Repeated execution shows a variance
of less than 0.1\,s in the execution time. Note that running on an
empty light curve takes less than 0.05\,s so the initialization of
\va{} is a negligible fraction of this time. Using the ``-l'' option
to execute on multiple copies of this same light curve shows that the
time scales linearly with the number of light curves. Reducing the
light curve size by half also reduces the time by half. If each of the
input columns has 8 digit-precision rather than 17 digit-precision,
then the process takes 1.5\,s. If we use the ``-inputlcformat
t:0,mag:0,err:0'' option, in which case the light curve is read-from
disk and the number of lines and columns are determined (this is done
whenever \va{} reads in a light curve and requires a scan over all
ASCII characters in the file), but the ASCII data are not converted to
binary, then the process takes 1.2\,s to execute. Finally, if we do
also include the {\bf -rms} command (i.e., the ``-if 0'' and ``-fi''
expressions are removed) the difference in time is negligible (less
than the variance between repeated executions of the command),
indicating that calculating the r.m.s.\ scatter of ASCII-format light
curves is an I/O limited process, dominated by the ASCII to binary
conversion, if not by the disk access itself.

To determine the time for ASCII output we carry out the following:
\begin{lstlisting}[frame=single,backgroundcolor=\color{lightgray},language=vartools]
prompt> time \
        vartools -i /run/shm/test_ascii.txt \
            -o test_ascii_out.txt > /dev/null

real    0m4.353s
user    0m4.216s
sys     0m0.116s
\end{lstlisting}
which, together with the 2.3\,s measured above for input, indicates
that it takes $\sim 2$\,s on this machine to output an ASCII light
curve (primarily step (3)) with 1,000,000 lines to shared memory.

To repeat the input-only test for a binary FITS light curve, we carry out the following:
\begin{lstlisting}[frame=single,backgroundcolor=\color{lightgray},language=vartools]
prompt> vartools -i test_ascii.txt \
            -o test_fits.fits

prompt> for i in $(seq 1 100) ; do \
            echo test_fits.fits \
        done > list_fits

prompt> time \
        vartools -l list_fits \
            -if 0 -rms -fi \
        > /dev/null

real    0m10.558s
user    0m7.092s
sys     0m3.436s
\end{lstlisting}
The first command is used to convert the ASCII light curve to a binary
FITS table. The second is used to generate a list with 100 repeated
entries of the same light curve (this is done because we found that
the input time for one of these light curves is not significantly
longer than the time required to start \va{}), and the third performs
the time test. Based on this test we find that it takes
$\sim 0.1$\,s on this machine to read-in a FITS light curve with 1,000,000
rows and 3 double-precision columns (roughly 2/3 of the time is user
time, the other third is system time). If we use the ``-inputlcformat
t:0,mag:0,err:0'' option, then only 20\,ms is required, with
negligible system time.

To determine the output time for FITS we carried out the following test:
\begin{lstlisting}[frame=single,backgroundcolor=\color{lightgray},language=vartools]
prompt> vartools -l list_fits \
            -o /run/shm/ \
             nameformat test_fits_out.fits \
             fits \
        > /dev/null

real    0m30.656s
user    0m17.133s
sys     0m13.469s
\end{lstlisting}
Where the ``nameformat'' option is used to output the light curve to a
different name than its input file. Subtracting the input time
determined above we find that it takes $\sim 0.2$\,s to output the
FITS light curve with 1,000,000 rows and 3 double-precision columns,
or roughly twice as long as input.

Altogether we estimate that neglecting additional differences in disk
access time when the data are stored on a physical hard disk, the
input time for an ASCII light curve is approxiately 20 times longer
than for a FITS-table, while the output time is approximately 10 times
longer for ASCII than for FITS. These scalings assume that the ASCII
data consists of 3 columns each stored to 17-digit precision. Changing
the precision, and/or the number of columns in the light curve will
adjust the scalings. In terms of storage, the ASCII light curve with
1,000,000 data points takes 56\,MB whereas the FITS-table with 3
double-precision columns takes 23\,MB.

Note that based on applying similar tests to the file
``EXAMPLES/2.fits'', which is used for the performance tests discussed
below, we conclude that it takes \va{} less than 0.3\,ms to load the
light curve.

\subsection{Test 2: -LS}

We now proceed to test the performance of individual \va{}
commands. For our first test we executed the {\bf -LS} command on 1000
copies of ``EXAMPLES/2.fits''. We choose the period range and
subsample factors such that the periodogram would be evaluated at
16,384 frequencies (due to the use of FFTs within the code the number
of frequencies sampled internally increases in powers of 2, and thus
the execuction time shows discrete jumps as ${\rm ceil}(\log_{2}(N_{f}))$
increments). Tests were carried out with commands similar to the
following:
\begin{lstlisting}[frame=single,backgroundcolor=\color{lightgray},language=vartools]
prompt> for i in $(seq 1 1000) ; do \
    echo /run/shm/2.fits; \
    done > /run/shm/lc_list

prompt> time \
        vartools -l /run/shm/lc_list \
                 -LS 0.018980 30.0 0.1 1 0 \
        > /dev/null

real    5m26.562s
user    5m20.624s
sys     0m5.700s
\end{lstlisting}
Repeated execution shows a variance of a few seconds in the time. The
command takes on average 330\,ms of user+system time per light curve
on this system. When executed in parallel on 10 processors the routine
takes 38\,ms per light curve processed. Note that here, and in the examples below, we are calculating the periodogram for 10 different light curves in parallel, parallel processing of a single light curve is not currently supported. The total time scales close to
linearly with the number of processors, however shared overhead and
locks used to prevent conflicts between threads causes the scaling to
be not exactly linear. In general the scaling is closer to linear with
the number of processors when the execution time for a single process
is longer. When 2048, 4096 or 8192 frequencies are scanned the process
takes 20\,ms, 48\,ms, and 110\,ms per light curve,
respectively. Adjusting the number of points in the light curve, while
keeping the number of frequencies fixed, has a negligible effect on
the processing time which is dominated by the calculation of FFTs over
the sampled frequencies. For very large light curves, however, there
may be a difference as the time required for the ``extirpolation''
routine, which depends on the number of points in the light curve,
becomes non-negligible.

\subsection{Test 3: -BLS}

We performed similar tests on {\bf -BLS} to those conducted for {\bf -LS}, using commands such as:
\begin{lstlisting}[frame=single,backgroundcolor=\color{lightgray},language=vartools]
prompt> for i in $(seq 1 100) ; do \
    echo /run/shm/2.fits; \
    done > /run/shm/lc_list_100

prompt> time \
        vartools -l /run/shm/lc_list_100 \
                 -BLS q 0.005 0.05 0.5 30. \
                     10000 200 0 1 0 0 0 \
        > /dev/null

real    1m54.527s
user    1m54.459s
sys     0m5.700s
\end{lstlisting}
The relevant numbers controlling the execution time in this case are
the number of frequencies sampled $N_{f}$ (the above listing uses
10,000), the number of phase bins $N_{\rm bin}$ (200 above), the
maximum number of phase bins for the transit duration $N_{\rm bin,tr}$
(for ``q 0.005 0.05'' and $N_{\rm bin} = 200$ above we have $N_{\rm
  bin,tr} = 0.05\times 200 = 10$), and the number of observations
$N_{\rm obs}$. We expect the execution time to scale roughly as
\begin{equation}
a\times N_{\rm f}\times N_{\rm obs} + b \times N_{\rm f}\times N_{\rm bin} \times N_{\rm bin,tr}
\end{equation}
for some constants $a$ and $b$. For the above example we find that the
execution takes 1.1\,s per light curve, or 120\,ms of user+system time
per light curve if we use 10 parallel processes (if the
``nobinnedrms'' keyword is used, then the execution is 920\,ms per
light curve, or 93\,ms when running with 10 parallel processes; the
difference in time is due to an extra square root being included in
the constant $b$ when the keyword is not used). Using 100,000
frequencies increases the execution time by a factor of ten as
expected (11\,s per light curve), while doubling $N_{\rm bin}$
increases the execution time by a factor of approximately two (or
$1.5$ for ``nobinnedrms'').

\subsection{Test 4: -aov\_harm}

We performed similar tests on {\bf -aov\_harm} to those conducted for the other commands. This was done using commands such as:
\begin{lstlisting}[frame=single,backgroundcolor=\color{lightgray},language=vartools]
prompt> for i in $(seq 1 100) ; do \
    echo /run/shm/2.fits; \
    done > /run/shm/lc_list_100

prompt> time \
        vartools -l /run/shm/lc_list_100 \
                 -aov_harm 1 0.018980 30.0 \
                     0.1 0.1 1 0 \
                 -parallel 10 \
        > /dev/null

real    1m33.439s
user    15m41.771s
sys     0m0.068s
\end{lstlisting}
Here we are scanning the same number of frequencies as for {\bf -LS}
and using a pure sinusoid for the signal. The processing however is
signficantly slower than for {\bf -LS}, with 16,384 frequencies taking
9.3\,s per light curve when not running in parallel (or 930\,ms per
light curve when running with 10 parallel processes). Using half as
many frequencies reduces the processing time by a factor of two. Using
2 or 3 harmonics (i.e.,the fundamental and first higher-order harmonic
in the first case, and the fundamental and first two higher-order
harmonics in the second case) increases the execution time to 11\,s
per light curve and 13\,s per light curve, respectively (when not running in parallel).

\subsection{Test 5: -wwz}

We used the following command to test the performance of {\bf
  -wwz}:
\begin{lstlisting}[frame=single,backgroundcolor=\color{lightgray},language=vartools]
prompt> for i in $(seq 1 100) ; do \
    echo /run/shm/2.fits; \
    done > /run/shm/lc_list_100

prompt> time \
        vartools -l /run/shm/lc_list_100 \
                 -wwz maxfreq 2.0 \
                     freqsamp 0.25 \
                     tau0 auto \
                     tau1 auto \
                     dtau 0.1 \
                 -parallel 10 \
        > /dev/null

real    2m10.350s
user    21m54.090s
sys     0m0.040s
\end{lstlisting}
Here we are scanning 249 frequencies each at 311 time-shifts. With 10
parallel processes this takes 1.3\,s per light curve, or 13\,s per
light curve per processor. Doubling the number of frequencies sampled
increases the execution time by a factor of two, as does doubling the
number of time-shifts.

\subsection{Test 6: -fastchi2}

We used the following command to test the performance of {\bf
  -fastchi2} which is included as an extension to \va{}:
\begin{lstlisting}[frame=single,backgroundcolor=\color{lightgray},language=vartools]
prompt> for i in $(seq 1 1000) ; do \
    echo /run/shm/2.fits; \
    done > /run/shm/lc_list

prompt> time \
        vartools -L $LIBDIR/fastchi2.la \
                 -l /run/shm/lc_list \
                 -fastchi2 \
                     Nharm fix 1 \
                     freqmax fix 52.687 \
                     freqmin fix 0.03333 \
                     oversample fix 4 \
        > /dev/null

real    0m54.122s
user    0m54.011s
sys     0m0.076s
\end{lstlisting}
where we explicitly load the library with the {\bf -L} option to avoid
spending time performing a disk search for the appropriate library,
and the parameters for {\bf -fastchi2} are chosen to calculate 16,384
frequencies using a simple sinusoid. We find that this process takes
54\,ms per light curve, or 5.6\,ms per light curve if run on 10
parallel processors. When 2048, 4096 or 8192 frequencies are scanned
the process takes 34\,ms, 36\,ms and 42\,ms, respectively. Thus for
large number of frequencies the {\bf -fastchi2} procedure is
substantially faster than {\bf -LS} (330\,ms for 16,384 frequencies),
but for 2048 or fewer frequencies the {\bf -LS} command is faster to
execute (20\,ms per light curve).

\section{Future Development}\label{sec:conclusion}

There are several areas in which \va{} may be improved to make it more
useful for astronomical time series analysis. Several new processing
commands are being actively developed, such as Fourier filtering
methods, and techniques for stitching together different light curves
for a given source. Other topics for future development, which require
more fundamental changes to the organization of the code, include the
parallelization of individual commands (currently only parallel processing of multiple light curves is supported), and support for parallel
operation in non-shared-memory mode (e.g., using MPICH). The methods
for handling analytic expressions may be extended to support vector
and matrix operations. We are also working on support for
user-developed commands and functions written in {\sc python}. As a
separate effort we are also working on developing a Graphical User
Interface (GUI) for \va{}, written in {\sc python}, which may be used
to construct processing pipelines and interact with the results.

\appendix

\onecolumn

\section{Command Syntax}\label{sec:syntax}

Here we describe in detail the expected syntax for each command, including a
brief explanation of the input parameters. Commands are listed in
alphabetical order. See section~\ref{sec:commands} for a description
of the algorithms and uses of each command. In general a command is
executed by typing the name of the command (including the ``-''
prefix, e.g., ``-rms'' for the command which calculates the
r.m.s.~scatter in a light curve) followed by various parameters and
options. Terms within angular brackets ``$<\,>$'' are required
parameters. Terms within square brackets ``[\,]'' are optional
parameters that may be ignored. Within a given set of brackets the
pipe ``$\vert$'' is used to distinguish between different choices that
are allowed (in general only one of the possible choices may be
used). Any term not placed within brackets is required. Terms within
quotations indicate keywords that should be typed out exactly as they
appear, but without the quotation marks (these are typically used to
control optional behavior, or to specify the name of a parameter which
then follows). The keywords are case sensitive. Unquoted terms
represent parameters for which the user should substitute an
appropriate value when calling the command. \va{} obeys a strict
ordering of parameters, you must provide them in the exact order
indicated. Parameters referred to below as ``flags'' are used to
control the behavior of the command. In these cases the user either
substitutes ``1'' for the flag to turn the option on, or ``0'' to turn
it off.

Many of the commands share syntax for controlling how parameter values
should be determined. The syntax 
\begin{lstlisting}[backgroundcolor=\color{white},language=vartools,basicstyle=\footnotesize]
<"fix" val | "list" ["column" col] | "fixcolumn" <colname | colnum>>
\end{lstlisting} 
is adopted to allow the user to choose between fixing a parameter to a
specific value for all light curves (the user types ``fix'' and then
substitutes the appropriate value for $val$), to read the parameter
value from the input list of light curve files allowing a different
value for each light curve (the user types ``list'' then optionally
indicates which column the parameter is to be read-in from by typing
``column'' and the column number; if the ``column'' keyword is not
used, then the values are read-in from the next unused column in the
file), or to use a value computed by a previously executed command
(the user types ``fixcolumn'' and then gives either the name of the
column, which is the column heading when the ``-header'' option is
used, or the number of the column in the output table). In some
cases only a subset of these options is allowed.

Below we list the expected syntax for each command, followed by a
table explaining the various parameters.

\begin{lstlisting}[backgroundcolor=\color{white},language=vartools,basicstyle=\footnotesize]
-addnoise
    <   "white"
            <"sig_white" <"fix" val | "list" ["column" col]>>
      | "squareexp"
            <"rho" <"fix" val | "list" ["column" col]>>
            <"sig_red" <"fix" val | "list" ["column" col]>>
            <"sig_white" <"fix" val | "list" ["column" col]>>
            ["bintime" <"fix" val | "list" ["column" col]>]
      | "exp"
            <"rho" <"fix" val | "list" ["column" col]>>
            <"sig_red" <"fix" val | "list" ["column" col]>>
            <"sig_white" <"fix" val | "list" ["column" col]>>
            ["bintime" <"fix" val | "list" ["column" col]>]
      | "matern"
            <"nu" <"fix" val | "list" ["column" col]>>
            <"rho" <"fix" val | "list" ["column" col]>>
            <"sig_red" <"fix" val | "list" ["column" col]>>
            <"sig_white" <"fix" val | "list" ["column" col]>>
            ["bintime" <"fix" val | "list" ["column" col]>]
      | "wavelet"
            <"gamma" <"fix" val | "list" ["column" col]>>
            <"sig_red" <"fix" val | "list" ["column" col]>>
            <"sig_white" <"fix" val | "list" ["column" col]>>
    >
\end{lstlisting}

\begin{center}
\tablecaption{Input Parameters for ``-addnoise'' Command (Section~\ref{cmd:addnoise})}
\footnotesize
\begin{xtabular}{|p{1in}p{5mm}p{5in}|}
\hline
"white" & \tabitem & Use the white-noise only noise model. \\
~~~~"sig\_white" & \tabitem & Standard deviation of white noise. \\ \hline
"squareexp" & \tabitem & Use the square-exponential correlated noise model (Eq.~\ref{eqn:squareexp}). \\
~~~~"rho" & \tabitem & The parameter $\rho$ in equation~\ref{eqn:squareexp}. \\
~~~~"sig\_red" & \tabitem & The standard deviation of the correlated component ($\sqrt{a}$ in equation~\ref{eqn:squareexp}). \\
~~~~"sig\_white" & \tabitem & The standard deviation of the white noise component ($\sigma_{i}$ in equation~\ref{eqn:squareexp}). \\ \hline
~~~~"bintime" & \tabitem & Optionally divide the input light curve into time bins and simulate the correlated noise independently in each bin. This option speeds up the simulation substantially in cases where the total time spanned by the light curve greatly exceeds the correlation timescale. \\ \hline
"exp" & \tabitem & Use the exponential correlated noise model (Eq.~\ref{eqn:expnoise}). \\
~~~~"rho" & \tabitem & The parameter $\rho$ in equation~\ref{eqn:expnoise}. \\
~~~~"sig\_red" & \tabitem & The standard deviation of the correlated component ($\sqrt{a}$ in equation~\ref{eqn:expnoise}). \\
~~~~"sig\_white" & \tabitem & The standard deviation of the white noise component ($\sigma_{i}$ in equation~\ref{eqn:expnoise}). \\ \hline
~~~~"bintime" & \tabitem & Optionally divide the input light curve into time bins and simulate the correlated noise independently in each bin. This option speeds up the simulation substantially in cases where the total time spanned by the light curve greatly exceeds the correlation timescale. \\ \hline
"matern" & \tabitem & Use the Mat\'ern correlated noise model (Eq.~\ref{eqn:matern}). \\
~~~~"nu" & \tabitem & The parameter $\nu$ in equation~\ref{eqn:matern}. \\
~~~~"rho" & \tabitem & The parameter $\rho$ in equation~\ref{eqn:matern}. \\
~~~~"sig\_red" & \tabitem & The standard deviation of the correlated component ($\sqrt{a}$ in equation~\ref{eqn:matern}). \\
~~~~"sig\_white" & \tabitem & The standard deviation of the white noise component ($\sigma_{i}$ in equation~\ref{eqn:matern}). \\ \hline
~~~~"bintime" & \tabitem & Optionally divide the input light curve into time bins and simulate the correlated noise independently in each bin. This option speeds up the simulation substantially in cases where the total time spanned by the light curve greatly exceeds the correlation timescale. \\ \hline
"wavelet" & \tabitem & Use a wavelet correlated noise model. \\
~~~~"gamma" & \tabitem & The parameter $\gamma$ discussed in section~\ref{sec:lcsim}. \\
~~~~"sig\_red" & \tabitem & The standard deviation of the correlated component. \\
~~~~"sig\_white" & \tabitem & The standard deviation of the white noise component. \\ \hline
\end{xtabular}
\end{center}

\begin{lstlisting}[backgroundcolor=\color{white},language=vartools,basicstyle=\footnotesize,breaklines=false,prebreak={}]
-alarm
\end{lstlisting}

No parameters or options are available.

\begin{lstlisting}[backgroundcolor=\color{white},language=vartools,basicstyle=\footnotesize,breaklines=false,prebreak={}]
-aov ["Nbin" Nbin] minp maxp subsample finetune Npeaks operiodogram
    [outdir] ["whiten"] ["clip" clip clipiter] ["uselog"]
    ["fixperiodSNR" <"aov" | "ls" | "injectharm" | "fix" period
        | "list" ["column" col]
        | "fixcolumn" <colname | colnum>>]
\end{lstlisting}

\begin{center}
\tablecaption{Input Parameters for ``-aov'' Command (Section~\ref{cmd:aov})}
\footnotesize
\begin{xtabular}{|p{1in}p{5mm}p{5in}|} \hline
"Nbin" & \tabitem & The number of phase bins to use. The default is $8$. \\ \hline
minp & \tabitem & The minimum period to search. \\ \hline
maxp & \tabitem & The maximum period to search. \\ \hline
subsample & \tabitem & The periodogram will be sampled at a resolution of subsample$/T$ where $T$ is the time baseline of the light curve. \\ \hline
finetune & \tabitem & Peaks found in the initial scan will then be resampled at a resolution of finetune$/T$. \\ \hline
Npeaks & \tabitem & The number of peaks in the periodogram to report. \\ \hline
operiodogram & \tabitem & Flag indicating if the periodogram for each light curve should be output to a separate file. \\
~~~~outdir & \tabitem & Required if operiodogram is set. The periodogram will be output to the file \$outdir/\$basename.aov where \$outdir is the value specified for outdir, and \$basename is the base filename of the light curve (any leading directory is stripped). \\ \hline
"whiten" & \tabitem & Keyword used to whiten the light curve and recalculate the periodogram for each peak. By default this is not done. \\ \hline
"clip" & \tabitem & Adjust the clipping performed on the periodogram in calculating the spectroscopic S/N of a peak. By default iterative 5$\sigma$ clipping is performed. \\ 
~~~~clip & \tabitem & The sigma-clipping factor to use. \\
~~~~clipiter & \tabitem & Flag which if set causes iterative clipping to be performed. \\ \hline
"uselog" & \tabitem & Keyword used to calculate the S/N from the natural logarithm of the AoV spectrum. By default the S/N is calculated directly from AoV. \\ \hline
"fixperiodSNR" & \tabitem & Report also AoV, the false alarm probability and the S/N for a fixed period. \\
~~~~"aov" & \tabitem & Use the peak period from the last executed -aov command for the fixperiodSNR. \\
~~~~"ls" & \tabitem & Use the peak period from the last executed -LS command. \\
~~~~"injectharm" & \tabitem & Use the period from the last executed -Injectharm command (the first one is used if multiple periods in a given command. \\ \hline
\end{xtabular}
\end{center}

\begin{lstlisting}[backgroundcolor=\color{white},language=vartools,basicstyle=\footnotesize,breaklines=false,prebreak={}]
-aov_harm Nharm minp maxp subsample finetune Npeaks operiodogram [outdir]
    ["whiten"] ["clip" clip clipiter]
    ["fixperiodSNR" <"aov" | "ls" | "injectharm" | "fix" period
        | "list" ["column" col]
        | "fixcolumn" <colname | colnum>>]
\end{lstlisting}

\begin{center}
\tablecaption{Input Parameters for ``-aov\_harm'' Command (Section~\ref{cmd:aovharm})}
\footnotesize
\begin{xtabular}{|p{1in}p{5mm}p{5in}|} \hline
Nharm & \tabitem & The number of harmonics to use. Set this to a value less than 1 to automatically optimize this number. The fundamental mode is counted as 1 harmonic here. \\ \hline
minp & \tabitem & The minimum period to search. \\ \hline
maxp & \tabitem & The maximum period to search. \\ \hline
subsample & \tabitem & The periodogram will be sampled at a resolution of subsample$/T$ where $T$ is the time baseline of the light curve. \\ \hline
finetune & \tabitem & Peaks found in the initial scan will then be resampled at a resolution of finetune$/T$. \\ \hline
Npeaks & \tabitem & The number of peaks in the periodogram to report. \\ \hline
operiodogram & \tabitem & Flag indicating if the periodogram for each light curve should be output to a separate file. \\
~~~~outdir & \tabitem & Required if operiodogram is set. The periodogram will be output to the file \$outdir/\$basename.aov\_harm where \$outdir is the value specified for outdir, and \$basename is the base filename of the light curve (any leading directory is stripped). \\ \hline
"whiten" & \tabitem & Keyword used to whiten the light curve and recalculate the periodogram for each peak. By default this is not done. \\ \hline
"clip" & \tabitem & Adjust the clipping performed on the periodogram in calculating the spectroscopic S/N of a peak. By default iterative 5$\sigma$ clipping is performed. \\ 
~~~~clip & \tabitem & The sigma-clipping factor to use. \\
~~~~clipiter & \tabitem & Flag which if set causes iterative clipping to be performed. \\ \hline
"fixperiodSNR" & \tabitem & Report also AoV, the false alarm probability and the S/N for a fixed period. \\
~~~~"aov" & \tabitem & Use the peak period from the last executed -aov command for the fixperiodSNR. \\
~~~~"ls" & \tabitem & Use the peak period from the last executed -LS command. \\
~~~~"injectharm" & \tabitem & Use the period from the last executed -Injectharm command (the first one is used if multiple periods in a given command. \\ \hline
\end{xtabular}
\end{center}

\begin{lstlisting}[backgroundcolor=\color{white},language=vartools,basicstyle=\footnotesize,breaklines=false,prebreak={}]
-autocorrelation start stop step outdir
\end{lstlisting}

\begin{center}
\tablecaption{Input Parameters for ``-autocorrelation'' Command (Section~\ref{cmd:autocorrelation})}
\footnotesize
\begin{xtabular}{|p{1in}p{5mm}p{5in}|}\hline
start & \tabitem & Starting time-lag for sampling the discrete autocorrelation function. \\ \hline
stop & \tabitem & Stopping time-lag  for sampling the discrete autocorrelation function. \\ \hline
step & \tabitem & Time-lag step-size. \\ \hline
outdir & \tabitem & Directory for outputting the autocorrelation files. The files will be written to \$outdir/\$basename.autocorr where \$outdir is the value specified for outdir, and \$basename is the base filename of the light curve (any leading directory is stripped). \\ \hline
\end{xtabular}
\end{center}

\begin{lstlisting}[backgroundcolor=\color{white},language=vartools,basicstyle=\footnotesize,breaklines=false,prebreak={}]
-binlc <"average" | "median" | "weightedaverage">
    <"binsize" binsize | "nbins" nbins>
    ["bincolumns" var1[:stats1][,var2[:stats2],...]]
    ["firstbinshift" firstbinshift]
    <"tcenter" | "taverage" | "tmedian" | "tnoshrink" ["bincolumnsonly"]>
\end{lstlisting}

\begin{center}
\tablecaption{Input Parameters for ``-binlc'' Command (Section~\ref{cmd:binlc})}
\footnotesize
\begin{xtabular}{|p{1in}p{5mm}p{5in}|}
\hline
"average" & \tabitem & Take the average of points in a bin \\ \hline
"median"  & \tabitem & Take the median of points in a bin \\ \hline
"weightedaverage" & \tabitem & Take the weighted average of points in a bin \\ \hline
"binsize" & \tabitem & Specify the size of the bin in the time unit of the light curve. Type the "binsize" keyword and then provide the value. \\ \hline
"nbins" & \tabitem & Specify the number of bins to use. In this case the binsize is equal to $T/nbins$ where $T$ is the time-span of the light curve. Type the "nbins" keyword and then provide the value. \\ \hline
"bincolumns" & \tabitem & By default all light curve vectors except for the time, and uncertainty, will be binned using the "average", "median", or "weightedaverage" as specified by the user. Any vectors that the user wishes to bin in a different manner may be specified with this option. Following the "bincolumns" keyword provide a comma separated list of light curve variables and associated statistics. After a given variable use a ":" and then indicate the statistic to calculate. The options for the statistics are the same as for the {\bf -stats} command. \\ \hline
"firstbinshift" & \tabitem & By default the first bin begins at the initial time in the light curve ($t_{0}$). Use this keyword to change this behavior, in which case the first bin will start at $t_{0} - {\rm firstbinshift}/{\rm binsize}$. \\ \hline
"tcenter" & \tabitem & Take the binned light curve time to be the time at the center of each bin. \\ \hline
"taverage" & \tabitem & Take the binned light curve time to be the average of the times of points that fall in a given bin. \\ \hline
"tmedian" & \tabitem & Take the binned light curve time to be the median of the times of points that fall in a given bin. \\ \hline
"tnoshrink" & \tabitem & Do not shrink the light curve. In this case all points in the light curve will be replaced by their binned value, but the times will not be changed. \\
"bincolumnsonly" & \tabitem & If this keyword is given, then only the columns specified after the "bincolumns" keyword will be binned. All other light curve vectors will be left unchanged. This option is only available when the "tnoshrink" option is used. \\ \hline
\end{xtabular}
\end{center}

\begin{lstlisting}[backgroundcolor=\color{white},language=vartools,basicstyle=\footnotesize,breaklines=false,prebreak={}]
-BLS < "r" rmin rmax | "q" qmin qmax > minper maxper nfreq nbins
    timezone Npeak outperiodogram [outdir] omodel [model_outdir]
    correctlc ["fittrap"] ["nobinnedrms"]
    ["ophcurve" outdir phmin phmax phstep]
    ["ojdcurve" outdir jdstep]
    ["stepP" | "steplogP"]
    ["adjust-qmin-by-mindt" ["reduce-nbins"]]
\end{lstlisting}

\begin{center}
\tablecaption{Input Parameters for ``-BLS'' Command (Section~\ref{cmd:bls})}
\footnotesize
\begin{xtabular}{|p{1in}p{5mm}p{5in}|}
\hline
``r'' & \tabitem & Allow for a period dependent limit on the transit durations to search. $r$ corresponds approximately to the stellar radius for lower main sequence stars assuming central transits, circular orbits, and that times are specified in days. \\ 
~~~~rmin & \tabitem & The minimum transit duration at period $P$ will be $0.076 {\rm rmin}^{2/3} P^{1/3}$. \\
~~~~rmax & \tabitem & The maximum transit duration at period $P$ will be $0.076 {\rm rmax}^{2/3} P^{1/3}$. \\ \hline
``q'' & \tabitem & Use a period-independent limit on the fractional transit durations considered. \\ 
~~~~qmin & \tabitem & The minimum transit duration at period $P$ will be ${\rm qmin} P$. \\
~~~~qmax & \tabitem & The maximum transit duration at period $P$ will be ${\rm qmax} P$. \\ \hline
minper & \tabitem & The minimum period to search for transits. \\ \hline
maxper & \tabitem & The maximum period to search for transits. \\ \hline
nfreq & \tabitem & The number of frequencies to search (the frequency step-size is $\Delta f = (1/{\rm minper} - 1/{\rm maxper})/{\rm nfreq}$). \\ \hline
nbins & \tabitem & The number of phase bins to use. \\ \hline
timezone & \tabitem & The timezone of the observatory (in hours from UTC). This only affects the grouping of points used to determine the fraction of $\Delta \chi^{2}$ from a single day. If multiple observatories were used, or if the observations were performed from space one can simply provide $0$ for this value and ignore the reported $\Delta \chi^{2}$ from one day. \\ \hline
Npeak & \tabitem & The number of peaks to identify in the BLS spectrum. \\ \hline
outperiodogram & \tabitem & A flag used to indicate whether or not the BLS spectra should be output. \\
~~~~outdir & \tabitem & Required if outperiodogram is set to $1$. The BLS spectra will be written to files with the name \$outdir/\$basename.bls \$outdir is the value specified for outdir, and \$basename is the base filename of the light curve (any leading directory is stripped). \\ \hline
omodel & \tabitem & A flag used to indicate whether or not the best-fit transit model light curves should be output. These light curves are evaluated at the observed times. \\
~~~~model\_outdir & \tabitem & Required if omodel is set to $1$. The models files will be written to \$model\_outdir/\$basename.bls.model. \\ \hline
correctlc & \tabitem & A flag used to indicate whether or not the best-fit transit model should be subtracted from the observations before passing the light curve on to the next command. \\ \hline
``fittrap'' & \tabitem & An optional keyword which, if used, causes a trapezoid-shape transit model to be fit to the light curve around each BLS peak identified. \\ \hline
``nobinnedrms'' & \tabitem & An optional keyword used to adjust the way in which the BLS spectroscopic S/N statistic is calculated. If this keyword is used then the average and standard deviation of the spectrum are calculated only from the maximum SR values at each trial frequency. If it is not used, then these are calculated from all SR values considered (including transit epochs and durations that do not optimize the fit at a given frequency). Using this keyword speeds up the BLS calculation, but the S/N will tend to be suppressed for high significance detections. \\ \hline
``ophcurve'' & \tabitem & A keyword used to output best-fit model light curves that are uniformly sampled in phase. \\
~~~~outdir & \tabitem & The models will be written out to files named \$outdir/\$basename.bls.phcurve. \\
~~~~phmin & \tabitem & The starting phase to use in the model. \\
~~~~phmax & \tabitem & The ending phase to use in the model. \\
~~~~phstep & \tabitem & The phase step-size to use in the model. \\ \hline
``ojdcurve'' & \tabitem & A keyword used to output best-fit model light curves that are uniformly sampled in time between the first and last observed times in the light curve. \\
~~~~outdir & \tabitem & The models will be written out to files named \$outdir/\$basename.bls.jdcurve. \\
~~~~jdstep & \tabitem & The time step-size to use in the model. \\ \hline
``stepP'' & \tabitem & Sample the BLS spectrum at uniform steps in period (by default uniform steps in frequency are used). \\ \hline
``steplogP'' & \tabitem & Sample the BLS spectrum at uniform steps in the logarithm of the period (by default uniform steps in frequency are used). \\ \hline
``adjust-qmin-by-mindt'' & \tabitem & Adaptively set the minimum $q$ value to the maximum of $qmin$ or $\Delta t_{\rm min} \times f$ where $\Delta t_{\rm min}$ is the minimum time difference between consecutive points in the light curve. \\
~~~~``reduce-bins'' & \tabitem & Adaptively reduce the number of phase bins at each frequency such that there are no more than two bins to sample a transit of phase duration $qmin$. \\ \hline
\end{xtabular}
\end{center}

\begin{lstlisting}[backgroundcolor=\color{white},language=vartools,basicstyle=\footnotesize,breaklines=false,prebreak={}]
-BLSFixPer <"aov" | "ls" | "list" ["column" col]
        | "fix" period | "fixcolumn" <colname | colnum>
        | "expr" expr>
    <"r" rmin rmax | "q" qmin qmax >
    nbins timezone omodel [model_outdir] correctlc ["fittrap"]
\end{lstlisting}

\begin{center}
\tablecaption{Input Parameters for ``-BLSFixPer'' Command (Section~\ref{cmd:blsfixper})}
\footnotesize
\begin{xtabular}{|p{1in}p{5mm}p{5in}|}
\hline
``aov'' $\vert$ ``ls'' $\vert$ ``list'' $\vert$ ``fix'' $\vert$ ``fixcolumn'' $\vert$ ``expr'' & \tabitem & Source for the period at which to calculate the BLS model (either from the last -aov command, the last -LS command, from the input light curve list file, fixed to a value given on the command line, set to the output from a previously executed command, or determined by evaluating an analytic expression). \\ \hline
``r'' & \tabitem & Allow for a period dependent limit on the transit durations to search. $r$ corresponds approximately to the stellar radius for lower main sequence stars assuming central transits, circular orbits, and that times are specified in days. \\ 
~~~~rmin & \tabitem & The minimum transit duration at period $P$ will be $0.076 {\rm rmin}^{2/3} P^{1/3}$. \\
~~~~rmax & \tabitem & The maximum transit duration at period $P$ will be $0.076 {\rm rmax}^{2/3} P^{1/3}$. \\ \hline
``q'' & \tabitem & Use a period-independent limit on the fractional transit durations considered. \\ 
~~~~qmin & \tabitem & The minimum transit duration at period $P$ will be ${\rm qmin} P$. \\
~~~~qmax & \tabitem & The maximum transit duration at period $P$ will be ${\rm qmax} P$. \\ \hline
nbins & \tabitem & The number of phase bins to use. \\ \hline
timezone & \tabitem & The timezone of the observatory (in hours from UTC). This only affects the grouping of points used to determine the fraction of $\Delta \chi^{2}$ from a single day. If multiple observatories were used, or if the observations were performed from space one can simply provide $0$ for this value and ignore the reported $\Delta \chi^{2}$ from one day. \\ \hline
omodel & \tabitem & A flag used to indicate whether or not the best-fit transit model light curves should be output. These light curves are evaluated at the observed times. \\
~~~~model\_outdir & \tabitem & Required if omodel is set to $1$. The models files will be written to \$model\_outdir/\$basename.blsfixper.model. \\ \hline
correctlc & \tabitem & A flag used to indicate whether or not the best-fit transit model should be subtracted from the observations before passing the light curve on to the next command. \\ \hline
``fittrap'' & \tabitem & An optional keyword which, if used, causes a trapezoid-shape transit model to be fit to the light curve around each BLS peak identified. \\ \hline
\end{xtabular}
\end{center}

\begin{lstlisting}[backgroundcolor=\color{white},language=vartools,basicstyle=\footnotesize,breaklines=false,prebreak={}]
-BLSFixDurTc
    <"duration" <"fix" dur | "fixcolumn" <colname | colnum>
        | "list" ["column" col]>>
    <"Tc" <"fix" Tc | "fixcolumn" <colname | colnum>
        | "list" ["column" col]>>
    ["fixdepth" <"fix" depth | "fixcolumn" <colname | colnum>
        | "list" ["column" col]>
        ["qgress" <"fix" qgress | "fixcolumn" <colname | colnum>
            | "list" ["column" col]>]]
    minper maxper nfreq timezone
    Npeak outperiodogram [outdir] omodel [model_outdir]
    correctlc ["fittrap"]
    ["ophcurve" outdir phmin phmax phstep]
    ["ojdcurve" outdir jdstep]
\end{lstlisting}

\begin{center}
\tablecaption{Input Parameters for ``-BLSFixDurTc'' Command (Section~\ref{cmd:blsfixdurtc})}
\footnotesize
\begin{xtabular}{|p{1in}p{5mm}p{5in}|}
\hline
``duration'' & \tabitem & Indicate how the fixed transit duration (in time-units of the light curve) should be determined. \\ \hline
``Tc'' & \tabitem & Indicate how the fixed transit epoch (in time-units of the light curve) should be determined. \\ \hline
``fixdepth'' & \tabitem & Optional keyword used to indicate that the transit depth should be fixed (by default it is allowed to vary). If this keyword is used, then one should follow it with an indication of how the fixed depth should be determined. \\ \hline
``qgress'' & \tabitem & Optional keyword used to indicate that the fractional duration of transit ingress (i.e., the ingress duration divided by the total transit duration) should be fixed (by default it is allowed to vary). This keyword may only be used if ``fixdepth'' is also used. One should follow the keyword with an indication of how the fixed fractional ingress duration should be determined. \\ \hline
minper & \tabitem & The minimum period to search for transits. \\ \hline
maxper & \tabitem & The maximum period to search for transits. \\ \hline
nfreq & \tabitem & The number of frequencies to search (the frequency step-size is $\Delta f = (1/{\rm minper} - 1/{\rm maxper})/{\rm nfreq}$). \\ \hline
timezone & \tabitem & The timezone of the observatory (in hours from UTC). This only affects the grouping of points used to determine the fraction of $\Delta \chi^{2}$ from a single day. If multiple observatories were used, or if the observations were performed from space one can simply provide $0$ for this value and ignore the reported $\Delta \chi^{2}$ from one day. \\ \hline
Npeak & \tabitem & The number of peaks to identify in the BLS spectrum. \\ \hline
outperiodogram & \tabitem & A flag used to indicate whether or not the BLS spectra should be output. \\
~~~~outdir & \tabitem & Required if outperiodogram is set to $1$. The BLS spectra will be written to files with the name \$outdir/\$basename.blsfixdurtc \$outdir is the value specified for outdir, and \$basename is the base filename of the light curve (any leading directory is stripped). \\ \hline
omodel & \tabitem & A flag used to indicate whether or not the best-fit transit model light curves should be output. These light curves are evaluated at the observed times. \\
~~~~model\_outdir & \tabitem & Required if omodel is set to $1$. The models files will be written to \$model\_outdir/\$basename.blsfixdurtc.model. \\ \hline
correctlc & \tabitem & A flag used to indicate whether or not the best-fit transit model should be subtracted from the observations before passing the light curve on to the next command. \\ \hline
``fittrap'' & \tabitem & An optional keyword which, if used, causes a trapezoid-shape transit model to be fit to the light curve around each BLS peak identified. \\ \hline
``ophcurve'' & \tabitem & A keyword used to output best-fit model light curves that are uniformly sampled in phase. \\
~~~~outdir & \tabitem & The models will be written out to files named \$outdir/\$basename.blsfixdurtc.phcurve. \\
~~~~phmin & \tabitem & The starting phase to use in the model. \\
~~~~phmax & \tabitem & The ending phase to use in the model. \\
~~~~phstep & \tabitem & The phase step-size to use in the model. \\ \hline
``ojdcurve'' & \tabitem & A keyword used to output best-fit model light curves that are uniformly sampled in time between the first and last observed times in the light curve. \\
~~~~outdir & \tabitem & The models will be written out to files named \$outdir/\$basename.blsfixdurtc.jdcurve. \\
~~~~jdstep & \tabitem & The time step-size to use in the model. \\ \hline
\end{xtabular}
\end{center}

\begin{lstlisting}[backgroundcolor=\color{white},language=vartools,basicstyle=\footnotesize,breaklines=false,prebreak={}]
-changeerror
\end{lstlisting}

No parameters or options are available.

\begin{lstlisting}[backgroundcolor=\color{white},language=vartools,basicstyle=\footnotesize,breaklines=false,prebreak={}]
-changevariable <"t" | "mag" | "err" | "id"> var
\end{lstlisting}

\begin{center}
\tablecaption{Input Parameters for ``-changevariable'' Command (Section~\ref{cmd:changevariable})}
\footnotesize
\begin{xtabular}{|p{1in}p{5mm}p{5in}|}
\hline
``t'' & \tabitem & Change the variable used for the light curve times in subsequent commands. \\ \hline
``mag'' & \tabitem & Change the variable used for the light curve magnitude values in subsequent commands. \\ \hline
``err'' & \tabitem & Change the variable used for the light curve uncertainties in subsequent commands. \\ \hline
``id'' & \tabitem & Change the variable used for the light curve image-ids in subsequent commands. \\ \hline
var & \tabitem & The name of the new variable to use for the indicated quantity. \\ \hline
\end{xtabular}
\end{center}

\begin{lstlisting}[backgroundcolor=\color{white},language=vartools,basicstyle=\footnotesize,breaklines=false,prebreak={}]
-chi2
\end{lstlisting}

No parameters or options are available.

\begin{lstlisting}[backgroundcolor=\color{white},language=vartools,basicstyle=\footnotesize,breaklines=false,prebreak={}]
-chi2bin Nbin bintime1...bintimeN
\end{lstlisting}

\begin{center}
\tablecaption{Input Parameters for ``-chi2bin'' Command (Section~\ref{cmd:chi2bin})}
\footnotesize
\begin{xtabular}{|p{1in}p{5mm}p{5in}|}
\hline
Nbin & \tabitem & The number of moving mean filters to use. \\ \hline
bintime1...bintimeN & \tabitem & A space-delimited list of filter widths, one for each of the Nbin filters used. The width of each filter is given by $2.0*{\rm bintime}$, where ${\rm bintime}$ is in minutes, assuming the times in the light curve are in days. \\ \hline
\end{xtabular}
\end{center}

\begin{lstlisting}[backgroundcolor=\color{white},language=vartools,basicstyle=\footnotesize,breaklines=false,prebreak={}]
-clip sigclip iter ["niter" n] ["median"]
\end{lstlisting}

\begin{center}
\tablecaption{Input Parameters for ``-clip'' Command (Section~\ref{cmd:clip})}
\footnotesize
\begin{xtabular}{|p{1in}p{5mm}p{5in}|}
\hline
sigclip & \tabitem & The $\sigma$-clipping factor to use. If a value $<=0$ is specified, then $\sigma$-clipping is not performed, but points with errors $<=0$ or NaN magnitude values will be clipped from the light curve. \\ \hline
iter & \tabitem & A flag that is 1 for iterative clipping (performed continuously until no further points are removed), or 0 to not do continuous iterative clipping. \\ \hline
``niter'' & \tabitem & An optional keyword used to specify a fixed number of iterations to use in performing the clipping. \\ \hline
``median'' & \tabitem & By default clipping is done with respect to the mean. Use this keyword to cause the clipping to be done instead with respect to the median. \\ \hline
\end{xtabular}
\end{center}

\begin{lstlisting}[backgroundcolor=\color{white},language=vartools,basicstyle=\footnotesize,breaklines=false,prebreak={}]
-converttime
    <"input" <"mjd" | "jd" | "hjd" | "bjd" >>
    ["inputsubtract" value] ["inputsys-tdb" | "inputsys-utc"]
    <"output" <"mjd" | "jd" | "hjd" | "bjd" >>
    ["outputsubtract" value] ["outputsys-tdb" | "outputsys-utc"]
    ["radec" <"list" ["column" col] | "fix" raval decval>
    ["epoch" epoch]]
    ["ppm" <"list" ["column" col] | "fix" mu_ra mu_dec>]
    ["input-radec" <"list" ["column" col] | "fix" raval decval>
    ["epoch" epoch]]
    ["input-ppm" <"list" ["column" col] | "fix" mu_ra mu_dec>]
    ["ephemfile" file] ["leapsecfile" file] ["planetdatafile" file]
    ["observatory" < code | "show-codes">
        | "coords"
            <"fix" latitude[deg] longitude[deg_E] altitude[m]
            | "list" ["column" collat collong colalt]
            | "fromlc" collat collong colalt>]
\end{lstlisting}

\begin{center}
\tablecaption{Input Parameters for ``-converttime'' Command (Section~\ref{cmd:converttime})}
\footnotesize
\begin{xtabular}{|p{1in}p{5mm}p{5in}|}
\hline
``input'' & \tabitem & Indicate the input time system used. The required ``input'' keyword must be followed by one of the following keywords. \\
~~~~``mjd'' & \tabitem & Modified Julian Date (${\rm MJD} = {\rm JD} - 2400000.5$). \\
~~~~``jd'' & \tabitem & Julian Date. \\
~~~~``hjd'' & \tabitem & Helio-centric Julian Date. \\
~~~~``bjd'' & \tabitem & Bary-centric Julian Date. \\ \hline
``inputsubtract'' & \tabitem & Optional keyword used to indicate a constant that has been subtracted from the input times (i.e., if the input time is ${\rm HJD}-2400000$, one would give ``input hjd inputsubtract 2400000'' on the command-line). \\ \hline
``inputsys-tdb'' & \tabitem & Optional keyword used to indicate that the input times have been corrected for leap-seconds. \\ \hline
``inputsys-utc'' & \tabitem & Optional keyword used to indicate that the input times have not been corrected for leap-seconds (this is assumed by default). \\ \hline
``output'' & \tabitem & Indicate the output time system used. The required ``output'' keyword must be followed by a keyword indicating the time system, options are the same as for ``input''. \\
``outputsubtract'' & \tabitem & Optional keyword used to indicate a constant that is to be subtracted from the output times. \\ \hline
``outputsys-tdb'' & \tabitem & Optional keyword used to indicate that the output times should be on a system corrected for leap-seconds. \\ \hline
``outputsys-utc'' & \tabitem & Optional keyword used to indicate that the output times should be on a system that is not corrected for leap-seconds (by default the same system as the input times is assumed). \\ \hline
``radec'' & \tabitem & Indicate the source for the RA and Dec coordinates of the source. This is required if converting to/from HJD or BJD. Both coordinates should be in decimal degrees. \\
~~~~``epoch'' & \tabitem & Optionally specify a time epoch in years on the Common Era system for which the RA and Dec coordinates are provided. The default is $2000.0$. \\ \hline
``ppm'' & \tabitem & Optionally indicate the source for the RA and Dec proper motions of the source. These should be given in units of milliarcseconds per year. By default sources are assumed to have no proper motion. \\ \hline
``input-radec'' & \tabitem & Optionally specify a different soruce for the RA and Dec coordinates that were assumed in calculating HJD or BJD for the input times. This may be useful, for example, if the times were converted to HJD or BJD assuming the coordinates for the center of an image, but one wishes now to correct them to the coordinates of the source itself. \\
~~~~``epoch'' & \tabitem & Optionally specify a time epoch in years on the Common Era system for which the input RA and Dec coordinates are provided. The default is $2000.0$. \\ \hline
``input-ppm'' & \tabitem & Optionally indicate the source for the input RA and Dec proper motions assumed for the source. These should be given in units of milliarcseconds per year. By the initial times are assumed to have been calculated with no proper motion, or with the proper motion specified using the ``ppm'' keyword. \\ \hline
``ephemfile'' & \tabitem & Provide the filename for the JPL NAIF ephemeris file used to determine the position of the Earth with respect to the Solar System Barycenter. If not specified \va{} will check if the environment variable CSPICE\_EPHEM\_FILE has been set, and use it if it has been set. \\
``leapsecfile'' & \tabitem & Provide the filename for the JPL NAIF leap-second file used to determine the number of leap-seconds since a reference epoch. This can also be set with the environment variable CSPICE\_LEAPSEC\_FILE. \\
``planetdatafile'' & \tabitem & Provide the filename for the JPL NAIF planetary physical data file used to determine the location of the observer with respect to the center of the Earth in the J2000.0 intertial frame. This can also be set with the environment variable CSPICE\_PLANETDATA\_FILE. \\
``observatory'' & \tabitem & Optionally provide this keyword to specify an observatory used to make the observations. If this is set then one should either give: \\
~~~~code & \tabitem & A short string used to indicate the observatory, e.g., ``maunakea'' for Mauna Kea Observatory, Hawaii, ``lasilla'' for La Silla Observatory, etc. \\
~~~~``show-codes'' & \tabitem & If this keyword is given, then the list of allowed observatory codes will be output, and \va{} will not perform any processing. \\ \hline
``coords'' & \tabitem & Optionally provide the latitude (in degrees), longitude (in degrees east of the prime meridian), and altitude (in meters) of the observatory at which the measurements were performed. One must then indicate the source for these quantities. In addition to the standard ``fix'' and ``list'' keywords, one may also use the ``fromlc'' keyword to read these quantities as columns in the light curve. This may be used if observations carried out from different locations were combined into a single light curve file. \\ \hline
\end{xtabular}
\end{center}

\begin{lstlisting}[backgroundcolor=\color{white},language=vartools,basicstyle=\footnotesize,breaklines=false,prebreak={}]
-copylc Ncopies
\end{lstlisting}

\begin{center}
\tablecaption{Input Parameters for ``-copylc'' Command (Section~\ref{cmd:copylc})}
\footnotesize
\begin{xtabular}{|p{1in}p{5mm}p{5in}|}
\hline
Ncopies & \tabitem & The number of light curve copies to make. \\ \hline
\end{xtabular}
\end{center}

\begin{lstlisting}[backgroundcolor=\color{white},language=vartools,basicstyle=\footnotesize,breaklines=false,prebreak={}]
-decorr correctlc zeropointterm subtractfirstterm Nglobalterms globalfile1
    order1 ... Nlcterms lccolumn1 lcorder1 ... omodel
    [model_outdir]
\end{lstlisting}

\begin{center}
\tablecaption{Input Parameters for ``-decorr'' Command (Section~\ref{cmd:decorr})}
\footnotesize
\begin{xtabular}{|p{1in}p{5mm}p{5in}|}
\hline
correctlc & \tabitem & A flag indicating whether or not the light curves passed on to the next command will be decorrelated. If it is set to $0$ then the value of $\chi^2$ from the decorrelation together with the decorrelation coefficients will be output to the table, but the light curves themselves will not be affected by the command. \\ \hline
zeropointterm & \tabitem & A flag indicating whether or not a zero-point offset should be included in the fit. \\ \hline
subtractfirstterm & \tabitem & A flag indicating whether or not the first terms in the parameter sequences used in the decorrelation should be subtracted. This may be useful, for example, if one is decorrelating against the JD, in which case one can set subtractfirstterm to $1$ to use $JD-JD0$ to prevent accumulating large round-off errors. \\ \hline
Nglobalterms & \tabitem & The number of global parameter sequences to read in from separate files. The two terms below must be repeated for each of the Nglobalterms files. \\ 
~~~~globalfile1 & \tabitem & The filename for the first global parameter sequencee to decorrelate against. The file should have the format: JD signal\_value, or ID signal\_value if the -matchstringid option to vartools has been set. \\
~~~~order1 & \tabitem & The order of the polynomial to use in decorrelating against this sequence. \\ \hline
Nlcterms & \tabitem & The number of sequences read from each light curve to use in the decorrelation. The two terms below must be repeated Nlcterms times. \\
~~~~lccolumn1 & \tabitem & The column in the light curve to use for the first parameter sequence. \\
~~~~lcorder1 & \tabitem & The order of the polynomial to use in decorrelating against this sequence. \\ \hline
omodel & \tabitem & A flag indicating whether or not the model decorrelation light curves should be output. \\
~~~~model\_outdir & \tabitem & This parameter is required if omodel is set to 1. The model light curves will be written to files named \$model\_outdir/\$basename.decorr.model where \$model\_outdir is the value supplied for model\_outdir, and \$basename is the base filename of the light curve (i.e., the name stripped of any directories). \\ \hline
\end{xtabular}
\end{center}

\begin{lstlisting}[backgroundcolor=\color{white},language=vartools,basicstyle=\footnotesize,breaklines=false,prebreak={}]
-dftclean nbeam ["maxfreq maxf"] ["outdspec" dspec_outdir]
    ["finddirtypeaks" Npeaks ["clip" clip clipiter]]
    ["outwfunc" wfunc_outdir]
    ["clean" gain SNlimit ["outcbeam" cbeam_outdir]
    ["outcspec" cspec_outdir]
    ["findcleanpeaks" Npeaks ["clip" clip clipiter]]]
    ["useampspec"] ["verboseout"]
\end{lstlisting}

\begin{center}
\tablecaption{Input Parameters for ``-dftclean'' Command (Section~\ref{cmd:dftclean})}
\footnotesize
\begin{xtabular}{|p{1in}p{5mm}p{5in}|}
\hline
nbeam & \tabitem & The number of points per $1/T$ frequency element to include in the calculated power spectrum, where $T$ is the time base-line of the input light curve. \\ \hline
``maxfreq'' & \tabitem & An optional keyword which should be followed by the maximum frequency (in cycles per day) at which to calculate the power spectrum. If not specified, the power spectrum will be calculated up to the Nyquist frequency. \\ \hline
``outdspec'' & \tabitem & An optional keyword used to output the raw Discrete Fourier Transform (i.e., before applying the CLEAN algorithm) to a file. \\
~~~~dspec\_outdir & \tabitem & Required if the ``outdspec'' keyword is set. The raw DFT will be output to a file named \$dspec\_outdir/\$basename.dftclean.dspec where \$basename is the base filename of the input light curve (stripped of any leading directory names). \\ \hline
``finddirtypeaks'' & \tabitem & An optional keyword used to search the raw DFT for peaks. \\
~~~~Npeaks & \tabitem & The number of peaks to report from the raw DFT power spectrum. This parameter is required if the ``finddirtypeaks'' keyword is given. \\
~~~~``clip'' & \tabitem & Optional keyword to change the clipping method used in calculating the average and standard deviation of the raw power spectrum for determining the spectroscopic S/N of each peak. If this keyword is given then follow it with the $\sigma$-clipping factor and a flag to indicate whether or not continuous iterative clipping should be performed. By default iterative 5$\sigma$ clipping is performed. \\ \hline
``outwfunc'' & \tabitem & An optional keyword used to output the window function to the file \$wfunc\_outdir/\$basename.dftclean.wfunc. \\ \hline
``clean'' & \tabitem & If this keyword is specified then the CLEAN deconvolution algorithm will be executed on the power spectrum. \\
~~~~gain & \tabitem & The gain factor used in cleaning the power spectrum. It should have a value between 0.1 and 1. \\
~~~~SNlimit & \tabitem & Cleaning will proceed until the last peak has a value that is less than SNlimit times the standard deviation. \\
~~~~outcbeam & \tabitem & Optional keyword to output the ``clean beam'' used in producing the final power spectrum. This will be output to the file named \$cbeam\_outdir/\$basename.dftclean.cbeam. \\
~~~~findcleanpeaks & \tabitem & An optional keyword used to search the cleaned DFT for peaks. It takes the same additional parameters as ``finddirtypeaks''. \\ \hline
``useampspec'' & \tabitem & Give this keyword to use the amplitude spectrum for calculating the S/N of peaks rather than the power spectrum, which is used by default. \\ \hline
``verboseout'' & \tabitem & Give this keyword to output the average and standard deviation of the spectrum before and after clipping, in additional to the final S/N value. \\ \hline
\end{xtabular}
\end{center}

\begin{lstlisting}[backgroundcolor=\color{white},language=vartools,basicstyle=\footnotesize,breaklines=false,prebreak={}]
-difffluxtomag mag_constant offset ["magcolumn" col]
\end{lstlisting}

\begin{center}
\tablecaption{Input Parameters for ``-difffluxtomag'' Command (Section~\ref{cmd:difffluxtomag})}
\footnotesize
\begin{xtabular}{|p{1in}p{5mm}p{5in}|}
\hline
mag\_constant & \tabitem & The magnitude of a source with a flux of 1\,ADU. \\ \hline
offset & \tabitem & An additive constant to apply to the output light curves. \\ \hline
``magcolumn'' & \tabitem & An optional keyword used to specify the column from the input light curve list with the reference magnitude of the source. If not given, then the next available column in the light curve list file will be used. \\ \hline
\end{xtabular}
\end{center}

\begin{lstlisting}[backgroundcolor=\color{white},language=vartools,basicstyle=\footnotesize,breaklines=false,prebreak={}]
-ensemblerescalesig sigclip
\end{lstlisting}

\begin{center}
\tablecaption{Input Parameters for ``-ensemblerescalesig'' Command (Section~\ref{cmd:ensemblerescalesig})}
\footnotesize
\begin{xtabular}{|p{1in}p{5mm}p{5in}|}
\hline
sigclip & \tabitem & $\sigma$-clipping factor used in clipping outliers from the ${\rm r.m.s.}_{\rm expected}^2$ vs.~$(\chi^{2}/{\rm d.o.f.}){\rm r.m.s.}_{\rm expected}^2$ distribution in determining the light curve uncertainty rescaling factor. \\ \hline
\end{xtabular}
\end{center}

\begin{lstlisting}[backgroundcolor=\color{white},language=vartools,basicstyle=\footnotesize,breaklines=false,prebreak={}]
-expr var"="expression
\end{lstlisting}

\begin{center}
\tablecaption{Input Parameters for ``-expr'' Command (Section~\ref{cmd:expr})}
\footnotesize
\begin{xtabular}{|p{1in}p{5mm}p{5in}|}
\hline
var''=''expression & \tabitem & Set the variable on the left hand side of the expression equal to the result of evaluating the analytic expression on the right hand side. \\ \hline
\end{xtabular}
\end{center}

\begin{lstlisting}[backgroundcolor=\color{white},language=vartools,basicstyle=\footnotesize,breaklines=false,prebreak={}]
-findblends matchrad ["radec"]
    ["xycol" xcol ycol]
    <"fix" period | "list" ["column" col]
        | "fixcolumn" <colname | colnum>>
    ["starlist" starlistfile] ["zeromag" zeromagval] ["nofluxconvert"]
    ["Nharm" Nharm] ["omatches" outputmatchfile]
\end{lstlisting}

\begin{center}
\tablecaption{Input Parameters for ``-findblends'' Command (Section~\ref{cmd:findblends})}
\footnotesize
\begin{xtabular}{|p{1in}p{5mm}p{5in}|}
\hline
matchrad & \tabitem & Matching radius used in determining if a given light curve might be blended with a source from the star list file. \\ \hline
``radec'' & \tabitem & If this keyword is specified then matchrad is assumed to be in arcseconds and the X and Y positions of the sources are RA and Dec.~in decimal degrees. If this keyword is not given, then rectangular matching will be performed on the X and Y coordinates, and matchrad should be in the same units. \\ \hline
``xycol'' & \tabitem & An optional keyword for specifying the columns from the input light curve list file to use for determining the X and Y coordinates of the each of the light curves. If not given, then the next unused columns in the list file will be assumed. \\ \hline
``fix'' $\vert$ ``list'' $\vert$ ``fixcolumn'' & \tabitem & Source for the period to use for the light curve. \\ \hline
``starlist'' & \tabitem & By default the input light curve list is matched to itself. Use this keyword to match instead to the list in the file starlistfile. The file should have 3 white-space delimited columns, the first being a string identifier, the second and third being the X and Y coordinates of the source.\\ \hline
``zeromag'' & \tabitem & Optional keyword for specifying the zero-point magnitude for converting from magnitudes into fluxes. The default value is 25.0. \\ \hline
``nofluxconvert'' & \tabitem & If this optional keyword is given then the magnitude to flux conversion is not performed. \\ \hline
``Nharm'' & \tabitem & Optional keyword to change the number of harmonics to include in the Fourier series fit to each light curve to determine its amplitude of variability. If it is 0 then only a sinusoid will be fit. The default value is 2. \\ \hline
``omatches'' & \tabitem & Optional keyword to output the names and flux amplitudes of all stars matching to each potential variable. outputmatchfile is the name of the file to output this information to. \\ \hline
\end{xtabular}
\end{center}

\begin{lstlisting}[backgroundcolor=\color{white},language=vartools,basicstyle=\footnotesize,breaklines=false,prebreak={}]
-fluxtomag mag_constant offset
\end{lstlisting}

\begin{center}
\tablecaption{Input Parameters for ``-fluxtomag'' Command (Section~\ref{cmd:fluxtomag})}
\footnotesize
\begin{xtabular}{|p{1in}p{5mm}p{5in}|}
\hline
mag\_constant & \tabitem & The magnitude of a source with a flux of 1 ADU. \\ \hline
offset & \tabitem & A constant to add to the output light curve magnitudes. \\ \hline
\end{xtabular}
\end{center}

\begin{lstlisting}[backgroundcolor=\color{white},language=vartools,basicstyle=\footnotesize,breaklines=false,prebreak={}]
-GetLSAmpThresh <"ls" | "list" ["column" col]> minp thresh
    <"harm" Nharm Nsubharm | "file" listfile> ["noGLS"]
\end{lstlisting}

\begin{center}
\tablecaption{Input Parameters for ``-GetLSAmpThresh'' Command (Section~\ref{cmd:getlsampthresh})}
\footnotesize
\begin{xtabular}{|p{1in}p{5mm}p{5in}|}
\hline
``ls'' $\vert$ ``list'' & \tabitem & The source for the period for which to calculate the minimally detectable amplitude. If the keyword ``ls'' is given then the source is the highest peak found by the most recent -LS command. If it is ``list'' then the period is read-in from the input light curve list. \\ \hline
minp & \tabitem & The minimum period considered by the L-S run. This is needed for determining the false alarm probability, which depends on the period range scanned. \\ \hline
thresh & \tabitem & The maximum value of $\log_{10}{\rm FAP}$ that the light curve have had and still have been considered a detection. \\ \hline
``harm'' & \tabitem & Use this keyword to fit a harmonic series to the light curve to determine its amplitude. \\
~~~~Nharm & \tabitem & The number of harmonics to include in the series. \\
~~~~Nsubharm & \tabitem & The number of sub-harmonics to include in the series. \\ \hline
``file'' & \tabitem & Use this keyword to read-in the model signal from a file (in general a separate file is used for each light curve). \\
~~~~listfile & \tabitem & The name of a file listing the files containing the model signals for each of the light curves. This list file has two columns of the form: signal\_file signal\_amp, with one line for each light curve being processed. Each signal\_file should contain the signal magnitude in the third column. signal\_amp is the amplitude in magnitudes of the signal (to allow for cases where the signal amplitude is greater than the difference between the minimum and maximum values in the file). \\ \hline
``noGLS'' & \tabitem & If this optional keyword is given then the traditional L-S periodogram is used for determining the false alarm probability of a signal. By default the Generalized L-S is used. \\ \hline
\end{xtabular}
\end{center}

\begin{lstlisting}[backgroundcolor=\color{white},language=vartools,basicstyle=\footnotesize,breaklines=false,prebreak={}]
-if <expression> [-command1 ... -commandN]
    [-elif <expression> [-command1 ... -commandN]]
    ...
    [-elif <expression> [-command1 ... -commandN]]
    [-else [-command1 ... -commandN]]
    -fi
\end{lstlisting}

\begin{center}
\tablecaption{Input Parameters for ``-if'' Command (Section~\ref{cmd:if})}
\footnotesize
\begin{xtabular}{|p{1in}p{5mm}p{5in}|}
\hline
expression & \tabitem & An analytic expression evaluated for a given light curve. If this evaulates to a number different from 0 the expression will be treated as ``true.'' \\ \hline
-command1 ... -commandN & \tabitem & A set of commands that will be executed conditional upon testing the expression. \\ \hline
\end{xtabular}
\end{center}

\begin{lstlisting}[backgroundcolor=\color{white},language=vartools,basicstyle=\footnotesize,breaklines=false,prebreak={}]
-Injectharm <"list" ["column" col] | "fix" per
    | "rand" minp maxp
    | "logrand" minp maxp | "randfreq" minf maxf
    | "lograndfreq" minf maxf>
    Nharm (<"amplist" ["column" col]
    | "ampfix" amp | "amprand" minamp maxamp 
    | "amplogrand" minamp maxamp> ["amprel"]
    <"phaselist" ["column" col]
    | "phasefix" phase | "phaserand"> ["phaserel"])0...Nharm Nsubharm
    (<"amplist" ["column" col] | "ampfix" amp
    | "amprand" minamp maxamp 
    | "amplogrand" minamp maxamp> ["amprel"]
    <"phaselist" ["column" col]
    | "phasefix" phase | "phaserand"> ["phaserel"])1...Nsubharm
    omodel [modeloutdir]
\end{lstlisting}

\begin{center}
\tablecaption{Input Parameters for ``-Injectharm'' Command (Section~\ref{cmd:injectharm})}
\footnotesize
\begin{xtabular}{|p{1in}p{5mm}p{5in}|}
\hline
``list'' & \tabitem & Take the period of the harmonic signal to add to the light curve from the input light curve list file, optionally specifying the column number. \\ \hline
``fix'' & \tabitem & Fix the period to the specified value. \\ \hline
``rand'' & \tabitem & Choose a random number for the period between minp and maxp. \\ \hline
``logrand'' & \tabitem & Choose a random number for the period from a distribution that is uniform in logarithm between minp and maxp. \\ \hline
``randfreq'' & \tabitem & Choose a random number for the period from a distribution that is uniform in frequency between minf and maxf (in cycles per day). \\ \hline
``lograndfreq'' & \tabitem & Choose a random number for the period from a distribution that is uniform in the logarithm of the frequency between minf and maxf (in cycles per day). \\ \hline
Nharm & \tabitem & The number of harmonics to include in the harmonic series. If this is 0 then only the fundamental mode is included (i.e., a simple sine curve). For each of the ${\rm Nharm}+1$ modes one must specify the source for the amplitude and phase. This done with the following keywords: \\
~~~~``amplist'' & \tabitem & Take the amplitude from the input light curve list file. Optionally specify the column to use with the ``column'' keyword. \\
~~~~``ampfix'' & \tabitem & Fix the amplitude to the specified value. \\
~~~~``amprand'' & \tabitem & Use a random number drawn from a uniform distribution between minamp and maxamp. \\
~~~~``amplogrand'' & \tabitem & Use a random number drawn from a uniform-log distribution between minamp and maxamp. \\
~~~~``amprel'' & \tabitem & If specified, amplitudes for harmonics 1 through ${\rm Nharm}$ are given relative to the amplitude of the fundamental mode. \\
~~~~``phaselist'' & \tabitem & Take the phase from the input light curve list file. Optionally specify the column to use with the ``column'' keyword. \\
~~~~``phasefix'' & \tabitem & Fix the phase to the specified value. \\
~~~~``phaserand'' & \tabitem & Draw a random number between 0 and 1. \\
~~~~``phaserel'' & \tabitem & If specified, phases for harmonics 1 through ${\rm Nharm}$ are given relative to the fundamental mode. \\ \hline
Nsubharm & \tabitem & The number of subharmonics to use. For each of the ${\rm Nsubharm}$ subharmonics one must specify the source for the amplitude and phase as above, the ``amprel'' and ``phaserel'' keywords may also be provided to make the amplitude and/or phase relative to the values for the fundamental mode. \\ \hline
omodel & \tabitem & Flag indicating whether or not to output the model harmonic series evaluated at the times of observation in the light curve. \\ 
~~~~modeloutdir & \tabitem & This parameter is required if the omodel flag is set. The model will be output to the file \$modeloutdir/\$basename.injectharm.model where \$basename is the base filename of the light curve (with directory names stripped). \\ \hline
\end{xtabular}
\end{center}

\begin{lstlisting}[backgroundcolor=\color{white},language=vartools,basicstyle=\footnotesize,breaklines=false,prebreak={}]
-Injecttransit <"Plist" ["column" col] | "Pfix" per
        | "Pexpr" expr | "Prand" minp maxp
        | "Plogrand" minp maxp | "randfreq" minf maxf
        | "lograndfreq" minf maxf>
    <"Rplist" ["column" col] | "Rpfix" Rp | "Rpexpr" expr
        | "Rprand" minRp maxRp | "Rplogrand" minRp maxRp>
    <"Mplist" ["column" col] | "Mpfix" Mp | "Mpexpr" expr
        | "Mprand" minMp maxMp | "Mplogrand" minMp maxMp>
    <"phaselist" ["column" col] | "phasefix" phase
        | "phasexpr" expr | "phaserand>
    <"sinilist" ["column" col] | "sinifix" sin_i
        | "siniexpr" expr | "sinirand">
    <"eomega" <"elist" ["column" col] | "efix" e | "eexpr" expr | "erand">
        <"olist" ["column" col] | "ofix" omega | "oexpr" expr | "orand">
    | "hk" <"hlist" ["column" col] | "hfix" h | "hexpr" expr | "hrand">
        <"klist" ["column" col] | "kfix" k | "kexpr" expr | "krand">>
    <"Mstarlist" ["column" col] | "Mstarfix" Mstar | "Mstarexpr" expr>
    <"Rstarlist" ["column" col] | "Rstarfix" Rstar | "Rstarexpr" expr>
    <"quad" | "nonlin"> <"ldlist" ["column" col]
    | "ldfix" ld1 ... ldn | "ldexpr" ld1 ... ldn>
    ["dilute" <"list" ["column" col] | "fix" dilute | "expr" diluteexpr>]
     omodel [modeloutdir]
\end{lstlisting}

\begin{center}
\tablecaption{Input Parameters for ``-Injecttransit'' Command (Section~\ref{cmd:injecttransit})}
\footnotesize
\begin{xtabular}{|p{1in}p{5mm}p{5in}|}
\hline
``Plist'' & \tabitem & Take the period of the injected transit from the input light curve list file, optionally specifying the column number. \\ 
``Pfix'' & \tabitem & Fix the period of the injected transit to the specified value. \\ 
``Pexpr'' & \tabitem & Evaluate an analytic expression for each light curve to determine the period. \\
``Prand'' & \tabitem & Set the period of the inject transit to a random number drawn from a uniform distribution between minp and maxp. \\ 
``Plogrand'' & \tabitem & Choose a random number for the period from a distribution that is uniform in logarithm between minp and maxp. \\ 
``randfreq'' & \tabitem & Choose a random number for the period from a distribution that is uniform in frequency between minf and maxf (in cycles per day). \\ 
``lograndfreq'' & \tabitem & Choose a random number for the period from a distribution that is uniform in the logarithm of the frequency between minf and maxf (in cycles per day). \\ \hline
``Rplist'' & \tabitem & Take the radius of the planet (in Jupiter radius units) from the input light curve list file, optionally specifying the column number. \\ 
``Rpfix'' & \tabitem & Fix the radius of the planet to the specified value. \\ 
``Rpexpr'' & \tabitem & Evaluate an analytic expression for each light curve to determine the planet radius. \\
``Rprand'' & \tabitem & Set the radius of the planet to a random number drawn from a uniform distribution between minRp and maxRp. \\ 
``Rplogrand'' & \tabitem & Set the radius of the planet to a random number drawn from a uniform-log distribution between minRp and maxRp. \\ \hline
``Mplist'' & \tabitem & Take the mass of the planet (in Jupiter mass units) from the input light curve list file, optionally specifying the column number. \\ 
``Mpfix'' & \tabitem & Fix the mass of the planet to the specified value. \\ 
``Pexpr'' & \tabitem & Evaluate an analytic expression for each light curve to determine the planet mass. \\
``Mprand'' & \tabitem & Set the mass of the planet to a random number drawn from a uniform distribution between minMp and maxMp. \\ 
``Mplogrand'' & \tabitem & Set the mass of the planet to a random number drawn from a uniform-log distribution between minMp and maxMp. \\ \hline
``phaselist'' & \tabitem & Take the phase of the orbit (phase$=0$ corresponds to transit center) at time $T=0$ from the input light curve list file, optionally specifying the column number. \\ 
``phasefix'' & \tabitem & Fix the phase at $T=0$ to the specified value. \\ 
``phaseexpr'' & \tabitem & Evaluate an analytic expression for each light curve to determine the phase. \\
``phaserand'' & \tabitem & Set the phase at $T=0$ to a random number between 0 and 1. \\ \hline
``sinilist'' & \tabitem & Take the $\sin(i)$, where $i$ is the inclination of the orbit, from the input light curve list file, optionally specifying the column number. \\ 
``sinifix'' & \tabitem & Fix $\sin(i)$ to the specified value. \\ 
``siniexpr'' & \tabitem & Evaluate an analytic expression for each light curve to determine $\sin(i)$. \\
``sinirand'' & \tabitem & Choose a random number for $\sin(i)$ drawn from a uniform orientation distribution with the constraint that there must be a transit. \\ \hline
``eomega'' & \tabitem & Keyword to indicate that the eccentricity and argument of periastron will be specified. \\
``elist'' & \tabitem & Take the eccentricity from the input light curve list file, optionally specifying the column number. \\
``efix'' & \tabitem & Fix the eccentricity to the specified value. \\
``eexpr'' & \tabitem & Evaluate an analytic expression for each light curve to determine the eccentricity. \\
``erand'' & \tabitem & Adopt a random number for the eccentricity. \\
``olist'' & \tabitem & Take the argument of periastron (in degrees) from the input light curve list file, optionally specifying the column number. \\
``ofix'' & \tabitem & Fix the argument of periastron to the specified value. \\
``oexpr'' & \tabitem & Evaluate an analytic expression for each light curve to determine the argument of periastron. \\
``orand'' & \tabitem & Adopt a random number for the argument of periastron. \\ \hline
``hk'' & \tabitem & Keyword to indicate that $e\sin\omega \equiv h$ and $e\cos\omega \equiv k$ will be specified rather than the eccentricity and argument of periastron. \\
``hlist'' & \tabitem & Take $e\sin\omega$ from the input light curve list file, optionally specifying the column number. \\
``hfix'' & \tabitem & Fix $e\sin\omega$ to the specified value. \\
``hexpr'' & \tabitem & Evaluate an analytic expression for each light curve to determine $e\sin\omega$. \\
``hrand'' & \tabitem & Adopt a random number for $e\sin\omega$. \\
``klist'' & \tabitem & Take $e\cos\omega$ from the input light curve list file, optionally specifying the column number. \\
``kfix'' & \tabitem & Fix $e\cos\omega$ to the specified value. \\
``kexpr'' & \tabitem & Evaluate an analytic expression for each light curve to determine $e\cos\omega$. \\
``krand'' & \tabitem & Adopt a random number for $e\cos\omega$. \\ \hline
``Mstarlist'' & \tabitem & Take the mass of the star (in solar masses) from the input light curve list file, optionally specifying the column number. \\
``Mstarfix'' & \tabitem & Fix the mass of the star to the specified value. \\
``Mstarexpr'' & \tabitem & Evaluate an analytic expression for each light curve to determine the stellar mass. \\ \hline
``Rstarlist'' & \tabitem & Take the radius of the star (in solar radii) from the input light curve list file, optionally specifying the column number. \\
``Rstarfix'' & \tabitem & Fix the radius of the star to the specified value. \\
``Rstarexpr'' & \tabitem & Evaluate an analytic expression for each light curve to determine the stellar period. \\ \hline
``quad'' $\vert$ ``nonlin'' & \tabitem & Keywords to indicate whether a 2-parameter quadratic limb darkening law or a 4-parameter non-linear limb darkening law should be used. \\
``ldlist'' & \tabitem & Take the limb darkening coefficients from the input light curve list file, optionally specifying the column number of the first coefficient (other coefficients must be in the subsequent columns). \\
``ldfix'' & \tabitem & Fix the limb darkening coefficients to the specified values. Two values must be supplied if the ``quad'' keyword was used, or four values if the ``nonlin'' keyword was used. \\
``ldexpr'' & \tabitem & Evaluate an analytic expression to determine the limb darkening coefficients. Two expressions must be supplied if the ``quad'' keyword was used, or four expressions if the ``nonlin'' keyword was used. \\ \hline
``dilute'' & \tabitem & An optional keyword indicating that the transit signal should be scaled by a value between 0 and 1. One must then either use ``list'' to indicate that the scaling factor should be read from the input light curve list, ``fix'' to specify on the command line the value to use for all light curves, or ``expr'' to provide an analytic expression to be evaluated for each light curve. \\ \hline
``omodel'' & \tabitem & An optional keyword used to output the model transit light curve, evaluated at the times of observation in the input light curve. The model will be written to a file of the name \$modeloutdir/\$basename.injecttransit.model where \$basename is the base filename of the input light curve, stripped of any directories. \\ \hline
\end{xtabular}
\end{center}

\begin{lstlisting}[backgroundcolor=\color{white},language=vartools,basicstyle=\footnotesize,breaklines=false,prebreak={}]
-Jstet timescale dates
\end{lstlisting}

\begin{center}
\tablecaption{Input Parameters for ``-Jstet'' Command (Section~\ref{cmd:jstet})}
\footnotesize
\begin{xtabular}{|p{1in}p{5mm}p{5in}|}
\hline
timescale & \tabitem & The time, in minutes (assuming input times are in days), that distinguishes between ``near'' and ``far'' observations. \\ \hline
dates & \tabitem & A file containing a list of all times that appear in the light curves in the first column. This is used to calculate the maximum possible weight for a light curve. \\ \hline
\end{xtabular}
\end{center}

\begin{lstlisting}[backgroundcolor=\color{white},language=vartools,basicstyle=\footnotesize,breaklines=false,prebreak={}]
-Killharm <"aov" | "ls" | "both" | "injectharm" 
    | "fix" Nper per1 ... perN
    | "list" Nper ["column" col1]> Nharm Nsubharm
    omodel [model_outdir] ["fitonly"]
    ["outampphase" | "outampradphase" | "outRphi" | "outRradphi"]
    ["clip" val]
\end{lstlisting}

\begin{center}
\tablecaption{Input Parameters for ``-Killharm'' Command (Section~\ref{cmd:killharm})}
\footnotesize
\begin{xtabular}{|p{1in}p{5mm}p{5in}|}
\hline
``aov'' & \tabitem & Take the period for the harmonic series from the most recently executed ``-aov'' or ``-aov\_harm'' command. \\ 
``ls'' & \tabitem & Take the period for the harmonic series from the most recently executed ``-LS'' command. \\ 
``both'' & \tabitem & Use two periods, taking one from the most recently executed ``-aov'' or ``-aov\_harm'' command and the other from the most recently executed ``-LS'' command. \\ 
``injectharm'' & \tabitem & Take the period for the harmonic series from the most recently executed ``-Injectharm'' command. \\
``fix'' & \tabitem & Fix the period(s) to the specified value(s). Nper is the number of periods to use, and per1 through perN is a space-delimited list of the Nper periods. \\
``list'' & \tabitem & Take the period(s) from the input light curve list. Nper is the number of periods to use, and they must be in consecutive columns within the list file. You may optionally specify the column for the first period using the ``column'' keyword. \\ \hline
Nharm & \tabitem & The number of higher-harmonics to use in the harmonic series that is fit for each period (frequencies of $2f_{0}$, $3f_{0}$, up to $(N_{\rm harm} + 1)f_{0}$ will be used for these harmonics, where $f_{0}$ is the fundamental frequency). Set this to zero to fit only a sinusoid. \\ \hline
Nsubharm & \tabitem & The number of sub-harmonics to use in the harmonic series that is fit for each period (frequencies of $f_{0}/2$, $f_{0}/3$ up to $f_{0}/(N_{\rm subharm}+1)$). Set this to zero to fit only a sinusoid. \\ \hline
omodel & \tabitem & A flag indicating whether or not to output the model harmonic series light curves. \\
model\_outdir & \tabitem & This parameter is required if omodel is set to 1. The model harmonic series light curves will be written to files named \$model\_outdir/\$basename.killharm.model where \$basename is the base filename of the input light curve (stripped of directories). \\ \hline
``fitonly'' & \tabitem & This optional keyword, if used, will cause \va{} to not subtract the harmonic series from the light curve, and instead only perform the fit reporting the coefficients. By default the harmonic series will be subtracted from the light curve before passing it on to the next command. \\ \hline
``outampphase'' & \tabitem & By default the $a_{k}$ and $b_{k}$ coefficients of the $\sin$ and $\cos$ functions in the harmonic series for harmonic number $k$ are output. If the ``outampphase'' keyword is given, then the amplitudes of the harmonics $A_{k} = \sqrt{a_{k}^2+b_{k}^2}$ and phases between 0 and 1 ($\phi_{k} = {\rm atan}2(-b_{k},a_{k})/2\pi$) will be output. \\
``outampradphase'' & \tabitem & If this keyword is given, then the amplitudes of the harmonics $A_{k} = \sqrt{a_{k}^2+b_{k}^2}$ and phases in radians $\phi_{k} = {\rm atan}2(-b_{k},a_{k})$ will be output. \\
``outRphi'' & \tabitem & If this keyword is given, then the amplitudes of the harmonics relative to the fundamental $R_{k1} = A_{k}/A_{1}$ and relative phases $\phi_{k1} = \phi_{k} - k\phi_{1}$ (between 0 and 1) will be output. Note for sub-harmonics $k = 1/2$, $1/3$, etc. For the fundamental mode the amplitude $A_{1}$ and phase $\phi_{1}$ will be given. \\
``outRradphi'' & \tabitem & Similar to ``outRphi'', in this case phases are in radians. \\ \hline
``clip'' & \tabitem & If this optional keyword is given then the model will be fit, $\sigma$-clipping with a clipping factor given by val will be applied to the residuals, and the model will be refit to the the points which passed the clipping. \\ \hline
\end{xtabular}
\end{center}

\begin{lstlisting}[backgroundcolor=\color{white},language=vartools,basicstyle=\footnotesize,breaklines=false,prebreak={}]
-linfit function paramlist ["modelvar" varname]
    ["correctlc"]
    ["omodel" model_outdir ["format" nameformat]]
\end{lstlisting}

\begin{center}
\tablecaption{Input Parameters for ``-linfit'' Command (Section~\ref{cmd:linfit})}
\footnotesize
\begin{xtabular}{|p{1in}p{5mm}p{5in}|}
\hline
function & \tabitem & Analytic expression to fit to the light curve magnitudes. This expression should be linear in the free parameters. An example is 'a*t$\wedge{}$2+b*t+c' to fit a quadratic function in time to the light curve, with 'a', 'b' and 'c' being free parameters.\\ \hline
paramlist & \tabitem & A comma-delimited list of variables which are the free parameters to optimize in the function. These must all enter linearly into the function. For the previous example one would supply 'a,b,c' for this parameter.\\ \hline
``modelvar'' & \tabitem & Optional keyword used to store the best-fit model light curve in a variable indicated by varname. \\ \hline
``correctlc'' & \tabitem & If this keyword is given then the best-fit model will be subtracted from the light curve before passing it on to the next command. By default the model is not subtracted from the light curve. \\ \hline
``omodel'' & \tabitem & Optional keyword used to indicate that the best-fit model light curve should be output to a file. The name of the file is given by \$model\_outdir/\$basename.linfit.model where \$basename is the base filename of the input light curve stripped of directories. The \"format\" keyword may be used to change the rule for naming the file as indicated below. \\
``format'' & \tabitem & Optional keyword used in conjunction with the ``omodel'' keyword to change the rule for naming the output model light curve files. In this case the file will be named \$model\_outdir/\$nameformat where instances of \%s in \$nameformat are replaced by \$basename, instances of \%d are replaced by the light curve number (starting with 1), instances of \%0nd where n is an integer are replaced with the formatted light curve number, and instances of \%\% are replaced with \%. \\ \hline
\end{xtabular}
\end{center}

\begin{lstlisting}[backgroundcolor=\color{white},language=vartools,basicstyle=\footnotesize,breaklines=false,prebreak={}]
-LS minp maxp subsample Npeaks operiodogram [outdir] ["noGLS"] ["whiten"]
    ["clip" clip clipiter] ["fixperiodSNR" <"aov" | "ls" | "injectharm"
        | "fix" period | "list" ["column" col]
        | "fixcolumn" <colname | colnum>>]
\end{lstlisting}

\begin{center}
\tablecaption{Input Parameters for ``-LS'' Command (Section~\ref{cmd:ls})}
\footnotesize
\begin{xtabular}{|p{1in}p{5mm}p{5in}|}
\hline
minp & \tabitem & The minimum period to search (in the same time units as the input light curve). \\ \hline
maxp & \tabitem & The maximum period to search (in the same time units as the input light curve). \\ \hline
subsample & \tabitem & The periodogram will be scanned at a frequency resolution of $\Delta f = {\rm subsample}/T$ where $T$ is the time base-line of the input light curve. \\ \hline
operiodogram & \tabitem & Flag used to indicate whether or not the periodograms should be output to separate files. \\
outdir & \tabitem & This parameter is supplied only if operiodogram is set to 1. The periodograms will be output to files named \$outdir/\$basename.ls where \$basename is the base filename of the input light curve stripped of any directories. \\ \hline
``noGLS'' & \tabitem & If this keyword is given then the traditional L-S periodogram will be calculated. By default the Generalized L-S periodogram is calculated. \\ \hline
"whiten" & \tabitem & Keyword used to whiten the light curve and recalculate the periodogram for each peak. By default this is not done. \\ \hline
"clip" & \tabitem & Adjust the clipping performed on the periodogram in calculating the spectroscopic S/N of a peak. By default iterative 5$\sigma$ clipping is performed. \\ 
~~~~clip & \tabitem & The sigma-clipping factor to use. \\
~~~~clipiter & \tabitem & Flag which if set causes iterative clipping to be performed. \\ \hline
"fixperiodSNR" & \tabitem & Report the periodogram value, the false alarm probability and the S/N for a fixed period in addition to the highest peaks found. \\
~~~~"aov" & \tabitem & Use the peak period from the last executed -aov or -aov\_harm command for the fixperiodSNR. \\
~~~~"ls" & \tabitem & Use the peak period from the last executed -LS command. \\
~~~~"injectharm" & \tabitem & Use the period from the last executed -Injectharm command (the first one is used if multiple periods were injected). \\ \hline
\end{xtabular}
\end{center}

\begin{lstlisting}[backgroundcolor=\color{white},language=vartools,basicstyle=\footnotesize,breaklines=false,prebreak={}]
-MandelAgolTransit <"bls" | "blsfixper"
        | P0 T00 r0 a0 <"i" inclination | "b" bimpact> e0 omega0 mconst0>
    <"quad" | "nonlin"> ldcoeff1_0 ... ldcoeffn_0 fitephem
    fitr fita fitinclterm fite fitomega fitmconst fitldcoeff1 ... fitldcoeffn
    fitRV [RVinputfile RVmodeloutfile K0 gamma0 fitK fitgamma]
    correctlc omodel [model_outdir]
    ["modelvar" var] ["ophcurve" curve\_outdir phmin phmax phstep]
    ["ojdcurve" curve\_outdir jdstep]
\end{lstlisting}

\begin{center}
\tablecaption{Input Parameters for ``-MandelAgolTransit'' Command (Section~\ref{cmd:mandelagoltransit})}
\footnotesize
\begin{xtabular}{|p{1in}p{5mm}p{5in}|}
\hline
``bls'' & \tabitem & Initialize the parameters based on the highest peak found with the most recent prior ``-BLS'' command. \\ \hline
``blsfixper'' & \tabitem & Initialize the parameters based on the highest peak found with the most recent prior ``-BLSFixPer'' command. \\ \hline
P0 & \tabitem & Initial orbital period. \\ \hline
T00 & \tabitem & Initial transit epoch (time at the center of a transit). \\ \hline
r0 & \tabitem & Initial ratio of the planet radius to the star radius ($R_{P}/R_{\star}$). \\ \hline
a0 & \tabitem & Initial ratio of the semi-major axis to the stellar radius ($a/R_{\star}$). \\ \hline
``i'' & \tabitem & Keyword used to indicate that the next parameter given is the initial orbital inclination angle, in degrees. \\ \hline
``b'' & \tabitem & Keyword used to indicate that the next parameter given is the initial normalized impact parameter. This is the distance between the projected centers of the planet and star at conjunction divided by the sum of the planet and stellar radii (note, not just the stellar radius). \\ \hline
``e0'' & \tabitem & Initial orbital eccentricity. \\ \hline
``omega0'' & \tabitem & Initial argument of periastron in degrees. \\ \hline
``mconst0'' & \tabitem & Initial out-of-transit magnitude. If a negative value is specified, then the optimal value will be determined automatically. \\ \hline
``quad'' & \tabitem & Keyword used to indicate that a two-parameter quadratic limb darkening law is used. \\ \hline
``nonlin'' & \tabitem & Keyword used to indicate that a four-parameter non-linear limb darkening law is used. \\ \hline
ldcoeff1\_0 $\ldots$ ldcoeffn\_0 & \tabitem & A space separated list of initial limb darkening parameters (2 for the quadratic law, 4 for the non-linear law). \\ \hline
fitephem & \tabitem & Flag indicating whether or not the ephemeris ($P$ and $T00$) should be varied in the fit. \\ \hline
fitr & \tabitem & Flag indicating whether $R_{P}/R_{\star}$ should be varied in the fit. \\ \hline
fita & \tabitem & Flag indicating whether $a/R_{\star}$ should be varied in the fit. \\ \hline
fitinclterm & \tabitem & Flag indicating whether the inclination term (either the inclination angle itself, or the impact parameter, whichever was specified) should be varied in the fit. \\ \hline
fite & \tabitem & Flag indicating whether the eccentricity should be varied in the fit. \\ \hline
fitomega & \tabitem & Flag indicating whether the argument of periastron should be varied in the fit. \\ \hline
fitmconst & \tabitem & Flag indicating whether the out-of-transit magnitude should be varied in the fit. \\ \hline
fitldcoeff1 $\ldots$ fitldcoeffn & \tabitem & Flags indicating whether the respective limb-darkening coefficients should be varied in the fit. \\ \hline
fitRV & \tabitem & Flag indicating whether a separate RV curve should be fit together with the light curve. \\ \hline
RVinputfile & \tabitem & The name of the file containing the RV data to fit. The expected file format is: JD, RV, RV uncertainty. This parameter should be given if and only if fitRV is set to 1. \\ \hline
RVmodeloutfile & \tabitem & The name of the file to which the best-fit RV orbit model (evaluated at the input times) should be output. This parameter should be given if and only if fitRV is set to 1. \\ \hline
K0 & \tabitem & The initial RV orbital semi-amplitude. This parameter should be given if and only if fitRV is set to 1. \\ \hline
gamma0 & \tabitem & The initial RV zero-point. This parameter should be given if and only if fitRV is set to 1. \\ \hline
fitK & \tabitem & Flag indicating whether the RV semi-amplitude should be varied in the fit. This parameter should be given if and only if fitRV is set to 1. \\ \hline
fitgamma & \tabitem & Flag indicating whether the RV zero-point should be varied in the fit. This parameter should be given if and only if fitRV is set to 1. \\ \hline
correctlc & \tabitem & Flag indicating whether the best-fit model should be subtracted from the light curve before passing it on to the next command. \\ \hline
omodel & \tabitem & Flag indicating whether the best-fit model light curve should be output to a file. The model will be evaluated at the times of observation. \\
model\_outdir & \tabitem & The model will be written to the file given by \$model\_outdir/\$basename.mandelagoltransit.model where \$basename is the base filename of the input light curve, stripped of directories. This parameter should be given if and only if omodel is set to 1. \\ \hline
``modelvar'' & \tabitem & Optional keyword used to store the best-fit model light curve in a variable indicated by $var$. \\ \hline
``ophcurve'' & \tabitem & A keyword used to output best-fit model light curves that are uniformly sampled in phase. \\
~~~~curve\_outdir & \tabitem & The models will be written out to files named \$outdir/\$basename.mandelagoltransit.phcurve. \\
~~~~phmin & \tabitem & The starting phase to use in the model. \\
~~~~phmax & \tabitem & The ending phase to use in the model. \\
~~~~phstep & \tabitem & The phase step-size to use in the model. \\ \hline
``ojdcurve'' & \tabitem & A keyword used to output best-fit model light curves that are uniformly sampled in time between the first and last observed times in the light curve. \\
~~~~curve\_outdir & \tabitem & The models will be written out to files named \$outdir/\$basename.mandelagoltransit.jdcurve. \\
~~~~jdstep & \tabitem & The time step-size to use in the model. \\ \hline
\end{xtabular}
\end{center}

\begin{lstlisting}[backgroundcolor=\color{white},language=vartools,basicstyle=\footnotesize,breaklines=false,prebreak={}]
-medianfilter time ["average" | "weightedaverage"] ["replace"]
\end{lstlisting}

\begin{center}
\tablecaption{Input Parameters for ``-medianfilter'' Command (Section~\ref{cmd:medianfilter})}
\footnotesize
\begin{xtabular}{|p{1in}p{5mm}p{5in}|}
\hline
time & \tabitem & The time separation used by the filter. By default, a high-pass filter is applied, whereby the median magnitude of all points within 'time' of a given observation is subtracted from the magnitude for that observation. \\ \hline
``average'' & \tabitem & Optional keyword used to take the average of the magnitudes, rather than the median. \\ \hline
``weightedaverage'' & \tabitem & Optional keyword used to take the weighted average of the magnitudes, rather than the median. \\ \hline
``replace'' & \tabitem & Optional keyword used to perform a low-pass filter, whereby the magnitude of a given observation will be replaced by the median (or average, or weighted average) magnitude of all points within 'time' of that observation. \\ \hline
\end{xtabular}
\end{center}

\begin{lstlisting}[backgroundcolor=\color{white},language=vartools,basicstyle=\footnotesize,breaklines=false,prebreak={}]
-microlens
    ["f0"
        ["fix" fixval | "list" ["column" col]
            | "fixcolumn" <colname | colnum>
            | "auto"]
        ["step" initialstepsize] ["novary"]]
    ["f1"
        ["fix" fixval | "list" ["column" col]
            | "fixcolumn" <colname | colnum>
            | "auto"]
        ["step" initialstepsize] ["novary"]]
    ["u0"
        ["fix" fixval | "list" ["column" col]
            | "fixcolumn" <colname | colnum>
            | "auto"]
        ["step" initialstepsize] ["novary"]]
    ["t0"
        ["fix" fixval | "list" ["column" col]
            | "fixcolumn" <colname | colnum>
            | "auto"]
        ["step" initialstepsize] ["novary"]]
    ["tmax"
        ["fix" fixval | "list" ["column" col]
            | "fixcolumn" <colname | colnum>
            | "auto"]
        ["step" initialstepsize] ["novary"]]
    ["correctlc"] ["omodel" outdir]
\end{lstlisting}

\begin{center}
\tablecaption{Input Parameters for ``-microlens'' Command (Section~\ref{cmd:microlens})}
\footnotesize
\begin{xtabular}{|p{1in}p{5mm}p{5in}|}
\hline
``f0'' & \tabitem & Keyword used to change the method for initializing and fitting the $f_{0}$ parameter in equation~\ref{eqn:microlens1}. If this keyword is not used, the parameter will be automatically initialized and then varied in the fit. \\
~~~~``fix'' & \tabitem & Keyword indicating that the initial value should be fixed to the value supplied for fixval. \\
~~~~``list'' & \tabitem & Keyword indicating that the initial value should be taken from the input light curve list file (the column to read it from may also be optionally specified). \\
~~~~``fixcolumn'' & \tabitem & Keyword indicating that the initial value should be taken from the output of a previously executed command. \\
~~~~``auto'' & \tabitem & Keyword indicating that the initial value should be determined automatically (the default behavior). \\
~~~~``step'' & \tabitem & Keyword used to change the initial step-size for varyig the parameter. By default the step-size is determined automatically. \\
~~~~``novary'' & \tabitem & If this keyword is specified than the parameter will not be varied in the fit. \\ \hline
``f1'' & \tabitem & Keyword used to change the method for initializing and fitting the $f_{1}$ parameter in equation~\ref{eqn:microlens1}. If this keyword is not used, the parameter will be automatically initialized and then varied in the fit. The syntax following this keyword is the same as for ``f0''.\\ \hline
``u0'' & \tabitem & Keyword used to change the method for initializing and fitting the $u_{0}$ parameter in equation~\ref{eqn:microlens3}. If this keyword is not used, the parameter will be automatically initialized and then varied in the fit. The syntax following this keyword is the same as for ``f0''.\\ \hline
``t0'' & \tabitem & Keyword used to change the method for initializing and fitting the $t_{0}$ parameter in equation~\ref{eqn:microlens3}. If this keyword is not used, the parameter will be automatically initialized and then varied in the fit. The syntax following this keyword is the same as for ``f0''.\\ \hline
``tmax'' & \tabitem & Keyword used to change the method for initializing and fitting the $t_{\rm max}$ parameter in equation~\ref{eqn:microlens3}. If this keyword is not used, the parameter will be automatically initialized and then varied in the fit. The syntax following this keyword is the same as for ``f0''.\\ \hline
``correctlc'' & \tabitem & If this keyword is given then the best-fit model will be subtracted from the light curve before passing it on to the next command. By default the light curve is not modified by this command.\\ \hline
``omodel'' & \tabitem & If this keyword is given the the best-fit model, evaluated at the observed times in the light curve, will be output to a file whose name is given by \$outdir/\$basename.microlens where \$basename is the base filename of the light curve, stripped of directories. \\ \hline
\end{xtabular}
\end{center}

\begin{lstlisting}[backgroundcolor=\color{white},language=vartools,basicstyle=\footnotesize,breaklines=false,prebreak={}]
-nonlinfit function paramlist ["linfit" linfitparams]
    ["errors" error_expr]
    ["covariance"
        <"squareexp" amp_var rho_var
         | "exp" amp_var rho_var
         | "matern" amp_var rho_var nu_var>]
    ["priors" priorlist] ["constraints" constraintlist]
    <"amoeba" ["tolerance" tol] ["maxsteps" steps]
     | "mcmc" ["Naccept" N | "Nlinkstotal" N]
            ["fracburnin" frac] ["eps" eps] ["skipamoeba"]
            ["chainstats" exprlist statslist]
            ["maxmemstore" maxmem]
            ["outchains" outdir ["format" format] ["printevery" N]] >
    ["modelvar" varname] ["correctlc"]
    ["omodel" model_outdir ["format" nameformat]]
\end{lstlisting}

\begin{center}
\tablecaption{Input Parameters for ``-nonlinfit'' Command (Section~\ref{cmd:nonlinfit})}
\footnotesize
\begin{xtabular}{|p{1in}p{5mm}p{5in}|}
\hline
function & \tabitem & Analytic expression to fit to the light curve magnitudes. For example this might be 'a*exp(-(t-t0)$\wedge$2/2/sigma$\wedge$2)+b' for a Gaussian function in time with free parameters 'a', 't0', 'sigma' and 'b'.\\ \hline
paramlist & \tabitem & A comma-delimited list of parameters to vary in the fit. For each parameter you must specify the initial guess and step-size using the format ``var=init:step''. For example, if the initial value for 't0' is 5.0, and its step-size is 2.0, and if the initial value for 'sigma' is 10.0 and its step-size is 8.0, you would use the expression 't0=5.0:2.0,sigma=10.0:8.0'. The initial values and step-sizes may also be more complicated analytic expressions using variables, functions, etc. Note that the free parameters should have names that are not used by any vector variables (e.g., 't', 'mag', 'err', or other variables defined by the {\bf -expr} command or {\bf -inputlcformat} option). Note also that these variables may be used by other commands as well (e.g., on the right-hand-side of the expression provided to the {\bf -expr} command). \\ \hline
``linfit'' & \tabitem & Optional keyword used to indicate nay parameters in the function which enter linearly and should be optimized using linear least squares. In this case linfitparams is a comma-separated list of the linear parameters. For example, in the Gaussian case this could be 'linear a,b'. \\ \hline
``errors'' & \tabitem & Optional keyword used to change the magnitude uncertainties used in calculating the likelihood function (Eq.~\ref{eqn:nonlinfitx2}). error\_expr is an analytic expression. Parameters which are varied in the fit may also be included in this expression. \\ \hline
``covariance'' & \tabitem & Optional keyword used to indicated correlated uncertainties (by default errors are assume to be uncorrelated). \\
~~~~``squareexp'' & \tabitem & Use a square-exponential covariance matrix (Eq.~\ref{eqn:squareexp}). \\
~~~~~~~~amp\_var & \tabitem & The name of the \va{} variable which should be used as $a$ in equation~\ref{eqn:squareexp}. If this variable does not appear in paramlist then it should be specified here as 'amp\_var=expr' where amp\_var is the name of the variable to use, and expr is an expression used to determine the fixed value to be used for this parameter. \\
~~~~~~~~rho\_var & \tabitem & The name of the \va{} variable which should be used as $\rho$ in equation~\ref{eqn:squareexp}. If this variable does not appear in paramlist then it should be specified here as 'rho\_var=expr' where rho\_var is the name of the variable to use, and expr is an expression used to determine the fixed value to be used for this parameter. \\
~~~~``exp'' & \tabitem & Use a exponential covariance matrix (Eq.~\ref{eqn:expnoise}). \\
~~~~~~~~amp\_var & \tabitem & The name of the \va{} variable which should be used as $a$ in equation~\ref{eqn:expnoise}. If this variable does not appear in paramlist then it should be specified here as 'amp\_var=expr' where amp\_var is the name of the variable to use, and expr is an expression used to determine the fixed value to be used for this parameter. \\
~~~~~~~~rho\_var & \tabitem & The name of the \va{} variable which should be used as $\rho$ in equation~\ref{eqn:expnoise}. If this variable does not appear in paramlist then it should be specified here as 'rho\_var=expr' where rho\_var is the name of the variable to use, and expr is an expression used to determine the fixed value to be used for this parameter. \\
~~~~``matern'' & \tabitem & Use a Mat\'ern covariance matrix (Eq.~\ref{eqn:matern}). \\
~~~~~~~~amp\_var & \tabitem & The name of the \va{} variable which should be used as $a$ in equation~\ref{eqn:matern}. If this variable does not appear in paramlist then it should be specified here as 'amp\_var=expr' where amp\_var is the name of the variable to use, and expr is an expression used to determine the fixed value to be used for this parameter. \\
~~~~~~~~rho\_var & \tabitem & The name of the \va{} variable which should be used as $\rho$ in equation~\ref{eqn:matern}. If this variable does not appear in paramlist then it should be specified here as 'rho\_var=expr' where rho\_var is the name of the variable to use, and expr is an expression used to determine the fixed value to be used for this parameter. \\
~~~~~~~~nu\_var & \tabitem & The name of the \va{} variable which should be used as $\nu$ in equation~\ref{eqn:matern}. If this variable does not appear in paramlist then it should be specified here as 'nu\_var=expr' where nu\_var is the name of the variable to use, and expr is an expression used to determine the fixed value to be used for this parameter. \\ \hline
``priors'' & \tabitem & Keyword used to indicate a list of priors to place on the variables. priorlist is a comma-separated list, where each entry in the list is an analytic expresion that should evaluate to $-2\ln(P)$ where $P$ is the prior probability for a variable given its value. For example, to place a Gaussian prior on 't0' with mean 4.0 and standard deviation 3.0 you would use 'prior (t0-4.0)$\wedge$2/3.0$\wedge$2'. \\ \hline
``constraints'' & \tabitem & Use this keyword to place constraints on the parameters. constraintlist is a comma-separated list of expressions. For example, to require a positive 'sigma' you can use 'constraints sigma$>$0'. \\ \hline
``amoeba'' & \tabitem & Use the down-hill simplex optimization algorithm. \\
~~~~``tolerance'' & \tabitem & Optional keyword to change the convergence tolerance. Here 'tol' corresponds to ${\rm TOL}$ in equation~\ref{eqn:amoebaconvergence}. \\
~~~~``maxsteps'' & \tabitem & Optional keyword to change the maximum number of amoeba iterations that are tried before giving up. \\ \hline
``mcmc'' & \tabitem & Run a Differential Evolution Markov Chain Monte Carlo procedure. \\
~~~~``Naccept'' & \tabitem & Optionally specify the number of accepted links to run in a given fit. Once this number is reached, the MCMC will terminate. \\
~~~~``Nlinkstotal'' & \tabitem & Optionally specify a maximum total number of links to run. Once this number is reached, the MCMC will terminal. This option is assumed by default, with a limit of 100,000 links. \\
~~~~``fracburnin'' & \tabitem & Optionally change the initial fraction of the chain that is ignored in computing statistics to be reported from the posterior distribution (the default is 0.1). \\
~~~~``eps'' & \tabitem & Optionally change the value of $\epsilon$ to use in equation~\ref{eqn:mcmcdef}. The default value is 0.001. \\
~~~~``skipamoeba'' & \tabitem & By default the down-hill simplex algirhtm is run initially before starting the mcmc procedure. This can be skipped by using this keyword. \\
~~~~``chainstats'' & \tabitem & By default the median and standard deviation for each of the parameters varied in the MCMC will be included in the output table. You may use this keyword to change the statistics, and/or the quantities that are used. Here 'exprlist' is a comma-separated list of analytic expressions to calculate from the chain, and 'statslist' is a comma-separated list of statistics to report for each of the expressions. The available statistics are the same as for the {\bf -stats} command. \\
~~~~``maxmemstore'' & \tabitem & Change the maximum limit on the total amount of memory to be used by this MCMC command. This will limit the length of the chain used for calculating the output statistics. 'maxmem' is the limit in GB. The default is 4.0. \\
~~~~``outchains'' & \tabitem & Use this keyword to output the MCMC chains. The chains will be output to files named \$model\_outdir/\$basename.mcmc, where \$basename is the base filename of the input light curve file, stripped of directories. \\
~~~~~~~~``format'' & \tabitem & Use this keyword to change the naming convention for the output MCMC files. In this case the file will be named \$model\_outdir/\$nameformat where instances of \%s in \$nameformat are replaced by \$basename, instances of \%d are replaced by the light curve number (starting with 1), instances of \%0nd, where n is an integer, are replaced with the formatted light curve number, and instances of \%\% are replaced with \%. \\ \hline
~~~~~~~~``printevery'' & \tabitem & Use this keyword to change the number of links that are written to the output file. By default every link is written out. If this keyword is used, then every 'N'th link is written out. \\ \hline
``modelvar'' & \tabitem & Optional keyword used to store the best-fit model light curve in a variable indicated by varname. \\ \hline
``correctlc'' & \tabitem & Use this keyword to subtract the best-fit model light curve from the input light curve before passing it on to the next command. \\ \hline
``omodel'' & \tabitem & If this keyword is given, the best-fit model, evaluated at the observed times in the light curve, will be output to a file whose name is given by \$model\_outdir/\$basename.nonlinfit.model where \$basename is the base filename of the light curve, stripped of directories. \\
~~~~``format'' & \tabitem & Use this keyword to change the naming convention for the output model light curve. In this case the file will be named \$model\_outdir/\$nameformat where instances of \%s in \$nameformat are replaced by \$basename, instances of \%d are replaced by the light curve number (starting with 1), instances of \%0nd, where n is an integer, are replaced with the formatted light curve number, and instances of \%\% are replaced with \%. \\ \hline
\end{xtabular}
\end{center}

\begin{lstlisting}[backgroundcolor=\color{white},language=vartools,basicstyle=\footnotesize,breaklines=false,prebreak={}]
-o <outdir | outname> ["nameformat" formatstring]
    ["columnformat" formatstring]
    ["fits"] ["noclobber"]
\end{lstlisting}

\begin{center}
\tablecaption{Input Parameters for ``-o'' Command (Section~\ref{cmd:o})}
\footnotesize
\begin{xtabular}{|p{1in}p{5mm}p{5in}|}
\hline
outdir & \tabitem & The directory to which the light curve files will be written. By default the output filenames will be \$outdir/\$basename where \$outdir is the value supplied for 'outdir' and \$basename is the base filename of the light curve, stripped of directories. The supplied parameter will be interpreted as the output directory only if a light curve list was used for input ({\bf -l}). \\ 
outname & \tabitem & The name of the output light curve file. This will be used if an individual light curve file was used for input ({\bf -i}). \\ \hline
``nameformat'' & \tabitem & Use this keyword to change the naming convention for the output light curves. In this case the file will be named \$outdir/\$formatstring where instances of \%s in \$formatstring are replaced by \$basename, instances of \%d are replaced by the light curve number (starting with 1), instances of \%0nd, where n is an integer, are replaced with the formatted light curve number, and instances of \%\% are replaced with \%. \\ \hline
``columnformat'' & \tabitem & Change the format of the output light curves. By default they will have three columns: time, mag and err. Here 'formatstring' is a comma-separated list of variable names to output, optionally with a ':' after each variable name to specify the ``printf''-style format to use for that variable. For example, 'columnformat t:\%.17g,mag:\%.5f,err:\%.5f,xpos:\%.3f' would output the variables t, mag, err, and xpos using 17-digit-precision for t, and floating points for mag, err and xpos, with 5, 5, and 3 digits after the decimal, respectively. Here 'xpos' is a non-default variable that one might have read-in with the {\bf -inputlcformat} option, or produced through one of the other commands. If the light curves are output in fits format, then terms after the ':' will be used to specify the units of the column to list in the light curve header. \\ \hline
``fits'' & \tabitem & Use this keyword to output the light curves in binary FITS table format. The output light curves will have the extension ``.fits'' appended to their filenames, if not already present.\\ \hline
``noclobber'' & \tabitem & Use this keyword to prevent over-writing any existing files. \va{} will terminate if it encounters an existing file when this keyword is used. \\ \hline
\end{xtabular}
\end{center}

\begin{lstlisting}[backgroundcolor=\color{white},language=vartools,basicstyle=\footnotesize,breaklines=false,prebreak={}]
-Phase <"aov" | "ls" | "bls" | "fixcolumn" <colname | colnum>
        | "list" ["column" col] | "fix" period>
    ["T0" <"bls" phaseTc | "fixcolumn" <colname | colnum>
        | "list" ["column" col] | "fix" T0>]
    ["phasevar" var] ["startphase" startphase]
\end{lstlisting}

\begin{center}
\tablecaption{Input Parameters for ``-Phase'' Command (Section~\ref{cmd:phase})}
\footnotesize
\begin{xtabular}{|p{1in}p{5mm}p{5in}|}
\hline
``aov'' & \tabitem & Use this keyword to take the period at which to phase-fold the light curve from the most recently issued {\bf -aov} or {\bf -aov\_harm} command. \\
``ls'' & \tabitem & Use this keyword to take the period at which to phase-fold the light curve from the most recently issued {\bf -LS} command. \\
``bls'' & \tabitem & Use this keyword to take the period at which to phase-fold the light curve from the most recently issued {\bf -BLS} command. \\ \hline
``T0'' & \tabitem & By default the light curves are folded such that phase zero corresponds to time $t=0$. Use this keyword to specify a different reference time for phase zero. \\ 
~~~~``bls'' & \tabitem & Take the reference time to be the time of central transit found in the most recent {\bf -BLS} command. Here 'phaseTc' is the phase to adopt for the central transit time. \\ \hline
``phasevar'' & \tabitem & Use this keyword to write the phases to the variable $var$ rather than overwriting the time variable. If the variable $var$ does not exist it will be created. If it does exist, it must be a double-precision light curve vector (e.g., it cannot be the name of an output column, or a parameter used in a fit, etc.).\\ \hline
``startphase'' & \tabitem & Use this keyword to change the range over which the phases are expressed. By default they run from 0 to 1. With this keyword they will run from $startphase$ to $startphase+1$.\\ \hline
\end{xtabular}
\end{center}

\begin{lstlisting}[backgroundcolor=\color{white},language=vartools,basicstyle=\footnotesize,breaklines=false,prebreak={}]
-resample
    <"nearest" |
      "linear" |
      "spline"  ["left" yp1] ["right" ypn] |
      "splinemonotonic" |
      "bspline" ["nbreaks" nbreaks] ["order" order] >
    ["file" <"fix" times_file ["column" time_column] |
        "list" ["listcolumn" col] ["tcolumn" time_column] > |
    ["tstart" <"fix" tstart | "fixcolumn" <colname | colnum> |
        "list" ["column" col] | "expr" expression > ]
    ["tstop" <"fix" tstop | "fixcolumn" <colname | colnum> |
        "list" ["column" col] | "expr" expression > ]
    [["delt" <"fix" delt | "fixcolumn" <colname | colnum> |
        "list" ["column" col] | "expr" expression > ]
     | ["Npoints" <"fix" Np | "fixcolumn" <colname | colnum> |
        "list" ["column" col] | "expr" expression > ]]]
    ["gaps" 
        <"fix" time_sep | "fixcolumn" <colname | colnum> |
            "list" ["column" col] | "expr" expression |
            "frac_min_sep" val | "frac_med_sep" val | "percentile_sep" val>
        <"nearest" |
          "linear"  |
          "spline"  ["left" yp1] ["right" ypn] |
          "splinemonotonic" |
          "bspline" ["nbreaks" nbreaks] ["order" order] >]
    ["extrap" 
        <"nearest" |
          "linear"  |
          "spline"  ["left" yp1] ["right" ypn] |
          "splinemonotonic" |
          "bspline" ["nbreaks" nbreaks] ["order" order] >]
\end{lstlisting}

\begin{center}
\tablecaption{Input Parameters for ``-resample'' Command (Section~\ref{cmd:resample})}
\footnotesize
\begin{xtabular}{|p{1in}p{5mm}p{5in}|}
\hline
``nearest'' & \tabitem & Set resampled values to the value of the observation that is closed in time to the resampled point. \\
``linear'' & \tabitem & Perform linear interpolation between points. \\
``spline'' & \tabitem & Perform cubic spline interpolation. \\
~~~~``left'' & \tabitem & Specify the left boundary condition for the spline. \\
~~~~``right'' & \tabitem & Specify the right boundary condition for the spline. \\
``splinemonotonic'' & \tabitem & Perform cubic spline interpolation with the interpolating function constrained to be monotonic between input observations. \\
``bspline'' & \tabitem & Interpolate with a Basis-spline function. \\
~~~~``nbreaks'' & \tabitem & The number of breaks to use in the B-spline (the default is 15). If the number given is $< 2$, then the routine will increase the number of breaks until $\chi^2$ per degree of freedom is less than one. Caution, this can be quite slow. \\
~~~~``order'' & \tabitem & The order of the spline function to use (the default is 3). \\ \hline
``file'' & \tabitem & Optional keyword to read the time-base on which to resample the light curve from a file. By default \va{} will resample the light curve onto a uniform time-base with the starting and stopping times equal to the minimum and maximum observed times in the light curve, respectively, and with a time step equal to the minimum time separation between consecutive points in the input light curve. \\
~~~~``fix'' & \tabitem & Use the same time-base file for all light curves. This file is given by 'times\_file'. Use the optional ``column'' keyword to specify the column in the file which stores the times to use.\\
~~~~``list'' & \tabitem & Allow a different time-base file for each light curve. The filenames in this case are taken from the input light curve list file. Use the optional ``listcolumn'' keyword to change the column in the input light curve list file to use, use the ``tcolumn'' keyword to change the columns in the files to use for the new times. \\ \hline
``tstart'' & \tabitem & Optional keyword to change the starting time for a regularly spaced grid of times to resample the light curve at. \\
~~~~``expr'' & \tabitem & In addition to the standard ``fix'', ``fixcolumn'', and ``list'' options for determining the starting time, one may also use this keyword to set it equal to the result from evaluating an analytic expression given by 'expression'. \\ \hline
``tstop'' & \tabitem & Optional keyword to change the starting time for a regularly spaced grid of times to resample the light curve at. Options for determining this parameter are the same as for ``tstart''. \\ \hline
``delt'' & \tabitem & Optional keyword to change the time step for a regularly spaced grid of times to resample the light curve at. Options for determining this parameter are the same as for ``tstart''. \\ \hline
``Npoints'' & \tabitem & Optional keyword to change the number of points to resample the light curve onto. Either ``delt'' or ``Npoints'' may be specified. Options for determining this parameter are the same as for ``tstart''. \\ \hline
``gaps'' & \tabitem & By default the same resampling method will be used for all points in the light curve. By giving the ``gaps'' keyword you can make the method depend on how far away a resampled time is from the closest observed time. You first need to indicate how the time separation used to distinguish between the near and far points will be determined, you then indicate the method for resampling the ``far'' points (the near points will be resampled using the method already indicated before providing the ``gaps'' keyword). Below are the options for determining the time separation. The options for determining the resampling method are the same as discussed above. \\
~~~~``fix'' & \tabitem & Fix the time separation to value specified by 'time\_sep'. \\
~~~~``fixcolumn'' & \tabitem & Set the time separation equal to a value calculated by a previously executed command. \\
~~~~``list'' & \tabitem & Take the value from the input light curve list file. \\
~~~~``expr'' & \tabitem & Set the value equal to the result from evaluating an analytic expression. \\
~~~~``frac\_min\_sep'' & \tabitem & Set the time separation to a fixed factor times the minimum separation between subsequent points in the input light curve (e.g., if you give ``frac\_min\_sep 5.0'' and the minimum separation between points in the light curve is 1 day, then the separation times scale will be set to 5 days). \\
~~~~``frac\_med\_sep'' & \tabitem & Same as ``frac\_min\_sep'', except using the median separation between subsequent points rather than the minimum separation. \\
~~~~``percentile\_sep'' & \tabitem & Set the time separation to a specified percentile of the separations in the input light curve (e.g., if you give ``percentile\_sep 70'', then the $N$ separations in the input light curve will be ordered, and the separation to be used for distinguishing between near and far points will be set equal to the $0.7\times N$ longest separation in the input light curve). \\ \hline
``extrap'' & \tabitem & Change the method for evaluating resampled points which are extrapolations rather than interpolations. Options are as already discussed. By default extrapolations and interpolations are performed using the same method. \\ \hline
\end{xtabular}
\end{center}

\begin{lstlisting}[backgroundcolor=\color{white},language=vartools,basicstyle=\footnotesize,breaklines=false,prebreak={}]
-rescalesig
\end{lstlisting}

No parameters or options are available.

\begin{lstlisting}[backgroundcolor=\color{white},language=vartools,basicstyle=\footnotesize,breaklines=false,prebreak={}]
-restorelc savenumber
\end{lstlisting}

\begin{center}
\tablecaption{Input Parameters for ``-restorelc'' Command (Section~\ref{cmd:restorelc})}
\footnotesize
\begin{xtabular}{|p{1in}p{5mm}p{5in}|}
\hline
savenumber & \tabitem & Restores the light curve to its state at the 'savenumber'-th call to {\bf -savelc} on the command line. \\ \hline
\end{xtabular}
\end{center}

\begin{lstlisting}[backgroundcolor=\color{white},language=vartools,basicstyle=\footnotesize,breaklines=false,prebreak={}]
-restricttimes ["exclude"]
    < "JDrange" minJD maxJD |
      "JDrangebylc"
        <"fix" minJD | "list" ["column" col] | "fixcolumn" <colname | colnum> |
         "expr" expression>
        <"fix" maxJD | "list" ["column" col] | "fixcolumn" <colname | colnum> |
         "expr" expression> |
      "JDlist" JDfilename | 
      "imagelist" imagefilename >
\end{lstlisting}

\begin{center}
\tablecaption{Input Parameters for ``-restricttimes'' Command (Section~\ref{cmd:restricttimes})}
\footnotesize
\begin{xtabular}{|p{1in}p{5mm}p{5in}|}
\hline
``exclude'' & \tabitem & If this optional keyword is given, then the times specified below will be removed from the light curve. By default times specified are included, and other times are excluded. \\ \hline
``JDrange'' & \tabitem & Provide a fixed range of times to keep (or exclude) with 'minJD' and 'maxJD'. \\ \hline
``JDrangebylc'' & \tabitem & Provide a range of times to keep (or exclude), allowing the limits on the range to change between light curves. If this keyword is given, then the method for determining the 'minJD' should be provided (either fixed to a constant value, read from the input light curve list, set to the result of a previously executed command, or set equal to the result from evaluating an analytic expression. The method for determining 'maxJD' should then be provided, with the same set of options. \\ \hline
``JDlist'' & \tabitem & Provide a list of individual JD values to keep (or exclude) in the file 'JDfilename'. The JD values should be given in the first column of this file. \\ \hline
``imagelist'' & \tabitem & Provide a list of image IDs to keep (or exclude) in the file 'imagefilename'. The IDs should be given in the first column of this file. \\ \hline
\end{xtabular}
\end{center}

\begin{lstlisting}[backgroundcolor=\color{white},language=vartools,basicstyle=\footnotesize,breaklines=false,prebreak={}]
-rms
\end{lstlisting}

No parameters or options are available.

\begin{lstlisting}[backgroundcolor=\color{white},language=vartools,basicstyle=\footnotesize,breaklines=false,prebreak={}]
-rmsbin Nbin bintime1...bintimeN
\end{lstlisting}

\begin{center}
\tablecaption{Input Parameters for ``-rmsbin'' Command (Section~\ref{cmd:rmsbin})}
\footnotesize
\begin{xtabular}{|p{1in}p{5mm}p{5in}|}
\hline
Nbin & \tabitem & The number of moving mean filters to use. \\ \hline
bintime1...bintimeN & \tabitem & A space-delimited list of filter widths, one for each of the Nbin filters used. The width of each filter is given by $2.0*{\rm bintime}$, where ${\rm bintime}$ is in minutes, assuming the times in the light curve are in days. \\ \hline
\end{xtabular}
\end{center}

\begin{lstlisting}[backgroundcolor=\color{white},language=vartools,basicstyle=\footnotesize,breaklines=false,prebreak={}]
-savelc
\end{lstlisting}

No parameters or options are available.

\begin{lstlisting}[backgroundcolor=\color{white},language=vartools,basicstyle=\footnotesize,breaklines=false,prebreak={}]
-SoftenedTransit <"bls" | "blsfixper" | P0 T00 eta0 delta0 mconst0 cval0>
    fitephem fiteta fitcval fitdelta fitmconst correctlc
    omodel [model_outdir] fit_harm [<"aov" | "ls" | "bls" 
    | "list" ["column" col] | "fix" Pharm> nharm nsubharm]
\end{lstlisting}

\begin{center}
\tablecaption{Input Parameters for ``-SoftenedTransit'' Command (Section~\ref{cmd:softenedtransit})}
\footnotesize
\begin{xtabular}{|p{1in}p{5mm}p{5in}|}
\hline
``bls'' & \tabitem & Use this keyword to initialize the parameters for this model based on the highest peak found in the most recently executed {\bf -BLS} command. \\ \hline
``blsfixper'' & \tabitem & Use this keyword to initialize the parameters for this model based on the most recently executed {\bf -BLSFixPer} command. \\ \hline
P0 & \tabitem & The initial period to use for the model ($P$ in Eq.~\ref{eqn:softenedtransit2}). \\
T00 & \tabitem & The initial transit epoch to use for the model ($T_{0}$ in Eq.~\ref{eqn:softenedtransit2}). \\
eta0 & \tabitem & The initial transit duration to use for the model ($\eta$ in Eq.~\ref{eqn:softenedtransit2}). \\
delta0 & \tabitem & The initial transit depth to use for the model ($\delta$ in Eq.~\ref{eqn:softenedtransit1}). \\
mconst0 & \tabitem & The initial out-of-transit magnitude to use for the model ($M_{0}$ in Eq.~\ref{eqn:softenedtransit1}). \\
cval0 & \tabitem & The initial value for the transit sharpness parameter ($c$ in Eq.~\ref{eqn:softenedtransit1}). \\ \hline
fitephem & \tabitem & Flag indicating whether or not $P$ and $T_{0}$ are to be varied in the fit. \\ \hline
fiteta & \tabitem & Flag indicating whether or not $\eta$ is to be varied in the fit. \\ \hline
fitcval & \tabitem & Flag indicating whether or not $c$ is to be varied in the fit. \\ \hline
fitdelta & \tabitem & Flag indicating whether or not $\delta$ is to be varied in the fit. \\ \hline
fitmconst & \tabitem & Flag indicating whether or not $M_{0}$ is to be varied in the fit. \\ \hline
correctlc & \tabitem & Flag indicating whether or not to subtract the best-fit model from the light curve before passing it on to the next command. \\ \hline
omodel & \tabitem & Flag indicating whether or not to output the best-fit model light curve, evaluated at the input times of observation, to a file. The model will be written to a file named \$model\_outdir/\$basename.softenedtransit.model where \$basename is the base filename of the input light curve file, stripped of directories. \\ \hline
fit\_harm & \tabitem & Flag indicating whether or not to simultaneously fit a harmonic series to the light curve together with the transit model. \\
~~~~``aov'' & \tabitem & Use this keyword to take the period for the harmonic series from the most recent {\bf -aov} or {\bf -aov\_harm} command. \\
~~~~``ls'' & \tabitem & Use this keyword to take the period for the harmonic series from the most recent {\bf -LS} command. \\
~~~~``bls'' & \tabitem & Use this keyword to take the period for the harmonic series from the most recent {\bf -BLS} command. \\
~~~~``list'' & \tabitem & Use this keyword to take the period for the harmonic series from the input light curve list file (optionally specify the column to use). \\
~~~~``fix'' & \tabitem & Use this keyword to fix the period for the harmonic series to the value specified by 'Pharm'. \\
~~~~nharm & \tabitem & The number of harmonics to use. \\
~~~~nsubharm & \tabitem & The number of sub-harmonics to use. \\ \hline
\end{xtabular}
\end{center}

\begin{lstlisting}[backgroundcolor=\color{white},language=vartools,basicstyle=\footnotesize,breaklines=false,prebreak={}]
-Starspot
    <"aov" | "ls" | "list" ["column" col] | "fix" period |
        "fixcolumn" <colname | colnum>>
    a0 b0 alpha0 i0 chi0 psi00 mconst0 fitP fita fitb
    fitalpha fiti fitchi fitpsi fitmconst correctlc omodel [model_outdir]
\end{lstlisting}

\begin{center}
\tablecaption{Input Parameters for ``-Starspot'' Command (Section~\ref{cmd:starspot})}
\footnotesize
\begin{xtabular}{|p{1in}p{5mm}p{5in}|}
\hline
``aov'' & \tabitem & Take the rotation period for the model from the most recent {\bf -aov} or {\bf -aov\_harm} command. \\ \hline
``ls'' & \tabitem & Take the rotation period for the model from the most recent {\bf -LS} command. \\ \hline
``list'' & \tabitem & Take the rotation period for the model from the input light curve list (optionally specifying the column to use). \\ \hline
``fix'' & \tabitem & Fix the rotation period for the model to the value specified by 'period'. \\ \hline
``fixcolumn'' & \tabitem & Set the rotation period for the model to the value from a previously executed command. \\ \hline
a0 & \tabitem & The initial value to use for $a$ in equation~\ref{eqn:starspot1}. \\ \hline
b0 & \tabitem & The initial value to use for $b$ in equation~\ref{eqn:starspot1}. \\ \hline
alpha0 & \tabitem & The initial value to use for the spot angular radius in degrees. \\ \hline
i0 & \tabitem & The initial value to use for the inclination of the stellar rotation axis in degrees ($90^{\circ}$ corresponds to the rotation axis being perpendicular to the line-of-sight). \\ \hline
chi0 & \tabitem & The initial value to use for the spot latitude in degrees ($0^{\circ}$ for a spot at the equator). \\ \hline
psi00 & \tabitem & The initial value to use for the longitude of the spot center (at the first time instance in the light curve) in degrees. \\ \hline
mconst0 & \tabitem & The initial value to use for the constant magnitude term ($M_0$ in Eq.~\ref{eqn:starspot1}). Set this to a negative value to have it determined automatically. \\ \hline
fitP & \tabitem & Flag indicating whether or not the rotation period should be varied in the fit. \\ \hline
fita & \tabitem & Flag indicating whether or not $a$ should be varied in the fit. \\ \hline
fitb & \tabitem & Flag indicating whether or not $b$ should be varied in the fit. \\ \hline
fitalpha & \tabitem & Flag indicating whether or not the spot angular radius should be varied in the fit. \\ \hline
fiti & \tabitem & Flag indicating whether or not the inclination of the stellar rotation axis should be varied in the fit. \\ \hline
fitchi & \tabitem & Flag indicating whether or not the spot latitude should be varied in the fit. \\ \hline
fitpsi & \tabitem & Flag indicating whether or not the spot longitude should be varied in the fit. \\ \hline
fitmconst & \tabitem & Flag indicating whether or not the constant magnitude should be varied in the fit. \\ \hline
correctlc & \tabitem & Flag indicating whether or not the best-fit model should be subtracted from the light curve before passing it on to the next command. \\ \hline
omodel & \tabitem & Flag indicating whether or not to write the best-fit model light curve, evaluated at the times in the input light curve, to a file. \\
~~~~model\_outdir & \tabitem & This parameter should be given if and only if omodel is set to 1. The filename for the output model light curve will be \$model\_outdir/\$basename.starspot.model where \$basename is the base filename of the input light curve file, stripped of directories. \\ \hline
\end{xtabular}
\end{center}

\begin{lstlisting}[backgroundcolor=\color{white},language=vartools,basicstyle=\footnotesize,breaklines=false,prebreak={}]
-stats var1,var2,... stats1,stats2,...
\end{lstlisting}

\begin{center}
\tablecaption{Input Parameters for ``-stats'' Command (Section~\ref{cmd:stats})}
\footnotesize
\begin{xtabular}{|p{1in}p{5mm}p{5in}|}
\hline
var1,var2,... & \tabitem & A comma-separated list of variable names to compute the statistics on. \\ \hline
stats1,stats2,... & \tabitem & A comma-separated list of one or more statistics to compute for each variable (every statistic is computed for every variable). Available statistics include: \\
~~~~mean & \tabitem & \\
~~~~weightedmean & \tabitem & mean of the vector weighted using the light curve uncertainties. \\
~~~~median & \tabitem & \\
~~~~wmedian & \tabitem & median of the vector weighted using the light curve uncertainties. \\
~~~~stddev & \tabitem & standard deviation calculated with respect to the mean. \\
~~~~meddev & \tabitem & standard deviation calculated with respect to the median. \\
~~~~medmeddev & \tabitem & median of the absolute deviations from the median. \\
~~~~MAD & \tabitem & $1.483 \times {\rm medmeddev}$. For a Gaussian distribution this equals the standard deviation in the limit of large $N$. \\
~~~~kurtosis & \tabitem & \\
~~~~skewness & \tabitem & \\
~~~~pct\%f & \tabitem & \%f percentile, where \%f is a floating point number between 0 and 100. Here 0 corresponds to the minimum value and 100 to the maximum value in the vector. \\
~~~~wpct\%f & \tabitem & percentile using the light curve uncertainties as weights. \\
~~~~max & \tabitem & maximum value, equivalent to pct100. \\
~~~~min & \tabitem & minimum value, equivalent to pct0. \\
~~~~sum & \tabitem & sum of all elements in the vector. \\ \hline
\end{xtabular}
\end{center}

\begin{lstlisting}[backgroundcolor=\color{white},language=vartools,basicstyle=\footnotesize,breaklines=false,prebreak={}]
-SYSREM Ninput_color ["column" col1] Ninput_airmass initial_airmass_file
    sigma_clip1 sigma_clip2 saturation correctlc omodel [model_outdir]
    otrends [trend_outfile] useweights
\end{lstlisting}

\begin{center}
\tablecaption{Input Parameters for ``-SYSREM'' Command (Section~\ref{cmd:sysrem})}
\footnotesize
\begin{xtabular}{|p{1in}p{5mm}p{5in}|}
\hline
Ninput\_color & \tabitem & The number of trend vectors to remove for which the color terms are to be specified initially. These will be read-in from the input light curve list file, by default starting from the next unused column. The terms must be in Ninput\_color consecutive columns. You can change the column to use for the first term with the ``column'' keyword. \\ \hline
Ninput\_airmass & \tabitem & The number of trend vectors to remove for which the airmass terms are to be specified initially. \\ \hline
initial\_airmass\_file & \tabitem & A file with the initial airmass trends to use. The first column in this file should be the JDs (or image IDs if the {\bf -matchstringid} option was given), and the subsequent Ninput\_airmass columns are the initial airmass trends. \\ \hline
sigma\_clip1 & \tabitem & $\sigma$-clipping factor used in calculating the mean magnitudes of the light curves. \\ \hline
sigma\_clip2 & \tabitem & $\sigma$-clipping factor used in determining whether or not points contribute to the airmass or color terms when performing the fit. \\ \hline
saturation & \tabitem & Any points with magnitude less than 'saturation' will not contribute to the fit. \\ \hline
correctlc & \tabitem & Flag indicating whether or not the best-fit trend model should be subtracted from the light curve before passing it on to the next command. \\ \hline
omodel & \tabitem & Flag indicating whether or not the model light curve files, evaluated at the observed times in the input light curve, should be output. \\
~~~~model\_outdir & \tabitem & This parameter should be given if and only if omodel is set to 1. The output model light curves will be written to files named \$model\_outdir/\$basename.sysrem.model where \$basename is the base filename of the input light curve, stripped of leading directories. \\ \hline
otrends & \tabitem & Flag indicating whether or not the final trends should be output to a file. \\
~~~~trend\_outfile & \tabitem & The name of the file to write the trends to. This parameter should be given if and only if otrends is set to 1. The output file will have JD (or image ID) in the first column, and the subsequent columns are for each trend signal. \\ \hline
useweights & \tabitem & Flag indicating whether or not the light curve uncertainties are used in performing the various fits. \\ \hline
\end{xtabular}
\end{center}

\begin{lstlisting}[backgroundcolor=\color{white},language=vartools,basicstyle=\footnotesize,breaklines=false,prebreak={}]
-TFA trendlist ["readformat" Nskip jdcol magcol]
    dates_file pixelsep ["xycol" xcol ycol]
    correctlc ocoeff [coeff_outdir] omodel [model_outdir]
\end{lstlisting}

\begin{center}
\tablecaption{Input Parameters for ``-TFA'' Command (Section~\ref{cmd:tfa})}
\footnotesize
\begin{xtabular}{|p{1in}p{5mm}p{5in}|}
\hline
trendlist & \tabitem & The name of the file containing the list of light curves to be used as trends. The first column in the file should contain the names of the light curve files, the second and third columns should be the X and Y coordinates of the stars. \\ \hline
``readformat'' & \tabitem & An optional keyword used to change the format assumed when reading in the trend light curves. \\
~~~~Nskip & \tabitem & The number of lines to skip from the top of the trend light curve files. The default is 0. Note that any line beginning with a '\#' is automatically skipped, and not counted in this number. \\
~~~~jdcol & \tabitem & The column storing the times in the trend light curve files. Note that if the {\bf -matchstringid} option is used with \va{}, then this should be the column with the image IDs. The default value is 1. \\
~~~~magcol & \tabitem & The column storing the magnitudes in the trend light curve files. The default is 2. \\ \hline
dates\_file & \tabitem & The name of a file providing a full list of times in the trend light curves or a full list of image IDs (if the {\bf -matchstringid} option is used). Times should be in the second column of this file, image IDs in the first column. \\ \hline
pixelsep & \tabitem & Trend stars within pixelsep of the light curve in question will not be used in detrending the light curve. \\ \hline
``xycol'' & \tabitem & Optional keyword to change the columns in the input light curve list file used to read-in the X and Y coordinates of the light curves. By default these are taken from the next unused columns. \\ \hline
correctlc & \tabitem & Flag indicating whether or not the best-fit trend model should be subtracted from the light curve before passing it on to the next command. \\ \hline
ocoeff & \tabitem & Flag indicating whether or not the trend coefficients should be output for each light curve. \\
~~~~coeff\_outdir & \tabitem & The trend coefficients will be written to files named \$coeff\_outdir/\$basename.tfa.coeff where \$basename is the base filename of the input light curve, stripped of directories. This parameter should be given if and only if ocoeff is set to 1. \\ \hline
omodel & \tabitem & Flag indicating whether or not to output the best-fit model trends, evaluated at the times in the input light curves. \\
~~~~model\_outdir & \tabitem & The trend models will be written to files named \$model\_outdir/\$basename.tfa.model where \$basename is the base filename of the input light curve, stripped of directories. This parameter should be given if and only if omodel is set to 1. \\ \hline
\end{xtabular}
\end{center}

\begin{lstlisting}[backgroundcolor=\color{white},language=vartools,basicstyle=\footnotesize,breaklines=false,prebreak={}]
-TFA_SR trendlist ["readformat" Nskip jdcol magcol] dates_file
    ["decorr" iterativeflag Nlcterms lccolumn1 lcorder1 ...] pixelsep
    ["xycol" colx coly]
    correctlc ocoeff [coeff_outdir] omodel [model_outdir] dotfafirst
    tfathresh maxiter <"bin" nbins ["period" <"aov" | "ls" 
    | "bls" | "list" ["column" col] | "fix" period>]
    | "signal" filename
    | "harm" Nharm Nsubharm ["period" <"aov" | "ls" 
    | "bls" | "list" ["column" col] | "fix" period>]>
\end{lstlisting}

\begin{center}
\tablecaption{Input Parameters for ``-TFA\_SR'' Command (Section~\ref{cmd:tfasr})}
\footnotesize
\begin{xtabular}{|p{1in}p{5mm}p{5in}|}
\hline
trendlist & \tabitem & The name of the file containing the list of light curves to be used as trends. The first column in the file should contain the names of the light curve files, the second and third columns should be the X and Y coordinates of the stars. \\ \hline
``readformat'' & \tabitem & An optional keyword used to change the format assumed when reading in the trend light curves. \\
~~~~Nskip & \tabitem & The number of lines to skip from the top of the trend light curve files. The default is 0. Note that any line beginning with a '\#' is automatically skipped, and not counted in this number. \\
~~~~jdcol & \tabitem & The column storing the times in the trend light curve files. Note that if the {\bf -matchstringid} option is used with \va{}, then this should be the column with the image IDs. The default value is 1. \\
~~~~magcol & \tabitem & The column storing the magnitudes in the trend light curve files. The default is 2. \\ \hline
dates\_file & \tabitem & The name of a file providing a full list of times in the trend light curves or a full list of image IDs (if the {\bf -matchstringid} option is used). Times should be in the second column of this file, image IDs in the first column. \\ \hline
``decorr'' & \tabitem & If this optional keyword is used, the the light curve will be simultaneously decorrelated against additional light-curve-specific signals. \\
~~~~iterativeflag & \tabitem & Flag indicating whether the decorrelation and TFA will be done iteratively (flag set to 1; this is faster) or if they will be done simultaneously (flag set to 0; slower but more correct). \\
~~~~Nlcterms & \tabitem & The number of decorrelation trends to use. \\
~~~~lccolumn1 & \tabitem & The column in the input light curve to use for the first light-curve-specific trend to decorrelate against. \\
~~~~lcorder1 & \tabitem & The polynomial order to use for the first trend. \\
~~~~$\ldots$ & \tabitem & The columns and orders should then be provided for the rest of the Nlcterms trends. \\ \hline
pixelsep & \tabitem & Trend stars within pixelsep of the light curve in question will not be used in detrending the light curve. \\ \hline
``xycol'' & \tabitem & Optional keyword to change the columns in the input light curve list file used to read-in the X and Y coordinates of the light curves. By default these are taken from the next unused columns. \\ \hline
correctlc & \tabitem & Flag indicating whether or not the best-fit trend model should be subtracted from the light curve before passing it on to the next command. \\ \hline
ocoeff & \tabitem & Flag indicating whether or not the trend coefficients should be output for each light curve. \\
~~~~coeff\_outdir & \tabitem & The trend coefficients will be written to files named \$coeff\_outdir/\$basename.tfa.coeff where \$basename is the base filename of the input light curve, stripped of directories. This parameter should be given if and only if ocoeff is set to 1. \\ \hline
omodel & \tabitem & Flag indicating whether or not to output the best-fit model trends, evaluated at the times in the input light curves. \\
~~~~model\_outdir & \tabitem & The trend models will be written to files named \$model\_outdir/\$basename.tfa.model where \$basename is the base filename of the input light curve, stripped of directories. This parameter should be given if and only if omodel is set to 1. \\ \hline
dotfafirst & \tabitem & Flag indicating whether TFA should be applied to the input light curve first, with the signal to preserve determined on the residual in each iteration (set the flag to 1), or if the signal is determined and subtracted from the light curve first and then TFA is applied to the residual in each iteration (set the flag to 0). \\ \hline
tfathresh & \tabitem & The iterations will stop if the fractional change in the r.m.s.\ is less than 'tfathresh'. \\ \hline
maxiter & \tabitem & The iterations will stop once 'maxiter' iterations have completed. \\ \hline
``bin'' & \tabitem & Use the binned light curve for the model signal to preserve. \\
~~~~nbins & \tabitem & The number of bins to use. \\
~~~~``period'' & \tabitem & Optional keyword to do phase-binning, rather than binning in time. \\
~~~~~~~~``aov'' & \tabitem & Take the period for phase-binning from the most recently executed {\bf -aov} of {\bf -aov\_harm} command. \\
~~~~~~~~``ls'' &  \tabitem & Take the period for phase-binning from the most recently executed {\bf -LS} command. \\
~~~~~~~~``bls'' &  \tabitem & Take the period for phase-binning from the most recently executed {\bf -BLS} command. \\
~~~~~~~~``list'' &  \tabitem & Take the period for phase-binning from the input light curve list file (optionally specifying the column). \\
~~~~~~~~``fix'' &  \tabitem & Fix the period for phase-binning to the value specified by 'period'. \\ \hline
``signal'' & \tabitem & Use a fixed signal form read-in from a file. Use 'filename' to specify a file listing the signal files, one for each light curve to process. Each signal file should contain the signal in the second column. The quantity $a\times S + b$ is fit to the light curve simultaneously with TFA, here $a$ and $b$ are free parameters and $S$ is the signal. \\ \hline
``harm'' & \tabitem & Use a harmonic series to model the signal. This will be fit simultaneously to the light curve with TFA. \\
~~~~Nharm & \tabitem & The number of harmonics to use. \\
~~~~Nsubharm & \tabitem & The number of sub-harmonics to use. \\
~~~~``period'' & \tabitem & Use this keyword to change the period for the harmonic series (the default is the time spanned by the light curve). \\
~~~~~~~~``aov'' & \tabitem & Use this keyword to take the period from the most recent {\bf -aov} or {\bf -aov\_harm} command. \\
~~~~~~~~``ls'' & \tabitem & Use this keyword to take the period from the most recent {\bf -LS} command. \\
~~~~~~~~``bls'' & \tabitem & Use this keyword to take the period from the most recent {\bf -BLS} command. \\
~~~~~~~~``list'' & \tabitem & Read the period from the light curve list file (optionally specify the column). \\
~~~~~~~~``fix'' & \tabitem & Fix the period to a specified value. \\ \hline
\end{xtabular}
\end{center}

\begin{lstlisting}[backgroundcolor=\color{white},language=vartools,basicstyle=\footnotesize,breaklines=false,prebreak={}]
-wwz <"maxfreq" <"auto" | maxfreq>> <"freqsamp" freqsamp>
    <"tau0" <"auto" | tau0>> <"tau1" <"auto" | tau1>>
    <"dtau" <"auto" | dtau>> ["c" cval]
    ["outfulltransform" outdir ["fits" | "pm3d"] ["format" format]]
    ["outmaxtransform" outdir ["format" format]]
\end{lstlisting}

\begin{center}
\tablecaption{Input Parameters for ``-wwz'' Command (Section~\ref{cmd:wwz})}
\footnotesize
\begin{xtabular}{|p{1in}p{5mm}p{5in}|}
\hline
``maxfreq'' & \tabitem & Indicate the maximum frequency for which the wavelet transform will be calculated. \\
~~~~``auto'' & \tabitem & If this keyword is given then the maximum frequency will be $1/(2\Delta t_{\rm min})$ where $\Delta t_{\rm min}$ is the minimum time seperation between consecutive points in the light curve. \\
~~~~maxfreq & \tabitem & The maximum frequency to use. \\ \hline
``freqsamp'' & \tabitem & Specify the frequency sampling as ${\rm freqsamp}/T$ where $T$ is the time baseline of the light curve. \\ \hline
``tau0'' & \tabitem & Indicate the minimum time-shift to test (the time-shift is $\tau$ in Eq.~\ref{eqn:waveletdef}). Use the ``auto'' keyword to set this to the minimum time in the light curve, otherwise provide a value. \\ \hline
``tau1'' & \tabitem & Indicate the maximum time-shift to test. Use the ``auto'' keyword to set this to the maximum time in the light curve, otherwise provide a value. \\ \hline
``dtau'' & \tabitem & Indicate the time-shift step to use. Use the ``auto'' keyword to set this to $\Delta t_{\rm min}$, otherwise provide a value. \\ \hline
``c'' & \tabitem & Optionally provide a value for the $c$ parameter in equation~\ref{eqn:wavelet2}. The default value is $c = (8\pi^2)^{-1}$. \\ \hline
``outfulltransform'' & \tabitem & Use this keyword to output the full wavelet transform for each light curve (i.e., calculated at every trial time-shift and frequency). \\
~~~~outdir & \tabitem & The directory to output the transforms to. The default naming convention is \$outdir/\$basename.wwz where \$basename is the base filename of the input light curve, stripped of directories. \\
~~~~``fits'' & \tabitem & Optionally output the transforms as multi-extension FITS image files. \\
~~~~``pm3d'' & \tabitem & Optionally output the transforms as ascii tables in a format suitable for plotting with the gnuplot pm3d plotting style. \\
~~~~``format'' & \tabitem & Modify the naming convention for the files. In this case the file will be named \$outdir/\$format where instances of \%s in \$format are replaced by \$basename, instances of \%d are replaced by the light curve number (starting with 1), instances of \%0nd, where n is an integer, are replaced with the formatted light curve number, and instances of \%\% are replaced with \%. \\ \hline
``outmaxtransform'' & \tabitem & Use this keyword to output the transform that maximizes $Z$ (Eq.~\ref{eqn:waveletZ}) over frequencies as a function of time-shift. \\
~~~~outdir & \tabitem & The directory to output these files to. The default naming convention is \$outdir/\$basename.mwwz where \$basename is the base filename of the input light curve, stripped of directories. \\
~~~~``format'' & \tabitem & Modify the naming convention for the files. In this case the file will be named \$outdir/\$format where instances of \%s in \$format are replaced by \$basename, instances of \%d are replaced by the light curve number (starting with 1), instances of \%0nd, where n is an integer, are replaced with the formatted light curve number, and instances of \%\% are replaced with \%. \\ \hline
\end{xtabular}
\end{center}

\section{Columns Output By Each Command}\label{sec:output}

The tables below list the columns added by each command to the output table.  The columns will be named \$colnum''\_''\$basecolname''\_''\$commandnum where \$colnum''\_'' is prepended only if the {\bf -numbercolumns} option is given to \va{}, \$colnum is the column number (starting at 1), \$basecolname is as listed below, and \$commandnum is the number of the command that this column was produced by (i.e., if one runs ``vartools -i input.txt -rms -chi2 -rms'' the columns produced by the first ``-rms'' call would have \$commandnum$=1$, those produced by the ``-chi2'' command would have \$commandnum$=2$, and those produced by the second ``-rms'' call would have \$commandnum$=3$). When using a ``fixcolumn'' option to one of the \va{} commands to set a parameter equal to the value of the command, one can either provide the column number (\$colnum), or the column name without the prepended column number (\$basecolname''\_''\$commandnum).  

\begin{center}
\tablecaption{Output Columns for ``-addnoise'' Command (Section~\ref{cmd:addnoise})}
\footnotesize
\begin{xtabular}{|p{1in}p{5mm}p{5in}|}
\hline
Addnoise\_Gamma\_1 & \tabitem & The value used for $\gamma$, as discussed in section~\ref{sec:lcsim}. This is only output if the ``wavelet'' noise model is used, and if ``list'' is used to initialize $\gamma$.\\ \hline
Addnoise\_Sig\_Red\_1 & \tabitem & The value used for the red-noise component standard deviation. This is only output if ``list'' is used to initialize red-noise standard deviation.\\ \hline
Addnoise\_Sig\_White\_1 & \tabitem & The value used for the white-noise component standard deviation.. This is only output if ``list'' is used to initialize the white noise standard deviation.\\ \hline
Addnoise\_Rho\_1 & \tabitem & The value used for $\rho$, as discussed in section~\ref{sec:lcsim}. This is only output if the ``squareexp'', ``exp'' or ``matern'' noise model is used, and if ``list'' is used to initialize $\rho$.\\ \hline
Addnoise\_Nu\_1 & \tabitem & The value used for $\nu$, as discussed in section~\ref{sec:lcsim}. This is only output if the ``matern'' noise model is used, and if ``list'' is used to initialize $\nu$.\\ \hline
\end{xtabular}
\end{center}

\begin{center}
\tablecaption{Output Columns for ``-alarm'' Command (Section~\ref{cmd:alarm})}
\footnotesize
\begin{xtabular}{|p{1in}p{5mm}p{5in}|}
\hline
Alarm & \tabitem & The computed alarm statistic. \\ \hline
\end{xtabular}
\end{center}

\begin{center}
\tablecaption{Output Columns for ``-aov'' Command (Section~\ref{cmd:aov})}
\footnotesize
\begin{xtabular}{|p{1.5in}p{5mm}p{4.5in}|}
\hline
Period\_\$peaknum & \tabitem & The period for peak number \$peaknum. \$peaknum runs from 1 for the highest peak in the periodogram to 'Npeak', the number of peaks to identify as given on the command line. Note that all columns for a given peak are output together, and then the same quantities are repeated for the next peak, until all 'Npeak' peaks have been reported.\\ \hline
AOV\_LOGSNR\_\$peaknum & \tabitem & The spectroscopic S/N measured on the logarithmic AoV periodogram for peak \$peaknum. This is output only if the ``uselog'' keyword was given to the {\bf -aov} command. \\ \hline
AOV\_\$peaknum & \tabitem & The value of the AoV periodogram for peak \$peaknum. This is output only if the ``uselog'' keyword was not given to the {\bf -aov} command. \\ \hline
AOV\_SNR\_\$peaknum & \tabitem & The spectroscopic S/N measured on the AoV periodogram for peak \$peaknum. This is output only if the ``uselog'' keyword was not given to the {\bf -aov} command. \\ \hline
AOV\_NEG\_LN\_FAP\_\$peaknum & \tabitem & The negative natural logarithm of the formal false alarm probability for peak \$peaknum. This is output only if the ``uselog'' keyword was not given to the {\bf -aov} command. \\ \hline
Mean\_lnAOV\_\$peaknum & \tabitem & The mean value of the logarithmic AoV periodogram at the whitening cycle used to find peak \$peaknum. This is output only if the ``uselog'' and the ``whiten'' keywords were both given to the {\bf -aov} command. \\ \hline
RMS\_lnAOV\_\$peaknum & \tabitem & The standard deviation of the logarithmic AoV periodogram at the whitening cycle used to find peak \$peaknum. This is output only if the ``uselog'' and the ``whiten'' keywords were both given to the {\bf -aov} command. \\ \hline
PeriodFix & \tabitem & The fixed period at which to report the periodogram properties. This is output only if the ``fixperiodSNR'' keyword was given to the {\bf -aov} command. \\ \hline
AOV\_LOGSNR\_PeriodFix & \tabitem & The spectroscopic S/N measured on the logarithmic AoV periodogram for the fixed period. This is output only if the ``fixperiodSNR'' and the ``uselog'' keywords were both given to the {\bf -aov} command. \\ \hline
AOV\_PeriodFix & \tabitem & The value of the AoV periodogram at the fixed period. This is output only if the ``fixperiodSNR'' keyword is given and the ``uselog'' keyword is not given. \\ \hline
AOV\_SNR\_PeriodFix & \tabitem & The spectroscopic S/N of the AoV periodogram at the fixed period. This is output only if the ``fixperiodSNR'' keyword is given and the ``uselog'' keyword is not given. \\ \hline
AOV\_NEG\_LN\_FAP\_PeriodFix & \tabitem & The negative natural logarithm of the formal false alarm probability at the fixed period. This is output only if the ``fixperiodSNR'' keyword is given and the ``uselog'' keyword is not given. \\ \hline
Mean\_lnAOV & \tabitem & The mean value of the logarithmic AoV periodogram. This is output only if the ``uselog'' keyword is given and the ``whiten'' keyword is not given. \\ \hline
RMS\_lnAOV & \tabitem & The standard deviation of the logarithmic AoV periodogram. This is output only if the ``uselog'' keyword is given and the ``whiten'' keyword is not given. \\ \hline
\end{xtabular}
\end{center}

\begin{center}
\tablecaption{Output Columns for ``-aov\_harm'' Command (Section~\ref{cmd:aovharm})}
\footnotesize
\begin{xtabular}{|p{2.0in}p{5mm}p{4.0in}|}
\hline
Period\_\$peaknum & \tabitem & The period for peak number \$peaknum. \$peaknum runs from 1 for the highest peak in the periodogram to 'Npeak', the number of peaks to identify as given on the command line. Note that all columns for a given peak are output together, and then the same quantities are repeated for the next peak, until all 'Npeak' peaks have been reported.\\ \hline
AOV\_HARM\_\$peaknum & \tabitem & The value of the AoV periodogram for peak \$peaknum. This is output only if a positive number was specified for the 'Nharm' parameter input to the {\bf -aov\_harm} command.\\ \hline
AOV\_HARM\_NEG\_LOG\_FAP\_\$peaknum & \tabitem & The negative logarithm of the formal false alarm probability for peak \$peaknum. This is output here only if a value less than or equal to zero was specified for the 'Nharm' parameter input to the {\bf -aov\_harm} command.\\ \hline
AOV\_HARM\_NHARM\_\$peaknum & \tabitem & The optimum number of harmonics used for peak \$peaknum. This is output here only if a value less than or equal to zero was specified for the 'Nharm' parameter input to the {\bf -aov\_harm} command.\\ \hline
AOV\_HARM\_SNR\_\$peaknum & \tabitem & The spectroscopic S/N measured on the AoV periodogram for peak number \$peaknum.\\ \hline
AOV\_HARM\_NEG\_LOG\_FAP\_\$peaknum & \tabitem & The negative logarithm of the formal false alarm probability for peak \$peaknum. This is output here only if a positive number was specified for the 'Nharm' parameter input to the {\bf -aov\_harm} command.\\ \hline
Mean\_AOV\_HARM\_\$peaknum & \tabitem & The mean value of the AoV periodogram for the whitening cycle in which peak \$peaknum was found. This is only output if the ``whiten'' keyword was given.\\ \hline
RMS\_AOV\_HARM\_\$peaknum & \tabitem & The standard deviation of the AoV periodogram for the whitening cycle in which peak \$peaknum was found. This is only output if the ``whiten'' keyword was given.\\ \hline
PeriodFix & \tabitem & The fixed period at which to report the periodogram properties. This is output only if the ``fixperiodSNR'' keyword was given to the {\bf -aov\_harm} command. \\ \hline
AOV\_HARM\_PeriodFix & \tabitem & The value of the AoV periodogram at the fixed period. This is output only if the ``fixperiodSNR'' keyword is given. \\ \hline
AOV\_HARM\_SNR\_PeriodFix & \tabitem & The spectroscopic S/N of the AoV periodogram at the fixed period. This is output only if the ``fixperiodSNR'' keyword is given. \\ \hline
AOV\_HARM\_NEG\_LN\_FAP\_PeriodFix & \tabitem & The negative logarithm of the formal false alarm probability for the fixed period. This is output here only if the ``fixperiodSNR'' keyword is given, and if a positive value was specified for the 'Nharm' parameter input to the {\bf -aov\_harm} command.\\ \hline
Mean\_AOV\_HARM & \tabitem & The mean of the AoV periodogram. This is output only if the ``whiten'' keyword was not provided.\\ \hline
RMS\_AOV\_HARM & \tabitem & The standard deviation of the AoV periodogram. This is output only if the ``whiten'' keyword was not provided.\\ \hline
\end{xtabular}
\end{center}

\begin{center}
\tablecaption{Output Columns for ``-autocorrelation'' Command (Section~\ref{cmd:autocorrelation})}
\footnotesize
\begin{xtabular}{|p{5.75in}p{0.1mm}p{0.25in}|}
\hline
No columns added to the output table. & & \\ \hline
\end{xtabular}
\end{center}

\begin{center}
\tablecaption{Output Columns for ``-binlc'' Command (Section~\ref{cmd:binlc})}
\footnotesize
\begin{xtabular}{|p{5.75in}p{0.1mm}p{0.25in}|}
\hline
No columns added to the output table. & & \\ \hline
\end{xtabular}
\end{center}

\begin{center}
\tablecaption{Output Columns for ``-BLS'' Command (Section~\ref{cmd:bls})}
\footnotesize
\begin{xtabular}{|p{2.0in}p{5mm}p{4.0in}|}
\hline
BLS\_Period\_\$peaknum & \tabitem & The period for peak number \$peaknum. \$peaknum runs from 1 for the highest peak in the periodogram to 'Npeak', the number of peaks to identify as given on the command line. Note that all columns for a given peak are output together, and then the same quantities are repeated for the next peak, until all 'Npeak' peaks have been reported.\\ \hline
BLS\_Tc\_\$peaknum & \tabitem & The time of central transit for peak number \$peaknum. \\ \hline
BLS\_SN\_\$peaknum & \tabitem & The spectroscopic S/N estimate for peak number \$peaknum calculated following equation~\ref{eqn:BLSSN}. If the ``nobinnedrms'' keyword is given, then this is instead equal to the ${\rm SDE}$ (Eq.~\ref{eqn:BLSSDE}), except that a $3\sigma$ clipping is first applied to the BLS spectrum before computing the average value and standard deviation to use in that equation. The BLS peaks are sorted by this quantity. \\ \hline
BLS\_SR\_\$peaknum & \tabitem & The signal residue for peak number \$peaknum (Eq.~\ref{eqn:BLSSR}).\\ \hline
BLS\_SDE\_\$peaknum & \tabitem & The signal detection efficiency as defined by \citet{kovacs:2002:BLS} for peak number \$peaknum. \\ \hline
BLS\_Depth\_\$peaknum & \tabitem & The transit depth in magnitudes (or whatever units are used for the input light curve) for peak number \$peaknum. \\ \hline
BLS\_Qtran\_\$peaknum & \tabitem & The fractional transit duration (duration divided by orbital period) for peak number \$peaknum. \\ \hline
BLS\_Qingress\_\$peaknum & \tabitem & The fractional ingress duration (duration of ingress divided by total transit duration) for peak number \$peaknum. This is output only if the ``fittrap'' keyword was used. \\ \hline
BLS\_OOTmag\_\$peaknum & \tabitem & The out-of-transit mangitude level for peak number \$peaknum. This is output here only if the ``fittrap'' keyword was used. \\ \hline
BLS\_i1\_\$peaknum & \tabitem & The phase of transit ingress for peak number \$peaknum when time $t=0$ is used as a reference time for zero phase. \\ \hline
BLS\_i2\_\$peaknum & \tabitem & The phase of transit egress for peak number \$peaknum when time $t=0$ is used as a reference time for zero phase. \\ \hline
BLS\_deltaChi2\_\$peaknum & \tabitem & The value of $\Delta \chi^2 = \chi^{2}_{0} - \chi^{2}_{\rm tr}$ where $\chi^{2}_{0}$ is for a model with no variation, and $\chi^{2}_{\rm tr}$ is for the transit model found for peak number \$peaknum. \\ \hline
BLS\_fraconenight\_\$peaknum & \tabitem & The fraction of $\Delta \chi^2$ that comes from a single night for peak number \$peaknum. \\ \hline
BLS\_Npointsintransit\_\$peaknum & \tabitem & The number of points in transit for peak number \$peaknum. \\ \hline
BLS\_Ntransits\_\$peaknum & \tabitem & The number of distinct transits with at least one point in transit for peak number \$peaknum. \\ \hline
BLS\_Npointsbeforetransit\_\$peaknum & \tabitem & The number of points in a phase bin of equal size to the transit and immediately preceding the transit, for peak number \$peaknum. \\ \hline
BLS\_Npointsaftertransit\_\$peaknum & \tabitem & The number of points in a phase bin of equal size to the transit and immediately following the transit, for peak number \$peaknum. \\ \hline
BLS\_Rednoise\_\$peaknum & \tabitem & An estimate of the red-noise of the light curve at the time-scale of a transit for peak number \$peaknum. This is calculated following equation~\ref{eqn:BLSrednoise}. \\ \hline
BLS\_Whitenoise\_\$peaknum & \tabitem & An estimate of the white-noise of the light curve for peak number \$peaknum, calculated as the standard deviation of the light curve after subtracting the transit. \\ \hline
BLS\_SignaltoPinknoise\_\$peaknum & \tabitem & An estimate of the S/N of the transit accounting for both red and white noise for peak number \$peaknum. This is calculated following equation~\ref{eqn:BLSSNpink}. \\ \hline
BLS\_Period\_invtransit & \tabitem & Period for the highest S/N inverse transit (periodic, box-shaped brightening rather than dimming) found in the light curve. \\ \hline
BLS\_deltaChi2\_invtransit & \tabitem & The value of $\Delta \chi^2 = \chi^{2}_{0} - \chi^{2}_{\rm invtr}$ where $\chi^{2}_{0}$ is for a model with no variation, and $\chi^{2}_{\rm invtr}$ is the best-fit inverse transit model found. \\ \hline
BLS\_MeanMag & \tabitem & The mean magnitude of the light curve which is subtracted before applying the BLS algorithm. \\ \hline
\end{xtabular}
\end{center}

\begin{center}
\tablecaption{Output Columns for ``-BLSFixPer'' Command (Section~\ref{cmd:blsfixper})}
\footnotesize
\begin{xtabular}{|p{2.0in}p{5mm}p{4.0in}|}
\hline
BLSFixPer\_Period & \tabitem & The period at which the BLS quantities are calculated. \\ \hline
BLSFixPer\_Tc & \tabitem & The time of transit center. \\ \hline
BLSFixPer\_SR & \tabitem & The signal residue at this period (Eq.~\ref{eqn:BLSSR}).\\ \hline
BLSFixPer\_Depth & \tabitem & The transit depth in magnitudes (or whatever units are used for the input light curve). \\ \hline
BLSFixPer\_Qtran & \tabitem & The fractional transit duration (duration divided by orbital period). \\ \hline
BLSFixPer\_Qingress & \tabitem & The fractional ingress duration (duration of ingress divided by total transit duration). This is output only if the ``fittrap'' keyword was used. \\ \hline
BLSFixPer\_OOTmag & \tabitem & The out-of-transit mangitude level. This is output here only if the ``fittrap'' keyword was used. \\ \hline
BLSFixPer\_i1 & \tabitem & The phase of transit ingress when time $t=0$ is used as a reference time for zero phase. \\ \hline
BLSFixPer\_i2 & \tabitem & The phase of transit egress when time $t=0$ is used as a reference time for zero phase. \\ \hline
BLSFixPer\_deltaChi2 & \tabitem & The value of $\Delta \chi^2 = \chi^{2}_{0} - \chi^{2}_{\rm tr}$ where $\chi^{2}_{0}$ is for a model with no variation, and $\chi^{2}_{\rm tr}$ is for the transit model. \\ \hline
BLSFixPer\_fraconenight & \tabitem & The fraction of $\Delta \chi^2$ that comes from a single night. \\ \hline
BLSFixPer\_Npointsintransit & \tabitem & The number of points in transit. \\ \hline
BLSFixPer\_Ntransits & \tabitem & The number of distinct transits with at least one point in transit. \\ \hline
BLSFixPer\_Npointsbeforetransit & \tabitem & The number of points in a phase bin of equal size to the transit and immediately preceding the transit. \\ \hline
BLSFixPer\_Npointsaftertransit & \tabitem & The number of points in a phase bin of equal size to the transit and immediately following the transit. \\ \hline
BLSFixPer\_Rednoise & \tabitem & An estimate of the red-noise of the light curve at the time-scale of a transit. This is calculated following equation~\ref{eqn:BLSrednoise}. \\ \hline
BLSFixPer\_Whitenoise & \tabitem & An estimate of the white-noise of the light curve, calculated as the standard deviation of the light curve after subtracting the transit. \\ \hline
BLSFixPer\_SignaltoPinknoise & \tabitem & An estimate of the S/N of the transit accounting for both red and white noise. This is calculated following equation~\ref{eqn:BLSSNpink}. \\ \hline
BLSFixPer\_deltaChi2\_invtransit & \tabitem & The value of $\Delta \chi^2 = \chi^{2}_{0} - \chi^{2}_{\rm invtr}$ where $\chi^{2}_{0}$ is for a model with no variation, and $\chi^{2}_{\rm invtr}$ is the best-fit inverse transit model found at the fixed period. \\ \hline
BLSFixPer\_MeanMag & \tabitem & The mean magnitude of the light curve which is subtracted before calculating the BLS statistics. \\ \hline
\end{xtabular}
\end{center}

\begin{center}
\tablecaption{Output Columns for ``-BLSFixDurTc'' Command (Section~\ref{cmd:blsfixdurtc})}
\footnotesize
\begin{xtabular}{|p{2.0in}p{5mm}p{4.0in}|}
\hline
BLSFixDurTc\_Duration & \tabitem & The fixed transit duration adopted for this command. \\ \hline
BLSFixDurTc\_Tc & \tabitem & The fixed time of transit center adopted for this command. \\ \hline
BLSFixDurTc\_Depth & \tabitem & The fixed transit depth adopted for this command. This is output here only if the ``fixdepth'' keyword was used. \\ \hline
BLSFixDurTc\_Qingress & \tabitem & The fixed fractional ingress duration (duration of ingress divided by total transit duration) adopted for this command. This is output only if the ``fixdepth'' keyword was used. If the ``qgress'' keyword was not given, this will have a value of 0. \\ \hline
BLSFixDurTc\_Period\_\$peaknum & \tabitem & The period for peak number \$peaknum. \$peaknum runs from 1 for the highest S/N peak in the periodogram to 'Npeak', the number of peaks to identify as given on the command line. Note that all columns for a given peak are output together, and then the same quantities are repeated for the next peak, until all 'Npeak' peaks have been reported.\\ \hline
BLSFixDurTc\_SN\_\$peaknum & \tabitem & The spectroscopic S/N estimate for peak number \$peaknum equal to the ${\rm SDE}$ (Eq.~\ref{eqn:BLSSDE}), except that a $3\sigma$ clipping is first applied to the BLS spectrum before computing the average value and standard deviation to use in that equation. The BLS peaks are sorted by this quantity. \\ \hline
BLSFixDurTc\_SR\_\$peaknum & \tabitem & The Signal Residue for peak \$peaknum (Eq.~\ref{eqn:BLSSR}).\\ \hline
BLSFixDurTc\_SDE\_\$peaknum & \tabitem & The Signal Detection Efficiency for peak \$peaknum (Eq.~\ref{eqn:BLSSDE}).\\ \hline
BLSFixDurTc\_Depth\_\$peaknum & \tabitem & The transit depth in magnitudes (or whatever units are used for the input light curve) for peak number \$peaknum. \\ \hline
BLSFixDurTc\_Qtran\_\$peaknum & \tabitem & The fractional transit duration (duration divided by orbital period) for peak number \$peaknum. \\ \hline
BLSFixDurTc\_Qingress\_\$peaknum & \tabitem & The fractional ingress duration (duration of ingress divided by total transit duration) for peak number \$peaknum. This is output only if the ``fittrap'' keyword was used and the ``fixdepth'' keyword was not used. \\ \hline
BLSFixDurTc\_OOTmag\_\$peaknum & \tabitem & The out-of-transit mangitude level for peak number \$peaknum. This is output here only if the ``fittrap'' keyword was used and the ``fixdepth'' keyword was not used. \\ \hline
BLSFixDurTc\_deltaChi2 & \tabitem & The value of $\Delta \chi^2 = \chi^{2}_{0} - \chi^{2}_{\rm tr}$ for peak number \$peaknum where $\chi^{2}_{0}$ is for a model with no variation, and $\chi^{2}_{\rm tr}$ is for the transit model. \\ \hline
BLSFixDurTc\_fraconenight\_\$peaknum & \tabitem & The fraction of $\Delta \chi^2$ that comes from a single night for peak number \$peaknum. \\ \hline
BLSFixDurTc\_Npointsintransit\_\$peaknum & \tabitem & The number of points in transit for peak number \$peaknum. \\ \hline
BLSFixDurTc\_Ntransits\_\$peaknum & \tabitem & The number of distinct transits with at least one point in transit for peak number \$peaknum. \\ \hline
BLSFixDurTc\_Npointsbeforetransit\_\$peaknum & \tabitem & The number of points in a phase bin of equal size to the transit and immediately preceding the transit for peak number \$peaknum. \\ \hline
BLSFixDurTc\_Npointsaftertransit\_\$peaknum & \tabitem & The number of points in a phase bin of equal size to the transit and immediately following the transit for peak number \$peaknum. \\ \hline
BLSFixDurTc\_Rednoise\_\$peaknum & \tabitem & An estimate of the red-noise of the light curve at the time-scale of a transit for peak number \$peaknum. This is calculated following equation~\ref{eqn:BLSrednoise}. \\ \hline
BLSFixDurTc\_Whitenoise\_\$peaknum & \tabitem & An estimate of the white-noise of the light curve for peak number \$peaknum, calculated as the standard deviation of the light curve after subtracting the transit. \\ \hline
BLSFixDurTc\_SignaltoPinknoise\_\$peaknum & \tabitem & An estimate of the S/N of the transit for peak number \$peaknum accounting for both red and white noise. This is calculated following equation~\ref{eqn:BLSSNpink}. \\ \hline
BLSFixDurTc\_Period\_invtransit & \tabitem & Period for the highest S/N inverse transit (periodic, box-shaped brightening rather than dimming) found with the same duration, $T_{c}$ and other constraints. \\ \hline
BLSFixDurTc\_deltaChi2\_invtransit & \tabitem & The value of $\Delta \chi^2 = \chi^{2}_{0} - \chi^{2}_{\rm invtr}$ where $\chi^{2}_{0}$ is for a model with no variation, and $\chi^{2}_{\rm invtr}$ is for the inverse transit model. \\ \hline
BLSFixDurTc\_MeanMag & \tabitem & The mean magnitude of the light curve which is subtracted before calculating the BLS statistics. \\ \hline
\end{xtabular}
\end{center}

\begin{center}
\tablecaption{Output Columns for ``-changeerror'' Command (Section~\ref{cmd:changeerror})}
\footnotesize
\begin{xtabular}{|p{1in}p{5mm}p{5in}|}
\hline
Mean\_Mag & \tabitem & The mean of the light curve
magnitudes. \\ \hline
RMS & \tabitem & The r.m.s.\ scatter of the light curve
magnitudes. All light curve uncertainties (including those with input
uncertainties $\sigma \leq 0$) are set equal to this quantity by this
command. \\ \hline
Npoints & \tabitem & The number of points in the light curve which
contributed to the r.m.s.\ scatter estimate. Points with input
uncertainty $\sigma \leq 0$ and points with NaN magnitude values do
not contribute to the scatter or mean estimates. \\ \hline
\end{xtabular}
\end{center}

\begin{center}
\tablecaption{Output Columns for ``-changevariable'' Command (Section~\ref{cmd:changevariable})}
\footnotesize
\begin{xtabular}{|p{5.75in}p{0.1mm}p{0.25in}|}
\hline
No columns added to the output table. & & \\ \hline
\end{xtabular}
\end{center}

\begin{center}
\tablecaption{Output Columns for ``-chi2'' Command (Section~\ref{cmd:chi2})}
\footnotesize
\begin{xtabular}{|p{1in}p{5mm}p{5in}|}
\hline
Chi2 & \tabitem & The value of $\chi^{2}_{0}/{\rm d.o.f.}$, the reduced $\chi^2$ of the light curve magnitudes. \\ \hline
Weighted\_Mean\_Mag & \tabitem & The mean magnitude of the light curve, using the input uncertainties to determine the weights. This is the value used in calculating the magnitude residuals for computing $\chi^2$. \\ \hline
\end{xtabular}
\end{center}

\begin{center}
\tablecaption{Output Columns for ``-chi2bin'' Command (Section~\ref{cmd:chi2bin})}
\footnotesize
\begin{xtabular}{|p{2in}p{5mm}p{4in}|}
\hline
Chi2Bin\_\$bintime & \tabitem & The reduced $\chi^2$ after binning the light curve with bintime \$bintime. The value of \$bintime is displayed to two decimal places.  Note that both columns for a given \$bintime (Chi2Bin and Weighted\_Mean\_Mag) are output together, and then the same quantities are repeated for the next \$bintime, until the results for all 'Nbin' values of \$bintime given on the command line have been reported.\\
Weighted\_Mean\_Mag\_Bin\_\$bintime & \tabitem & The mean magnitude of the light curve, using the input uncertainties to determine the weights, after binning with by \$bintime. \\ \hline
\end{xtabular}
\end{center}

\begin{center}
\tablecaption{Output Columns for ``-clip'' Command (Section~\ref{cmd:clip})}
\footnotesize
\begin{xtabular}{|p{1in}p{5mm}p{5in}|}
\hline
Nclip & \tabitem & The number of points clipped from the light curve. \\ \hline
\end{xtabular}
\end{center}

\begin{center}
\tablecaption{Output Columns for ``-converttime'' Command (Section~\ref{cmd:converttime})}
\footnotesize
\begin{xtabular}{|p{5.75in}p{0.1mm}p{0.25in}|}
\hline
No columns added to the output table. & & \\ \hline
\end{xtabular}
\end{center}

\begin{center}
\tablecaption{Output Columns for ``-copylc'' Command (Section~\ref{cmd:copylc})}
\footnotesize
\begin{xtabular}{|p{5.75in}p{0.1mm}p{0.25in}|}
\hline
No columns added to the output table. & & \\ \hline
\end{xtabular}
\end{center}

\begin{center}
\tablecaption{Output Columns for ``-decorr'' Command (Section~\ref{cmd:decorr})}
\footnotesize
\begin{xtabular}{|p{2in}p{5mm}p{4in}|}
\hline
Decorr\_constant\_term & \tabitem & The best-fit value for the zero-point term in the decorrelation model. This is output only if the 'zeropointterm' flag was set to 1.\\ \hline
Decorr\_constant\_term\_err & \tabitem & The uncertainty on the zero-point term in the decorrelation model. This is output only if the 'zeropointterm' flag was set to 1. The uncertainties output from this command are correct only if the input magnitude uncertainties are accurate, and if the noise in the light curve is uncorrelated and follows a Gaussian distribution.\\ \hline
Global\_\$i\_coeff\_\$j & \tabitem & The best-fit value of the coefficient for global parameter sequence number \$i (running from 1 to the value of 'Nglobalterms' given on the command line) raised to the power \$j (running from 1 to the polynomial 'order' specified for this sequence on the command line). A separate column is output for each combination of \$i and \$j, with the loop over \$j nested within the loop over \$i. The best-fit value, and uncertainty for a given \$i and \$j combination are output together.\\ \hline
Global\_\$i\_coeff\_err\_\$j & \tabitem & The uncertainty on Global\_\$i\_coeff\_\$j.\\ \hline
LCColumn\_\$lccolumn$[$\$i$]$\_coeff\_\$j & \tabitem & The best-fit value of the coefficient for the parameter sequence read from column \$lccolumn$[$\$i$]$ in the input light curve (\$i runs from 1 to the value of 'Nlcterms' given on the command line, and \$lccolumn$[$\$i$]$ is the value of 'lccolumn' given on the command line for term \$i) raised to the power \$j (running from 1 to the polynomial 'order' specified for this light curve column). A separate column is output for each combination of \$i and \$j, with the loop over \$j nested within the loop over \$i. The best-fit value, and uncertainty for a given \$i and \$j combination are output together.\\ \hline
LCColumn\_\$lccolumn$[$\$i$]$\_coeff\_err\_\$j & \tabitem & The uncertainty on LCColumn\_\$lccolumn$[$\$i$]$\_coeff\_\$j.\\ \hline
Decorr\_chi2 & \tabitem & The $\chi^2$ of the fit (not reduced).\\ \hline
\end{xtabular}
\end{center}

\begin{center}
\tablecaption{Output Columns for ``-dftclean'' Command (Section~\ref{cmd:dftclean})}
\footnotesize
\begin{xtabular}{|p{2.5in}p{5mm}p{3.5in}|}
\hline
DFTCLEAN\_DSPEC\_PEAK\_FREQ\_\$peaknum & \tabitem & The frequency in cycles per day (assuming input times are in days) for peak number \$peaknum identified in the ``dirty'' DFT spectrum (i.e., the spectrum before application of the ``CLEAN'' deconvolution algorithm). \$peaknum runs from 1 for the highest peak in the spectrum to 'Npeaks', the number of peaks to identify in the ``dirty'' spectrum as given on the command line. Note that all columns for a given peak are output together, and then the same quantities are repeated for the next peak, until all 'Npeaks' peaks have been reported. These columns are only reported if the ``findddirtypeaks'' keyword is given.\\ \hline
DFTCLEAN\_DSPEC\_PEAK\_POW\_\$peaknum & \tabitem & The value of the power spectrum at peak number \$peaknum. This is equal to $P(f) = F_{r}^2(f) + F_{i}^2(f)$ where $F_{r}$ and $F_{i}$ are the real and imaginary components, respectively, of $F(f)$ given in equation~\ref{eqn:dft}. If the ``useampspec'' keyword is given on the command line, then it is the square root of this quantity that is reported.\\ \hline
DFTCLEAN\_DSPEC\_PEAK\_SNR\_\$peaknum & \tabitem & The spectroscopic S/N ratio for peak number \$peaknum. This equal to $(P(f) - \bar{P})/\sigma_{P}$, where $P(f)$ is the value of the power spectrum at peak number \$peaknum (or amplitude spectrum if the ``useampspec'' keyword is given), and $\bar{P}$ and $\sigma_{P}$ are the average and standard deviation of the spectrum, respectively, after $\sigma$-clipping.\\ \hline
DFTCLEAN\_DSPEC\_AVESPEC & \tabitem & The value of $\bar{P}$ used in computing the S/N ratio(s) for the dirty power spectrum. This is output only if the ``verboseout'' keyword is given, and if the ``finddirtypeaks'' keyword is given with 'Npeaks' greater than 0.\\ \hline
DFTCLEAN\_DSPEC\_STDSPEC & \tabitem & The value of $\sigma_{P}$ used in computing the S/N ratio(s) for the dirty power spectrum. This is output only if the ``verboseout'' keyword is given, and if the ``finddirtypeaks'' keyword is given with 'Npeaks' greater than 0.\\ \hline
DFTCLEAN\_DSPEC\_AVESPEC\_NOCLIP & \tabitem & The value of $\bar{P}$ for the dirty spectrum when no clipping is applied. This is output only if the ``verboseout'' keyword is given, and if the ``finddirtypeaks'' keyword is given with 'Npeaks' greater than 0.\\ \hline
DFTCLEAN\_DSPEC\_STDSPEC\_NOCLIP & \tabitem & The value of $\sigma_{P}$ for the dirty spectrum when no clipping is applied. This is output only if the ``verboseout'' keyword is given, and if the ``finddirtypeaks'' keyword is given with 'Npeaks' greater than 0.\\ \hline
DFTCLEAN\_CSPEC\_PEAK\_FREQ\_\$peaknum & \tabitem & The frequency in cycles per day (assuming input times are in days) for peak number \$peaknum identified in the ``CLEAN'' DFT spectrum. \$peaknum runs from 1 for the highest peak in the spectrum to 'Npeaks', the number of peaks to identify in the ``CLEAN'' spectrum as given on the command line. Note that all columns for a given peak are output together, and then the same quantities are repeated for the next peak, until all 'Npeaks' peaks have been reported. These columns are only reported if the CLEAN algirthm is performed, and the ``findcleanpeaks'' keyword is given.\\ \hline
DFTCLEAN\_CSPEC\_PEAK\_POW\_\$peaknum & \tabitem & The value of the ``CLEAN'' power spectrum at peak number \$peaknum.\\ \hline
DFTCLEAN\_CSPEC\_PEAK\_SNR\_\$peaknum & \tabitem & The spectroscopic S/N ratio for peak number \$peaknum in the ``CLEAN'' spectrum.\\ \hline
DFTCLEAN\_CSPEC\_AVESPEC & \tabitem & The value of $\bar{P}$ used in computing the S/N ratio(s) for the clean power spectrum. This is output only if the ``verboseout'' keyword is given, and if the ``findcleanpeaks'' keyword is given with 'Npeaks' greater than 0.\\ \hline
DFTCLEAN\_CSPEC\_STDSPEC & \tabitem & The value of $\sigma_{P}$ used in computing the S/N ratio(s) for the clean power spectrum. This is output only if the ``verboseout'' keyword is given, and if the ``findcleanpeaks'' keyword is given with 'Npeaks' greater than 0.\\ \hline
DFTCLEAN\_CSPEC\_AVESPEC\_NOCLIP & \tabitem & The value of $\bar{P}$ for the clean spectrum when no clipping is applied. This is output only if the ``verboseout'' keyword is given, and if the ``findcleanpeaks'' keyword is given with 'Npeaks' greater than 0.\\ \hline
DFTCLEAN\_CSPEC\_STDSPEC\_NOCLIP & \tabitem & The value of $\sigma_{P}$ for the clean spectrum when no clipping is applied. This is output only if the ``verboseout'' keyword is given, and if the ``findcleanpeaks'' keyword is given with 'Npeaks' greater than 0.\\ \hline
\end{xtabular}
\end{center}

\begin{center}
\tablecaption{Output Columns for ``-difffluxtomag'' Command (Section~\ref{cmd:difffluxtomag})}
\footnotesize
\begin{xtabular}{|p{5.75in}p{0.1mm}p{0.25in}|}
\hline
No columns added to the output table. & & \\ \hline
\end{xtabular}
\end{center}

\begin{center}
\tablecaption{Output Columns for ``-ensemblerescalesig'' Command (Section~\ref{cmd:ensemblerescalesig})}
\footnotesize
\begin{xtabular}{|p{1in}p{5mm}p{5in}|}
\hline
SigmaRescaleFactor & \tabitem & The value of $\sqrt{\chi^{2}_{\rm new}/\chi^{2}_{\rm old}}$ where $\chi^{2}_{\rm old}$ and $\chi^{2}_{\rm new}$ are the reduced $\chi^2$ values of the light curve before and after transforming the magnitude uncertainties, respectively. \\ \hline
\end{xtabular}
\end{center}

\begin{center}
\tablecaption{Output Columns for ``-expr'' Command (Section~\ref{cmd:expr})}
\footnotesize
\begin{xtabular}{|p{5.75in}p{0.1mm}p{0.25in}|}
\hline
No columns added to the output table. & & \\ \hline
\end{xtabular}
\end{center}

\begin{center}
\tablecaption{Output Columns for ``-findblends'' Command (Section~\ref{cmd:findblends})}
\footnotesize
\begin{xtabular}{|p{1in}p{5mm}p{5in}|}
\hline
FindBlends\_Period & \tabitem & The period of the harmonic series fitted to the light curves. \\ \hline
FindBlends\_LCname & \tabitem & The name of the light curve with the largest amplitude (in flux) harmonic series at the indicated period. This light curve is the most likely source for the measured variability.\\ \hline
FindBlends\_FluxAmp & \tabitem & The amplitude (in flux) of the highest-amplitude light curve.\\ \hline
\end{xtabular}
\end{center}

\begin{center}
\tablecaption{Output Columns for ``-fluxtomag'' Command (Section~\ref{cmd:fluxtomag})}
\footnotesize
\begin{xtabular}{|p{5.75in}p{0.1mm}p{0.25in}|}
\hline
No columns added to the output table. & & \\ \hline
\end{xtabular}
\end{center}

\begin{center}
\tablecaption{Output Columns for ``-GetLSAmpThresh'' Command (Section~\ref{cmd:getlsampthresh})}
\footnotesize
\begin{xtabular}{|p{1in}p{5mm}p{5in}|}
\hline
LS\_AmplitudeScaleFactor & \tabitem & The scaling factor $\alpha$ for which the detected FAP would be equal to the threshold.\\ \hline
LS\_MinimumAmplitude & \tabitem & The minimum peak-to-peak amplitude that the light curve could have had, and still have been detected by L-S with a FAP below the threshold.\\ \hline
\end{xtabular}
\end{center}

\begin{center}
\tablecaption{Output Columns for ``-if'' Command (Section~\ref{cmd:if})}
\footnotesize
\begin{xtabular}{|p{5.75in}p{0.1mm}p{0.25in}|}
\hline
No columns added to the output table. & & \\ \hline
\end{xtabular}
\end{center}

\begin{center}
\tablecaption{Output Columns for ``-Injectharm'' Command (Section~\ref{cmd:injectharm})}
\footnotesize
\begin{xtabular}{|p{1.5in}p{5mm}p{4.5in}|}
\hline
Injectharm\_Period & \tabitem & The period of the harmonic series added to the light curve.\\ \hline
Injectharm\_Subharm\_\$k\_Amp & \tabitem & The value of $B_{k}$ in equation~\ref{eqn:injectharm} for $k = $\$k. Here \$k varies from 2 to 'Nsubharm'$+1$, where 'Nsubharm' is the number of sub-harmonics to use, as indicated on the command line. The amplitude and phase for sub-harmonic \$k are output together, then the amplitude and phase of the next sub-harmonic are output, up to 'Nsubharm'$+1$.\\ \hline
Injectharm\_Subharm\_\$k\_Phase & \tabitem & The value of $\psi_{k}$ in equation~\ref{eqn:injectharm} for $k = $\$k.\\ \hline
Injectharm\_Fundamental\_Amp & \tabitem & The value of $A_{1}$ in equation~\ref{eqn:injectharm}.\\ \hline
Injectharm\_Fundamental\_Phase & \tabitem & The value of $\phi_{1}$ in equation~\ref{eqn:injectharm}.\\ \hline
Injectharm\_Harm\_\$k\_Amp & \tabitem & The value of $A_{k}$ in equation~\ref{eqn:injectharm} for $k = $\$k. Here \$k varies from 2 to 'Nharm'$+1$, where 'Nharm' is the number of harmonics to use, as indicated on the command line. The amplitude and phase for harmonic \$k are output together, then the amplitude and phase of the next harmonic are output, up to 'Nharm'$+1$.\\ \hline
Injectharm\_Harm\_\$k\_Phase & \tabitem & The value of $\phi_{k}$ in equation~\ref{eqn:injectharm} for $k = $\$k.\\ \hline
\end{xtabular}
\end{center}

\begin{center}
\tablecaption{Output Columns for ``-Injecttransit'' Command (Section~\ref{cmd:injecttransit})}
\footnotesize
\begin{xtabular}{|p{1.5in}p{5mm}p{4.5in}|}
\hline
Injecttransit\_Period & \tabitem & The period of the transit signal added to the light curve.\\ \hline
Injecttransit\_Rp & \tabitem & The radius of the planet in Jupiter radii for the transit signal added to the light curve.\\ \hline
Injecttransit\_Mp & \tabitem & The mass of the planet in Jupiter masses for the transit signal added to the light curve.\\ \hline
Injecttransit\_phase & \tabitem & The orbital phase at time $t = 0$ (phase$=0$ corresponds to transit center) for the transit signal added to the light curve.\\ \hline
Injecttransit\_sin\_i & \tabitem & The sine of the orbital inclination angle for the transit signal added to the light curve.\\ \hline
Injecttransit\_e & \tabitem & The orbital eccentricity $e$ for the transit signal added to the light curve. This is output only if the ``eomega'' keyword is used.\\ \hline
Injecttransit\_omega & \tabitem & The argument of periastron $\omega$ (in degrees) for the transit signal added to the light curve. This is output only if the ``eomega'' keyword is used.\\ \hline
Injecttransit\_h & \tabitem & The value of $e\sin\omega$ for the transit signal added to the light curve. This is output only if the ``hk'' keyword is used.\\ \hline
Injecttransit\_k & \tabitem & The value of $e\cos\omega$ for the transit signal added to the light curve. This is output only if the ``hk'' keyword is used.\\ \hline
Injecttransit\_Mstar & \tabitem & The mass of the star in Solar masses for the transit signal added to the light curve.\\ \hline
Injecttransit\_Rstar & \tabitem & The radius of the star in Solar radii for the transit signal added to the light curve.\\ \hline
Injecttransit\_ld\_\$i & \tabitem & The value of limb darkening coefficient \$i (\$i takes values from 1 to 2 for the quadratic limb-darkening law, or from 1 to 4 for the non-linear limb-darkening law) for the transit signal added to the light curve.\\ \hline
\end{xtabular}
\end{center}

\begin{center}
\tablecaption{Output Columns for ``-Jstet'' Command (Section~\ref{cmd:jstet})}
\footnotesize
\begin{xtabular}{|p{1in}p{5mm}p{5in}|}
\hline
Jstet & \tabitem & The computed value of Stetson's $J$ statistic (Eq.~\ref{eqn:jstet}).\\ \hline
Kurtosis & \tabitem & The estimated kurtosis of the light curve (Eq.~\ref{eqn:kurtosisstet}.\\ \hline
Lstet & \tabitem & The computed value of Stetson's $L$ statistic (Eq.~\ref{eqn:lstet}).\\ \hline
\end{xtabular}
\end{center}

\begin{center}
\tablecaption{Output Columns for ``-Killharm'' Command (Section~\ref{cmd:killharm})}
\footnotesize
\begin{xtabular}{|p{2.5in}p{5mm}p{3.5in}|}
\hline
Killharm\_Mean\_Mag & \tabitem & The fitted value of $m_{0}$ from equation~\ref{eqn:harm}.\\ \hline
Killharm\_Period\_\$i & \tabitem & The value of the period for harmonic series $i=$\$i$+1$ ($1/f_{i}$ in Eq.~\ref{eqn:harm}). Here \$i takes values from 1 to 'Nper', where 'Nper' is the number of periods indicated on the command-line ('Nper' is 1 for the ``aov'', ``ls'' and ``injectharm'' options, it is 2 for the ``both'' option).  All columns for a given value of \$i are output before the columns for the next value of \$i are output.\\ \hline
Killharm\_Per\$i\_Subharm\_\$k\_Sincoeff & \tabitem & The fitted value of $c_{i,k}$ from equation~\ref{eqn:harm} where $k=$\$k. \$k takes values from 2 to 'Nsubharm'$+1$, where 'Nsubharm' is the number of sub-harmonics to use as specified on the command line. This is output only if none of the keywords ``outampphase'', ``outampradphase'', ``outRphi'' or ``outRradphi'' are given.\\ \hline
Killharm\_Per\$i\_Subharm\_\$k\_Coscoeff & \tabitem & The fitted value of $d_{i,k}$ from equation~\ref{eqn:harm} where $k=$\$k. \$k takes values from 2 to 'Nsubharm'$+1$, where 'Nsubharm' is the number of sub-harmonics to use as specified on the command line. This is output only if none of the keywords ``outampphase'', ``outampradphase'', ``outRphi'' or ``outRradphi'' are given.\\ \hline
Killharm\_Per\$i\_Subharm\_Amp\_\$k & \tabitem & The fitted value of $B_{i,k}$ from equation~\ref{eqn:subharmamp}. This is output only if one of the keywords ``outampphase'' or ``outampradphase'' is given.\\ \hline
Killharm\_Per\$i\_Subharm\_Phi\_\$k & \tabitem & The fitted value of $\psi_{i,k}$ from equation~\ref{eqn:subharmphase}. This is output only if one of the keywords ``outampphase'' or ``outampradphase'' is given. If ``outampphase'' is given, then $\psi_{i,k}/(2\pi)$ is reported.\\ \hline
Killharm\_Per\$i\_Subharm\_R\_\$k\_1 & \tabitem & The fitted value of the relative amplitude $Q_{i,k,1}$ from equation~\ref{eqn:subharmrelamp}. This is output only if the ``outRphi'' or ``outRradphi'' keyword is given.\\ \hline
Killharm\_Per\$i\_Subharm\_Phi\_\$k\_1 & \tabitem & The fitted value of the relative phase $\psi_{i,k,1}$ from equation~\ref{eqn:subharmrelphase}. This is output only if the ``outRphi'' or ``outRradphi'' keyword is given. If ``outRphi'' is given, then $\psi_{i,k,1}/(2\pi)$ is reported.\\ \hline
Killharm\_Per\$i\_Fundamental\_Sincoeff & \tabitem & The fitted value of $a_{i,1}$ from equation~\ref{eqn:harm}. This is output only if none of the keywords ``outampphase'', ``outampradphase'', ``outRphi'' or ``outRradphi'' are given.\\ \hline
Killharm\_Per\$i\_Fundamental\_Coscoeff & \tabitem & The fitted value of $b_{i,1}$ from equation~\ref{eqn:harm}. This is output only if none of the keywords ``outampphase'', ``outampradphase'', ``outRphi'' or ``outRradphi'' are given.\\ \hline
Killharm\_Per\$i\_Fundamental\_Amp & \tabitem & The fitted value of $A_{i,1}$ from equation~\ref{eqn:harmamp}. This is output only if one of the keywords ``outampphase'', ``outampradphase'', ``outRphi'' or ``outRradphi'' is given.\\ \hline
Killharm\_Per\$i\_Fundamental\_Phi & \tabitem & The fitted value of $\phi_{i,1}$ from equation~\ref{eqn:harmphase}. This is output only if one of the keywords ``outampphase'', ``outampradphase'', ``outRphi'' or ``outRradphi'' is given. If ``outampphase'' or ``outRphi'' is given, then `$\phi_{i,1}/(2\pi)$ is reported.\\ \hline
Killharm\_Per\$i\_Harm\_\$k\_Sincoeff & \tabitem & The fitted value of $a_{i,k}$ from equation~\ref{eqn:harm} where $k=$\$k. \$k takes values from 2 to 'Nharm'$+1$, where 'Nharm' is the number of harmonics to use as specified on the command line. This is output only if none of the keywords ``outampphase'', ``outampradphase'', ``outRphi'' or ``outRradphi'' are given.\\ \hline
Killharm\_Per\$i\_Harm\_\$k\_Coscoeff & \tabitem & The fitted value of $b_{i,k}$ from equation~\ref{eqn:harm} where $k=$\$k. \$k takes values from 2 to 'Nharm'$+1$, where 'Nharm' is the number of harmonics to use as specified on the command line. This is output only if none of the keywords ``outampphase'', ``outampradphase'', ``outRphi'' or ``outRradphi'' are given.\\ \hline
Killharm\_Per\$i\_Harm\_Amp\_\$k & \tabitem & The fitted value of $A_{i,k}$ from equation~\ref{eqn:harmamp}. This is output only if one of the keywords ``outampphase'' or ``outampradphase'' is given.\\ \hline
Killharm\_Per\$i\_Harm\_Phi\_\$k & \tabitem & The fitted value of $\phi_{i,k}$ from equation~\ref{eqn:harmphase}. This is output only if one of the keywords ``outampphase'' or ``outampradphase'' is given. If ``outampphase'' is given, then $\phi_{i,k}/(2\pi)$ is reported.\\ \hline
Killharm\_Per\$i\_Harm\_R\_\$k\_1 & \tabitem & The fitted value of the relative amplitude $R_{i,k,1}$ from equation~\ref{eqn:harmrelamp}. This is output only if the ``outRphi'' or ``outRradphi'' keyword is given.\\ \hline
Killharm\_Per\$i\_Harm\_Phi\_\$k\_1 & \tabitem & The fitted value of the relative phase $\phi_{i,k,1}$ from equation~\ref{eqn:harmrelphase}. This is output only if the ``outRphi'' or ``outRradphi'' keyword is given. If ``outRphi'' is given, then $\phi_{i,k,1}/(2\pi)$ is reported.\\ \hline
Killharm\_Per\$i\_Amplitude & \tabitem & The peak-to-peak amplitude of the full Fourier series (fundamental, sub-harmonics and harmonics) for period \$i.\\ \hline
\end{xtabular}
\end{center}

\begin{center}
\tablecaption{Output Columns for ``-linfit'' Command (Section~\ref{cmd:linfit})}
\footnotesize
\begin{xtabular}{|p{1in}p{5mm}p{5in}|}
\hline
Linfit\_\$paramname & \tabitem & The fitted value for parameter \$paramname, where \$paramname is the name of a variable in 'paramlist' supplied on the command line. The Linfit\_\$paramname and Linfit\_err\$paramname columns for each parameter \$paramname are output together, followed by the respective columns for the next value of \$paramname, until the results have been shown for all varied parameters.\\ \hline
Linfit\_err\$paramname & \tabitem & The formal uncertainty for parameter \$paramname, this is accurate only if the noise in the light curve is Gaussian and uncorrelated in time, and if the measurement uncertainties supplied in the light curve are correct.\\ \hline
\end{xtabular}
\end{center}

\begin{center}
\tablecaption{Output Columns for ``-LS'' Command (Section~\ref{cmd:ls})}
\footnotesize
\begin{xtabular}{|p{2in}p{5mm}p{4in}|}
\hline
LS\_Period\_\$peaknum & \tabitem & The period for peak number \$peaknum. \$peaknum runs from 1 for the highest peak in the periodogram to 'Npeaks', the number of peaks to identify as given on the command line. Note that all columns for a given peak are output together, and then the same quantities are repeated for the next peak, until all 'Npeaks' peaks have been reported.\\ \hline
Log10\_LS\_Prob\_\$peaknum & \tabitem & The base-10 logarithm of the false alarm probability for peak number \$peaknum, calculated using equation~\ref{eqn:LSFAP}, or following \citet{press:1992} if the ``noGLS'' keyword is given.\\ \hline
LS\_Periodogram\_Value\_\$peaknum & \tabitem & The value of the Generalized L-S periodogram (Eq.~\ref{eqn:LSdef}) at peak number \$peaknum. If the ``noGLS'' keyword is given, this is instead the traditional normalized L-S periodogram value.\\ \hline
LS\_SNR\_\$peaknum & \tabitem & The spectroscopic ${\rm S/N}$ ratio for peak number \$peaknum (Eq.~\ref{eqn:LSSNR}).\\ \hline
LS\_PeriodFix & \tabitem & The fixed period at which the Generalized L-S periodogram statistics are to be evaluated. This is output only if the ``fixperiodSNR'' keyword is given.\\ \hline
Log10\_LS\_Prob\_PeriodFix & \tabitem & The base-10 logarithm of the false alarm probability at the fixed period. This is output only if the ``fixperiodSNR'' keyword is given.\\ \hline
LS\_Periodogram\_Value\_PeriodFix & \tabitem & The value of the Generalized L-S periodogram at the fixed period. This is output only if the ``fixperiodSNR'' keyword is given.\\ \hline
LS\_SNR\_PeriodFix & \tabitem & The spectroscopic ${\rm S/N}$ at the fixed period. This is output only if the ``fixperiodSNR'' keyword is given.\\ \hline
\end{xtabular}
\end{center}

\begin{center}
\tablecaption{Output Columns for ``-MandelAgolTransit'' Command (Section~\ref{cmd:mandelagoltransit})}
\footnotesize
\begin{xtabular}{|p{2in}p{5mm}p{4in}|}
\hline
MandelAgolTransit\_Period & \tabitem & The period of the transit signal. This is either the input value if the parameter is not varied, or the fitted parameter if it is. This also holds for the other parameters output by this command.\\ \hline
MandelAgolTransit\_T0 & \tabitem & The time of transit center.\\ \hline
MandelAgolTransit\_r & \tabitem & The ratio of the planetary and stellar radii ($R_{P}/R_{\star}$).\\ \hline
MandelAgolTransit\_a & \tabitem & The ratio of the semimajor axis to the stellar radius ($a/R_{\star}$).\\ \hline
MandelAgolTransit\_bimpact & \tabitem & The normalized impact parameter of the transit (in units of $R_{\star}+R_{P}$).\\ \hline
MandelAgolTransit\_inc & \tabitem & The orbital inclination angle of the transiting system ($90^{\circ}$ is an edge on system giving central transits).\\ \hline
MandelAgolTransit\_e & \tabitem & The orbital eccentricity.\\ \hline
MandelAgolTransit\_omega & \tabitem & The argument of periastron in degrees.\\ \hline
MandelAgolTransit\_mconst & \tabitem & The out-of-transit magnitude.\\ \hline
MandelAgolTransit\_ldcoeff1 & \tabitem & The first limb darkening coefficient.\\ \hline
MandelAgolTransit\_ldcoeff2 & \tabitem & The second limb darkening coefficient.\\ \hline
MandelAgolTransit\_ldcoeff3 & \tabitem & The third limb darkening coefficient. This is output only if the nonlinear limb darkening law is used.\\ \hline
MandelAgolTransit\_ldcoeff4 & \tabitem & The fourth limb darkening coefficient. This is output only if the nonlinear limb darkening law is used.\\ \hline
MandelAgolTransit\_K & \tabitem & The semi-amplitude of the RV orbit. This is output only if the fitRV flag is set to 1.\\ \hline
MandelAgolTransit\_gamma & \tabitem & The zero-point velocity of the RV orbit. This is output only if the fitRV flag is set to 1.\\ \hline
MandelAgolTransit\_chi2 & \tabitem & $\chi^2$ per degree of freedom about the best-fit model. This includes the RVs if they are fitted.\\ \hline
\end{xtabular}
\end{center}

\begin{center}
\tablecaption{Output Columns for ``-medianfilter'' Command (Section~\ref{cmd:medianfilter})}
\footnotesize
\begin{xtabular}{|p{5.75in}p{0.1mm}p{0.25in}|}
\hline
No columns added to the output table. & & \\ \hline
\end{xtabular}
\end{center}

\begin{center}
\tablecaption{Output Columns for ``-microlens'' Command (Section~\ref{cmd:microlens})}
\footnotesize
\begin{xtabular}{|p{2in}p{5mm}p{4in}|}
\hline
Microlens\_f0 & \tabitem & The parameter $f_{0}$ in equation~\ref{eqn:microlens1}. This is either the input value if the parameter is not varied (the ``novary'' keyword was used for this parameter), or the fitted parameter if it is. This also holds for the other parameters output by this command.\\ \hline
Microlens\_f1 & \tabitem & The parameter $f_{1}$ in equation~\ref{eqn:microlens1}\\ \hline
Microlens\_u0 & \tabitem & The parameter $u_{0}$ in equation~\ref{eqn:microlens3}.\\ \hline
Microlens\_t0 & \tabitem & The parameter $t_{0}$ in equation~\ref{eqn:microlens3}.\\ \hline
Microlens\_tmax & \tabitem & The parameter $t_{\rm max}$ in equation~\ref{eqn:microlens3}.\\ \hline
Microlens\_chi2perdof & \tabitem & $\chi^2$ per degree of freedom about the best-fit microlens model.\\ \hline
\end{xtabular}
\end{center}

\begin{center}
\tablecaption{Output Columns for ``-nonlinfit'' Command (Section~\ref{cmd:nonlinfit})}
\footnotesize
\begin{xtabular}{|p{2in}p{5mm}p{4in}|}
\hline
Nonlinfit\_HasConverged & \tabitem & This is ``1'' if the DHSX fitting procedure converged to a solution, or ``0'' if the routine terminated before convergence had been achieved. The parameter is output only if the ``amoeba'' fitting mode is used.\\ \hline
Nonlinfit\_\$paramname$[$\$i$]$\_BestFit & \tabitem & The best-fit (minimum $X^{2}$; Eq.~\ref{eqn:nonlinfitx2}) value for \$paramname$[$\$i$]$, the name of the \$i-th parameter listed in 'paramlist' (together with 'linfitparams' if the ``linfit'' kyword is used) given on the command line. Here \$i runs from 1 to the total number of parameters listed in 'paramlist' and 'linfitparams', and all columns for a given parameter are output before continuing to the next parameter in the sequence.\\ \hline
Nonlinfit\_\$paramname$[$\$i$]$\_Err & \tabitem & The formal uncertainty on \$paramname$[$\$i$]$ determined by finding the size of the perturbation to the parameter which leads to $\Delta X^{2} = 1$ from the best-fit solution, when all other parameters are held constant. This is output only if the ``amoeba'' fitting mode is used.\\ \hline
Nonlinfit\_BestFit\_Chi2 & \tabitem & The value of $X^{2}$ for the best-fit model.\\ \hline
Nonlinfit\_\$expression$[$\$i$]$\_\$stats$[$\$j$]$ & \tabitem & The result of calculating the statistic \$stats$[$\$j$]$ on the expression \$expression$[$\$i$]$ evaluated on the DEMCMC chains. This is output only if the ``mcmc'' fitting mode is used. Here \$expression$[$\$i$]$ is the \$i-th expression listed in 'exprlist' on the command-line if the ``chainstats'' keyword is used, or the \$i-th parameter name in 'paramlist' if the ``chainstats'' keyword is not used. The term \$stats$[$\$j$]$ is a string based on the \$j-th statistic in 'statslist' (if the ``chainstats'' keyword is not given then 'median,stddev' is assumed for 'statslist'). A separate column is output for each combination of \$i and \$j, with the loop over \$j nested within the loop over \$i. The possible values for \$stats$[$\$j$]$ are: ``MEAN'', ``WEIGHTEDMEAN'', ``MEDIAN'', ``WEIGHTEDMEDIAN'', ``STDDEV'', ``MEDDEV'', ``MEDMEDDEV'', ``MAD'', ``KURTOSIS'', ``SKEWNESS'', ``PCT\$pctval'' (where \$pctval is the percentile value to two decimal places), ``WPCT\$pctval'', ``MAX'', ``MIN'' and ``SUM''. These terms are output only if the ``mcmc'' fitting mode is used.\\ \hline
\end{xtabular}
\end{center}

\begin{center}
\tablecaption{Output Columns for ``-o'' Command (Section~\ref{cmd:o})}
\footnotesize
\begin{xtabular}{|p{5.75in}p{0.1mm}p{0.25in}|}
\hline
No columns added to the output table. & & \\ \hline
\end{xtabular}
\end{center}

\begin{center}
\tablecaption{Output Columns for ``-Phase'' Command (Section~\ref{cmd:phase})}
\footnotesize
\begin{xtabular}{|p{5.75in}p{0.1mm}p{0.25in}|}
\hline
No columns added to the output table. & & \\ \hline
\end{xtabular}
\end{center}

\begin{center}
\tablecaption{Output Columns for ``-resample'' Command (Section~\ref{cmd:resample})}
\footnotesize
\begin{xtabular}{|p{5.75in}p{0.1mm}p{0.25in}|}
\hline
No columns added to the output table. & & \\ \hline
\end{xtabular}
\end{center}

\begin{center}
\tablecaption{Output Columns for ``-rescalesig'' Command (Section~\ref{cmd:rescalesig})}
\footnotesize
\begin{xtabular}{|p{1in}p{5mm}p{5in}|}
\hline
SigmaRescaleFactor & \tabitem & The factor by which the magnitude uncertainties are scaled to achieve $\chi^2/{\rm dof} = 1$ for that light curve.\\ \hline
\end{xtabular}
\end{center}

\begin{center}
\tablecaption{Output Columns for ``-restorelc'' Command (Section~\ref{cmd:restorelc})}
\footnotesize
\begin{xtabular}{|p{5.75in}p{0.1mm}p{0.25in}|}
\hline
No columns added to the output table. & & \\ \hline
\end{xtabular}
\end{center}

\begin{center}
\tablecaption{Output Columns for ``-restricttimes'' Command (Section~\ref{cmd:restricttimes})}
\footnotesize
\begin{xtabular}{|p{1in}p{5mm}p{5in}|}
\hline
RestrictTimes\_MinJD & \tabitem & The minimum JD used for restricting the light curve times. This is output only if the ``JDrange'' or ``JDrangebylc'' keyword is used. \\ \hline
RestrictTimes\_MaxJD & \tabitem & The maximum JD used for restricting the light curve times. This is output only if the ``JDrange'' or ``JDrangebylc'' keyword is used. \\ \hline
\end{xtabular}
\end{center}

\begin{center}
\tablecaption{Output Columns for ``-rms'' Command (Section~\ref{cmd:rms})}
\footnotesize
\begin{xtabular}{|p{1in}p{5mm}p{5in}|}
\hline
Mean\_Mag & \tabitem & The mean of the light curve magnitudes. \\ \hline
RMS & \tabitem & The r.m.s.\ scatter of the light curve magnitudes. \\  \hline
Expected\_RMS & \tabitem & The expected r.m.s.\ scatter of the light curve magnitudes based on the magnitude uncertainties (Eq.~\ref{eqn:expectedrms}).\\ \hline
Npoints & \tabitem & The number of points in the light curve which contribute to r.m.s.\ measurement. Only points with magnitude values that are not NaN, and which have magnitude uncertainties greater than zero contribute.\\ \hline
\end{xtabular}
\end{center}

\begin{center}
\tablecaption{Output Columns for ``-rmsbin'' Command (Section~\ref{cmd:rmsbin})}
\footnotesize
\begin{xtabular}{|p{2in}p{5mm}p{4in}|}
\hline
RMSBin\_\$bintime & \tabitem & The r.m.s.\ scatter of the light curve magnitudes after binning the light curve with bintime \$bintime. The value of \$bintime is displayed to two decimal places.  Note that both columns for a given \$bintime (RMSBin and Expected\_RMS\_Bin) are output together, and then the same quantities are repeated for the next \$bintime, until the results for all 'Nbin' values of \$bintime given on the command line have been reported.\\
Expected\_RMS\_Bin\_\$bintime & \tabitem & The expected r.m.s.\ scatter of the light curve magnitudes, based on the magnitude uncertainties, after binning with by \$bintime. \\ \hline
\end{xtabular}
\end{center}

\begin{center}
\tablecaption{Output Columns for ``-savelc'' Command (Section~\ref{cmd:restorelc})}
\footnotesize
\begin{xtabular}{|p{5.75in}p{0.1mm}p{0.25in}|}
\hline
No columns added to the output table. & & \\ \hline
\end{xtabular}
\end{center}

\begin{center}
\tablecaption{Output Columns for ``-SoftenedTransit'' Command (Section~\ref{cmd:softenedtransit})}
\footnotesize
\begin{xtabular}{|p{2in}p{5mm}p{4in}|}
\hline
SoftenedTransit\_Period & \tabitem & The period for the transit model ($P$ in Eq.~\ref{eqn:softenedtransit2}).  This is either the input value if the parameter is not varied, or the fitted parameter if it is. This also holds for the other parameters output by this command.\\ \hline
SoftenedTransit\_T0 & \tabitem & The transit epoch ($T_{0}$ in Eq.~\ref{eqn:softenedtransit2}).\\ \hline
SoftenedTransit\_eta & \tabitem & The transit duration ($\eta$ in Eq.~\ref{eqn:softenedtransit2}).\\ \hline
SoftenedTransit\_cval & \tabitem & The transit sharpness ($c$ in Eq.~\ref{eqn:softenedtransit1}).\\ \hline
SoftenedTransit\_delta & \tabitem & The transit depth ($\delta$ in Eq.~\ref{eqn:softenedtransit1}).\\ \hline
SoftenedTransit\_mconst & \tabitem & The out-of-transit magnitude ($M_{0}$ in Eq.~\ref{eqn:softenedtransit1}).\\ \hline
SoftenedTransit\_perharm & \tabitem & The period for the harmonic series fit simultaneously to the light curve with the transit model (this is included only if the 'fit\_harm' flag is set to '1').\\ \hline
SoftenedTransit\_Subharm\_\$k\_Sincoeff & \tabitem & Coefficient of the sine function for sub-harmonic \$k where \$k varies from 2 to 'nsubharm'$+1$ (i.e., this is $c_{1,\$k}$ in Eq.~\ref{eqn:harm} with $N_{P} = 1$). This is output only if the 'fit\_harm' flag is set to '1'.\\ \hline
SoftenedTransit\_Subharm\_\$k\_Coscoeff & \tabitem & Coefficient of the cosine function for sub-harmonic \$k where \$k varies from 2 to 'nsubharm'$+1$ (i.e., this is $d_{1,\$k}$ in Eq.~\ref{eqn:harm} with $N_{P} = 1$). This is output only if the 'fit\_harm' flag is set to '1'.\\ \hline
SoftenedTransit\_Fundamental\_Sincoeff & \tabitem & Coefficient of the sine function for the fundamental mode in the harmonic series fit simultaneously to the transit model (i.e., this is $a_{1,1}$ in Eq.~\ref{eqn:harm} with $N_{P} = 1$). This is output only if the 'fit\_harm' flag is set to '1'.\\ \hline
SoftenedTransit\_Fundamental\_Coscoeff & \tabitem & Coefficient of the cosine function for the fundamental mode in the harmonic series fit simultaneously to the transit model (i.e., this is $b_{1,1}$ in Eq.~\ref{eqn:harm} with $N_{P} = 1$). This is output only if the 'fit\_harm' flag is set to '1'.\\ \hline
SoftenedTransit\_Harm\_\$k\_Sincoeff & \tabitem & Coefficient of the sine function for harmonic \$k where \$k varies from 2 to 'nharm'$+1$ (i.e., this is $a_{1,\$k}$ in Eq.~\ref{eqn:harm} with $N_{P} = 1$). This is output only if the 'fit\_harm' flag is set to '1'.\\ \hline
SoftenedTransit\_Harm\_\$k\_Coscoeff & \tabitem & Coefficient of the cosine function for harmonic \$k where \$k varies from 2 to 'nharm'$+1$ (i.e., this is $b_{1,\$k}$ in Eq.~\ref{eqn:harm} with $N_{P} = 1$). This is output only if the 'fit\_harm' flag is set to '1'.\\ \hline
SoftenedTransit\_chi2perdof & \tabitem & $\chi^2$ per degree of freedom for the best-fit model.\\ \hline
\end{xtabular}
\end{center}

\begin{center}
\tablecaption{Output Columns for ``-Starspot'' Command (Section~\ref{cmd:starspot})}
\footnotesize
\begin{xtabular}{|p{2in}p{5mm}p{4in}|}
\hline
Starspot\_Period & \tabitem & The rotation period of the star. This is either the input value if the parameter is not varied, or the fitted parameter if it is. This also holds for the other parameters output by this command.\\ \hline
Starspot\_a & \tabitem & The parameter $a$ in equation~\ref{eqn:starspot1}. \\ \hline
Starspot\_b & \tabitem & The parameter $b$ in equation~\ref{eqn:starspot1}. \\ \hline
Starspot\_alpha & \tabitem & The angular radius of the starspot in degrees. \\ \hline
Starspot\_inclination & \tabitem & The inclination angle of the stellar rotation axis in degrees ($90^{\circ}$ corresponds to the rotation axis being perpendicular to the line-of-sight). \\ \hline
Starspot\_chi & \tabitem & The spot latitude in degrees ($0^{\circ}$ for a spot at the equator).\\ \hline
Starspot\_psi0 & \tabitem & The longitude of the spot center at the first time instance in the light curve, in degrees. \\ \hline
Starspot\_mconst & \tabitem & The constant magnitude term ($M_{0}$ in Eq.~\ref{eqn:starspot1}).\\ \hline
Starspot\_chi2perdof & \tabitem & $\chi^2$ per degree of freedom for the best-fit model.\\ \hline
\end{xtabular}
\end{center}

\begin{center}
\tablecaption{Output Columns for ``-stats'' Command (Section~\ref{cmd:stats})}
\footnotesize
\begin{xtabular}{|p{2in}p{5mm}p{4in}|}
\hline
STATS\_\$varname$[$\$i$]$\_\$stats$[$\$j$]$ & \tabitem & The statistic \$stats$[$\$j$]$ calculated for variable \$varname$[$\$i$]$. Here \$varname$[$\$i$]$ is the \$i-th variable listed in 'var1,var2,...' on the command-line. The term \$stats$[$\$j$]$ is a string based on the \$j-th statistic in 'stats1,stats2,...'. A separate column is output for each combination of \$i and \$j, with the loop over \$j nested within the loop over \$i. The possible values for \$stats$[$\$j$]$ are: ``MEAN'', ``WEIGHTEDMEAN'', ``MEDIAN'', ``WEIGHTEDMEDIAN'', ``STDDEV'', ``MEDDEV'', ``MEDMEDDEV'', ``MAD'', ``KURTOSIS'', ``SKEWNESS'', ``PCT\$pctval'' (where \$pctval is the percentile value to two decimal places), ``WPCT\$pctval'', ``MAX'', ``MIN'' and ``SUM''.\\ \hline
\end{xtabular}
\end{center}

\begin{center}
\tablecaption{Output Columns for ``-SYSREM'' Command (Section~\ref{cmd:sysrem})}
\footnotesize
\begin{xtabular}{|p{2in}p{5mm}p{4in}|}
\hline
SYSREM\_MeanMag & \tabitem & The mean magnitude of the light curve.\\ \hline
SYSREM\_Trend\_\$i\_Coeff & \tabitem & The output value for the color term for the \$i-th corrected trend. Here \$i ranges from 0 to 'Ninput\_color'$+$'Ninput\_airmass'$-1$, with the coefficients from the trends with airmass terms specified initially listed before those from the trends with color terms specified initially.\\ \hline
SYSREM\_RMS & \tabitem & The r.m.s.\ scatter of the light curve after removing the trends.\\ \hline
\end{xtabular}
\end{center}

\begin{center}
\tablecaption{Output Columns for ``-TFA'' Command (Section~\ref{cmd:tfa})}
\footnotesize
\begin{xtabular}{|p{2in}p{5mm}p{4in}|}
\hline
TFA\_MeanMag & \tabitem & The mean magnitude of the light curve.\\ \hline
TFA\_RMS & \tabitem & The r.m.s.\ scatter of the light curve after removing the trends.\\ \hline
\end{xtabular}
\end{center}

\begin{center}
\tablecaption{Output Columns for ``-TFA\_SR'' Command (Section~\ref{cmd:tfasr})}
\footnotesize
\begin{xtabular}{|p{2in}p{5mm}p{4in}|}
\hline
TFA\_SR\_MeanMag & \tabitem & The mean magnitude of the light curve.\\ \hline
TFA\_SR\_RMS & \tabitem & The r.m.s.\ scatter of the light curve after removing the trends.\\ \hline
\end{xtabular}
\end{center}

\begin{center}
\tablecaption{Output Columns for ``-wwz'' Command (Section~\ref{cmd:wwz})}
\footnotesize
\begin{xtabular}{|p{2in}p{5mm}p{4in}|}
\hline
MaxWWZ & \tabitem & The maximum value of $Z$ (Eq.~\ref{eqn:waveletZ}) over all values of time-shifts $\tau$ and frequencies $f$ scanned.\\ \hline
MaxWWZ\_Freq & \tabitem & The value of $f$ at which $Z$ is maximized.\\ \hline
MaxWWZ\_TShift & \tabitem & The value of $\tau$ at which $Z$ is maximized.\\ \hline
MaxWWZ\_Power & \tabitem & The value of the $WWT$ statistic (Eq.~\ref{eqn:waveletT}) at the maximum $Z$ peak.\\ \hline
MaxWWZ\_Amplitude & \tabitem & The value of the Weighted-Wavelet-Amplitude ($WWA$; Eq.~\ref{eqn:waveletA}) at the maximum $Z$ peak.\\ \hline
MaxWWZ\_Neffective & \tabitem & The effective number of data points contributing to the signal at the maximum $Z$ peak (Eq.~\ref{eqn:waveletNeff}).\\ \hline
MaxWWZ\_AverageMag & \tabitem & The mean magnitude of the signal at the maximum $Z$ peak.\\ \hline
Med\_WWZ & \tabitem & The median value of $Z_{\rm max}(\tau)$, where $Z_{\rm max}(\tau)$ is the maximum value of $Z$ over all frequencies, at a fixed value of $\tau$, and the median is then taken over all $\tau$ values.\\ \hline
Med\_Freq & \tabitem & The median value of $f_{\rm Zmax}(\tau)$, where $f_{\rm Zmax}(\tau)$ is the frequency at which $Z$ is maximum at a fixed value of $\tau$, and the median is then taken over all $\tau$ values.\\ \hline
Med\_Power & \tabitem & The median value of $WWT_{\rm Zmax}(\tau)$, where $WWT_{\rm Zmax}(\tau)$ is the value of $WWT$ (Eq.~\ref{eqn:waveletT}) calculated at the $Z$ maximum for fixed $\tau$, and the median is then taken over all $\tau$ values.\\ \hline
Med\_Amplitude & \tabitem & The median value of $WWA_{\rm Zmax}(\tau)$, where $WWA_{\rm Zmax}(\tau)$ is the value of $WWA$ (Eq.~\ref{eqn:waveletA}) calculated at the $Z$ maximum for fixed $\tau$, and the median is then taken over all $\tau$ values.\\ \hline
Med\_Neffective & \tabitem & The median value of $N_{\rm eff,Zmax}(\tau)$, where $N_{\rm eff,Zmax}(\tau)$ is the value of $N_{\rm eff}$ (Eq.~\ref{eqn:waveletNeff}) calculated at the $Z$ maximum for fixed $\tau$, and the median is then taken over all $\tau$ values.\\ \hline
Med\_AverageMag & \tabitem & The median value of $M_{\rm ave,Zmax}(\tau)$, where $M_{\rm ave,Zmax}(\tau)$ is the average magnitude of the signal calculated at the $Z$ maximum for fixed $\tau$, and the median is then taken over all $\tau$ values.\\ \hline
\end{xtabular}
\end{center}

\section{Analytic Functions, Constants, and Operators}\label{sec:reservednames}

Below is the list of analytic functions, constants, and operators
recognized by the \va{} analytic expression interpreted. These strings
are reserved and cannot be used for naming variables.

\begin{center}
\footnotesize
\begin{xtabular}{|p{1in}p{5mm}p{5in}|}\hline
Operators & & \\ \hline
~~~~$a+b$ & \tabitem & Addition \\
~~~~$a-b$ & \tabitem & Subtraction \\
~~~~$a*b$ & \tabitem & Multiplication \\
~~~~$a/b$ & \tabitem & Division \\
~~~~$a\%b$ & \tabitem & Floating point reminder (fmod function in c) \\
~~~~$a\wedge\,b$ & \tabitem & Exponentiation \\
~~~~$a>b$ & \tabitem & Greater than comparison \\
~~~~$a>=b$ & \tabitem & Greater than or equal to comparison \\
~~~~$a<b$ & \tabitem & Less than comparison \\
~~~~$a<=b$ & \tabitem & Less than or equal to comparison \\
~~~~$a==b$ & \tabitem & Logical equals \\
~~~~$a!=b$ & \tabitem & Logical not equal \\
~~~~$a\&\&b$ & \tabitem & Logical "and" comparison \\
~~~~$a\Vert b$ & \tabitem & Logical "or" comparison \\
~~~~$!a$ & \tabitem & Logical "not" \\ \hline
Functions & & \\ \hline
~~~~$\exp(x)$ & \tabitem & exponential of $x$. \\
~~~~$\log(x)$ & \tabitem & natural logarithm of $x$. \\
~~~~$\log 10(x)$ & \tabitem & base 10 logarithm of $x$. \\
~~~~${\rm sqrt}(x)$ & \tabitem & square root of $x$. \\
~~~~${\rm abs}(x)$ & \tabitem & absolute value of $x$. \\
~~~~${\rm max}(x,y)$ & \tabitem & the larger of $x$ or $y$. \\
~~~~${\rm min}(x,y)$ & \tabitem & the smaller of $x$ or $y$. \\
~~~~${\rm hypot}(x,y)$ & \tabitem & $\sqrt{x^{2}+y^{2}}$. \\
~~~~${\rm sin}(x)$ & \tabitem & trigonometric sine of $x$. Input in radians. \\
~~~~${\rm cos}(x)$ & \tabitem & trigonometric cosine of $x$. Input in radians. \\
~~~~${\rm tan}(x)$ & \tabitem & trigonometric tangent of $x$. Input in radians. \\
~~~~${\rm sindegr}(x)$ & \tabitem & trigonometric sine of $x$. Input in degrees. \\
~~~~${\rm cosdegr}(x)$ & \tabitem & trigonometric cosine of $x$. Input in degrees. \\
~~~~${\rm tandegr}(x)$ & \tabitem & trigonometric tangent of $x$. Input in degrees. \\
~~~~${\rm asin}(x)$ & \tabitem & inverse sine of $x$. Output in radians. \\
~~~~${\rm acos}(x)$ & \tabitem & inverse cosine of $x$. Output in radians. \\
~~~~${\rm atan2}(y,x)$ & \tabitem & $4$ quadrant inverse tangent of $y/x$. Output in radians. \\
~~~~${\rm asindegr}(x)$ & \tabitem & inverse sine of $x$. Output in degrees. \\
~~~~${\rm acosdegr}(x)$ & \tabitem & inverse cosine of $x$. Output in degrees. \\
~~~~${\rm atan2degr}(y,x)$ & \tabitem & $4$ quadrant inverse tangent of $y/x$. Output in degrees. \\
~~~~${\rm sinh}(x)$ & \tabitem & hyperbolic sine of $x$. \\
~~~~${\rm cosh}(x)$ & \tabitem & hyperbolic cosine of $x$. \\
~~~~${\rm tanh}(x)$ & \tabitem & hyperbolic tangent of $x$. \\
~~~~${\rm asinh}(x)$ & \tabitem & inverse hyperbolic sine of $x$. \\
~~~~${\rm acosh}(x)$ & \tabitem & inverse hyperbolic cosine of $x$. \\
~~~~${\rm atanh}(x)$ & \tabitem & inverse hyperbolic tangent of $x$. \\
~~~~${\rm erf}(x)$ & \tabitem & error function of $x$ ****TBD denote convention***. \\
~~~~${\rm erfc}(x)$ & \tabitem & complimentary error function of $x$. \\
~~~~${\rm lgamma}(x)$ & \tabitem & natural logarithm of $\Gamma(x)$. \\
~~~~${\rm gamma}(x)$ & \tabitem & $\Gamma(x)$. \\
~~~~${\rm theta}(x)$ & \tabitem & $1$ for $x >= 0$, $0$ for $x < 0$. \\
~~~~${\rm round}(x)$ & \tabitem & round $x$ to the nearest integer. \\
~~~~${\rm ceil}(x)$ & \tabitem & smallest integer that is greater than or equal to $x$. \\
~~~~${\rm floor}(x)$ & \tabitem & largest integer that is less than or equal to $x$. \\
~~~~${\rm rand}()$ & \tabitem & random number drawn from a uniform distribution between $0$ and $1$. \\
~~~~${\rm gauss}()$ & \tabitem & random number drawn from a normal distribution with $0$ mean and unit variance. \\ \hline
Constants & & \\ \hline
~~~~${\rm pi}$ & \tabitem & Evaluates to the M\_PI constant defined in the ``math.h'' {\sc c} header file. \\
~~~~${\rm e}$ & \tabitem & Evaluates to the M\_E constant defined in the ``math.h'' {\sc c} header file. \\ \hline
Special Variables: & & \\ \hline
~~~~${\rm NR}$ & \tabitem & image index in the light curve starting from 0. \\
~~~~${\rm NF}$ & \tabitem & light curve index starting from 0. \\ \hline
\end{xtabular}
\end{center}

\section{Justification for the {\bf -ensemblerescalesig} procedure}\label{sec:ensemblerescaletransjust}

Here we justify the procedure used by the {\bf
  -ensemblerescalesig} command to find parameters $a$ and $b$ for the
transformation in equation~\ref{eqn:ensemblerescaletrans} such that
$\chi^{2}/{\rm dof} = 1$ for the typical light curve after the
transformation. The procedure is to solve the linear least squares
problem in
equations~\ref{eqn:ensemblerescalefitrelation}--\ref{eqn:ensemblerescalefitrelationxi}
for $a$ and $b$.

Let \begin{equation}
\chi_{i}^{\prime 2}/{\rm dof} = \sum_{j=1}^{N_{\rm JD,i}}(m_{i,j} - \bar{m_{i}})^2/(N_{\rm JD,i}-1)/\sigma^{\prime 2}_{i,j}
\label{eqn:chi2trans}
\end{equation}
be the reduced $\chi^2$ for light curve $i$ after applying the
transformation in equation~\ref{eqn:ensemblerescaletrans}. We want $a$ and $b$ such that
\begin{equation}
\chi_{i}^{\prime 2}/{\rm dof} = 1
\label{eqn:chi2transgoal}
\end{equation}
for the ``typical'' light curve. One way to do this is to perform a least-squares fit for $a$ and $b$. In other words, to find the values of $a$ and $b$ which minimizes
\begin{equation}
X^{2} = \sum_{i=1}^{N_{\rm LC}}\frac{(\chi_{i}^{\prime 2}/{\rm dof} - 1)^{2}}{2/{\rm dof}}
\label{eqn:chi2transchi2}
\end{equation}
where the expected value of $\chi_{i}^{\prime 2}/{\rm dof}$ is $1$ and
its variance is $2/{\rm dof}$. However, substituting
equation~\ref{eqn:ensemblerescaletrans} directly into
equations~\ref{eqn:chi2trans} and~\ref{eqn:chi2transchi2} results in a non-linear least squares
problem for $a$ and $b$ whose solution using standard non-linear least
squares algorithms would be very time consuming due to the need to
perform a sum over every observation in every light curve for each
evaluation of $X^2$. An alternative approach is to find a simple approximation which reduces this to a linear least-squares problem. The {\bf -ensemblerescalesig} command makes the following approximation for $\sigma_{i,j}^{\prime 2}$ in equation~\ref{eqn:chi2trans}:
\begin{equation}
\sigma_{i,j}^{\prime 2} \approx \sigma_{i,j}^2 \frac{a\bar{\sigma_{i}}^{2}+b}{\bar{\sigma_{i}}^{2}}.
\end{equation}
In other words, rather than using the transformed value of $\sigma_{i,j}^{2}$ directy, the individual uncertainty is scaled by the ratio of
the expected variance after the transformation, to the expected
variance before the transformation. Substituting this expression into equation~\ref{eqn:chi2transgoal}, we then have
\begin{equation}
\sum_{j=1}^{N_{\rm JD,i}}\frac{(m_{i,j} - \bar{m_{i}})^2\bar{\sigma_{i}}^2}{(N_{\rm JD,i}-1)\sigma_{i,j}^2(a\bar{\sigma_{i}}^2+b)} = 1
\end{equation}
or
\begin{equation}
\sum_{j=1}^{N_{\rm JD,i}}\frac{(m_{i,j} - \bar{m_{i}})^2\bar{\sigma_{i}}^2}{(N_{\rm JD,i}-1)\sigma_{i,j}^2} = a\bar{\sigma_{i}}^2 + b
\end{equation}
or
\begin{equation}
\bar{RMS_{i}}^2\chi_{i}^{2}/{\rm dof} = a\bar{RMS_{i}}^2 + b
\end{equation}
which may then be treated as a linear least squares problem for $a$
and $b$ (Eq.~\ref{eqn:ensemblerescalefitrelation}).

\twocolumn

\section*{Acknowledgements}
The development of \va{} has been supported by NASA grant
NNX14AE87G. We also acknowledge support from NSF/AST-1108686 and NASA
grants NNX12AH91H and NNX13AJ15G. We would like to thank the anonymous
referee for their review of this paper. We would also like to thank
the users of \va{} who have contributed bug reports and other
feedback, especially T.~Beatty, A.~Bonanos, T.~Bovaird, D.~Flateau,
L.~Macri, D.~Nataf, J.~Pepper, J.~Rasor, B.~Sip\"ocz, R.~Siverd,
A.~Schwarzenberg-Czerny, K.~Stanek, and A.~Sweeney.

%% Bibliography

%bibliography{vartools.bib}


\begin{thebibliography}{65}
\expandafter\ifx\csname natexlab\endcsname\relax\def\natexlab#1{#1}\fi
\providecommand{\url}[1]{\texttt{#1}}
\providecommand{\href}[2]{#2}
\providecommand{\path}[1]{#1}
\providecommand{\DOIprefix}{doi:}
\providecommand{\ArXivprefix}{arXiv:}
\providecommand{\URLprefix}{URL: }
\providecommand{\Pubmedprefix}{pmid:}
\providecommand{\doi}[1]{\href{http://dx.doi.org/#1}{\path{#1}}}
\providecommand{\Pubmed}[1]{\href{pmid:#1}{\path{#1}}}
\providecommand{\bibinfo}[2]{#2}
\ifx\xfnm\relax \def\xfnm[#1]{\unskip,\space#1}\fi
%Type = Article
\bibitem[{{Acton}(1996)}]{acton:1996}
\bibinfo{author}{{Acton}, C.H.}, \bibinfo{year}{1996}.
\newblock \bibinfo{title}{{Ancillary data services of NASA's Navigation and
  Ancillary Information Facility}}.
\newblock \bibinfo{journal}{\planss} \bibinfo{volume}{44},
  \bibinfo{pages}{65--70}.
\newblock \DOIprefix\doi{10.1016/0032-0633(95)00107-7}.
%Type = Article
\bibitem[{{Alard} and {Lupton}(1998)}]{alard:1998}
\bibinfo{author}{{Alard}, C.}, \bibinfo{author}{{Lupton}, R.H.},
  \bibinfo{year}{1998}.
\newblock \bibinfo{title}{{A Method for Optimal Image Subtraction}}.
\newblock \bibinfo{journal}{\apj} \bibinfo{volume}{503}, \bibinfo{pages}{325}.
\newblock \DOIprefix\doi{10.1086/305984},
  \href{http://arxiv.org/abs/arXiv:astro-ph/9712287}{\tt
  arXiv:arXiv:astro-ph/9712287}.
%Type = Article
\bibitem[{{Bakos} et~al.(2012){Bakos}, {Csubry}, {Penev}, {Bayliss},
  {Jord{\'a}n}, {Afonso}, {Hartman}, {Henning}, {Kov{\'a}cs}, {Noyes},
  {B{\'e}ky}, {Suc}, {Cs{\'a}k}, {Rabus}, {L{\'a}z{\'a}r}, {Papp}, {S{\'a}ri},
  {Conroy}, {Zhou}, {Sackett}, {Schmidt}, {Mancini}, {Sasselov} and
  {Ueltzhoeffer}}]{bakos:2012:hatsouth}
\bibinfo{author}{{Bakos}, G.{\'A}.}, \bibinfo{author}{{Csubry}, Z.},
  \bibinfo{author}{{Penev}, K.}, \bibinfo{author}{{Bayliss}, D.},
  \bibinfo{author}{{Jord{\'a}n}, A.}, \bibinfo{author}{{Afonso}, C.},
  \bibinfo{author}{{Hartman}, J.D.}, \bibinfo{author}{{Henning}, T.},
  \bibinfo{author}{{Kov{\'a}cs}, G.}, \bibinfo{author}{{Noyes}, R.W.},
  \bibinfo{author}{{B{\'e}ky}, B.}, \bibinfo{author}{{Suc}, V.},
  \bibinfo{author}{{Cs{\'a}k}, B.}, \bibinfo{author}{{Rabus}, M.},
  \bibinfo{author}{{L{\'a}z{\'a}r}, J.}, \bibinfo{author}{{Papp}, I.},
  \bibinfo{author}{{S{\'a}ri}, P.}, \bibinfo{author}{{Conroy}, P.},
  \bibinfo{author}{{Zhou}, G.}, \bibinfo{author}{{Sackett}, P.D.},
  \bibinfo{author}{{Schmidt}, B.}, \bibinfo{author}{{Mancini}, L.},
  \bibinfo{author}{{Sasselov}, D.D.}, \bibinfo{author}{{Ueltzhoeffer}, K.},
  \bibinfo{year}{2012}.
\newblock \bibinfo{title}{{HATSouth: a global network of fully automated
  identical wide-field telescopes}}.
\newblock \bibinfo{journal}{ArXiv e-prints}
  \href{http://arxiv.org/abs/1206.1391}{\tt arXiv:1206.1391}.
%Type = Article
\bibitem[{{Bond} et~al.(2004){Bond}, {Udalski}, {Jaroszy{\'n}ski},
  {Rattenbury}, {Paczy{\'n}ski}, {Soszy{\'n}ski}, {Wyrzykowski},
  {Szyma{\'n}ski}, {Kubiak}, {Szewczyk}, {{\.Z}ebru{\'n}}, {Pietrzy{\'n}ski},
  {Abe}, {Bennett}, {Eguchi}, {Furuta}, {Hearnshaw}, {Kamiya}, {Kilmartin},
  {Kurata}, {Masuda}, {Matsubara}, {Muraki}, {Noda}, {Okajima}, {Sako},
  {Sekiguchi}, {Sullivan}, {Sumi}, {Tristram}, {Yanagisawa}, {Yock} and {OGLE
  Collaboration}}]{bond:2004}
\bibinfo{author}{{Bond}, I.A.}, \bibinfo{author}{{Udalski}, A.},
  \bibinfo{author}{{Jaroszy{\'n}ski}, M.}, \bibinfo{author}{{Rattenbury},
  N.J.}, \bibinfo{author}{{Paczy{\'n}ski}, B.},
  \bibinfo{author}{{Soszy{\'n}ski}, I.}, \bibinfo{author}{{Wyrzykowski}, L.},
  \bibinfo{author}{{Szyma{\'n}ski}, M.K.}, \bibinfo{author}{{Kubiak}, M.},
  \bibinfo{author}{{Szewczyk}, O.}, \bibinfo{author}{{{\.Z}ebru{\'n}}, K.},
  \bibinfo{author}{{Pietrzy{\'n}ski}, G.}, \bibinfo{author}{{Abe}, F.},
  \bibinfo{author}{{Bennett}, D.P.}, \bibinfo{author}{{Eguchi}, S.},
  \bibinfo{author}{{Furuta}, Y.}, \bibinfo{author}{{Hearnshaw}, J.B.},
  \bibinfo{author}{{Kamiya}, K.}, \bibinfo{author}{{Kilmartin}, P.M.},
  \bibinfo{author}{{Kurata}, Y.}, \bibinfo{author}{{Masuda}, K.},
  \bibinfo{author}{{Matsubara}, Y.}, \bibinfo{author}{{Muraki}, Y.},
  \bibinfo{author}{{Noda}, S.}, \bibinfo{author}{{Okajima}, K.},
  \bibinfo{author}{{Sako}, T.}, \bibinfo{author}{{Sekiguchi}, T.},
  \bibinfo{author}{{Sullivan}, D.J.}, \bibinfo{author}{{Sumi}, T.},
  \bibinfo{author}{{Tristram}, P.J.}, \bibinfo{author}{{Yanagisawa}, T.},
  \bibinfo{author}{{Yock}, P.C.M.}, \bibinfo{author}{{OGLE Collaboration}},
  \bibinfo{year}{2004}.
\newblock \bibinfo{title}{{OGLE 2003-BLG-235/MOA 2003-BLG-53: A Planetary
  Microlensing Event}}.
\newblock \bibinfo{journal}{\apjl} \bibinfo{volume}{606},
  \bibinfo{pages}{L155--L158}.
\newblock \DOIprefix\doi{10.1086/420928},
  \href{http://arxiv.org/abs/arXiv:astro-ph/0404309}{\tt
  arXiv:arXiv:astro-ph/0404309}.
%Type = Article
\bibitem[{{Borucki} et~al.(2010){Borucki}, {Koch}, {Basri}, {Batalha}, {Brown},
  {Caldwell}, {Caldwell}, {Christensen-Dalsgaard}, {Cochran}, {DeVore},
  {Dunham}, {Dupree}, {Gautier}, {Geary}, {Gilliland}, {Gould}, {Howell},
  {Jenkins}, {Kondo}, {Latham}, {Marcy}, {Meibom}, {Kjeldsen}, {Lissauer},
  {Monet}, {Morrison}, {Sasselov}, {Tarter}, {Boss}, {Brownlee}, {Owen},
  {Buzasi}, {Charbonneau}, {Doyle}, {Fortney}, {Ford}, {Holman}, {Seager},
  {Steffen}, {Welsh}, {Rowe}, {Anderson}, {Buchhave}, {Ciardi}, {Walkowicz},
  {Sherry}, {Horch}, {Isaacson}, {Everett}, {Fischer}, {Torres}, {Johnson},
  {Endl}, {MacQueen}, {Bryson}, {Dotson}, {Haas}, {Kolodziejczak}, {Van Cleve},
  {Chandrasekaran}, {Twicken}, {Quintana}, {Clarke}, {Allen}, {Li}, {Wu},
  {Tenenbaum}, {Verner}, {Bruhweiler}, {Barnes} and {Prsa}}]{borucki:2010}
\bibinfo{author}{{Borucki}, W.J.}, \bibinfo{author}{{Koch}, D.},
  \bibinfo{author}{{Basri}, G.}, \bibinfo{author}{{Batalha}, N.},
  \bibinfo{author}{{Brown}, T.}, \bibinfo{author}{{Caldwell}, D.},
  \bibinfo{author}{{Caldwell}, J.}, \bibinfo{author}{{Christensen-Dalsgaard},
  J.}, \bibinfo{author}{{Cochran}, W.D.}, \bibinfo{author}{{DeVore}, E.},
  \bibinfo{author}{{Dunham}, E.W.}, \bibinfo{author}{{Dupree}, A.K.},
  \bibinfo{author}{{Gautier}, T.N.}, \bibinfo{author}{{Geary}, J.C.},
  \bibinfo{author}{{Gilliland}, R.}, \bibinfo{author}{{Gould}, A.},
  \bibinfo{author}{{Howell}, S.B.}, \bibinfo{author}{{Jenkins}, J.M.},
  \bibinfo{author}{{Kondo}, Y.}, \bibinfo{author}{{Latham}, D.W.},
  \bibinfo{author}{{Marcy}, G.W.}, \bibinfo{author}{{Meibom}, S.},
  \bibinfo{author}{{Kjeldsen}, H.}, \bibinfo{author}{{Lissauer}, J.J.},
  \bibinfo{author}{{Monet}, D.G.}, \bibinfo{author}{{Morrison}, D.},
  \bibinfo{author}{{Sasselov}, D.}, \bibinfo{author}{{Tarter}, J.},
  \bibinfo{author}{{Boss}, A.}, \bibinfo{author}{{Brownlee}, D.},
  \bibinfo{author}{{Owen}, T.}, \bibinfo{author}{{Buzasi}, D.},
  \bibinfo{author}{{Charbonneau}, D.}, \bibinfo{author}{{Doyle}, L.},
  \bibinfo{author}{{Fortney}, J.}, \bibinfo{author}{{Ford}, E.B.},
  \bibinfo{author}{{Holman}, M.J.}, \bibinfo{author}{{Seager}, S.},
  \bibinfo{author}{{Steffen}, J.H.}, \bibinfo{author}{{Welsh}, W.F.},
  \bibinfo{author}{{Rowe}, J.}, \bibinfo{author}{{Anderson}, H.},
  \bibinfo{author}{{Buchhave}, L.}, \bibinfo{author}{{Ciardi}, D.},
  \bibinfo{author}{{Walkowicz}, L.}, \bibinfo{author}{{Sherry}, W.},
  \bibinfo{author}{{Horch}, E.}, \bibinfo{author}{{Isaacson}, H.},
  \bibinfo{author}{{Everett}, M.E.}, \bibinfo{author}{{Fischer}, D.},
  \bibinfo{author}{{Torres}, G.}, \bibinfo{author}{{Johnson}, J.A.},
  \bibinfo{author}{{Endl}, M.}, \bibinfo{author}{{MacQueen}, P.},
  \bibinfo{author}{{Bryson}, S.T.}, \bibinfo{author}{{Dotson}, J.},
  \bibinfo{author}{{Haas}, M.}, \bibinfo{author}{{Kolodziejczak}, J.},
  \bibinfo{author}{{Van Cleve}, J.}, \bibinfo{author}{{Chandrasekaran}, H.},
  \bibinfo{author}{{Twicken}, J.D.}, \bibinfo{author}{{Quintana}, E.V.},
  \bibinfo{author}{{Clarke}, B.D.}, \bibinfo{author}{{Allen}, C.},
  \bibinfo{author}{{Li}, J.}, \bibinfo{author}{{Wu}, H.},
  \bibinfo{author}{{Tenenbaum}, P.}, \bibinfo{author}{{Verner}, E.},
  \bibinfo{author}{{Bruhweiler}, F.}, \bibinfo{author}{{Barnes}, J.},
  \bibinfo{author}{{Prsa}, A.}, \bibinfo{year}{2010}.
\newblock \bibinfo{title}{{Kepler Planet-Detection Mission: Introduction and
  First Results}}.
\newblock \bibinfo{journal}{Science} \bibinfo{volume}{327},
  \bibinfo{pages}{977--}.
\newblock \DOIprefix\doi{10.1126/science.1185402}.
%Type = Article
\bibitem[{{Charbonneau} et~al.(2000){Charbonneau}, {Brown}, {Latham} and
  {Mayor}}]{charbonneau:2000}
\bibinfo{author}{{Charbonneau}, D.}, \bibinfo{author}{{Brown}, T.M.},
  \bibinfo{author}{{Latham}, D.W.}, \bibinfo{author}{{Mayor}, M.},
  \bibinfo{year}{2000}.
\newblock \bibinfo{title}{{Detection of Planetary Transits Across a Sun-like
  Star}}.
\newblock \bibinfo{journal}{\apjl} \bibinfo{volume}{529},
  \bibinfo{pages}{L45--L48}.
\newblock \DOIprefix\doi{10.1086/312457},
  \href{http://arxiv.org/abs/arXiv:astro-ph/9911436}{\tt
  arXiv:arXiv:astro-ph/9911436}.
%Type = Article
\bibitem[{{Cumming} et~al.(1999){Cumming}, {Marcy} and {Butler}}]{cumming:1999}
\bibinfo{author}{{Cumming}, A.}, \bibinfo{author}{{Marcy}, G.W.},
  \bibinfo{author}{{Butler}, R.P.}, \bibinfo{year}{1999}.
\newblock \bibinfo{title}{{The Lick Planet Search: Detectability and Mass
  Thresholds}}.
\newblock \bibinfo{journal}{\apj} \bibinfo{volume}{526},
  \bibinfo{pages}{890--915}.
\newblock \DOIprefix\doi{10.1086/308020},
  \href{http://arxiv.org/abs/astro-ph/9906466}{\tt arXiv:astro-ph/9906466}.
%Type = Article
\bibitem[{{Devor}(2005)}]{devor:2005}
\bibinfo{author}{{Devor}, J.}, \bibinfo{year}{2005}.
\newblock \bibinfo{title}{{Solutions for 10,000 Eclipsing Binaries in the Bulge
  Fields of OGLE II Using DEBiL}}.
\newblock \bibinfo{journal}{\apj} \bibinfo{volume}{628},
  \bibinfo{pages}{411--425}.
\newblock \DOIprefix\doi{10.1086/431170},
  \href{http://arxiv.org/abs/arXiv:astro-ph/0504399}{\tt
  arXiv:arXiv:astro-ph/0504399}.
%Type = Article
\bibitem[{{Dorren}(1987)}]{dorren:1987}
\bibinfo{author}{{Dorren}, J.D.}, \bibinfo{year}{1987}.
\newblock \bibinfo{title}{{A new formulation of the starspot model, and the
  consequences of starspot structure}}.
\newblock \bibinfo{journal}{\apj} \bibinfo{volume}{320},
  \bibinfo{pages}{756--767}.
\newblock \DOIprefix\doi{10.1086/165593}.
%Type = Article
\bibitem[{{Eastman} et~al.(2010){Eastman}, {Siverd} and {Gaudi}}]{eastman:2010}
\bibinfo{author}{{Eastman}, J.}, \bibinfo{author}{{Siverd}, R.},
  \bibinfo{author}{{Gaudi}, B.S.}, \bibinfo{year}{2010}.
\newblock \bibinfo{title}{{Achieving Better Than 1 Minute Accuracy in the
  Heliocentric and Barycentric Julian Dates}}.
\newblock \bibinfo{journal}{\pasp} \bibinfo{volume}{122},
  \bibinfo{pages}{935--946}.
\newblock \DOIprefix\doi{10.1086/655938},
  \href{http://arxiv.org/abs/1005.4415}{\tt arXiv:1005.4415}.
%Type = Article
\bibitem[{{Edelson} and {Krolik}(1988)}]{edelson:1988}
\bibinfo{author}{{Edelson}, R.A.}, \bibinfo{author}{{Krolik}, J.H.},
  \bibinfo{year}{1988}.
\newblock \bibinfo{title}{{The discrete correlation function - A new method for
  analyzing unevenly sampled variability data}}.
\newblock \bibinfo{journal}{\apj} \bibinfo{volume}{333},
  \bibinfo{pages}{646--659}.
\newblock \DOIprefix\doi{10.1086/166773}.
%Type = Inproceedings
\bibitem[{{Etzel}(1981)}]{etzel:1981}
\bibinfo{author}{{Etzel}, P.B.}, \bibinfo{year}{1981}.
\newblock \bibinfo{title}{{A Simple Synthesis Method for Solving the Elements
  of Well-Detached Eclipsing Systems}}, in: \bibinfo{editor}{{Carling}, E.B.},
  \bibinfo{editor}{{Kopal}, Z.} (Eds.), \bibinfo{booktitle}{Photometric and
  Spectroscopic Binary Systems}, p. \bibinfo{pages}{111}.
%Type = Article
\bibitem[{{Foster}(1996)}]{foster:1996}
\bibinfo{author}{{Foster}, G.}, \bibinfo{year}{1996}.
\newblock \bibinfo{title}{{Wavelets for period analysis of unevenly sampled
  time series}}.
\newblock \bibinfo{journal}{\aj} \bibinfo{volume}{112}, \bibinfo{pages}{1709}.
\newblock \DOIprefix\doi{10.1086/118137}.
%Type = Book
\bibitem[{Galassi et~al.(2009)Galassi, Davies, Theiler, Gough, Jungman, Alken,
  Booth and Rossi}]{galassi:2009}
\bibinfo{author}{Galassi, M.}, \bibinfo{author}{Davies, J.},
  \bibinfo{author}{Theiler, J.}, \bibinfo{author}{Gough, B.},
  \bibinfo{author}{Jungman, G.}, \bibinfo{author}{Alken, P.},
  \bibinfo{author}{Booth, M.}, \bibinfo{author}{Rossi, F.},
  \bibinfo{year}{2009}.
\newblock \bibinfo{title}{GNU Scientific Library Reference Manual}.
\newblock \bibinfo{edition}{Third} ed., \bibinfo{publisher}{Network Theory
  Ltd.}
%Type = Article
\bibitem[{{Goupillaud} et~al.(1984){Goupillaud}, {Grossmann} and
  {Morlet}}]{goupillaud:1984}
\bibinfo{author}{{Goupillaud}, P.}, \bibinfo{author}{{Grossmann}, A.},
  \bibinfo{author}{{Morlet}, J.}, \bibinfo{year}{1984}.
\newblock \bibinfo{title}{{Cycle-octave and related transforms in seismic
  signal analysis}}.
\newblock \bibinfo{journal}{Geoexploration} \bibinfo{volume}{23},
  \bibinfo{pages}{85--102}.
\newblock \DOIprefix\doi{10.1016/0016-7142(84)90025-5}.
%Type = Article
\bibitem[{Handcock and Stein(1993)}]{handcock:1993}
\bibinfo{author}{Handcock, M.S.}, \bibinfo{author}{Stein, M.L.},
  \bibinfo{year}{1993}.
\newblock \bibinfo{title}{A bayesian analysis of kriging}.
\newblock \bibinfo{journal}{Technometrics} \bibinfo{volume}{35},
  \bibinfo{pages}{403--410}.
\newblock \URLprefix
  \url{http://amstat.tandfonline.com/doi/abs/10.1080/00401706.1993.10485354},
  \DOIprefix\doi{10.1080/00401706.1993.10485354},
  \href{http://arxiv.org/abs/http://amstat.tandfonline.com/doi/pdf/10.1080/00401706.1993.10485354}{\tt
  arXiv:http://amstat.tandfonline.com/doi/pdf/10.1080/00401706.1993.10485354}.
%Type = Article
\bibitem[{{Hartman} et~al.(2011){Hartman}, {Bakos}, {Noyes}, {Sip{\H o}cz},
  {Kov{\'a}cs}, {Mazeh}, {Shporer} and {P{\'a}l}}]{hartman:2011:kmdwarf}
\bibinfo{author}{{Hartman}, J.D.}, \bibinfo{author}{{Bakos}, G.{\'A}.},
  \bibinfo{author}{{Noyes}, R.W.}, \bibinfo{author}{{Sip{\H o}cz}, B.},
  \bibinfo{author}{{Kov{\'a}cs}, G.}, \bibinfo{author}{{Mazeh}, T.},
  \bibinfo{author}{{Shporer}, A.}, \bibinfo{author}{{P{\'a}l}, A.},
  \bibinfo{year}{2011}.
\newblock \bibinfo{title}{{A Photometric Variability Survey of Field K and M
  Dwarf Stars with HATNet}}.
\newblock \bibinfo{journal}{\aj} \bibinfo{volume}{141}, \bibinfo{pages}{166}.
\newblock \DOIprefix\doi{10.1088/0004-6256/141/5/166}.
%Type = Article
\bibitem[{{Hartman} et~al.(2008){Hartman}, {Gaudi}, {Holman}, {McLeod},
  {Stanek}, {Barranco}, {Pinsonneault} and {Kalirai}}]{hartman:m37:2}
\bibinfo{author}{{Hartman}, J.D.}, \bibinfo{author}{{Gaudi}, B.S.},
  \bibinfo{author}{{Holman}, M.J.}, \bibinfo{author}{{McLeod}, B.A.},
  \bibinfo{author}{{Stanek}, K.Z.}, \bibinfo{author}{{Barranco}, J.A.},
  \bibinfo{author}{{Pinsonneault}, M.H.}, \bibinfo{author}{{Kalirai}, J.S.},
  \bibinfo{year}{2008}.
\newblock \bibinfo{title}{{Deep MMT Transit Survey of the Open Cluster M37. II.
  Variable Stars}}.
\newblock \bibinfo{journal}{\apj} \bibinfo{volume}{675},
  \bibinfo{pages}{1254--1277}.
\newblock \DOIprefix\doi{10.1086/527460},
  \href{http://arxiv.org/abs/0709.3484}{\tt arXiv:0709.3484}.
%Type = Article
\bibitem[{{Hartman} et~al.(2005){Hartman}, {Stanek}, {Gaudi}, {Holman} and
  {McLeod}}]{hartman:2005:ngc6791}
\bibinfo{author}{{Hartman}, J.D.}, \bibinfo{author}{{Stanek}, K.Z.},
  \bibinfo{author}{{Gaudi}, B.S.}, \bibinfo{author}{{Holman}, M.J.},
  \bibinfo{author}{{McLeod}, B.A.}, \bibinfo{year}{2005}.
\newblock \bibinfo{title}{{Pushing the Limits of Ground-based Photometric
  Precision: Submillimagnitude Time-Series Photometry of the Open Cluster NGC
  6791}}.
\newblock \bibinfo{journal}{\aj} \bibinfo{volume}{130},
  \bibinfo{pages}{2241--2251}.
\newblock \DOIprefix\doi{10.1086/462405},
  \href{http://arxiv.org/abs/arXiv:astro-ph/0504487}{\tt
  arXiv:arXiv:astro-ph/0504487}.
%Type = Article
\bibitem[{{Henry} et~al.(1999){Henry}, {Marcy}, {Butler} and
  {Vogt}}]{henry:1999}
\bibinfo{author}{{Henry}, G.W.}, \bibinfo{author}{{Marcy}, G.},
  \bibinfo{author}{{Butler}, R.P.}, \bibinfo{author}{{Vogt}, S.S.},
  \bibinfo{year}{1999}.
\newblock \bibinfo{title}{{HD 209458}}.
\newblock \bibinfo{journal}{\iaucirc} \bibinfo{volume}{7307},
  \bibinfo{pages}{1}.
%Type = Article
\bibitem[{{Horne} and {Baliunas}(1986)}]{horne:1986}
\bibinfo{author}{{Horne}, J.H.}, \bibinfo{author}{{Baliunas}, S.L.},
  \bibinfo{year}{1986}.
\newblock \bibinfo{title}{{A prescription for period analysis of unevenly
  sampled time series}}.
\newblock \bibinfo{journal}{\apj} \bibinfo{volume}{302},
  \bibinfo{pages}{757--763}.
\newblock \DOIprefix\doi{10.1086/164037}.
%Type = Article
\bibitem[{{Hubble}(1929)}]{hubble:1929:m31}
\bibinfo{author}{{Hubble}, E.P.}, \bibinfo{year}{1929}.
\newblock \bibinfo{title}{{A spiral nebula as a stellar system, Messier 31.}}
\newblock \bibinfo{journal}{\apj} \bibinfo{volume}{69},
  \bibinfo{pages}{103--158}.
\newblock \DOIprefix\doi{10.1086/143167}.
%Type = Article
\bibitem[{{Kipping}(2012)}]{kipping:2012:macula}
\bibinfo{author}{{Kipping}, D.M.}, \bibinfo{year}{2012}.
\newblock \bibinfo{title}{{An analytic model for rotational modulations in the
  photometry of spotted stars}}.
\newblock \bibinfo{journal}{\mnras} \bibinfo{volume}{427},
  \bibinfo{pages}{2487--2511}.
\newblock \DOIprefix\doi{10.1111/j.1365-2966.2012.22124.x},
  \href{http://arxiv.org/abs/1209.2985}{\tt arXiv:1209.2985}.
%Type = Article
\bibitem[{{Konacki} et~al.(2003){Konacki}, {Torres}, {Jha} and
  {Sasselov}}]{konacki:2003}
\bibinfo{author}{{Konacki}, M.}, \bibinfo{author}{{Torres}, G.},
  \bibinfo{author}{{Jha}, S.}, \bibinfo{author}{{Sasselov}, D.D.},
  \bibinfo{year}{2003}.
\newblock \bibinfo{title}{{An extrasolar planet that transits the disk of its
  parent star}}.
\newblock \bibinfo{journal}{\nat} \bibinfo{volume}{421},
  \bibinfo{pages}{507--509}.
\newblock \DOIprefix\doi{10.1038/nature01379}.
%Type = Book
\bibitem[{{Kopal}(1959)}]{kopal:1959}
\bibinfo{author}{{Kopal}, Z.}, \bibinfo{year}{1959}.
\newblock \bibinfo{title}{{Close binary systems}}.
\newblock \bibinfo{publisher}{The International Astrophysics Series, London:
  Chapman and Hall, 1959}.
%Type = Article
\bibitem[{{Kovacs}(1980)}]{kovacs:1980}
\bibinfo{author}{{Kovacs}, G.}, \bibinfo{year}{1980}.
\newblock \bibinfo{title}{{Period analysis at high noise level}}.
\newblock \bibinfo{journal}{\apss} \bibinfo{volume}{69},
  \bibinfo{pages}{485--493}.
\newblock \DOIprefix\doi{10.1007/BF00661932}.
%Type = Article
\bibitem[{{Kov{\'a}cs} et~al.(2005){Kov{\'a}cs}, {Bakos} and
  {Noyes}}]{kovacs:2005:TFA}
\bibinfo{author}{{Kov{\'a}cs}, G.}, \bibinfo{author}{{Bakos}, G.},
  \bibinfo{author}{{Noyes}, R.W.}, \bibinfo{year}{2005}.
\newblock \bibinfo{title}{{A trend filtering algorithm for wide-field
  variability surveys}}.
\newblock \bibinfo{journal}{\mnras} \bibinfo{volume}{356},
  \bibinfo{pages}{557--567}.
\newblock \DOIprefix\doi{10.1111/j.1365-2966.2004.08479.x},
  \href{http://arxiv.org/abs/arXiv:astro-ph/0411724}{\tt
  arXiv:arXiv:astro-ph/0411724}.
%Type = Article
\bibitem[{{Kov{\'a}cs} et~al.(2002){Kov{\'a}cs}, {Zucker} and
  {Mazeh}}]{kovacs:2002:BLS}
\bibinfo{author}{{Kov{\'a}cs}, G.}, \bibinfo{author}{{Zucker}, S.},
  \bibinfo{author}{{Mazeh}, T.}, \bibinfo{year}{2002}.
\newblock \bibinfo{title}{{A box-fitting algorithm in the search for periodic
  transits}}.
\newblock \bibinfo{journal}{\aap} \bibinfo{volume}{391},
  \bibinfo{pages}{369--377}.
\newblock \DOIprefix\doi{10.1051/0004-6361:20020802},
  \href{http://arxiv.org/abs/arXiv:astro-ph/0206099}{\tt
  arXiv:arXiv:astro-ph/0206099}.
%Type = Article
\bibitem[{{Kurtz}(1985)}]{kurtz:1985}
\bibinfo{author}{{Kurtz}, D.W.}, \bibinfo{year}{1985}.
\newblock \bibinfo{title}{{An algorithm for significantly reducing the time
  necessary to compute a Discrete Fourier Transform periodogram of unequally
  spaced data}}.
\newblock \bibinfo{journal}{\mnras} \bibinfo{volume}{213},
  \bibinfo{pages}{773--776}.
%Type = Article
\bibitem[{{Lenz} and {Breger}(2005)}]{lenz:2005}
\bibinfo{author}{{Lenz}, P.}, \bibinfo{author}{{Breger}, M.},
  \bibinfo{year}{2005}.
\newblock \bibinfo{title}{{Period04 User Guide}}.
\newblock \bibinfo{journal}{Communications in Asteroseismology}
  \bibinfo{volume}{146}, \bibinfo{pages}{53--136}.
\newblock \DOIprefix\doi{10.1553/cia146s53}.
%Type = Article
\bibitem[{{Lomb}(1976)}]{lomb:1976}
\bibinfo{author}{{Lomb}, N.R.}, \bibinfo{year}{1976}.
\newblock \bibinfo{title}{{Least-squares frequency analysis of unequally spaced
  data}}.
\newblock \bibinfo{journal}{\apss} \bibinfo{volume}{39},
  \bibinfo{pages}{447--462}.
\newblock \DOIprefix\doi{10.1007/BF00648343}.
%Type = Article
\bibitem[{{Mandel} and {Agol}(2002)}]{mandel:2002}
\bibinfo{author}{{Mandel}, K.}, \bibinfo{author}{{Agol}, E.},
  \bibinfo{year}{2002}.
\newblock \bibinfo{title}{{Analytic Light Curves for Planetary Transit
  Searches}}.
\newblock \bibinfo{journal}{\apjl} \bibinfo{volume}{580},
  \bibinfo{pages}{L171--L175}.
\newblock \DOIprefix\doi{10.1086/345520},
  \href{http://arxiv.org/abs/arXiv:astro-ph/0210099}{\tt
  arXiv:arXiv:astro-ph/0210099}.
%Type = Article
\bibitem[{{McCoy} and {Walden}(1996)}]{mccoy:1996}
\bibinfo{author}{{McCoy}, E.J.}, \bibinfo{author}{{Walden}, A.T.},
  \bibinfo{year}{1996}.
\newblock \bibinfo{title}{{Wavelet Analysis and Synthesis of Stationary
  Long-Memory Processes}}.
\newblock \bibinfo{journal}{Journal of Computational and Graphical Statistics}
  \bibinfo{volume}{5}, \bibinfo{pages}{26--56}.
\newblock \URLprefix \url{http://www.jstor.org/stable/1390751}.
%Type = Article
\bibitem[{{Nelder} and {Mead}(1965)}]{nelder:1965}
\bibinfo{author}{{Nelder}, J.A.}, \bibinfo{author}{{Mead}, R.},
  \bibinfo{year}{1965}.
\newblock \bibinfo{title}{{A Simplex Method for Function Minimization}}.
\newblock \bibinfo{journal}{The Computer Journal} \bibinfo{volume}{7},
  \bibinfo{pages}{308--313}.
\newblock \DOIprefix\doi{10.1093/comjnl/7.4.308}.
%Type = Article
\bibitem[{{Nelson} and {Davis}(1972)}]{nelson:1972}
\bibinfo{author}{{Nelson}, B.}, \bibinfo{author}{{Davis}, W.D.},
  \bibinfo{year}{1972}.
\newblock \bibinfo{title}{{Eclipsing-Binary Solutions by Sequential
  Optimization of the Parameters}}.
\newblock \bibinfo{journal}{\apj} \bibinfo{volume}{174}, \bibinfo{pages}{617}.
\newblock \DOIprefix\doi{10.1086/151524}.
%Type = Article
\bibitem[{{Paczynski}(1986)}]{paczynski:1986}
\bibinfo{author}{{Paczynski}, B.}, \bibinfo{year}{1986}.
\newblock \bibinfo{title}{{Gravitational microlensing by the galactic halo}}.
\newblock \bibinfo{journal}{\apj} \bibinfo{volume}{304}, \bibinfo{pages}{1--5}.
\newblock \DOIprefix\doi{10.1086/164140}.
%Type = Article
\bibitem[{{Palmer}(2009)}]{palmer:2009}
\bibinfo{author}{{Palmer}, D.M.}, \bibinfo{year}{2009}.
\newblock \bibinfo{title}{{A Fast Chi-Squared Technique for Period Search of
  Irregularly Sampled Data}}.
\newblock \bibinfo{journal}{\apj} \bibinfo{volume}{695},
  \bibinfo{pages}{496--502}.
\newblock \DOIprefix\doi{10.1088/0004-637X/695/1/496},
  \href{http://arxiv.org/abs/0901.1913}{\tt arXiv:0901.1913}.
%Type = Article
\bibitem[{{Paltani}(2004)}]{paltani:2004}
\bibinfo{author}{{Paltani}, S.}, \bibinfo{year}{2004}.
\newblock \bibinfo{title}{{Searching for periods in X-ray observations using
  Kuiper's test. Application to the ROSAT PSPC archive}}.
\newblock \bibinfo{journal}{\aap} \bibinfo{volume}{420},
  \bibinfo{pages}{789--797}.
\newblock \DOIprefix\doi{10.1051/0004-6361:20034220},
  \href{http://arxiv.org/abs/arXiv:astro-ph/0403186}{\tt
  arXiv:arXiv:astro-ph/0403186}.
%Type = Article
\bibitem[{{Paunzen} and {Vanmunster}(2016)}]{paunzen:2016}
\bibinfo{author}{{Paunzen}, E.}, \bibinfo{author}{{Vanmunster}, T.},
  \bibinfo{year}{2016}.
\newblock \bibinfo{title}{{Peranso - Light Curve and Period Analysis
  Software}}.
\newblock \bibinfo{journal}{ArXiv e-prints}
  \href{http://arxiv.org/abs/1602.05329}{\tt arXiv:1602.05329}.
%Type = Inproceedings
\bibitem[{{Pence}(1999)}]{pence:1999}
\bibinfo{author}{{Pence}, W.}, \bibinfo{year}{1999}.
\newblock \bibinfo{title}{{CFITSIO, v2.0: A New Full-Featured Data Interface}},
  in: \bibinfo{editor}{{Mehringer}, D.M.}, \bibinfo{editor}{{Plante}, R.L.},
  \bibinfo{editor}{{Roberts}, D.A.} (Eds.), \bibinfo{booktitle}{Astronomical
  Data Analysis Software and Systems VIII}, p. \bibinfo{pages}{487}.
%Type = Article
\bibitem[{{Perlmutter} et~al.(1999){Perlmutter}, {Aldering}, {Goldhaber},
  {Knop}, {Nugent}, {Castro}, {Deustua}, {Fabbro}, {Goobar}, {Groom}, {Hook},
  {Kim}, {Kim}, {Lee}, {Nunes}, {Pain}, {Pennypacker}, {Quimby}, {Lidman},
  {Ellis}, {Irwin}, {McMahon}, {Ruiz-Lapuente}, {Walton}, {Schaefer}, {Boyle},
  {Filippenko}, {Matheson}, {Fruchter}, {Panagia}, {Newberg}, {Couch} and
  {Supernova Cosmology Project}}]{perlmutter:1999}
\bibinfo{author}{{Perlmutter}, S.}, \bibinfo{author}{{Aldering}, G.},
  \bibinfo{author}{{Goldhaber}, G.}, \bibinfo{author}{{Knop}, R.A.},
  \bibinfo{author}{{Nugent}, P.}, \bibinfo{author}{{Castro}, P.G.},
  \bibinfo{author}{{Deustua}, S.}, \bibinfo{author}{{Fabbro}, S.},
  \bibinfo{author}{{Goobar}, A.}, \bibinfo{author}{{Groom}, D.E.},
  \bibinfo{author}{{Hook}, I.M.}, \bibinfo{author}{{Kim}, A.G.},
  \bibinfo{author}{{Kim}, M.Y.}, \bibinfo{author}{{Lee}, J.C.},
  \bibinfo{author}{{Nunes}, N.J.}, \bibinfo{author}{{Pain}, R.},
  \bibinfo{author}{{Pennypacker}, C.R.}, \bibinfo{author}{{Quimby}, R.},
  \bibinfo{author}{{Lidman}, C.}, \bibinfo{author}{{Ellis}, R.S.},
  \bibinfo{author}{{Irwin}, M.}, \bibinfo{author}{{McMahon}, R.G.},
  \bibinfo{author}{{Ruiz-Lapuente}, P.}, \bibinfo{author}{{Walton}, N.},
  \bibinfo{author}{{Schaefer}, B.}, \bibinfo{author}{{Boyle}, B.J.},
  \bibinfo{author}{{Filippenko}, A.V.}, \bibinfo{author}{{Matheson}, T.},
  \bibinfo{author}{{Fruchter}, A.S.}, \bibinfo{author}{{Panagia}, N.},
  \bibinfo{author}{{Newberg}, H.J.M.}, \bibinfo{author}{{Couch}, W.J.},
  \bibinfo{author}{{Supernova Cosmology Project}}, \bibinfo{year}{1999}.
\newblock \bibinfo{title}{{Measurements of Omega and Lambda from 42
  High-Redshift Supernovae}}.
\newblock \bibinfo{journal}{\apj} \bibinfo{volume}{517},
  \bibinfo{pages}{565--586}.
\newblock \DOIprefix\doi{10.1086/307221},
  \href{http://arxiv.org/abs/arXiv:astro-ph/9812133}{\tt
  arXiv:arXiv:astro-ph/9812133}.
%Type = Article
\bibitem[{{Pont} et~al.(2006){Pont}, {Zucker} and {Queloz}}]{pont:2006}
\bibinfo{author}{{Pont}, F.}, \bibinfo{author}{{Zucker}, S.},
  \bibinfo{author}{{Queloz}, D.}, \bibinfo{year}{2006}.
\newblock \bibinfo{title}{{The effect of red noise on planetary transit
  detection}}.
\newblock \bibinfo{journal}{\mnras} \bibinfo{volume}{373},
  \bibinfo{pages}{231--242}.
\newblock \DOIprefix\doi{10.1111/j.1365-2966.2006.11012.x},
  \href{http://arxiv.org/abs/astro-ph/0608597}{\tt arXiv:astro-ph/0608597}.
%Type = Article
\bibitem[{{Popper} and {Etzel}(1981)}]{popper:1981}
\bibinfo{author}{{Popper}, D.M.}, \bibinfo{author}{{Etzel}, P.B.},
  \bibinfo{year}{1981}.
\newblock \bibinfo{title}{{Photometric orbits of seven detached eclipsing
  binaries}}.
\newblock \bibinfo{journal}{\aj} \bibinfo{volume}{86},
  \bibinfo{pages}{102--120}.
\newblock \DOIprefix\doi{10.1086/112862}.
%Type = Article
\bibitem[{{Press} and {Rybicki}(1989)}]{press:1989}
\bibinfo{author}{{Press}, W.H.}, \bibinfo{author}{{Rybicki}, G.B.},
  \bibinfo{year}{1989}.
\newblock \bibinfo{title}{{Fast algorithm for spectral analysis of unevenly
  sampled data}}.
\newblock \bibinfo{journal}{\apj} \bibinfo{volume}{338},
  \bibinfo{pages}{277--280}.
\newblock \DOIprefix\doi{10.1086/167197}.
%Type = Book
\bibitem[{{Press} et~al.(1992){Press}, {Teukolsky}, {Vetterling} and
  {Flannery}}]{press:1992}
\bibinfo{author}{{Press}, W.H.}, \bibinfo{author}{{Teukolsky}, S.A.},
  \bibinfo{author}{{Vetterling}, W.T.}, \bibinfo{author}{{Flannery}, B.P.},
  \bibinfo{year}{1992}.
\newblock \bibinfo{title}{{Numerical recipes in C. The art of scientific
  computing}}.
\newblock \bibinfo{publisher}{Cambridge: University Press, |c1992, 2nd ed.}
%Type = Article
\bibitem[{{Protopapas} et~al.(2005){Protopapas}, {Jimenez} and
  {Alcock}}]{protopapas:2005}
\bibinfo{author}{{Protopapas}, P.}, \bibinfo{author}{{Jimenez}, R.},
  \bibinfo{author}{{Alcock}, C.}, \bibinfo{year}{2005}.
\newblock \bibinfo{title}{{Fast identification of transits from light-curves}}.
\newblock \bibinfo{journal}{\mnras} \bibinfo{volume}{362},
  \bibinfo{pages}{460--468}.
\newblock \DOIprefix\doi{10.1111/j.1365-2966.2005.09305.x},
  \href{http://arxiv.org/abs/arXiv:astro-ph/0502301}{\tt
  arXiv:arXiv:astro-ph/0502301}.
%Type = Book
\bibitem[{{Richards} and {Whitby-Strevens}(1979)}]{richards:1979}
\bibinfo{author}{{Richards}, M.}, \bibinfo{author}{{Whitby-Strevens}, C.},
  \bibinfo{year}{1979}.
\newblock \bibinfo{title}{{BCPL, the Language and Its Compiler}}.
\newblock \bibinfo{publisher}{Cambridge: Cambridge University Press, 1979}.
%Type = Article
\bibitem[{{Riess} et~al.(1998){Riess}, {Filippenko}, {Challis}, {Clocchiatti},
  {Diercks}, {Garnavich}, {Gilliland}, {Hogan}, {Jha}, {Kirshner},
  {Leibundgut}, {Phillips}, {Reiss}, {Schmidt}, {Schommer}, {Smith},
  {Spyromilio}, {Stubbs}, {Suntzeff} and {Tonry}}]{riess:1998}
\bibinfo{author}{{Riess}, A.G.}, \bibinfo{author}{{Filippenko}, A.V.},
  \bibinfo{author}{{Challis}, P.}, \bibinfo{author}{{Clocchiatti}, A.},
  \bibinfo{author}{{Diercks}, A.}, \bibinfo{author}{{Garnavich}, P.M.},
  \bibinfo{author}{{Gilliland}, R.L.}, \bibinfo{author}{{Hogan}, C.J.},
  \bibinfo{author}{{Jha}, S.}, \bibinfo{author}{{Kirshner}, R.P.},
  \bibinfo{author}{{Leibundgut}, B.}, \bibinfo{author}{{Phillips}, M.M.},
  \bibinfo{author}{{Reiss}, D.}, \bibinfo{author}{{Schmidt}, B.P.},
  \bibinfo{author}{{Schommer}, R.A.}, \bibinfo{author}{{Smith}, R.C.},
  \bibinfo{author}{{Spyromilio}, J.}, \bibinfo{author}{{Stubbs}, C.},
  \bibinfo{author}{{Suntzeff}, N.B.}, \bibinfo{author}{{Tonry}, J.},
  \bibinfo{year}{1998}.
\newblock \bibinfo{title}{{Observational Evidence from Supernovae for an
  Accelerating Universe and a Cosmological Constant}}.
\newblock \bibinfo{journal}{\aj} \bibinfo{volume}{116},
  \bibinfo{pages}{1009--1038}.
\newblock \DOIprefix\doi{10.1086/300499},
  \href{http://arxiv.org/abs/arXiv:astro-ph/9805201}{\tt
  arXiv:arXiv:astro-ph/9805201}.
%Type = Article
\bibitem[{{Roberts} et~al.(1987){Roberts}, {Lehar} and {Dreher}}]{roberts:1987}
\bibinfo{author}{{Roberts}, D.H.}, \bibinfo{author}{{Lehar}, J.},
  \bibinfo{author}{{Dreher}, J.W.}, \bibinfo{year}{1987}.
\newblock \bibinfo{title}{{Time Series Analysis with Clean - Part One -
  Derivation of a Spectrum}}.
\newblock \bibinfo{journal}{\aj} \bibinfo{volume}{93}, \bibinfo{pages}{968}.
\newblock \DOIprefix\doi{10.1086/114383}.
%Type = Article
\bibitem[{{Scargle}(1982)}]{scargle:1982}
\bibinfo{author}{{Scargle}, J.D.}, \bibinfo{year}{1982}.
\newblock \bibinfo{title}{{Studies in astronomical time series analysis. II -
  Statistical aspects of spectral analysis of unevenly spaced data}}.
\newblock \bibinfo{journal}{\apj} \bibinfo{volume}{263},
  \bibinfo{pages}{835--853}.
\newblock \DOIprefix\doi{10.1086/160554}.
%Type = Article
\bibitem[{{Schwarzenberg-Czerny}(1989)}]{schwarzenbergczerny:1989}
\bibinfo{author}{{Schwarzenberg-Czerny}, A.}, \bibinfo{year}{1989}.
\newblock \bibinfo{title}{{On the advantage of using analysis of variance for
  period search}}.
\newblock \bibinfo{journal}{\mnras} \bibinfo{volume}{241},
  \bibinfo{pages}{153--165}.
%Type = Article
\bibitem[{{Schwarzenberg-Czerny}(1996)}]{schwarzenbergczerny:1996}
\bibinfo{author}{{Schwarzenberg-Czerny}, A.}, \bibinfo{year}{1996}.
\newblock \bibinfo{title}{{Fast and Statistically Optimal Period Search in
  Uneven Sampled Observations}}.
\newblock \bibinfo{journal}{\apjl} \bibinfo{volume}{460},
  \bibinfo{pages}{L107}.
\newblock \DOIprefix\doi{10.1086/309985}.
%Type = Inproceedings
\bibitem[{{Schwarzenberg-Czerny}(2012)}]{schwarzenbergczerny:2012}
\bibinfo{author}{{Schwarzenberg-Czerny}, A.}, \bibinfo{year}{2012}.
\newblock \bibinfo{title}{{On the Sensitivity of Period Searches}}, in:
  \bibinfo{editor}{{Griffin}, E.}, \bibinfo{editor}{{Hanisch}, R.},
  \bibinfo{editor}{{Seaman}, R.} (Eds.), \bibinfo{booktitle}{IAU Symposium},
  pp. \bibinfo{pages}{81--86}.
\newblock \DOIprefix\doi{10.1017/S1743921312000294}.
%Type = Article
\bibitem[{{Southworth} et~al.(2004){Southworth}, {Maxted} and
  {Smalley}}]{southworth:2004}
\bibinfo{author}{{Southworth}, J.}, \bibinfo{author}{{Maxted}, P.F.L.},
  \bibinfo{author}{{Smalley}, B.}, \bibinfo{year}{2004}.
\newblock \bibinfo{title}{{Eclipsing binaries in open clusters - II. V453 Cyg
  in NGC 6871}}.
\newblock \bibinfo{journal}{\mnras} \bibinfo{volume}{351},
  \bibinfo{pages}{1277--1289}.
\newblock \DOIprefix\doi{10.1111/j.1365-2966.2004.07871.x},
  \href{http://arxiv.org/abs/astro-ph/0403572}{\tt arXiv:astro-ph/0403572}.
%Type = Article
\bibitem[{{Steffen}(1990)}]{steffen:1990}
\bibinfo{author}{{Steffen}, M.}, \bibinfo{year}{1990}.
\newblock \bibinfo{title}{{A Simple Method for Monotonic Interpolation in One
  Dimension}}.
\newblock \bibinfo{journal}{\aap} \bibinfo{volume}{239}, \bibinfo{pages}{443}.
%Type = Article
\bibitem[{{Stellingwerf}(1978)}]{stellingwerf:1978}
\bibinfo{author}{{Stellingwerf}, R.F.}, \bibinfo{year}{1978}.
\newblock \bibinfo{title}{{Period determination using phase dispersion
  minimization}}.
\newblock \bibinfo{journal}{\apj} \bibinfo{volume}{224},
  \bibinfo{pages}{953--960}.
\newblock \DOIprefix\doi{10.1086/156444}.
%Type = Article
\bibitem[{{Stetson}(1996)}]{stetson:1996}
\bibinfo{author}{{Stetson}, P.B.}, \bibinfo{year}{1996}.
\newblock \bibinfo{title}{{On the Automatic Determination of Light-Curve
  Parameters for Cepheid Variables}}.
\newblock \bibinfo{journal}{\pasp} \bibinfo{volume}{108}, \bibinfo{pages}{851}.
\newblock \DOIprefix\doi{10.1086/133808}.
%Type = Misc
\bibitem[{{Still} and {Barclay}(2012)}]{still:2012}
\bibinfo{author}{{Still}, M.}, \bibinfo{author}{{Barclay}, T.},
  \bibinfo{year}{2012}.
\newblock \bibinfo{title}{{PyKE: Reduction and analysis of Kepler Simple
  Aperture Photometry data}}.
\newblock \href{http://arxiv.org/abs/1208.004}{\tt arXiv:1208.004}.
  \bibinfo{note}{astrophysics Source Code Library}.
%Type = Article
\bibitem[{{Tamuz} et~al.(2006){Tamuz}, {Mazeh} and {North}}]{tamuz:2006}
\bibinfo{author}{{Tamuz}, O.}, \bibinfo{author}{{Mazeh}, T.},
  \bibinfo{author}{{North}, P.}, \bibinfo{year}{2006}.
\newblock \bibinfo{title}{{Automated analysis of eclipsing binary light curves
  - I. EBAS - a new Eclipsing Binary Automated Solver with EBOP}}.
\newblock \bibinfo{journal}{\mnras} \bibinfo{volume}{367},
  \bibinfo{pages}{1521--1530}.
\newblock \DOIprefix\doi{10.1111/j.1365-2966.2006.10049.x},
  \href{http://arxiv.org/abs/arXiv:astro-ph/0601199}{\tt
  arXiv:arXiv:astro-ph/0601199}.
%Type = Article
\bibitem[{{Tamuz} et~al.(2005){Tamuz}, {Mazeh} and {Zucker}}]{tamuz:2005}
\bibinfo{author}{{Tamuz}, O.}, \bibinfo{author}{{Mazeh}, T.},
  \bibinfo{author}{{Zucker}, S.}, \bibinfo{year}{2005}.
\newblock \bibinfo{title}{{Correcting systematic effects in a large set of
  photometric light curves}}.
\newblock \bibinfo{journal}{\mnras} \bibinfo{volume}{356},
  \bibinfo{pages}{1466--1470}.
\newblock \DOIprefix\doi{10.1111/j.1365-2966.2004.08585.x},
  \href{http://arxiv.org/abs/arXiv:astro-ph/0502056}{\tt
  arXiv:arXiv:astro-ph/0502056}.
%Type = Article
\bibitem[{{ter Braak}(2006)}]{terbraak:2006}
\bibinfo{author}{{ter Braak}, C.J.F.}, \bibinfo{year}{2006}.
\newblock \bibinfo{title}{{A Markov Chain Monte Carlo version of the genetic
  algorithm Differential Evolution: easy Bayesian computing for real parameter
  spaces}}.
\newblock \bibinfo{journal}{Statistics and Computing} \bibinfo{volume}{16},
  \bibinfo{pages}{239--249}.
%Type = Article
\bibitem[{{Thompson} and {Mullally}(2009)}]{thompson:2009}
\bibinfo{author}{{Thompson}, S.E.}, \bibinfo{author}{{Mullally}, F.},
  \bibinfo{year}{2009}.
\newblock \bibinfo{title}{{Wqed: A lightcurve analysis suite}}.
\newblock \bibinfo{journal}{Journal of Physics Conference Series}
  \bibinfo{volume}{172}, \bibinfo{pages}{012081}.
\newblock \DOIprefix\doi{10.1088/1742-6596/172/1/012081}.
%Type = Article
\bibitem[{{Torres} et~al.(2010){Torres}, {Andersen} and
  {Gim{\'e}nez}}]{torres:2010:EBreview}
\bibinfo{author}{{Torres}, G.}, \bibinfo{author}{{Andersen}, J.},
  \bibinfo{author}{{Gim{\'e}nez}, A.}, \bibinfo{year}{2010}.
\newblock \bibinfo{title}{{Accurate masses and radii of normal stars: modern
  results and applications}}.
\newblock \bibinfo{journal}{\aapr} \bibinfo{volume}{18},
  \bibinfo{pages}{67--126}.
\newblock \DOIprefix\doi{10.1007/s00159-009-0025-1},
  \href{http://arxiv.org/abs/0908.2624}{\tt arXiv:0908.2624}.
%Type = Article
\bibitem[{{Wozniak} et~al.(2001){Wozniak}, {Udalski}, {Szymanski}, {Kubiak},
  {Pietrzynski}, {Soszynski} and {Zebrun}}]{wozniak:2001}
\bibinfo{author}{{Wozniak}, P.R.}, \bibinfo{author}{{Udalski}, A.},
  \bibinfo{author}{{Szymanski}, M.}, \bibinfo{author}{{Kubiak}, M.},
  \bibinfo{author}{{Pietrzynski}, G.}, \bibinfo{author}{{Soszynski}, I.},
  \bibinfo{author}{{Zebrun}, K.}, \bibinfo{year}{2001}.
\newblock \bibinfo{title}{{Difference Image Analysis of the OGLE-II Bulge Data.
  II. Microlensing Events}}.
\newblock \bibinfo{journal}{Acta Astronomica} \bibinfo{volume}{51},
  \bibinfo{pages}{175--219}.
\newblock \href{http://arxiv.org/abs/arXiv:astro-ph/0106474}{\tt
  arXiv:arXiv:astro-ph/0106474}.
%Type = Article
\bibitem[{{Zechmeister} and {K{\"u}rster}(2009)}]{zechmeister:2009}
\bibinfo{author}{{Zechmeister}, M.}, \bibinfo{author}{{K{\"u}rster}, M.},
  \bibinfo{year}{2009}.
\newblock \bibinfo{title}{{The generalised Lomb-Scargle periodogram. A new
  formalism for the floating-mean and Keplerian periodograms}}.
\newblock \bibinfo{journal}{\aap} \bibinfo{volume}{496},
  \bibinfo{pages}{577--584}.
\newblock \DOIprefix\doi{10.1051/0004-6361:200811296},
  \href{http://arxiv.org/abs/0901.2573}{\tt arXiv:0901.2573}.

\end{thebibliography}
\end{document}